\documentclass[aps,prd,showpacs,preprintnumbers,twocolumn,groupedaddress,nofootinbib]{revtex4-1}
\usepackage{graphicx}
\usepackage{color}
\usepackage{latexsym}
\usepackage{multirow}
\usepackage{amsmath, footnote, epsfig}

\usepackage{paralist}
\usepackage{booktabs}
\usepackage{aas_macros}
\usepackage{subfigure}
\usepackage{hyperref}
\usepackage{placeins}
\usepackage[toc,page]{appendix}
\usepackage{multirow}
\usepackage{array} % for defining a new column type
\usepackage{varwidth} %for the varwidth minipage environment

\usepackage{rotating}

\pdfoutput=1
\def\beq{\begin{equation}}
\def\eeq{\end{equation}}
\def\bey{\begin{eqnarray}}
\def\eey{\end{eqnarray}}

\def\lsim{\mathrel{\raise.3ex\hbox{$<$\kern-.75em\lower1ex\hbox{$\sim$}}}}
\def\gsim{\mathrel{\raise.3ex\hbox{$>$\kern-.75em\lower1ex\hbox{$\sim$}}}}

\newcommand{\alf}{Alfv\'en}

\newcommand{\fh}{$f_{\rm H2}~$}
\newcommand{\Dpp}{D_{pp}}
\newcommand{\Dxx}{D_{xx}}

\newcommand{\ddp}{\frac{\partial}{\partial p}}

\newcommand{\xco}{$X_{\rm CO}~$}
\newcommand{\htwo}{$\rm H_2~$}

\AtBeginDocument{
\heavyrulewidth=.08em
\lightrulewidth=.05em
\cmidrulewidth=.03em
\belowrulesep=.65ex
\belowbottomsep=0pt
\aboverulesep=.4ex
\abovetopsep=0pt
\cmidrulesep=\doublerulesep
\cmidrulekern=.5em
\defaultaddspace=.5em
}

\begin{document}

\title{Improved Cosmic-Ray Injection Models and the Galactic Center Gamma-Ray Excess}
\author{Eric Carlson$^{1,2}$, Tim Linden$^{3,4}$, Stefano Profumo$^{1,2}$}
\affiliation{$^1$ Department of Physics, University of California, Santa Cruz 1156 High St, Santa Cruz, CA 95064}
\affiliation{$^2$ Santa Cruz Institute for Particle Physics, 1156 High St, Santa Cruz, CA 95064}
\affiliation{$^3$ Kavli Institute for Cosmological Physics, University of Chicago, Chicago, IL}
\affiliation{$^4$ Center for Cosmology and AstroParticle Physics (CCAPP) and Department of Physics, The Ohio State University Columbus, OH }
\begin{abstract}
Fermi-LAT observations of the Milky Way Galactic Center (GC) have revealed a spherically-symmetric excess of GeV $\gamma$ rays extending to at least 10$^\circ$ from the dynamical center of the Galaxy. A critical uncertainty in extracting the intensity, spectrum, and morphology of this excess concerns the accuracy of astrophysical diffuse $\gamma$-ray emission models near the GC. Recently, it has been noted that many diffuse emission models utilize a cosmic-ray injection rate far below that predicted based on the observed star formation rate in the Central Molecular Zone. In this study, we add a cosmic-ray injection component which non-linearly traces the Galactic H$_{2}$ density determined in three-dimensions, and find that the associated $\gamma$-ray emission is degenerate with many properties of the GC $\gamma$-ray excess. Specifically, in models that utilize a large sideband ($40^\circ\times40^\circ$ surrounding the GC) to normalize the best-fitting diffuse emission models, the intensity of the GC excess decreases by approximately a factor of 2, and the morphology of the excess becomes less peaked and less spherically symmetric. In models which utilize a smaller region of interest  ($15^\circ\times 15^\circ$) the addition of an excess template instead suppresses the intensity of the best-fit astrophysical diffuse emission, and the GC excess is rather resilient to changes in the details of the astrophysical diffuse modeling. In both analyses, the addition of a GC excess template still provides a statistically significant improvement to the overall fit to the $\gamma$-ray data.  We also implement advective winds at the GC, and find that the Fermi-LAT data strongly prefer outflows of order several hundred km/s, whose role is to efficiently advect low-energy cosmic rays from the inner few kpc of the Galaxy.  Finally, we perform numerous tests of our diffuse emission models, and conclude that they provide a significant improvement in the physical modeling of the multi-wavelength non-thermal emission from the GC region.

\end{abstract}
\maketitle

\section{Introduction}
\label{sec:introduction}

%{\bf EC: Intro needs to be reviewed/details added to structure after paper complete.}

Gamma-ray data from observations with the Large Area Telescope on board the Fermi Gamma-Ray Satellite (Fermi-LAT) have consistently indicated the presence of a bright, extended, and spherically symmetric excess coincident with the position of the dynamical center of the Milky Way (Galactic Center, hereafter GC)~\citep{Goodenough:2009gk, Hooper:2010mq, Hooper:2011ti, Abazajian:2012pn, gordon_macias:2013, Hooper:2013rwa, Abazajian:2014fta, Daylan:2014rsa, Zhou:2014lva, Calore:2015, TheFermi-LAT:2015kwa}. Most notably, Ref.~\citep{Daylan:2014rsa} and~\citep{Calore:2015} have obtained detailed determinations of the $\gamma$-ray excess, significantly improving our understanding of its spectrum and morphology, and including an assessment of possible systematic effects in extracting the spectral and morphological details of the excess. Using the {\em current set of Galactic foreground models}, the key features of the GC excess are as follows:

(i) A spectrum which peaks at a $\gamma$-ray energy of $\sim$2~GeV, with a low-energy tail harder than generically expected from astrophysical $\pi^0$-decay;

(ii) A morphology which extends out to at least 10$^\circ$ away from the GC, with a 3D intensity falling as r$^{-2.2}$~to~r$^{-2.8}$;

(iii) An approximate spherical symmetry throughout its spatial extent, and

(iv) An intensity peak centered on the dynamical position of Sgr A* to within 0.05$^\circ$.

%Dark Matter/MSP/Hadronic/Leptonic Models of these observations
The GC excess features listed above have attracted significant attention, in part because of their consistency with predictions from simple models where the excess is explained by the pair-annihilation of dark matter particles~\citep[see e.g.][]{Berlin:2014tja, Agrawal:2014una, Alves:2014yha, Abdullah:2014lla,Ipek:2014gua}. Most notably, the excess is well fit by generic dark matter models with a dark matter particle mass of around 35---50~GeV and pair-annihilates to a quark-antiquark final state with a cross-section similar to that expected for a simple thermal relic, and a density profile similar to that expected from a Navarro-Frenk-White (NFW)~\cite{NFW:1996} dark matter density profile that has undergone moderate adiabatic contraction due to baryonic effects~~\citep{Hooper:2010mq, Hooper:2011ti, Calore:2014nla, Berlin:2014tja}.

In addition to dark matter models of the GC excess, several astrophysical scenarios have also been posited as counterparts to the excess. These include the emission from a yet-undetected population of milli-second pulsars densely concentrated near the GC and throughout the Galactic bulge~\citep{2011JCAP...03..010A,Abazajian:2012pn, Yuan:2014yda, Petrovic:2014xra, O'Leary:2015gfa}, or an outburst originating from the position of Sgr A*, of either hadronic~\cite{Carlson:2014} or leptonic~\cite{Petrovic:2014,Cholis:2015} origin. At present, these astrophysical models have produced fits to the $\gamma$-ray data which are of poorer quality than dark matter models \cite{Daylan:2014rsa, Calore:2015, Hooper:2013nhl}, or which appear to be in strong tension with existing constraints, especially in the case of pulsar emission models~\citep{Hooper:2013nhl, Cholis:2014lta}. However, further investigation of these models is highly warranted, due to the high Bayesian prior on the existence of unknown astrophysical emission components in the unique GC environment. Intriguingly, recent studies have tentatively shown that the excess appears compatible with a collection of (unresolved) point sources rather than with a genuinely diffuse emission~\citep{Lee:2015fea, Bartels:2015aea}, an observation which would favor e.g. a pulsar interpretations of the GC excess. Further analyses are clearly needed in order to assess the robustness of this conclusion. In reality, an astrophysical interpretation of the Galactic center excess is likely to involve both cosmic-ray and pulsar contributions.

%Reliance on similar diffuse models
The detailed features of the GC excess are prone to large systematic uncertainties stemming from bright astrophysical diffuse emission which must be removed in order to determine the excess. It is important to note that {\em the GC excess only accounts for approximately 10--20\% of total emission within 10$^\circ$ of the GC, where the excess can be statistically observed.} The majority of the $\gamma$-ray emission in this region of interest (ROI) instead stems from the collision of high-energy cosmic rays with the diffuse Galactic medium. Moreover, {\em all existing studies of the GC excess rely on very similar (or in some cases identical) models for the diffuse Galactic background}, thus increasing the likelihood of systematic errors associated with incorrect modeling of the astrophysical diffuse emission. In particular, the majority of studies~\cite{Abazajian:2012pn,gordon_macias:2013,Hooper:2013rwa,Daylan:2014rsa,Abazajian:2014fta} have utilized the diffuse emission models provided by the Fermi-LAT collaboration\footnote{The analyses of \citep{Abazajian:2012pn, gordon_macias:2013, Daylan:2014rsa, Abazajian:2014fta} employ gal\_2yearp7v6\_v0.fits, while \citep{Hooper:2013rwa, Daylan:2014rsa} employ the gll\_iem\_v02\_P6\_V11\_DIFFUSE.fit diffuse model for an Inner Galaxy type analysis.}. While these models are based on a physically motivated model of the $\gamma$-ray sky, and should in principle provide accurate models for the diffuse Galactic $\gamma$-ray emission, these models are intended primarily to maximize the sensitivity of Fermi-LAT searches for $\gamma$-ray point sources in regions far from the GC. Thus, the Fermi-LAT team has, in some cases, added non-physical extended emission templates in order to reduce extended astrophysical excesses.

We note that several studies of the GC excess  have utilized alternative models for the astrophysical diffuse $\gamma$-ray emission. The earliest studies, e.g. Ref.~\citep{Goodenough:2009gk, Hooper:2010mq, Hooper:2011ti} utilized simple background subtraction models which assumed that the astrophysical $\gamma$-ray emission was either planer, or directly traces the molecular gas density in the GC region. More recent analyses, e.g. Ref.~\citep{Calore:2015, TheFermi-LAT:2015kwa} have utilized diffuse emission models based on the {\tt Galprop} cosmic-ray propagation code, but have typically assumed that the cosmic-ray propagation parameters near the GC region are identical to those which best fit cosmic-ray data in the solar neighborhood. This introduces a similar problem as models utilizing the Fermi-LAT diffuse emission models, where the physical modeling of the GC is based on analyses tuned to fit residuals far from the GC region. 

In Ref.~\cite{paper_one} we showed that improving the physical modeling of the Galactic diffuse emission can dramatically affect the nature, and possibly the very existence,  of the GC excess. In particular, in Ref.~\cite{paper_one} we pushed the envelope of current physical models for diffuse Galactic $\gamma$ rays in two distinct but complementary directions, producing: 

(i) a 3-dimensional modeling of cosmic-ray propagation, and 

(ii) a 3-dimensional and up-to-date choice for the Galactic gas density distribution combined with physical models for the morphology of cosmic-ray injection sources. 

Specifically, we postulated that cosmic-ray injection traces regions of star formation, which is, in turn, traced by the observed molecular hydrogen (H$_2$) density distribution via Schmidt laws.

The key results of our study are that: 

(1) diffuse emission models of the full sky strongly favor a cosmic-ray injection distribution that includes a counterpart to star-forming regions, and 

(2) the features of the GC excess are significantly affected by the choice of the cosmic-ray source distribution. In particular, we found that postulating 20-25\% of the cosmic-ray injection to trace the distribution of H$_2$ regions improves the global fit to the observed $\gamma$-ray data, while also suppressing the GC excess and distorting its spherical symmetry. However, we note that the GC excess is still present in the best-fit models focused toward the inner Galaxy. 

\smallskip 

In this \emph{paper}, we examine the parameter space of models which utilize H$_2$ as a tracer of cosmic-ray injection, and determine the degeneracies between the ensemble of diffusion scenarios and the properties of the $\gamma$-ray excess in various regions of interest. Our key results show that in all cases, $\gamma$-ray data toward the inner Galaxy and Galactic center statistically favor models with 10\% to 15\% of cosmic-ray sources tracing H$_2$. Notably, all fits to the $\gamma$-ray data near the GC prefer the existence of a GC excess component. However, both all-sky $\gamma$-ray fits and the observed star formation rate near the Galactic center prefer a larger fraction of H$_2$ distributed sources.  Depending on one's confidence in these as Bayesian priors on the source distribution, we find that the intensity, morphology and spectrum of this emission component change considerably under different assumptions for the astrophysical diffuse $\gamma$-ray emission model.  Finally, the preferred Galactic diffuse emission models that include physically-motivated cosmic-ray injection sources are observed to produce a significant population of low energy ($\lesssim 30$ GeV) electrons and protons, giving rise to a bright sub-GeV $\gamma$-ray emission. We find that such emission is naturally suppressed in the presence of high-velocity winds emanating from the Galactic center region -- a result which brings $\gamma$-ray observations into considerably better agreement with multi-wavelength observations indicating the existence of a strong Galactic wind~\citep{Crocker:2011,2011MNRAS.411L..11C,1984Natur.310..568S,1992ApJ...397L..39C,2003ApJ...582..246B,2011MNRAS.411L..11C,2000ApJ...540..224S,2016ApJ...817..171Z,Yoast-Hull2014}. We explore the interplay between the many diffuse emission processes and scenarios in the GC region in great detail, and produce an astrophysical diffuse emission model which substantially enhances our understanding of this complex region of the sky.

The outline of this study is as follows. In Section~\ref{sec:galprop} we summarize the current state of the art of public tools that compute Galactic cosmic-ray propagation; we then discuss current gas models in Section~\ref{sec:gas}, and the distribution of cosmic-ray injection sources in Section~\ref{sec:sources}. Within the latter section we address the key pitfalls of previously employed source distributions for cosmic rays in the Galactic center region, and describe a physically-motivated new prescription based on the notion that cosmic-ray injection traces star forming regions (Sec.~\ref{sec:sfr_prescription}). We then describe a set of reference benchmark models (Section~\ref{sec:benchmark_models}) before delving into a detailed comparison of our newly-proposed diffuse models with $\gamma$-ray data from the Fermi telescope (sec.~\ref{sec:gammarays}). The results of our study, especially in connection with the features of the Galactic center excess, are given in Section \ref{sec:results}, while Section \ref{sec:conclusions} summarizes and concludes. The Appendices present additional details on gas distributions (App.~\ref{app:gas_extra}), the \xco conversion factor toward the inner Galaxy (\ref{sec:X_CO}), the ROI dependence of the fit results (\ref{sec:scan_ROI}), a comparison to the Gaussian CMZ models of Ref.~\cite{Gaggero2015}, fits of the Galactic center excess over 10 sky segments from Ref.~\cite{Calore:2015}, and stability of the Galactic center results when using the 1FIG point source catalog~\cite{TheFermi-LAT:2015kwa}.

\section{Diffuse Emission Modeling}

\subsection{Galprop Propagation Models}
\label{sec:galprop}

At the heart of several, modern attempts at Galactic diffuse emission modeling lies the cosmic-ray propagation code {\tt Galprop v54r2504}~\cite{galprop_sourceforge, galprop_stanford, galprop0,galprop1,galprop2}.  Here, we briefly review cosmic-ray propagation in {\tt Galprop}. Cosmic-ray transport is modeled by the following differential equation:

\begin{equation}
\begin{aligned}
\label{eqn:transport}
\frac{\partial \psi}{\partial t}
&= q(\vec r, p)
+ \vec{\nabla} \cdot ( \Dxx \vec{\nabla} \psi - \vec{V}\psi )
- \frac{1}{\tau_f}\psi - \frac{1}{\tau_r}\psi\ \\
&+ \ddp\, p^2 \Dpp \ddp\, \frac{1}{p^2}\, \psi
- \frac{\partial}{\partial p} \left[\dot{p}
- \frac{p}{3} \, (\vec\nabla \cdot \vec V )\psi\right],
\end{aligned}
\end{equation}
where $\psi=\psi (\vec r,p,t)$ is the density per unit of total
particle momentum, i.e. $\psi(p)dp = 4\pi p^2 f(\vec p)$
where $f(\vec p)$ indicates the phase-space density, $q(\vec r, p)$ is a cosmic-ray injection source term,
$D_{xx}$ is the spatial diffusion coefficient, $\vec V$ is the convection
velocity, re-acceleration is described as diffusion in momentum space,
and is determined by a diffusion coefficient in momentum space $\Dpp$, and $\dot{p}\equiv dp/dt$
is the momentum loss rate.  Finally, $\tau_f$ and $\tau_r$ are the time scales for
fragmentation and radioactive decay, both of which we keep fixed to their default values in {\tt Galprop}.

For a given source distribution, {\tt Galprop} solves the above transport equation numerically, assuming free escape boundary conditions at the edges of a cylindrical diffusion halo with half-height $z_{\rm halo}$ and radius ${R_{\rm halo}}$. In most Galactic diffuse emission (hereafter, GDE) models, including the present one, the numerical solution proceeds until a steady state cosmic-ray density is reached (for time-dependent effects see e.g. \cite{Carlson:2014,Petrovic:2014,Cholis:2015}). Diffusion is normally assumed to be homogeneous and isotropic throughout the Galaxy, and exclusively dependent on the particle rigidity via a power law with an index $\delta \in 0.3 \div 0.6$,
\begin{eqnarray}
\Dxx(\mathcal{R}) \equiv D_0 \left( \frac{\mathcal{R}}{4~\rm GV}\right)^\delta.
\end{eqnarray}
The validity of this assumption is difficult to test due to astrophysical uncertainties and the significant degeneracy between transport parameters. In general, one expects that the diffusion properties can vary considerably in different regions of the Galactic, in particular within the Galactic center environment where strong turbulence and large scale poloidal magnetic fields likely induce strongly anisotropic diffusion~\citep{Morris:2007jk}.  Here, we will typically assume isotropic diffusion, but will also examine enhanced diffusion perpendicular to the plane, i.e $D_{zz} \geq \Dxx$, in Section~\ref{sec:global_sensitivity}.

Galactic winds convect cosmic rays out of the plane of the Galaxy at a velocity $\vec{V}=dv/dz \times \vec{z}$ where the gradient $dv/dz$ is a free parameter and is assumed (likely unphysically) to be independent of radius.  Cosmic rays can also scatter off of propagating interstellar \alf~waves, leading to stochastic diffusive reacceleration. This can be effectively described using momentum space diffusion~\cite{1994ApJ...431..705S} and is related to the spatial diffusion coefficient and \alf~velocity, $v_a$, via
\begin{eqnarray}
\Dpp(\mathcal{R}) \equiv \frac{4}{ 3\delta (2-\delta)(4-\delta)(2+\delta)} \frac{\mathcal{R}^2 v_a^2}{\Dxx(\mathcal{R})}.
\label{eqn:reacceleration}
\end{eqnarray}

Large scale Galactic magnetic fields dominate synchrotron energy losses for cosmic-ray electrons.  Throughout this paper we employ a cylindrically symmetric magnetic field model
\begin{eqnarray}
B(r,z) = B_0 e^{(R_\odot-r)/r_B} e^{-|z|/z_B},
\end{eqnarray}
where $R_\odot=8.5$ kpc is the solar radius, and $r_B$ and $z_B$ are the radial and vertical scale-lengths. Very strong fields, which we do not model here, are known to be present within the inner few hundred parsecs of Galactic center~\cite{Crocker:2010xc}.  Although these will impact the resulting spatial and spectral distribution of cosmic-ray leptons, as well as the output synchrotron emission, the properties of these magnetic fields are poorly constrained, occur at the resolution limit of our simulations, and complicate any analysis of leptonic CR injection near the Galactic center~\cite{Cholis:2015}. On a Galactic scale, it is worth noting that more elaborate magnetic field models (see e.g. Ref.~\cite{Jansson2012}) have been incorporated into the latest {\tt Galprop} release~\cite{2015arXiv150705020S}, which may provide a promising avenue for describing more realistic spatially-varying and anisotropic diffusion. These effects include an enhanced diffusion constant along the magnetic field direction, as has been done before in studies of isotropic $\gamma$-ray emission~\cite{isotropic}. Variations on the magnetic field strength and morphology can significantly impact the GCE spectrum~\cite{Calore:2015}, particularly for the new centrally concentrated source distributions described below.  We present details of such variations in Section~\ref{sec:global_sensitivity}.

The interstellar radiation field is comprised of three main components, as described in detail in Refs.~\cite{2006ApJ...640L.155M,2008ApJ...682..400P}:  

(1) cosmic microwave background radiation (CMB), 

(2) starlight, which peaks in the optical band (Opt.), and 

(3) a far-infrared component (FIR) arising from dust-reprocessed starlight.\\
The tightly correlated origin of the optical and FIR components motivates linking their normalizations and allowing for variations with respect to the fixed CMB radiation density.  We have tested Opt.+FIR normalizations from $0.5-3$ compared to their {\tt Galprop} default of 1, finding that values near $.75-1.5$ are slightly preferred and that the significance and spectrum of the GCE is hardly affected, consistent with energy loss timescales near the Galactic center being dominated by synctrotron losses.  Additional details are presented in Sec.~\ref{sec:global_sensitivity}.

It is important to note the limitations of {\tt Galprop's} current ISRF model near the Galactic center.  Current models are oriented toward reproducing the global ICS component and have not yet evolved to the point of adding in individual small scale Galactic structures.  This includes contributions from the stellar populations and the corresponding dust reprocessed photons in the Central Molecular Zone, where additional radiation fields will directly steepen the radial profile of ICS emission over the inner few degrees of the Galactic center.  Unfortunately, state-of-the-art ISRF models are complex, placing improved implementations near the GC beyond the scope of the present work. Some additional details are presented in Section~\ref{sec:discussion}, but we note that these caveats should be considered when assessing the quality of diffuse models very close to the Galactic center.

For the purposes of generating $\gamma$-ray sky-maps, we only calculate primary spectra of nuclear species up to $A=4$, additionally neglecting secondary hadrons which contribute less than $10^{-4}$ of the total nuclear abundance.  This very significantly reduces the required computation time and memory footprint.  For leptonic species, the contribution of secondaries is significant (5-15\% of the total between 1 GeV and 1 TeV~\cite{PhysRevLett.111.081102,PhysRevLett.110.141102}) and must be included.  The injected spectrum of primary cosmic rays is assumed to be spatially homogeneous and distributed as a broken power-law in rigidity with lower and higher indices fixed to the values of the reference models from Refs.~\cite{fermi_diffuse,Calore:2015} and listed in Table~\ref{tab:galprop_params}.  If treated as spatially homogeneous, the exact spectrum of cosmic-rays does not significantly impact our inner Galaxy or Galactic center analyses since the $\gamma$-ray spectrum is fit bin-by-bin.   We generate the photon spectrum arising from $\pi^0$ decays using the formalism of Kamae et al~\cite{2006ApJ...647..692K} to facilitate comparison to previous results, although improved calculations are available~\cite{Kachelriess:2012} which can result in non-negligible differences~\cite{Dermer2013,Carlson:2014}. However, we again note that allowing the spectrum of $\pi^0$-decay emission to float independently in each energy bin should alleviate errors in the $\pi^0$ spectrum calculation. We fix the relative electron and proton normalizations to their locally measured ratio noting that variations should only have minor effects, given the degeneracy between the $\pi^0$ \& bremsstrahlung components, and the existence of a freely floating ICS template.

In contrast to previous (2-dimensional) studies of the Galactic center, we run {\tt Galprop} on a fully three-dimensional cosmic-ray grid with a spacing of 500 pc in the plane of the Galaxy and 125 pc in the vertical direction.  In the planar axes, this represents a doubling of the resolution compared with previous studies.  Further resolution increases are computationally expensive, and do not noticeably impact our results as can be seen in Section~\ref{sec:winds} where we add Galactic center winds and double the planar resolution.

Below we describe a new gas model used to map cosmic-ray sources.  Since this gas cube varies significantly on $\sim$100 pc scales, we subsample the source density over each grid cell using trilinear interpolation.  Line of sight integrals are also subsampled during the generation of $\gamma-$ray skymaps.  We use the Crank-Nicholson method to forward evolve the numerical solution in {\tt Galprop}\footnote{Specifically, we use {\tt start (end) timestep}=$10^9$ ($10^2$) yr, {\tt timestep factor}=0.25 and {\tt timestep repeat}=20 following Ref.~\cite{fermi_diffuse}. These settings are much more computationally efficient and are found to yield stable, convergent solutions compared with much finer time-stepping, and compared with explicit time-step solutions.}.

Moving to 3D cosmic-ray propagation not only allows for new modeling possibilities, but it also, remarkably, removes numerical artifacts present in 2D solutions.  Specifically, the implicit boundary condition at the Galactic center prevents cosmic-rays from propagating through $r=0$ and systematically underestimates the cosmic-ray population at the GC.  Using otherwise identical models, we do find a few-percent enhancement to the CR density near the GC (larger for electrons), but the resulting diffuse model and GCE properties are essentially indistinguishable.  In light of the similarities, we do not further study this effect in isolation, and focus instead on the much larger impact of fully 3D source distributions.

\subsection{Gas Models}
\label{sec:gas}

{\tt Galprop} traditionally uses gas density models for two purposes: {\em propagation} and {\em $\gamma$-ray generation}.  During propagation, the spatial distribution of gas is used to calculate both ionization and hadronic energy loss rates, and to generate secondary particles produced through cosmic-ray spallation.  Below, we employ a new and improved model of the Galactic CO distribution (a tracer for H$_2$) in order to distribute cosmic-ray sources in dense star-forming giant molecular cloud complexes. In each of these cases, the cosmic-ray distribution function is ultimately smoothed out through diffusion, and thus our results are not sensitive to the precise cosmic-ray injection distribution on scales smaller than $\sim$500~pc. An important exception is the extremely dense Galactic center region where leptonic cosmic-rays are confined by very rapid energy losses. For the $\gamma$-ray generation phase, however, the $\gamma$-ray signal is directly impacted by the morphology of Galactic gas, and the precise gas distribution is important.  In this case, the $\gamma$-ray intensity along each line-of-sight is renormalized using survey column densities.

Previous studies of the Galactic center using {\tt Galprop} have relied on azimuthally symmetric gas models for cosmic-ray propagation, renormalizing the $\gamma$-ray intensities using 21cm~\cite{LAB} and CO$_{J=1\to0}$~\cite{Dame:2001} (used as a tracer for H$_2$) line surveys.  Assuming circular motion, a Galactic rotation curve~\cite{Clemens:1985} is used to bin gas in 10 to 20 `Galactocentric annuli' along each line-of-sight, providing a pseudo-three-dimensional distribution of gas in the Galaxy.  This procedure is described in full detail in Refs.~\cite{galprop1,fermi_diffuse}, but contains several important caveats toward the Galactic center which are reviewed in Appendix~\ref{app:gas_extra}.

We incorporate a novel gas model into the {\tt Galprop} code as far as the source injection is concerned, referring to this new model as the `PEB' gas model, based on the improved velocity deconvolution performed by Pohl, Englemier, and Bissantz (2008)~\cite{PEB}.  Gas flow in the inner Galaxy gives rise to large non-circular motions that are not correctly reconstructed by standard (circular) velocity de-convolutions. PEB resolves this issue by employing smoothed particle hydrodynamic simulations of gas flow in a gravitational potential~\cite{MNR:MNR6358} containing 2 spiral arms and a strong central bar which feeds gas into the Galactic center.  The simulated gas velocity field can then be used rather than assuming pure circular motion.  Not only does this more accurately resolve cloud orbits, spiral arms, and the Galactic bar, but it provides kinematic resolution toward the Galactic center where there was previously none.  Beyond the solar circle (8.5 kpc), the Galaxy is essentially in pure circular motion and simulations become less reliable.  PEB therefore linearly interpolates between the gas flow model and pure circular motion for Galactic radii between 7 and 9 kpc.  Additional details are described in Appendix~\ref{app:gas_extra}. 

It is also possible to utilize the PEB gas model for {\em $\gamma$-ray generation}, i.e. the production of the $\gamma$-ray emission map through a convolution of the steady state cosmic-ray population and the target gas density.  However, we find that (i) the impact of alternative gas models on the properties of the GCE is minimal, (ii) the parametric dependence of the GCE properties on the source model parameters is qualitatively similar, and (iii) the overall $\chi^2$ near the Galactic center is worse compared with the usual {\tt Galprop} treatment.  As the conclusions of this {\em paper} are unaffected by the choice of gas model for $\gamma$-ray generation we thus proceed using the usual {\tt Galprop} models when generating $\gamma$-rays.

\subsection{Primary Cosmic Ray Sources}
\label{sec:sources}

Primary cosmic rays are believed to arise from {\em in situ} shock acceleration of supra-thermal precursor ions and electrons.  While, in principle, the source distribution depends on the species of interest, in practice, the vast majority of primary Galactic cosmic-rays are believed to originate in supernova remnants (SNR) and the source distributions of all species are typically taken to be identical, neglecting the significantly smaller contributions from millisecond pulsars and pulsar wind nebulae.

Currently, primary source distributions used in diffuse $\gamma$-ray emission modeling assume cylindrical symmetry about the Galactic rotation axis. Specifically, they (i) are spatially smooth and continuous distributions, (ii) utilize isotropic injection spectra, and (iii) are not time-dependent.  It is remarkable that despite these drastic simplifications, these models reproduce most properties of the locally measured cosmic-ray nuclei and electron spectra as well as the global spectrum and intensity of $\gamma$-rays.  This provides good evidence that, on average, the cosmic-ray density in the Galaxy is reasonably smooth and uniform.  On finer scales, however, these simplifying assumptions break down, as is likely to be true in the case of the significant rise in the local positron spectrum above 10 GeV~\cite{PhysRevLett.111.081102,PhysRevLett.110.141102}, %, and a possible excess of antiprotons recently measured by AMS-02~\cite{ams02:2015}, both of 
which might point toward stochastic discrete sources in the vicinity of the solar system. Simple three dimensional source distributions containing logarithmic spiral arms~\cite{PhysRevD.89.083007,2013PhRvL.111b1102G} and a central bar~\cite{Werner201518,Kissmann201539} have already been investigated in the context of cosmic-ray data. However, local cosmic-ray measurements are not sensitive to sources at the Galactic center.  Gamma-rays provide a more granular probe of the three-dimensional cosmic-ray density, and the observed abundance of diffuse residuals along the Galactic plane highlight not only the difficulty of foreground modeling, but also rich ecosystem of astrophysical environments whose cosmic-ray physics deviate from the Galactic norm, with the Galactic center being the ultimate example.

\subsubsection{Pitfalls of Previous Source Distributions at the Galactic Center}

As a starting point, we delineate the four traditional cosmic-ray source distributions used in {\tt Galprop}.  Each is intended to approximate the true azimuthally averaged surface density\footnote{The surface density here is defined as the three dimensional density integrated over the height (z-axis) of the Galaxy.} of supernovae remnants and consist of either direct observations of SNR, Pulsars, or OB type stars.  Most importantly for the current study, {\em three of them are strictly zero at the Galactic center, and all four underestimate the CMZ injection rate by more than a factor 20}.

The first distribution, `SNR', uses the direct observation of 178 SNR~\cite{Case:1998}, of which 36 had reliable distance estimates.  Distances to the remaining SNRs were estimated by fitting a model of radio surface-brightness to apparent diameter. The Galaxy was then divided into 2~kpc wide radial bins which were used to fit a gamma function,
\begin{eqnarray}
f(r)=\left(\frac{r}{r_\odot} \right)^\alpha \exp\left(-\beta \frac{r-r_\odot}{r_\odot}\right).
\label{eqn:source_param}
\end{eqnarray}
In all {\tt Galprop} models, the surface density is assumed to fall off as $e^{-|z|/z_0}$ with $z_0=0.2$~kpc.  Ref.~\cite{Case:1998} notes that although this form suggests the surface density is zero at $r=0$, the data indicate that this is not correct, proposing an alternative functional form to describe the inner 16.8 kpc (this distribution is not used in {\tt Galprop}).
\begin{eqnarray}
f(r)=A \sin\left(\frac{\pi r}{r_0} + \theta_0 \right)e^{-\beta r}.
\label{eqn:SNR_sources}
\end{eqnarray}
Notably, the statistical and systematic uncertainties on the central 0-2~kpc bin used to fit these distributions are 75\% of the nominal value.

Since this 1998 study, the number of known SNR has now grown by almost 50\% as have the number of reliable distance indicators, the calibration of surface brightness-distance relations and the understanding of selection effects.  New distributions~\cite{Green:2015} have very recently become available, which are much more concentrated toward the inner Galaxy.  However, as the Eq.~(\ref{eqn:source_param}) parameterization is still used, the models again neglect the Galactic center and still remain unsuitable for studies of the Galactic center region.

The next two distributions rely on the observed surface density of pulsars from Lorimer et al~\cite{Lorimer:2004,Lorimer:2006} and Yusifov \& K{\"u}{\c c}{\"u}k~\cite{Yusifov:2004}.  Pulsars offer both a factor 5-10 improved statistics as well as improved distance estimates via radio pulse dispersion, which, used in combination with a model of the free electron density provide a more reliable estimate compared with the SNR surface-brightness/diameter relations. The Lorimer model attempts to derive the underlying pulsar population by matching surveys to Monte-Carlo simulations which include survey selection effects, free electron model uncertainties, and pulsar timing.  Unfortunately, the radial distribution of pulsars is still very strongly correlated with the assumed free electron density~\cite{Lorimer:2006}.  In the inner 1 kpc, this leads to an uncertainty of more than a factor two, but contains a nonzero surface density at a $2-3\sigma$ level.  However, the resulting data is nonetheless fit to the functional form of Eq.~(\ref{eqn:source_param}), which again artificially forces the distribution to zero at the Galactic center.  Similarly, the Yusifov distribution~\cite{Yusifov:2004} is a fitted gamma function (Equation~\ref{eqn:source_param}).  In this case, however, $r$ and $r_\odot$ are shifted to $r\to r-r_{\rm off}$ in order to explicitly preserve a non-zero value at the origin. This is the {\em only} {\tt Galprop} source distribution which is non-zero at the Galactic center. As we will see below, our improved cosmic-ray source models still imply more than an order of magnitude more sources within 500 parsecs of the GC with respect to the Yusifov distribution, suggesting that the innermost regions of the Galaxy have a cosmic-ray injection rate which is dramatically underestimated.

The final standard cosmic-ray source distribution is based on the observed surface density of 748 regions of OB star formation regions~\cite{Bronfman:2000} and motivated by the long-standing connection~\cite{Montmerle1979,Montmerle2009} between cosmic-ray sources (i.e. Type II supernovae) and OB star formation. Detection and distance measurements of these regions rely on CS($2\to1$) molecular line surveys.  Near the Galactic center this method not only suffers from poor kinematic resolution due to the vanishing LSR velocity, but CS($2\to1$) provides an unreliable tracer of massive star forming regions (SFR) in the unique GC environment~\cite{Bally:1987}.  For these reasons, the OB distribution of Ref.~\cite{Bronfman:2000} explicitly indicates that their focus is on the Galactic disk and neglects all sources within $10^\circ$ of the Galactic center.  On its own, this distribution should therefore be excluded from future studies of the $\gamma$-rays at the Galactic center, though it is included here for completeness.

\begin{figure*}[tbp]
  \centering
\begin{subfigure}%{.5\textwidth}
  \centering
  \includegraphics[width=.475\textwidth]{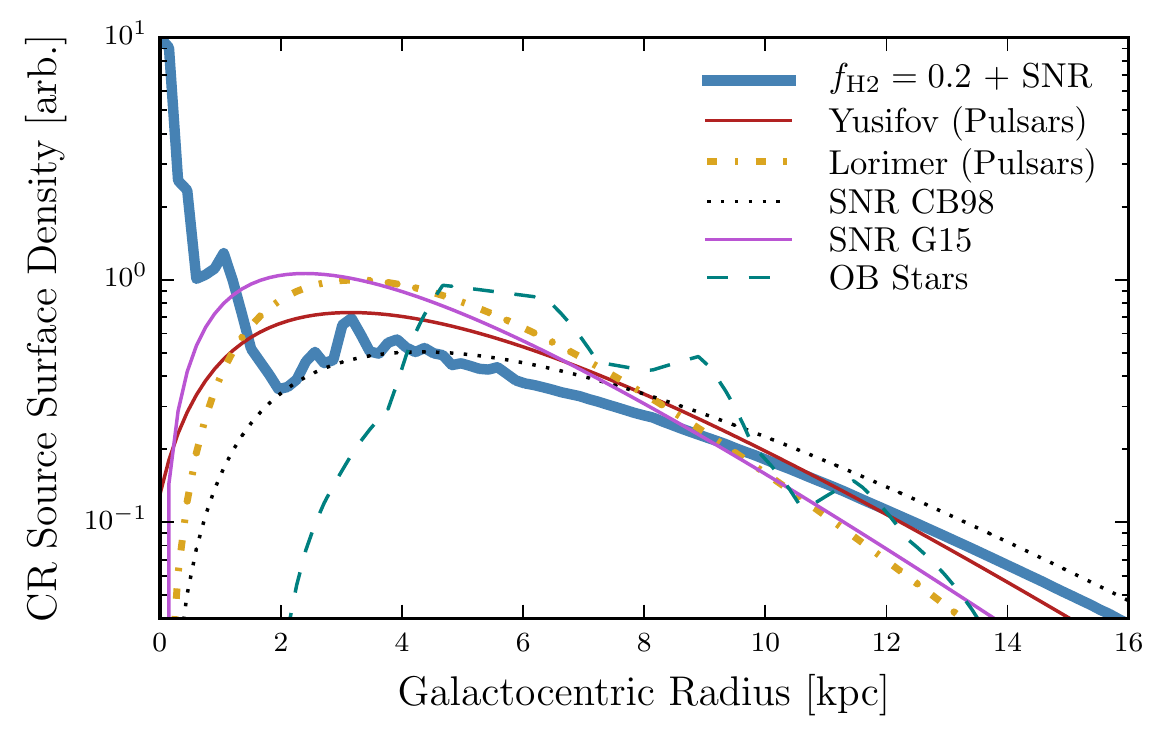}
\end{subfigure}%
\begin{subfigure}%{.5\textwidth}
  \centering
  \includegraphics[width=.475\textwidth]{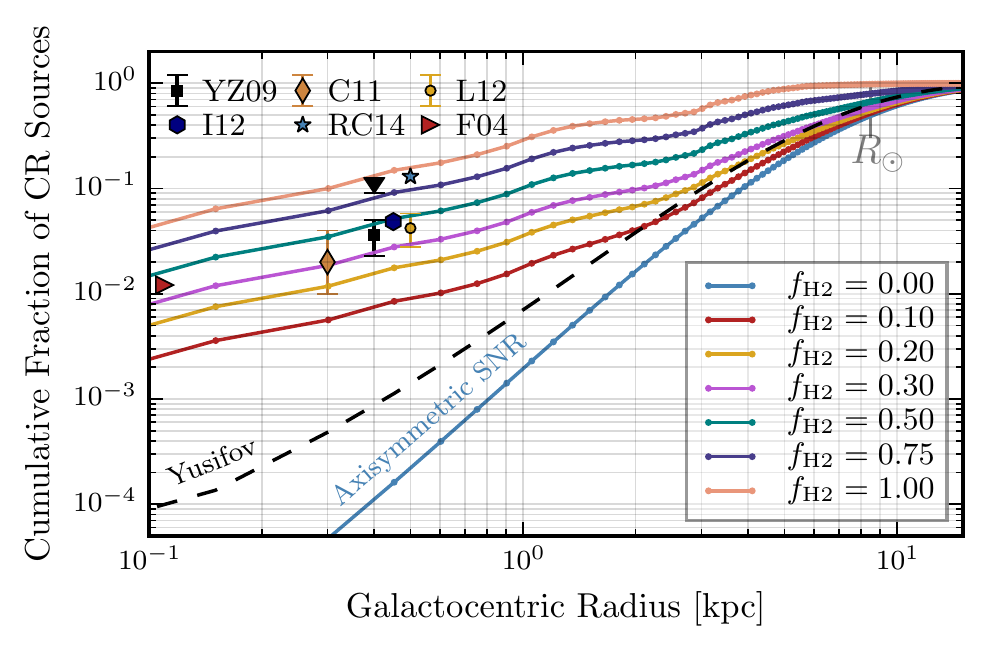}
\end{subfigure}  
  \begin{subfigure}%{.5\textwidth}
  \centering
  \includegraphics[width=.95\textwidth]{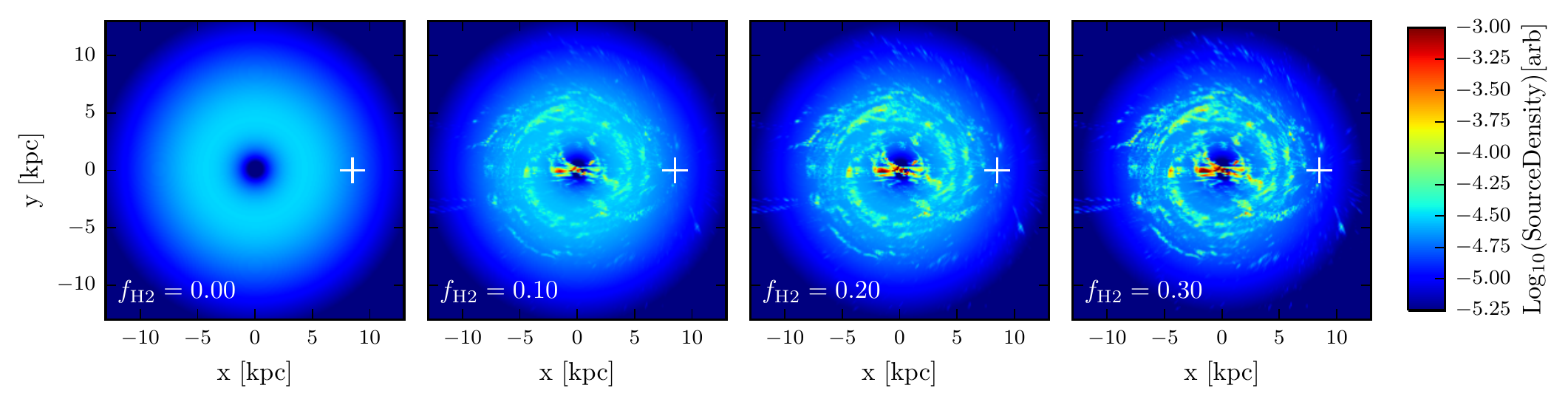}
\end{subfigure}  
  \caption{{\bf Top Left:} The azimuthally averaged surface density of cosmic-ray source distributions used in this analysis. The distribution of supernova remnants is taken from Ref.~\cite{Case:1998} (SNR CB98) and Ref.~\cite{Green:2015} (SNR G15). The Yusifov~\cite{Yusifov:2004} and Lorimer~\cite{Lorimer:2004,Lorimer:2006} distributions use pulsars as a proxy for supernovae remnants while the cosmic-ray injection morphology tracing OB Stars is taken from Ref.~\cite{Bronfman:2000}.   The best fitting cosmic-ray injection rate globally and in the inner Galaxy (with no GCE template) is an admixture with 80\% of cosmic-rays tracing SNR and 20\% tracing the molecular gas density (\fh=0.2) according to the star formation prescription presented in Section~\ref{sec:sfr_prescription} (with $n_s=1.5$ and $\rho_c=0.1 \rm\ cm^{-3}$). All distributions have dimension of length$^{-2}$ and are normalized in arbitrary units to have the same integrated source count.  Note that the $\rm H_2$ distribution contains a strong Galactic bar and spiral arms making it highly azimuthally {\em asymmetric}, as seen in the bottom panel.
{\bf Top Right:} Cumulative source count versus radius from the Galactic center for a variety of \fh values, as well as the axisymmetric SNR-CB98 and Yusifov pulsar models.  Also shown are observations of the fraction of the Milky Way's total cosmic-ray injection rate produced within the CMZ as computed from either the average star formation rate within the CMZ (F04~\cite{2004ApJ...601..319F}, YZ09 including an upper limit~\cite{Yusef-Zadeh2009}, I12~\cite{Immer2012}, L12~\cite{Longmore2013}) relative to the total Galactic SFR of $1.65\pm0.19\rm\ M_{\odot}\ yr^{-1}$~\cite{Licquia2015}, estimates of the fractional SNR occurring within the CMZ (C11~\cite{Crocker:2011}), or the fraction of Wolf-Rayet stars contained in the CMZ (R14~\cite{Rosslowe2015}).  The F04 marker should be placed at $r=50$ pc, though this is below the resolution of our model.
{\bf Bottom:} The primary cosmic ray source distribution derived from our star formation model for increasing values of $f_{\rm H2}$ assuming a Schmidt index $n_s=1.5$ and a critical density $\rho_c=0.1 \rm ~cm^{-3}$.  The leftmost panel corresponds to the pure SNR~\cite{Case:1998} source model while the center panels are typical of models providing improved fits to the full-sky $\gamma$-ray data.}  
 \label{fig:CR_sources}
\end{figure*}

These traditional {\tt Galprop} source distributions are plagued by two additional problems that are specific to the Galactic center region: 
\begin{itemize} 
\item The first issue is due to course binning and azimuthal averaging which blend together three structures: the CMZ, central bar, and gas depleted regions on either side of the bar (see Fig~\ref{fig:CR_sources} bottom panel, discussed below).  Radial surface densities are derived by binning counts of SNR, pulsars, or OB stars in bins of $\Delta r\approx 1-2$ kpc.  Between the CMZ ($r\lesssim 250 pc$) and the inner spiral arms (0.25-3 kpc) the Galaxy is largely devoid of gas and significant star formation.  By area, the CMZ only represents 1-2\% of the central radial bin and this depletion gap at larger radii strongly suppresses the resulting fitted source density at the GC.  In reality, the source density should be sharply peaked over the CMZ and bar, whose semi-major axis is $\sim$3 kpc and oriented within 20-40$^\circ$ to our line of sight~\cite{Dame:2001,Ferriere:2007,PEB}.  This results in a projected central density of cosmic-ray sources which is much higher than reflected in current models. 
\item Second, high extinction and distance uncertainties toward the GC make both selection effects and systematic uncertainties large.  For example, pulsar dispersion measures only provide a reliable distance estimate when the free electron density along the line of sight is well known, and distance measures to OB star forming regions require kinematic resolution.
\end{itemize}

We have shown that the radial distributions of all current tracers of cosmic-ray sources are systematically biased toward zero in the Galactic center, primarily due to the fact that these studies have focused on large scale properties of the disk, and thus choose parametrization that explicitly force the source density to zero at $r=0$. Additionally, the inherent observational, systematic, and statistical difficulties surrounding the Galactic center region make the reliable determination of the cosmic-ray injection rate from each existing tracer difficult. Furthermore, axial symmetry in the Galaxy is strongly broken by the central bar and spiral arms. These issues prompt the exploration of alternative models of primary cosmic-ray source distributions which more realistically reflect the geometry of the Milky Way and resolve scales below 1 kpc.

\subsubsection{Primary Cosmic-Rays from Star Forming $\rm H_2$ Regions}
\label{sec:sfr_prescription}
The connection between supernova and star forming regions is well known~\cite{Montmerle1979,Montmerle2009}.  The high-mass OB stars which precede Type II supernovae evolve on time scales of 10-30 Myr and thus produce cosmic rays on timescales similar to the typical $17\pm4$ Myr lifetimes of giant molecular clouds~\cite{2011ApJ...729..133M}.  We therefore expect that a significant fraction of cosmic-rays in the Galaxy are accelerated in the vicinity of collapsed molecular clouds, and that the rate of cosmic-ray injection should be approximately proportional to the rate of star formation.  Under these assumptions, we present a new primary source distribution based on high-resolution ($\sim 100$pc) three dimensional density maps of $\rm H_2$~\cite{PEB} and a simple model of star formation based on the local volumetric gas density.

At the most basic level, star formation rates are thought to be governed by the gravitational collapse of these clouds, with a characteristic free fall-time $\tau_{\rm f.f.}\propto (\rho_{\rm gas})^{-1}$.  In the absence of feedback the star formation-rate should then be proportional to the gas infall rate $\dot{\rho}_{*} \propto \rho_{\rm gas}/t_{\rm f.f.} \propto G^{1/2}\rho_{\rm gas}^{3/2}$.  This relation is known as the Schmidt law~\cite{1959ApJ...129..243S}.

On the other hand, the Kennicutt-Schmidt law~\cite{1959ApJ...129..243S} encapsulates the empirical observation that the surface density of star formation scales as a power law in the gas surface density with index $1.4\pm .15$\cite{1998ApJ...498..541K}. If one assumes a constant scale-height for the gas disk, then the surface density and volume density are linearly proportional and the Kennicutt-Schmidt law is reproduced within 1$\sigma$ error bars on the power law index.  Of course, this scenario is highly oversimplified and one can devise much more advanced models, particularly in the context of full magnetohydrodynamic simulations~\cite{Schaye2008} where not only thermodynamic quantities are known, but effects such as radiative feedback, cooling, turbulence, processes can be included~\cite{0004-637X-630-1-250,0004-637X-740-2-107}.  This can significantly alter the relationship between the Kennicutt-Schmidt (surface density) power-law index and the Schmidt (volume) index $n_s$.  In addition to the power-law relationship between the local gas density and the star-formation rate, star-formation is observed to terminate below a critical gas density $\rho_c$.  In our cosmic-ray simulations, we adopt a phenomenological prescription which reproduces these essential features, setting the cosmic-ray injection proportional to
\begin{eqnarray}
\dot{\rho}_{\rm CR} \propto \dot{\rho}_{*} \propto
\begin{cases}
      0 & \rho_{\rm H2} < \rho_c \\
      \rho_{\rm H2}^{n_s} & \rho_{\rm H2} \geq \rho_c \\
   \end{cases},
\label{eqn:SFR_source_model}
\end{eqnarray}
where the Schmidt-index $n_s$ is allowed to vary between 1 and 2, and $\rho_c= 0.1\ \rm cm^{-3}$, consistent with numerical simulations and theoretical expectations~\cite{Schaye2008}. In Section~\ref{sec:SFR_tuning}, we will show that our results are relatively independent of $\rho_c$, except for extremely high values which are disfavored by our $\gamma$-ray results.  One must also consider that the gas density in the Galaxy is evolving and may not reflect the distribution of cosmic-ray sources at past epochs.  A benchmark diffusion rate for cosmic rays at the energies of interest for {\em Fermi} $\gamma$-ray studies is,
$R_{\rm diff}=2\sqrt{D_0 t} \approx 3 ~ {\rm kpc} ~t_{\rm Myr}^{1/2}$ (assuming $D_0\sim10^{28}\rm \ cm^2\ s^{-1} $),
where a few kpc is the scale at which non-axisymmetric structure in the Galaxy becomes washed out. A few Myr can also compared against the typical residence time of Galactic cosmic-ray protons, $10^7-10^8$ yr (depending on the height of the Galaxy's diffusion halo)~\cite{0004-637X-528-2-789,1999ICRC....4..314C}.  We should therefore expect that only a portion of cosmic rays should be represented by the current gas distribution and introduce a single additional parameter, $f_{\rm H2}$, that controls the fraction of cosmic-rays injected according to the star formation prescription.  The remaining fraction (1-$f_{\rm H2}$) is then distributed according to the SNR-CB98 model, which provides the best fit of the four traditional primary source distributions.

Because H$_2$ molecules possess no permanent dipole moment, observations of the local H$_2$ density instead employ observations of a tracer molecule, chosen here to be the $^{12}{\rm CO}_{j=1\to 0}$ transition line temperature, where the brightness temperature $\rm W_{CO}$ is related to the H$_2$ column density via a conversion factor $X_{\rm CO}=\rm N_{H2}/W_{CO}$. Theoretical and observational results indicate that $X_{\rm CO}$ is subject to significant spatial and environmental variations, especially in the centers of star forming galaxies in the local group~\cite{Sandstrom:2012ni}.  For simplicity, and to reduce the computational complexity of iterating solutions, we will assume a uniform value of $X_{\rm CO}=2\times10^{20}~\rm cm^{-2}\ (K\ km\ s^{-1})^{-1}$ for propagation and cosmic-ray injection, while fitting the radial profile of $X_{\rm CO}$ in $\gamma$-ray fits (see sec.~\ref{sec:global_analysis}). There is reasonable evidence for a significant suppression of $X_{\rm CO}$ in the central few kpc~\cite{Sandstrom:2012ni,MS:2004} and thus a suppression in $\rm H_2$.  In addition, CO does not provide an optimal tracer of H$_2$ in many cases.  Nonetheless, it is the only H$_2$ related tracer with full sky coverage. These issues are discussed further in Appendix~\ref{sec:X_CO} and in several reviews~\cite{2015ARA&A..53..583H,Bolatto:2013}. Here we will defer further study to future work, but one should keep in mind that the $\rm H_2$ density in the inner few kpc of the Galaxy may be lower than modeled here, and thus our cosmic-ray injection rate may be suppressed.  In contrast, we show that our best fitting models still slightly under-predict star formation in the CMZ relative to observations, a contradiction that points toward additional transport mechanisms near the Galactic center as explored in Section~\ref{sec:winds} and Appendix~\ref{sec:X_CO}.

In the top-left panel of Figure~\ref{fig:CR_sources} we show each of the cosmic-ray source distributions described above as well as the more recent SNR G15 parametrization~\cite{Green:2015}.  We also show the azimuthal average of our proposed star-formation based model with 20\% of sources distributed as $\rho_{H2}^{1.5}$ and 80\% following SNR CB98.  This dramatically enhances the source density within the central few hundred parsecs and generally concentrates more sources in the inner Galaxy. Additionally, this method adds significant structures like the clearly visible spiral arm from 3-5 kpc and gas depleted region from 1-3 kpc. Our later global and inner Galaxy analyses will reveal this source distribution be strongly preferred compared with any of the previous four models.

The top-right panel shows the cumulative source count for our new models as a function of Galactic radius, for different values of $f_{\rm H2}$, with the $f_{\rm H2}=0$ case corresponding to the SNR model.  Also shown is the Yusifov pulsar model which contains the largest central source density of the axisymmetric distributions.  Three of the markers shown (F04, YZ09 and I12, L12) are estimates of the CMZ star formation rate averaged over periods of at least several Myr~\cite{2004ApJ...601..319F,Yusef-Zadeh2009,Immer2012,Longmore2013}.  These have then been divided by recent estimates of the total Galactic star formation rate~\cite{Licquia2015} to yield the fractional SFR of the CMZ.  For the C11 marker, Ref.~\cite{Crocker:2011} estimates the supernova rate of the CMZ to be $2\%_{-1}^{+2}$ of the Galactic total.  Finally, the R14 marker estimates the fraction of Wolf-Rayet stars contained in the inner 500 pc of the Galaxy~\cite{Rosslowe2015}.

It is clear that the traditional models of primary cosmic-ray source injection {\em severely underestimate} the supernova rate of the inner Galaxy by at least a factor 20, while our best fitting proposed model ($f_{\rm H2}\approx0.2$) lies just below the measured rate.  In current diffusion models, the Galactic center region is highly calorimetric for both protons and electrons, and very high values of \fh ($\gtrsim 0.50$) lead to over-luminous $\pi^0$ and inverse-Compton fluxes that are strongly disfavored by the $\gamma$-ray data. This seems to imply that cosmic-ray transport must be much more efficient near the Galactic center via some combination of strongly anisotopic diffusion driven by poloidal magnetic fields, or intense advective winds driven by either starbursts or activity from Sgr. A*~\cite{2003ApJ...582..246B,Crocker2010,Crocker:2011}.  We explore such scenarios in Sec.~\ref{sec:global_sensitivity} and Sec.~\ref{sec:winds} below.  In any case, only values of $f_{\rm H2}\approx 0.3-0.5$ are compatible with observation, providing significant prior evidence that the cosmic-ray injection rate may be even higher than most of the models explored below.  In what follows, we will show that models with large values of $f_{\rm H2}$ are also favored by $\gamma$-ray observations toward the inner Galaxy ($|b|<8^\circ$ and $|l|<80^\circ$) and across the high-latitude sky ($|b|>8\deg$).

\begin{figure*}[tbp]
  \centering 
\begin{subfigure}%{.5\textwidth}
  \centering
	\includegraphics[width=.85\textwidth]{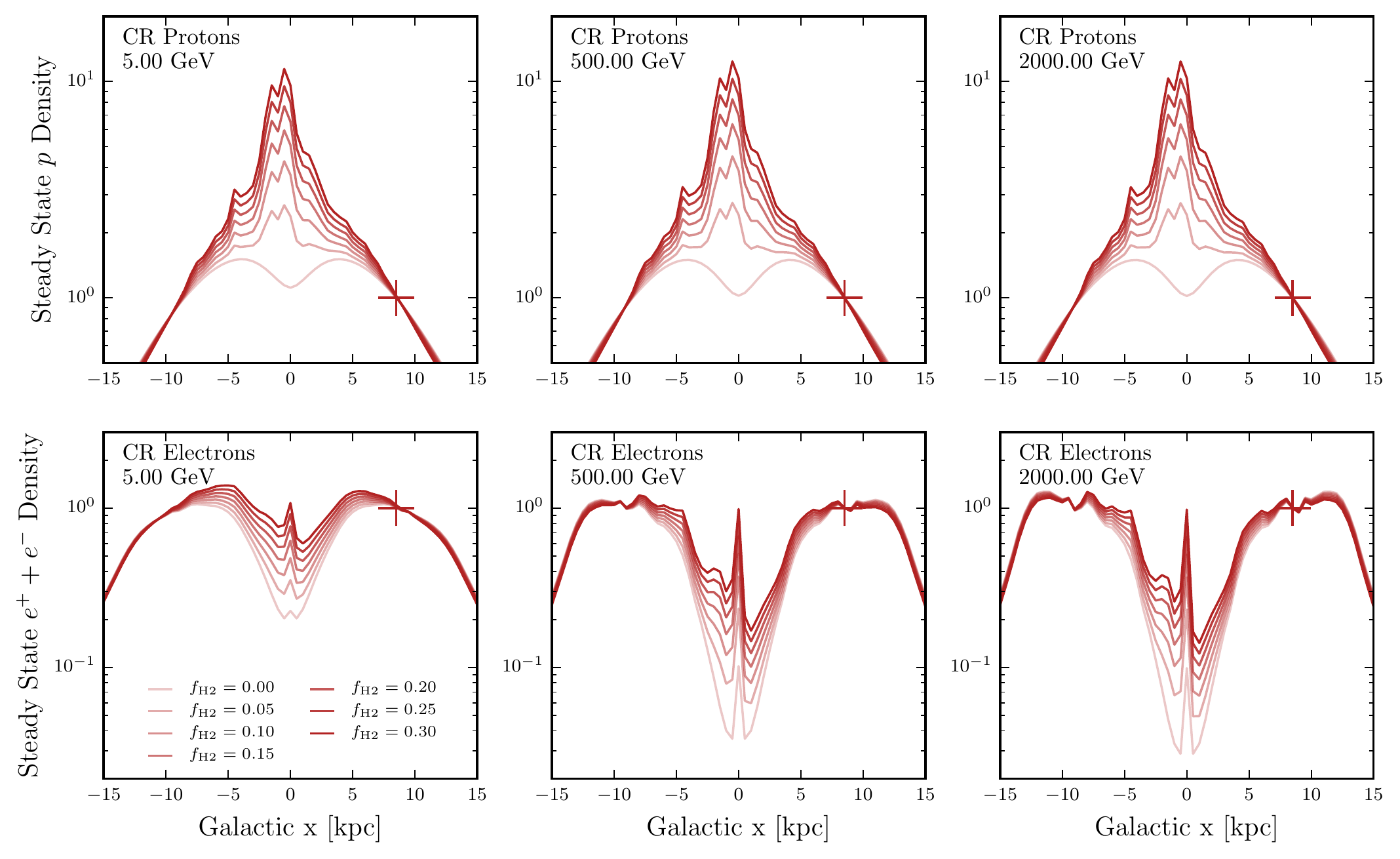}
\end{subfigure}\\% 
\begin{subfigure}%{.5\textwidth}
  \centering
	\includegraphics[width=.95\textwidth]{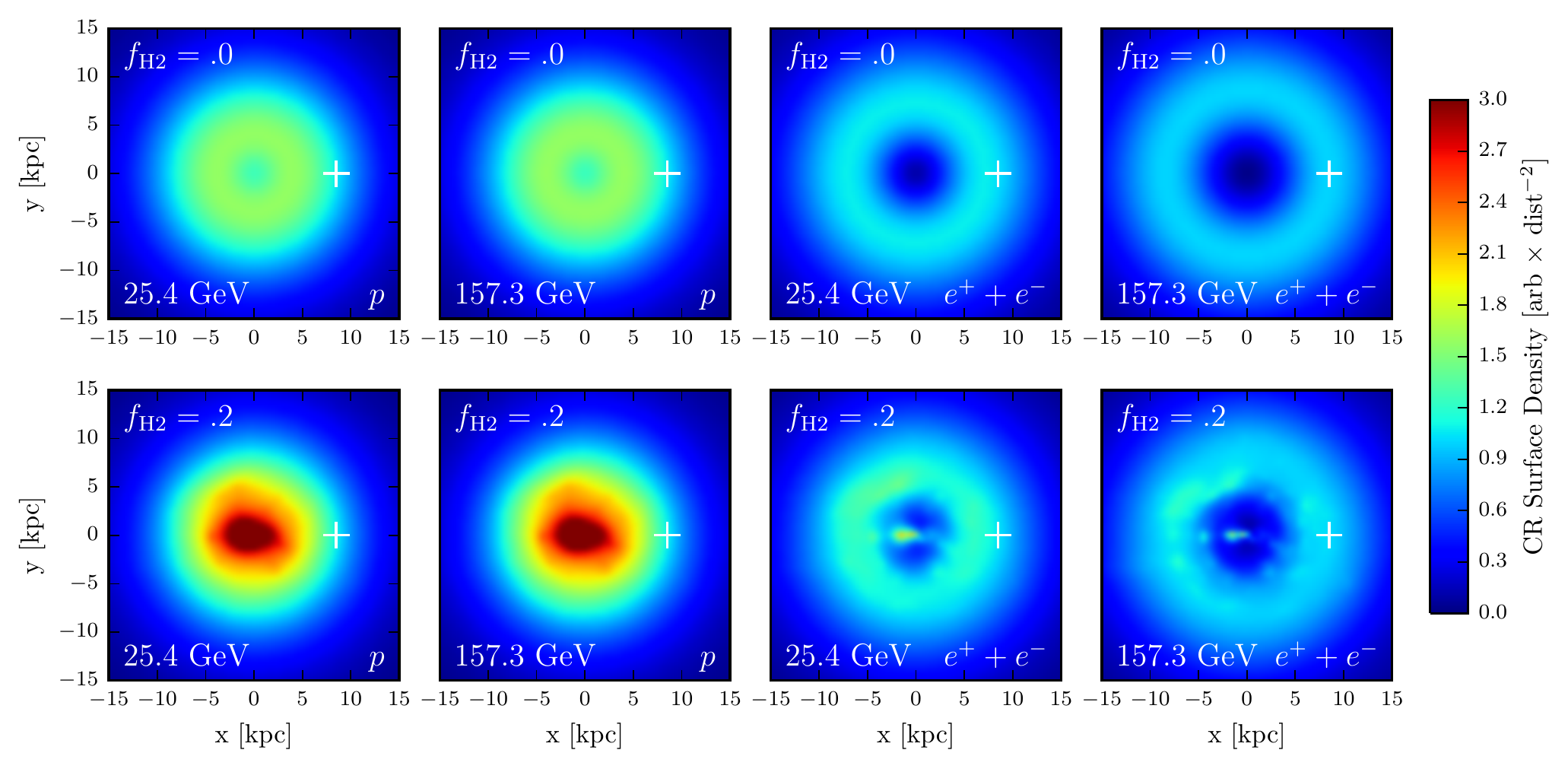}  	
\end{subfigure}  
  \caption{{\bf Top}:  The cosmic-ray proton and electron+positron fluxes, for several representative energies, along the line-of-sight to the Galactic center ($l=b=0^\circ$) after propagation in {\tt Galprop}. Light to dark lines show increasing $f_{\rm H2}$.  The distributions are normalized at the solar position $x=8.5$ kpc, $y=0, z\approx 0$ as indicated by a red `$+$'  {\bf Bottom}: Steady state cosmic-ray surface density for protons (left two columns) and electrons (right two columns) at representative energies for generating $\gamma$-rays over the Fermi-LAT band.  The top row shows the case of $f_{\rm H2 }=0$ corresponding to the traditional axisymmetric SNR CB98~\cite{Case:1998} source density while the bottom row shows the case of $f_{\rm H2 }=0.2$ which includes the proposed new source density model.  A white `+' indicates the solar position, where the cosmic-ray densities have been normalized to unity.}
 \label{fig:CR_steady_state}
\end{figure*}

We have implicitly assumed that the star-formation rate (in $M_\odot\ \rm yr^{-1}$) is proportional to the supernova rate.  However, some observations of stellar clusters near the Galactic center~\cite{2010ApJ...708..834B,2010MNRAS.402..519L,2013ApJ...764..155L} suggest that region may favor a top-heavy initial mass function (IMF). For IMF slopes $1<\alpha<2.35$, the number of Type II SNe per unit star-formation can increase by up to a factor $\approx 2$, moving many of our SFR based lines upward.  The RC14 marker is furthermore based on the relative population of Wolf-Rayet stars in the CMZ and Galactic disk.  If the CMZ IMF has a slope $\alpha=$1 (versus a Salpeter disk $\alpha=2.35$), the number of Type II SNe per WR star is reduced by a factor 2, moving the marker down by $\sim$50\%.

\smallskip

In the bottom panel of Figure~\ref{fig:CR_sources} we show a top-down view of the Galaxy, plotting the surface density of cosmic-ray injection as $f_{\rm H2}$ is increased. The left panel corresponds to a pure SNR distribution, where the lack of cosmic-ray sources in the central Galaxy is readily apparent.  As we increase $f_{\rm H2}$, the Galactic center becomes populated by CMZ sources while preserving the observed gas depleted regions above and below the Galactic bar ($|y|\lesssim 3$~kpc).  The Galactic bar and spiral arms become visible as well as many of the largest giant molecular clouds which harbor star-formation.  In each case, azimuthal symmetry is strongly broken and the model introduces a wealth of new structure into the simulations. 

\medskip

These new cosmic-ray injection models strongly alter the steady-state (after diffusion, energy losses and secondary production are accounted for) cosmic-ray populations in the Milky Way. In the top panel of Figure~\ref{fig:CR_steady_state}, we show the resulting steady-state cosmic-ray density from {\tt Galprop} along the line of sight to the Galactic center. For both protons and electrons, we show representative energies as \fh is increased in our canonical diffusion model. In the bottom panel, we show the steady state surface density (integrated top-down through the Galactic disk) of cosmic-ray protons and leptons for a subset of cases with $f_{\rm H2}=0, 0.2$ at 25 and 157 GeV.

Although much of the fine structure of the new source model is smoothed out by diffusion, the impact of an increasing \fh on the final cosmic-ray density is dramatic.  In the case of protons, the energy dependence is predictably very small from 5 GeV up to 2 TeV due to the logarithmic energy dependence of proton cooling times. The central source density is very strongly enhanced over the traditional SNR model ($f_{\rm H2}=0$), which shows a significant `deficit' in the central Galaxy.  As was discussed in the above description of Figure~\ref{fig:CR_sources}, it is important to note that this does not represent a dramatic change in the total cosmic-ray power of the Galaxy due to the small relative volume of the inner Galaxy region. Also visible in the $f_{\rm H2}=0.2$ case are features which break axial symmetry, including an elongated region of enhanced cosmic-ray density from the central bar and several large clouds to the north-west and south-east quadrants.  More localized features are visible when looking at planar cross-sections of the Galaxy rather than the full column density.

The distribution of leptons is more interesting.  High energy electrons and positrons experience strong energy dependent ($dE/dt\propto E^2$) energy losses due to inverse-Compton and synchrotron cooling, which limit the diffusion radius to a size dependent on the local strength of the magnetic and interstellar radiation fields. These are strongest at the Galactic center and confine leptonic cosmic-rays on scales well below 2 kpc (under standard diffusion assumptions). The range of CRe densities at the Galactic center is stark, due to the peaked cosmic-ray injection sources within the CMZ and Galactic bar. For the SNR models, this region is systematically vacant due to the vanishing of cosmic-ray sources toward the Galactic center. Additional distinct clouds are also visible throughout the Galactic plane, particularly inside the solar circle.

The results above have two important implications for diffuse $\gamma$-ray emission modeling.  First, many $\gamma$-ray analyses opt to use gas column-density templates as a proxy for the gas-correlated $\gamma$-ray emission. However, given that the cosmic-ray density can vary by much more than a factor of 2 along the line of sight (even in the $f_{\rm H2}=0$ case) a simple gas column density template will {\em not} produce the correct emission morphology, and large residuals are likely to occur.  Instead one should either allow for gas templates to vary in annuli around the Galactic center, or use a model of the cosmic-ray density along the line of sight to compute the full three-dimensional convolution of gas and cosmic-rays.

Secondly, one cannot arbitrarily increase the model's cosmic-ray density without over-brightening the gas-correlated emission from the disk.  This `cosmic-ray gradient problem' is well known~\cite{galprop0,galprop_x_co,MS:2004}, and even traditional source+diffusion models require $X_{\rm CO}$ in the inner $\approx$ 2 kpc to be a factor 5 lower than in the disk. Our new source distributions exacerbate this problem and we find that plausible CMZ injection rates are not reconcilable with the range of $X_{\rm CO}$ measured in the centers of nearby spiral galaxies.  Such issues can also be alleviated by efficient evacuation cosmic-rays from the inner few kpc via either enhanced anisotopic diffusion perpendicular to the disk or perhaps through high-velocity convective winds in the Galactic center (see Sections~\ref{sec:global_sensitivity}, \ref{sec:winds}, and Appendix~\ref{sec:X_CO}).

An additional limitation of past and present models is that the spectrum of primary cosmic-ray injection is traditionally assumed to be homogeneous throughout the Galaxy, depending only on whether the injected particle is an electron or nuclear species.  We emphasize that this is both theoretically and observationally known to be incorrect. Fermi measurements of several SNR~\cite{fermi_snr1,fermi_snr2,fermi_snr3,fermi_snr4,Dermer2013c,fermi_snr5} expanding into dense molecular clouds provide direct evidence that cosmic-ray injection spectra at the energies of interest (1-100 GeV) can be very sensitive to environmental factors.  Of particular importance at the Galactic center is the likely presence of ion-neutral damping~\cite{1996A&A...309.1002O,Malkov2005,Malkov2011,Blasi2012,uchiyama2010}. When the upstream edge of supernovae shocks interacts with molecular clouds, ion-neutral collisions effectively damp a range of otherwise resonant \alf~waves, severely deteriorating particle confinement within a slab of momentum space and steepening the spectral index of protons and electrons by precisely one. The energy break for this softening depends strongly on environmental parameters~\cite{Malkov2005}.  This mechanism was studied in detail in the context of the Galactic center environment by two of the authors in Ref.~\cite{Carlson:2014}, where it was shown that the corresponding $\pi^0$ emission can potentially reproduce the observed spectral features and intensity of the GCE.  While we do test hardened CMZ injection spectra in Sec~\ref{sec:winds}, we leave a detailed study of {\em broken} CMZ spectra to future work. 

As a final note, we discuss the impact of our new source models on the predicted local cosmic-ray spectrum.  As noted by Refs~\cite{2013PhRvL.111b1102G, PhysRevD.89.083007, Werner201518}, both the primary and secondary cosmic-ray spectra depend mildly on the Solar System's proximity to Galactic spiral arms, particularly for leptons where the spectra are hardened as one moves closer to the CR source.  Comparing the traditional SNR distribution against our new model (with $f_{\rm H2}=0.2$), the proton and antiproton spectral indices are negligibly changed above 1 GeV.  Below 1 GeV, measurements of the local interstellar spectra are strongly modulated by the heliosphere, but our models predict a $\approx 1\%$ hardening, well within measurement errors.   The ratio $\bar{p}/p$ is enhanced by an about 7\% between 10 MeV and 100 TeV, likely owing to the enhanced source density in the inner Galaxy which increases antiproton production through spallation.  While this is not negligible, it is well within the systematic uncertainties since the ratio here will be directly sensitive to both the \xco profile of the inner Galaxy and to propagation conditions (namely the convection gradient, diffusion coefficient, and halo height).  The positron fraction is hardened above 1 GeV, with a 1\% enhancement at 1 GeV up to a 5\% enhancement at 10 GeV. Thus our changes to the injection morphologies are fully compatible with local cosmic-ray measurements.  We prefer also to remain agnostic about propagation conditions throughout the rest of the Galaxy, particularly in the Galactic center region.  For both of these reasons we do not further constrain our models based on cosmic-ray measurements.

\subsection{Benchmark Models}
\label{sec:benchmark_models}

Throughout this paper we consider three benchmark models and the effects of varying individual parameters within those models.  First, we adopt reference Model A (hereafter `Mod A') from Ref.~\cite{Calore:2015} which performs better than the {\tt P7V6} and {\tt P6V11} Fermi diffuse models over a $40^\circ \times 40^\circ$ `inner Galaxy' region of interest centered on the GC. The {\tt Galprop} parameters are quite typical for diffusion models, with the possible exception of an elevated convection gradient $dv/dz=50\rm ~km s^{-1}$.  Given the intense star formation toward the inner Galaxy, such values are not unreasonable when focusing on the Galactic center region. Furthermore, the cosmic-ray electron population is approximately doubled compared to the locally measured $e^-/p$ flux, and the Opt.+FIR ISRF density is enhanced by $\approx 40\%$ over the {\tt Galprop} value.  Originally this model uses a step function for the \xco gradient taken from Ref.~\cite{MS:2004}, though here we have refit the radial \xco profile to more fairly compare against our modified models.  This results in a significantly better overall $\chi^2$, but does not strongly impact any of the other GCE results below, including our profiles of fit quality versus $f_{\rm H2}$.

Next, we consider a set of ``Canonical'' models which take advantage of the improved features discussed above.  Most importantly, the models incorporate our new source distribution with a fraction of the cosmic-ray injection tracing $f_{\rm H2}$.
  
For propagation (energy losses and generation of secondary species) we use Galprop's analytic gas model.  When generating $\gamma$-rays from $\pi^0$ or bremsstrahlung, we use Galprop's standard (survey renormalized) gas maps assuming a hydrogen spin temperature $\rm T_s=150\ K$ and a reddening cut such that $\rm E(B-V)\le 5$ magnitudes.  We have also verified that varying these assumptions within the model space of Ref.~\cite{fermi_diffuse} does not change the primary results below. Compared with Mod A, these models also possess a slightly smaller diffusion halo height, no convective wind, and a larger diffusion constant.  Of less importance is the higher spatial resolution in the plane of the Galaxy, 3D diffusion, lower Opt.+FIR ISRF and a smaller vertical magnetic field scale-height.  The Canonical models were roughly optimized by hand and provide a better fit globally compared to Mod A.  The model parameters are summarized in Table~\ref{tab:galprop_params}.  Below we will consider the \fh, $n_s$, and $\rho_c$ parameter space which most intensely impacts the properties of the Galactic center excess. In Section~\ref{sec:global_sensitivity}, we will also study the effect of varying global diffusion parameters on both the global fits and on the Galactic center excess. %We will also show the impact of varying transport properties and briefly examine specialized models such as a strong radial wind, anisotropic diffusion, and the cosmic-ray spike model of Ref.~\cite{Gaggero:2015}.

\begin{table*}[tbhp]
\begin{center}
%\scriptsize
\begin{tabular}{llllp{3in}}
\toprule

Parameter & Units & Canonical & Mod A & Description\\
\midrule

$D_0$ & $\rm cm^2~ s^{-1} $ & $7.2 \times 10^{28} $ & $5.0 \times 10^{28} $ & Diffusion constant at $\mathcal{R}=4$~GV\\

$\delta$ & -- & 0.33 & 0.33 & Index of diffusion constant energy dependence\\

$z_{\rm halo}$ & kpc & 3 & 4 & Half-height of diffusion halo\\

$R_{\rm halo}$ & kpc & 20 & 20 & Radius diffusion halo\\

$v_a$ & $\rm km~ s^{-1}$ & 35 & 32.7 & \alf~velocity\\
$dv/dz$ & $\rm km~ s^{-1}~ kpc^{-1}$ & 0 & 50 & Vertical convection gradient\\
\midrule

$\alpha_{\rm p}$ & -- & 1.88 (2.39) & 1.88 (2.47) &  $p$ injection index below (above) $\mathcal{R}=11.5$~GV \\

$\alpha_{\rm e}$ & -- & 1.6 (2.42) & 1.6 (2.43) & $e^-$ injection index below (above) $\mathcal{R}=2$~GV \\

Source & -- & SNR & SNR & Distribution of $(1-f_{\rm H2})$ primary sources$^*$\\

\fh & -- & .20 & N/A & Fraction of sources in star formation model$^*$\\

$n_s$ & -- &1.5 & N/A & Schmidt Index$^*$\\

$\rho_c$ & $\rm cm^{-3}$ & 0.1 & N/A  & Critical \htwo density for star formation$^*$\\
\midrule

$B_0$ & $\mu$G & 7.2 & 9.0 & Local ($r=R_\odot$) magnetic field strength\\

$r_B, z_B$ & kpc & 5, 1 & 5, 2 & Scaling radius and height for magnetic field\\

ISRF & -- & (1.0,.86,.86) & (1.0,.86,.86) & Relative CMB, Optical, FIR density\\

$T_s$ & K & 150 & 150 & Hydrogen spin temperature \\ 
$\rm E(B-V)$ & mag & 5 & 5 & Reddening cutoff for SFD correction\\ 

\midrule

$dx, dy$ &  kpc & 0.5, 0.5 & 1 (2D) & x, y (3D) or radial (2D) cosmic-ray grid spacing\\
$dz$ &  kpc & 0.125 & .1 & z-axis cosmic-ray grid spacing\\

\bottomrule
\end{tabular}
\end{center}
\caption{Summary of {\tt Galprop} parameters for our Canonical model and Mod A from Ref.~\cite{Calore:2015}. $^*$See Section~\ref{sec:sfr_prescription} for additional details on the Star formation parameters.}
\label{tab:galprop_params}
\end{table*}

\section{Gamma-Ray Analyses}\label{sec:gammarays}

In order to compare our new diffuse models against {\em Fermi} $\gamma$-ray data, we employ three distinct maximum likelihood template regressions.  We first perform a `global' $\gamma$-ray analysis over three regions (inner, outer, and local Galaxy) which collectively cover the entire sky and are used to fit the radial dependence of the $\rm CO \to H_2$ conversion factors $X_{\rm CO}(r)$.  It is necessary to refit $X_{\rm CO}$ in this analysis due to the re-distribution of cosmic-rays and due to the variations in propagation parameters. A major benefit of global fits is the ability to statistically assess the quality of the global diffusion model rather than focusing solely on the Galactic center region.  Although it is possible that diffusion in the Galactic center deviates radically from the rest of the Galaxy, it can be instructive to interpret the global likelihood (as well as the CMZ star formation rate) as a Bayesian prior toward conditions near the Galactic center, and to check that our new models remain compatible with the broader Galactic $\gamma$-ray emission.  

The second analysis concerns the inner Galaxy, a $40^\circ$ square region of interest centered on the Galactic center with the plane ($|b|<2^\circ$) masked.  For this purpose we have precisely reproduced the `inner Galaxy' analysis of Calore et al (2015)~\cite{Calore:2015} which is used to characterize the extended GCE emission without significant bias from the Galactic plane. 

Finally, the immediate vicinity around the Galactic center is very complex and depends sensitively on bright point sources which must be simultaneously fit to the diffuse GCE component.  This `Galactic center' analysis is based on that of Daylan et al (2015)~\cite{Daylan:2014rsa}, but extends the window to a larger $15^\circ$ square ROI.  

We notice that the key differences with respect to previous studies of the Galactic center excess are (i) the inclusion of $X_{\rm CO}$ fitting, (ii) global likelihood results, (iii) the first analysis of the GCE using {\em Fermi's} Pass 8 dataset, and (iv) the use of the new {\em Fermi} 3FGL source catalog~\cite{3FGL}.  Below we describe the individual analyses in succession before presenting the overall analysis results.

\subsection{Data Selection}
\label{sec:data_selection}
In our analysis we employ 360 weeks worth of {\em Fermi} data using the recent Pass 8 release.  We select front+back converting photons in the {\tt P8R2\_CLEAN} event class ({\tt evclass=256, evtype=3}) using {\tt Fermi ScienceTools v10r0p5}. Earth limb contamination is mitigated using a zenith angle cut $\theta\leq90^\circ$, which has been updated from the Pass 7 standard $\theta\leq100^\circ$.  We use {\tt gtmktime} with the standard filters {\tt DATA QUAL>0 \&\& LAT CONFIG==1 \&\& ABS(ROCK ANGLE)<52}.  

We note that Pass 8 provides an approximately 25\% increase in effective area over Pass 7. Combined with additional exposure time this provides $\sim 50\%$ increased statistics compared with Ref.~\cite{Calore:2015}, as well as more accurate instrumental response functions.

For analyses of the extremely dense Galactic Center ROI our event selection is identical except that we examine only events which convert in the front of the Fermi-LAT instrument, providing an enhanced angular resolution for this analysis ({\tt evclass=256, evtype=1}). 

\subsection{Additional $\gamma$-ray Templates}
In Section~\ref{sec:galprop}, we described {\tt Galprop's} diffuse emission components ($\pi^0$, bremsstrahlung, and ICS) arising from the sea of Galactic cosmic-rays interacting with interstellar matter and radiation. However, there are several additional $\gamma$-ray components arising from individual point and extended sources, collections of sub-threshold extragalactic sources making up the Isotropic Gamma-Ray Background (IGRB), as well as new diffuse components such as the Fermi Bubbles~\cite{0004-637X-724-2-1044}, and possibly from an additional Galactic center excess (GCE) template.

\begin{itemize}
\item {\em Point Sources (PSC):} The contribution of point sources in each of our $\gamma$-ray analyses is based on {\em Fermi's} 4 year third point source catalog, 3FGL~\cite{3FGL}, including the 13 extended sources.  In the Global and Inner Galaxy analyses the spectrum and normalization of each source is fixed to the 3FGL values and the finite angular resolution of the {\em Fermi-LAT} is taken into account by smearing photons according to the precise energy dependent point spread function (PSF).  Although changes to the fore/background diffuse model will inevitably change the spectrum and flux of 3FGL sources, refitting sources with the new diffuse model introduces a problematic number of degrees of freedom over the large regions of interest here.  We instead fix sources to their 3FGL values and rely on adaptive masking~\cite{Calore:2015} to reduce bias from mis-modeled point sources.

In the Galactic center analysis, we utilize the 3FGL catalog, and include all point sources within 18$^\circ$ of the Galactic Center. We allow any point source to vary freely in normalization if $\{|\ell|$, $|b|$\}~$<$~8$^\circ$, which combined with our diffuse models leaves us with 81 degrees of freedom. We note that allowing the point source normalizations to float freely in each small energy bin makes our fits independent of the global 3FGL spectral shape. We have additionally tested point source distributions based on the recently released 1FIG catalog~\citep{TheFermi-LAT:2015kwa}, and found that the addition of these point sources has no impact on the results presented in this paper, a result we show in Appendix~\ref{appendix:1FIG}.

\item {\em Isotropic Gamma-Ray Background (IGRB):} In theory, the IGRB template is composed only of the population of unresolved, extra-galactic $\gamma$-ray emitters, a distribution which should be roughly isotropic throughout both the IG and GC ROIs. However, due to relatively small ROIs employed in each analysis, the IGRB template may also absorb any diffuse emission of Galactic origin which appears relatively isotropic through out the inner Milky Way (e.g. diffuse Galactic $\gamma$-ray emission from nearby sources). While the spectrum and intensity of the ``true" IGRB is well constrained by observations far from the Galactic center region, the same is not true of the effective IGRB (which includes Galactic contributions), and the degeneracy between these components must be carefully treated in order to correctly model and subtract the IGRB component. In the Global analysis, the spectrum of all components is fixed and we use the {\em Fermi collaboration's} most recent determination of the IGRB spectrum~\cite{isotropic}, choosing Galactic foreground Model A\footnote{ This is not to be confused with our benchmark Galactic diffuse emission model ModA from Calore et al (2015)~\cite{Calore:2015}. In all following sections, mention of ModA will always refer to the latter GDE case.}.  Like our constructed GDE models, this Model A assumes isotropic diffusion parameters throughout the Galaxy, but does not differ substantially from the IGRB spectrum inferred using alternative foreground models.  As detailed in previous GCE studies~\cite{Daylan:2014rsa,Calore:2015}, the isotropic spectrum and flux are poorly constrained over the small ROI of the inner Galaxy analysis. We therefore opt to use the prescription of Ref.~\cite{Calore:2015} whereby an external $\chi^2_{\rm ext}$ is imposed to constrain the isotropic spectrum within it's uncertainties~\cite{isotropic} as determined from the larger regions of interest (the entire sky in this case).  This takes the form
\begin{eqnarray}
\chi^2_{\rm ext} = \sum_i \left( \frac{\phi_i-\bar{\phi_i}}{\Delta\phi_i} \right)^2
\label{eqn:ext_chi2}
\end{eqnarray}
where $\phi_i$ is the flux in energy bin $i$ and $\bar{\phi}_i$ and $\Delta\phi_i$ are the mean and standard deviation of Ref.~\cite{isotropic}'s IGRB Model A. However, {\em in the Galactic center analysis}, we employ a much smaller ROI, making it difficult to avoid contamination of the isotropic background by numerous diffuse Galactic sources.  This is particularly true at low energies where the instrumental point spread function is large. In this case, we allow the normalization of the isotropic background to vary freely in each energy bin without an external $\chi^2$ cost imposed. However, we will also show results produced when this template is fixed to the parameters of the physical isotropic background emission. 

\item {\em Fermi Bubbles:} The flux and spectrum of the Fermi bubbles is assumed to be spatially uniform over the region defined in Ref.~\cite{0004-637X-724-2-1044} with more recent corrections made near the Galactic plane~\cite{bubbles_correction} which are shown in Figure~\ref{fig:bubbles}.  We do not model the North and South Lobes independently.  Similarly to the isotropic template above, we impose the spectrum from the latest {\em Fermi} Collaboration paper~\cite{2014ApJ...793...64A}.  For the Global analysis, the spectrum is fixed (but not the overall normalization), while in the inner Galaxy, the spectrum is constrained to the form of Eq.~(\ref{eqn:ext_chi2}), In the Galactic Center analysis the spectrum is allowed to float freely due to our relatively poor understanding of the Fermi bubbles in regions close to the Galactic center. We also examine scenarios where the Fermi bubble spectrum and intensity are fixed to their values far from the Galactic center, and find that our treatment of the Fermi bubbles has a negligible impact on our results of the Galactic Center analysis.

\begin{figure}
\centering
\includegraphics[width=.45\textwidth]{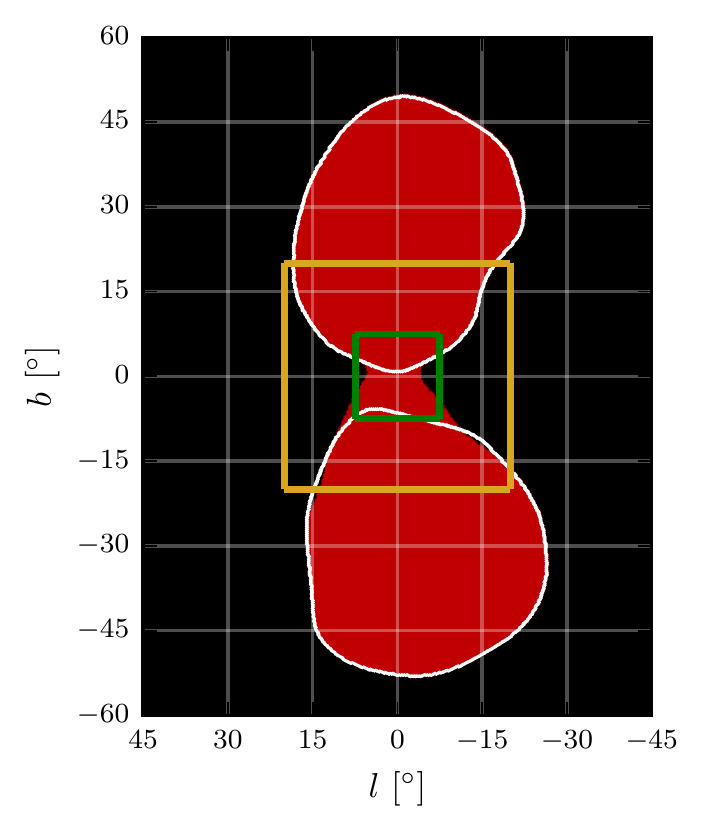}
\caption{The new Fermi bubbles template~\cite{bubbles_correction} used in this analysis is shown in red, and extends throughout the Galactic center region.  The previous template versions~\cite{0004-637X-724-2-1044} are shown by white contours. Also shown are the bounding windows for the inner Galaxy analysis (gold) and Galactic center analysis (green).}
\label{fig:bubbles}
\end{figure}

\item {\em Galactic Center Excess:} Motivated by numerous studies observing an excess in $\gamma$-rays spherically concentrated around the Galactic center~\citep{Goodenough:2009gk, Hooper:2010mq, Hooper:2011ti, Abazajian:2012pn, gordon_macias:2013, Hooper:2013rwa, Abazajian:2014fta, Daylan:2014rsa, Zhou:2014lva, Calore:2015, TheFermi-LAT:2015kwa}, we add and examine the properties of an additional template built to model this emission. Motivated by the reasonable fit provided by dark matter models to the morphology of the excess, we produce an additional template with a global morphology described by the integral over the line of sight of the squared Navarro-Frenk-White (NFW) density profile~\cite{NFW:1996}, which has a three dimensional density profile given by:

\begin{eqnarray}
\rho(r) = \rho_0\left( \frac{r_s}{r} \right)^\alpha \frac{1}{(1+r/r_s)^{3-\alpha}}.
\end{eqnarray}

Based on several fits to the data, our Canonical model employs an inner slope of $\alpha=1.05$, which is shallower than the NFW profile used in previous studies~\cite{Calore:2015,Daylan:2014rsa}, as well as a scale radius of $r_s=20$~kpc.  In Section~\ref{sec:radial_profiles} we scan over different values of the inner slopes and ellipticity, in order to determine the resilience of this excess component to changes in the diffuse modeling. 

\end{itemize}

\subsection{Global Analysis and \xco Fitting}
\label{sec:global_analysis}

After utilizing {\tt Galprop} to generate the energy-dependent $\gamma$-ray morphology of each astrophysical model component (excepting the NFW profile and Fermi-bubbles), we construct the diffuse model by fitting the radial variations of $X_{\rm CO}$.  {\tt Galprop} outputs $\pi^0$ and bremsstrahlung templates in 9 radial annuli (defined in Tab.~\ref{tab:annuli}) for each of the gas components HI, HII, and $\rm H_2$.  Given the strong degeneracy between the gas-correlated $\pi^0$ and bremsstrahlung components, we merge the two into a single $\pi^0 + 1.25~\times$ bremsstrahlung template for each annulus and each gas component\footnote{For the diffuse models in Ref.~\cite{Calore:2015}, a $\pi^0$ to bremsstrahlung ratio of 1:1.25 was found to minimize the absolute value the residuals.  This precise ratio is not important even when fixed in the Galactic center analysis where the bremsstrahlung template is normally allowed to float independently.  For the Global and Inner Galaxy analyses, combining these templates both improves convergence and allows us to fit a single value of \xco for each \htwo ring.}. Next, we merge the HI and HII annuli into a single template whose total normalization is freely varied.  The \htwo rings are kept separate in order to fit the \xco conversion factor for each annulus.
ICS emission in {\tt Galprop} is comprised of the cosmic microwave background, optical, and far-infrared components which have their relative normalizations fixed by the model under consideration.  The total ICS normalization, however, is left free.  The point source template has normalizations and spectra fixed to 3FGL values.  The overall normalizations for the Fermi-bubbles and isotropic component are allowed to vary with constraints from the external $\chi^2$ described above.

Photons are spatially binned into an equal area {\tt Healpix}~\cite{2005ApJ...622..759G} grid with $n_{\rm side}=256$, providing a spatial resolution of $\sim0.23^\circ$.  Spectral binning follows the recipe of Ref.~\cite{Calore:2015} consisting of four linearly spaced bins between 300-500 MeV, with $n_{\rm bins}=20$ additional bins between $E_{\rm min}=500$~MeV and $E_{\rm max}=500$ GeV whose edges are defined recursively by,
\begin{eqnarray}
E_{j+1}=\left(E_j^{1-\Gamma} - \frac{E_{\rm min}^{1-\Gamma}-E_{\rm max}^{1-\Gamma}}{n_{\rm bins}} \right)^{\frac{1}{1-\Gamma}},
\end{eqnarray}
for $j\in[0,1,..n_{\rm bins}]$. This spacing provides an equal number of photons in each bin for a power-law spectrum with index $\Gamma$.  A hard index of $\Gamma=1.45$ is chosen to balance loss of statistics and unreasonably large bin widths at high energies~\cite{Calore:2015}.

Next, each diffuse emission component is smoothed by a Gaussian kernel to approximate the LAT PSF in a computationally efficient manner\footnote{{\em Fermi's} PSF has substantially longer tails than a Gaussian function and one can more accurately implement the PSF by performing a spherical harmonic transform, re-weighting the coefficients, and performing the inverse transform. Unfortunately, this is only accurate if the PSF is much larger than the Healpix angular size ($\approx 0.23^\circ$) or the Healpix resolution is first up-sampled, reweighted, and downsampled, which is computationally expensive.  Maps for individual point sources are still calculated using the precise Fermi PSF.}.  For each energy bin, the width of the Gaussian is set to the 68\% containment radius of the actual PSF, computed by averaging the PSF over the bin, weighted by the spectrum of {\em Fermi's} P8R2 diffuse Galactic background model.  We confirm previous findings~\cite{Calore:2015,Gaggero:2015} that the details of smoothing the diffuse emission components are not important for the inner Galaxy analysis.

For a given region of interest we use {\tt Minuit}\footnote{See \href{https://seal.web.cern.ch/seal/MathLibs/Minuit2/html/}{https://seal.web.cern.ch/seal/MathLibs/Minuit2/html/} and its Python interface, iMinuit \href{http://iminuit.readthedocs.org/en/latest/}{http://iminuit.readthedocs.org/en/latest/} for further information.} to minimize the $\chi^2$ in three fitting regions as defined by,
\begin{eqnarray}
\chi^2 &\equiv& -2 \ln \mathcal{L} + \chi^2_{\rm ext}\\ &=& 2 \sum_{i,j} w_{i,j} (\mu_{i,j}-\theta_{i,j} \ln \mu_{i,j}) + \chi^2_{\rm bub}+\chi^2_{\rm IGRB}\nonumber.
\label{eqn:chi2}
\end{eqnarray}
Here $\theta_{i,j}$ is the observed number of photons in energy bin $i$ and pixel $j$. The model flux $\mu_{i,j}$ is the normalization weighted sum of all model templates.  In the Global analysis, the spectrum of all {\tt Galprop} derived templates is fixed and only the total normalization over all energy bins is varied. The $\chi^2_{\rm ext}$ terms constrain the spectrum of the Fermi bubbles and isotropic components. Ref.~\cite{Calore:2015} developed the weighting coefficient $w_{i,j}$ which allows for adaptive source masking based on the ratio of the point (or extended) source flux $\mu^{\rm PSC}_{i,j}$ to the diffuse flux $\mu^{\rm BG}_{i,j}$,
\begin{eqnarray}
w_{i,j} \equiv \left[\left( \frac{\mu^{\rm PSC}_{i,j}}{f_{\rm PSC} \mu^{\rm BG}_{i,j}}\right)^{\alpha_{\rm PSC}}+1 \right]^{-1}.
\label{eqn:psc_weights}
\end{eqnarray}
Here $f_{\rm PSC}=0.1$ and $\alpha_{\rm PSC}=5$ determine the point source masking threshold and transition rate from masked to unmasked pixels~\cite{Calore:2015}.

With the statistical framework and templates defined, we now determine the value of \xco for each molecular ring. In order to avoid bias from the bright Galactic plane emission, we preform subsequent fits over three sky regions as was done in Ref.~\cite{fermi_diffuse}.  A major benefit of this method is the ability to assess (via likelihood ratio tests) the quality of our new diffuse models not just in the Galactic center, but also in independent regions of the global sky.  We refer to these regions as global-local, global-outer, and global-inner as defined below and summarized in Table~\ref{tab:analyses}.  Annuli are defined in Table~\ref{tab:annuli}.

\begin{enumerate}
\item {\em Local Ring:}  As we are embedded inside the local \htwo gas ring, the local value of \xco is well determined by high-latitude emission which is not influenced by emission along Galactic plane.  We therefore fix all of the $\rm H_2$ rings except for the local ring 7 to their {\tt Galprop} defaults and fit to the full high-latitude sky $|b|>8^\circ$, allowing the IGRB, Bubble, $\pi^0$+bremsstrahlung, and ICS templates to vary.

\item {\em Outer Rings:}  There are two \htwo rings in the outer Galaxy. We fix the normalizations of the Ring 7 and isotropic templates to the global-local values and proceed to fit \xco in the outer two annuli over the region $|b|<8^\circ|, |l|>80^\circ$.

\item {\em Inner Rings:} Rings 7-9 are now fixed and the 6 remaining \htwo rings may be fit in the `Global Inner Galaxy', defined by the region $|b|<8^\circ|, |l|<80^\circ$.  The Fermi bubbles extend to $|b|\approx 50^\circ$ and are better constrained by high-latitude fits than in the Galactic plane. We therefore fix their normalization to the value determined in the global-local fit.
\end{enumerate}

\label{sec:X_CO_global}
\begin{table}[h!]
\begin{center}
\small
\begin{tabular}{ccccc}
\toprule
Ring Number & Radius & Fit Region & $\rm X_{\rm CO}$ \\
& [kpc] &   & [$\rm cm^{-2}\ (K\ km\ s^{-1})^{-1}$ ]\\
\midrule
1 & $0-2.0$ & Inner  & $1.00\times 10^{19\dagger}$\\
2 & $2.0-3.0$& Inner & $8.42\times 10^{19}$\\
3 & $3.0-4.0$& Inner & $1.61\times 10^{20}$\\
4 & $4.0-5.0$& Inner & $1.73\times 10^{20}$\\
5 & $5.0-6.5$& Inner & $1.72\times 10^{20}$\\
6 & $6.5-8.0$& Inner & $1.74\times 10^{20}$\\
7 & $8.0-10.0$& Local & $8.61\times 10^{19}$\\
8 & $10.0-16.5$& Outer & $4.29 \times 10^{20}$\\
9 & $16.5-50.0$& Outer & $2.01\times 10^{21}$\\
\bottomrule
\end{tabular}
\end{center}
\caption{Definitions of gas annuli used during the 3-stage global $X_{\rm CO}$ fitting described in Section~\ref{sec:global_analysis}.  The right column shows the iteratively determined `seed' \xco values which are input to {\tt Galprop} before discretely $\gamma$-ray fitting the rings to each model.}  
\label{tab:annuli}
\end{table}

When generating $\gamma$-ray skymaps, {\tt Galprop} uses power-law interpolation to determine $X_{\rm CO}(r)$ from a discrete set of \xco values.  The final model is therefore somewhat sensitive to the initial input values unless one iteratively determines \xco by feeding the fitted values back into {\tt Galprop} until convergence is reached~\cite{fermi_diffuse}.  In our high-resolution three-dimensional simulations, however, this approach is computationally expensive.  Because \xco is most strongly effected by the gas and source distributions, we choose to iterate our Canonical model four times.  These values are then used as the input \xco for all other models and are listed in Table~\ref{tab:annuli}.   Finally, we sum the \xco renormalized \htwo rings and add them to the HI+HII template (whose normalization is determined by the global-inner fit) to obtain a single map for $\pi^0$ and bremsstrahlung which can be used in the Inner Galaxy or Galactic Center analyses below.

\subsection{Inner Galaxy Analysis}
\label{sec:inner_analysis}

In order to model the region of the sky near the Galactic center, we define an {\em inner Galaxy} analysis comprised of a $40^\circ\times40^\circ$ window centered on $l=b=0^\circ$, but with latitudes $|b|<2^\circ$ masked out in order to avoid bias from the bright and complex Galactic plane. The choice of a $40^\circ\times40^\circ$ ROI balances the specificity of the model for the Galactic center against the ability to fit model components far away from regions where there may be a significant GCE contribution. 

The analysis details for the inner Galaxy are identical to the global analysis with several important exceptions: 

(i) each free component is now fit bin-by-bin in energy, allowing for a model independent determination of the spectrum based solely on the emission morphology; 

(ii) a single $\pi^0 + 1.25\times$ bremsstrahlung template is used for all the gas correlated emission; and 

(iii) we consider fits with and without a GCE template to assess the significance of a new spherical component.\\ 
 
The inner Galaxy analysis is thus an exact replica of Ref.~\cite{Calore:2015}, but using Pass 8 data and instrument response functions (versus Pass 7) as well as the 3FGL catalog (versus 2FGL).  Other than point sources, every template is free to vary in each energy bin, with the Fermi bubbles and IGRB constrained by $\chi^2_{\rm ext}$.

\begin{table*}[tb]
\begin{center}
%\scriptsize
\begin{tabular}{llcc}
\toprule
Analysis & Template & Fixed\\ \midrule

\multirow{7}{20em}{{\bf Global-Local}\\ ROI -- $|b|>8^\circ$\\ Pixel Type -- Healpix Nside=256\\ Binning -- 24 bins .3-500 GeV Ref.\cite{Calore:2015}\\ Events -- P8 Clean Front+Back \\ Spectra Fixed to Galprop/3FGL}  & $\pi^0+$ 1.25 Bremss. (HI+HII) & No & \\
& $\pi^0+$ 1.25 Bremss. ($\rm H_2$ Ring 7) & No & \\
& $\pi^0+$ 1.25 Bremss. ($\rm H_2$ Ring 1-6,8,9) & Yes & \\
& Inverse Compton Scattering & No & \\
& Isotropic (Ext. $\chi^2$) & No & \\
& Fermi Bubbles (Ext. $\chi^2$) & No & \\
& Point Sources (3FGL) & Yes (To 3FGL) & \\
\midrule
\multirow{7}{20em}{{\bf Global-Outer}\\ ROI -- $|l|>80^\circ, |b|<8^\circ$\\ Pixel Type -- Healpix Nside=256\\ Binning -- 24 bins .3-500 GeV Ref.\cite{Calore:2015}\\ Events -- P8 Clean Front+Back\\ Spectra Fixed to Galprop/3FGL}  & $\pi^0+$ 1.25 Bremss. (HI+HII) & No & \\
& $\pi^0+$ 1.25 Bremss. ($\rm H_2$ Ring 1-6) & Yes & \\
& $\pi^0+$ 1.25 Bremss. ($\rm H_2$ Ring 7) & Yes (From Local) & \\
& $\pi^0+$ 1.25 Bremss. ($\rm H_2$ Ring 8,9) & No & \\
& Inverse Compton Scattering & No & \\
& Isotropic & Yes (From Local) & \\
& Point Sources (3FGL) & Yes (To 3FGL) & \\
\midrule
\multirow{8}{20em}{{\bf Global-Inner}\\ ROI -- $|l|<80^\circ, |b|<8^\circ $\\ Pixel Type -- Healpix Nside=256\\ Binning -- 24 bins .3-500 GeV Ref.\cite{Calore:2015}\\ Events -- P8 Clean Front+Back\\ Spectra Fixed to Galprop/3FGL}  & $\pi^0+$ 1.25 Bremss. (HI+HII) & No & \\
& $\pi^0+$ 1.25 Bremss. ($\rm H_2$ Ring 1-6) & No & \\
& $\pi^0+$ 1.25 Bremss. ($\rm H_2$ Ring 7) & Yes (From Local)\\
& $\pi^0+$ 1.25 Bremss. ($\rm H_2$ Ring 8,9) & Yes (From Outer)\\
& Inverse Compton Scattering & No & \\
& Isotropic  & Yes (From Local) & \\
& Fermi Bubbles  & Yes (From Local) & \\
& Point Sources (3FGL) & Yes (To 3FGL) & \\
\midrule
\multirow{6}{20em}{{\bf Inner Galaxy}\\ ROI -- $|l|<20^\circ, 2^\circ<|b|<20^\circ $\\ Pixel Type -- Healpix Nside=256 \\ Binning -- 24 bins .3-500 GeV Ref.\cite{Calore:2015}\\ Events -- P8 Clean Front+Back\\ Bin-by-Bin Spectral Fit}  & $\pi^0+$ 1.25 Bremss. & No & \\
& Inverse Compton Scattering & No & \\
& Isotropic (Ext. $\chi^2$) & No & \\
& Fermi Bubbles (Ext. $\chi^2$) & No & \\
& Point Sources (3FGL) & Yes (To 3FGL)& \\
& GCE NFW$_{\alpha=1.05}$ & No &\\
\midrule
\multirow{6}{20em}{{\bf Galactic Center}\\ ROI -- $|l|<7.5^\circ, |b|<7.5^\circ $\\ Pixel Type -- Cartesian $0.05^\circ\times 0.05^\circ$ \\ Binning -- 0.119-300 GeV 34 Log-Spaced \\ Events -- P8 Clean Front\\Bin-by-Bin Spectral Fit}  & $\pi^0+$ 1.25 Bremss. & No \\
& Inverse Compton Scattering & No & \\
& Isotropic & No & \\
& Fermi Bubbles & No & \\
& Point Sources (3FGL) & No & \\
& GCE NFW$_{\alpha=1.05}$ & No &\\
\bottomrule
\end{tabular}
\end{center}
\caption{Summary of $\gamma$-ray analyses used in this study.}
\label{tab:analyses}
\end{table*}

\subsection{Galactic Center Analysis}
\label{sec:gc_analysis}
To model the $\gamma$-ray intensity and spectrum in regions very close to the GC ($15^\circ\times 15^\circ$ centered on the GC with no latitude mask), we utilize the Fermi-LAT tools to bin photons into 300$\times$300 angular bins and 30 logarithmically spaced energy bins between 300~MeV and 300 GeV. We place photon selection cuts which are identical to all previous analyses in this paper, with the exception that we select only events which convert in the front of the Fermi-LAT detector. We utilize the {\tt gtsrcmaps} toolset to convolve all 81 model components with the Fermi-LAT PSF in each energy range, using a minimum bin size of 0.01$^\circ$ to calculate each source model. By producing the sourcemaps on an angular scale much smaller than our analysis scale, we avoid errors in the determination of steeply sloped emission profiles such as the NFW template in regions very close to the GC. 

We then utilize the {\tt gtlike} algorithm to calculate the best fitting normalization of each model component, fixing the spectra of each source at their default values within the very small energy bins chosen for this study. We then calculate the resulting spectra of each emission template from the ensemble of normalizations calculated for each energy bin, and utilize {\tt gtmodel} to determine the emission model and calculate the LG(L) of our fit to the $\gamma$-ray data. In some simulations (noted throughout the text) we add a 2$^\circ$ latitude mask into the Galactic Center analysis. This is done by masking the output of {\tt gtsrcmaps}, setting both the $\gamma$-ray data and model fluxes to 0 within a given ROI. During the calculation of the likelihood function by {\tt gtlike}, the pixels within the mask have no weight in determining the best fitting model parameters. We have tested that this strategy produces consistent results and introduces no errors into the fitting procedure.  In simulations constraining the dark matter density profile and ellipticity of the NFW profile, we bin the Fermi-LAT data into 150$\times$150 angular bins in order to decrease the computational time, and have tested that this change has no significant effect on our results.

\section{Results}
\label{sec:results}

The diffuse emission models and methodology adopted here can be employed to address a wide variety of questions. For example, the specificity of our models to the Galactic center region make them ideal for studies of the Fermi bubbles, and the three-dimensional nature of our models makes them ideal for studies of the contribution of the spiral arms to the locally observed cosmic-ray population.

However, for the remainder of this paper, we will study the impact of our improved diffuse emission models on the existence, spectrum, and morphology of the Galactic center $\gamma$-ray excess (GCE). In this context, we also present the relevant results from the full-sky global analysis, which can be useful to inform our parameter choices, and to establish the quality of these new models.

In the following sections, we first examine the parameter space of our star-formation model, studying the resulting changes to the global diffuse $\gamma$-ray emission as well as the impact of these models on the Inner Galaxy and Galactic center ROIs and GCE properties. For the interested Reader, we quickly note the key results of our analysis: 

\begin{enumerate}

\item Larger values of \fh enhance the central population of cosmic rays. The CMZ electron population in particular, produces an approximately spherical, extended, and sharply peaked ICS halo surrounding the Galactic center.  Depending on the value of $f_{\rm H2}$, this feature is highly degenerate with the bulk properties of the GCE.

\item When we consider only the Galactic diffuse emission model in the analysis (i.e. no GCE template), a value of $f_{\rm H2}\approx$ 0.1 -- 0.2 is strongly preferred by the data.  Notably, larger ROIs prefer larger values of $f_{\rm H2}$; the best fit is $\sim$0.1 in the Galactic Center analysis, and 0.2  in both the Inner Galaxy analysis and the full-sky analysis. 

\item Models with $f_{\rm H2}\approx$ 0.1 -- 0.2 still substantially underpredict the observed CMZ star formation rate ({\rm cf.} Sec.~\ref{sec:sources}). In addition, the the global-inner and global-local analyses very strongly prefer $f_{\rm H2}\approx$~0.2 -- 0.4.  {\em How the results below are interpreted depends strongly on the relative weights of these priors (toward large $f_{\rm H2}$) pitted against the statistical preference toward lower $f_{\rm H2}$ provided by the narrower inner Galaxy and Galactic center ROIs}. At present, the large unknown systematics of the region, and the potential for missing model elements obfuscates an objective statistical assessment of the two possibilities. We therefore present both interpretations:

\noindent (a)  When a diffuse model utilizing a value $f_{\rm H2}\approx$ 0.2 is {\em imposed} in the Inner Galaxy analysis, it greatly affects the spectrum and morphology of the GCE, and decreases the intensity of the GCE component by approximately a factor of 3. In particular, for the Inner Galaxy ROI, we observe a marked degeneracy between the emission attributable to the $\gamma$-ray excess, and diffuse emission models. On the other hand, the GCE is relatively robust in the Galactic center analysis due to two major effects: 

(i) the bright residual component within two degrees of the Galactic center which is not well fit by diffuse emission models (but is masked from the Inner Galaxy analysis), and 

(ii) the smaller ROI allows the normalization of diffuse emission components to float more freely in the Galactic center analysis compared to analyses of the Inner Galaxy ROI. 

\noindent (b) For all models, both the Inner Galaxy and the Galactic Center, the {\em inclusion of a GCE template remains statistically preferred} compared to diffuse emission models that do not include a GCE component. In these models the best fit value of \fh is reduced to approximately 0.1 in both the Inner Galaxy and Galactic Center analyses, and the normalization of the excess is reduced by only a factor of $\sim$33-50\% in the Inner Galaxy analysis (and remains unchanged in the Galactic Center analysis). However, the spectrum and morphology of the GCE template can still be significantly altered by these relatively modest values of \fh, in some cases producing an unphysically hard spectrum. This appears to be a result of the GDE model becoming too bright below 1 GeV near the Galactic plane, and indicates the need for further enhancements in the diffuse emission modeling.

\item It is difficult to further reduce the residual GCE emission by varying standard diffusion parameters. In order to reconcile the large expected cosmic-ray injection rate of the CMZ with the observed $\gamma$-ray data, one must reduce the number of cosmic-ray electrons below $\sim30$ GeV. We find that a hardened CMZ injection spectrum cannot explain the troublesome low-energy spectrum and morphology. The remaining option is advection-dominated transport out of the CMZ. The addition of a strong Galactic center wind (i) improves the low-energy $\gamma$-ray fit, (ii) helps to reconcile CMZ injection rates with already oversaturated $\pi^0$ emission near the Galactic center (see Appendix~\ref{sec:X_CO}), and (iii) prefers larger values of \fh -- which better match observed CMZ injection rates -- in models with and without a GCE component.

\end{enumerate}

At this point our analysis offers two distinct possible interpretations -- one with a significant GCE component and one without. If one applies the full sky analysis of \fh and adopts a strong a priori preference for $f_{\rm H2}\approx$~0.2 or higher in the Inner Galaxy and Galactic center analysis, then the large scale emission from the GCE component is significantly mitigated, and the GCE may be interpreted as a symptom of mismodeling of the diffuse emission outside the inner-few degrees surrounding the Galactic Center. In this interpretation, some residual component is still necessary in the inner few degrees surrounding the GC, but as significant systematic uncertainties exist in this region, its interpretation would be unclear.  On the other hand, one may argue that the best-fit value of \fh in the spiral arms should not be correlated to the value of \fh near the Galactic center and that the current CMZ injection rate lies below that of the long term average. In that case, the fit should allow the value of \fh and the normalization of the GCE template to float freely in the fit, and our results indicate that the full log-likelihood fitting strongly prefers a lower value of \fh along with a significant GCE component.  

Ultimately, Galactic diffuse emission modeling toward the Galactic center remains a difficult and open problem whose complete, physical solution is only in its infancy.  We aim here to highlight the degeneracies between the astrophysical diffuse emission and any putative dark matter emission sources, present an up-to-date analysis of the GCE in this context, and discuss avenues for improved GDE models in the Galactic center region. In these state-of-the art, yet still ``simplified'' models, the GCE remains an important emission component, and we will study the morphology and spectrum of the GCE in great detail using both the Galactic center and Inner Galaxy analyses. We will then study the impact of global diffusion parameters on the GCE residual, focusing especially on how they reshape the new CMZ electron cloud.  CMZ specific solutions are then discussed including hardening the injection spectrum and adding strong outflowing winds from the GC.  In the appendices we also discuss the implications on $X_{\rm CO}$ and gas calorimetry at the GC, the ROI dependence of fit components, GCE fits across the 10 sky segments used in Ref.~\cite{Calore:2015}, comparisons to the CMZ `cosmic-ray spike' models of Ref.~\cite{Gaggero:2015}, and robustness of the Galactic center reults against the 1FIG point source catalog~\cite{TheFermi-LAT:2015kwa}.

\subsection{Tuning the Star Formation Model}
\label{sec:SFR_tuning}

Our star formation prescription contains three parameters: the Schmidt power-law index $n_s$, the fraction \fh of sources distributed according to Eq.~(\ref{eqn:SFR_source_model}), and $\rho_c$, the critical density needed to initiate star formation. In Section~\ref{sec:sfr_prescription} we provided physical arguments describing the importance and range of each of these parameters. However, the spatial resolution of our gas map ($\sim$ 100~pc) is much lower than typical single cloud hydrodynamic simulations, making our prescription necessarily phenomenological in its modeling of the sub-grid physics. We therefore opt to explore a broad range of the parameter space initially and choose our Canonical model values as those which fit the data well over the full sky.  

In this section, we show that the star formation parameters $n_s$ and $\rho_c$ only weakly impact the statistical fits with respect to the existence and properties of the Galactic center excess. Our star formation parameter space is approximately reduced to a single dimension aligned with $f_{\rm H2}$. We will then examine our Canonical model in great detail, focusing here on how the new GDE models ($f_{\rm H2}\neq0$) impact the properties of the GCE.  

\begin{figure*}[thb]
  \centering
  \includegraphics[width=\textwidth]{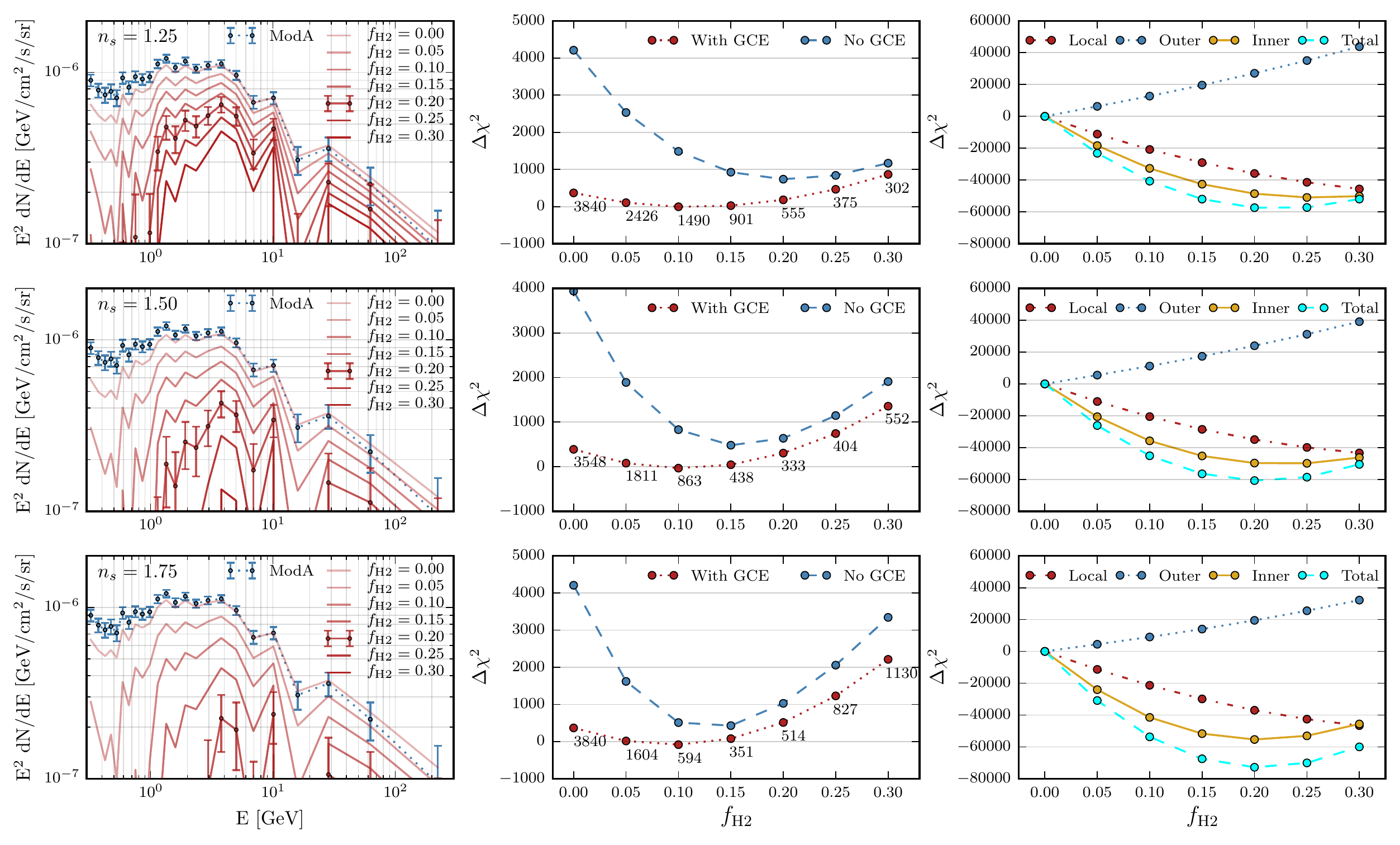}
  \caption{{\bf Left:} Flux of the Galactic Center Excess in the inner Galaxy analysis using an NFW$_{\gamma=1.05}$ GCE template. From top to bottom rows we also vary the Schmidt index ($n_s=1.25, 1.5, 1.75$) and the fraction \fh of primary cosmic-ray sources distributed according to molecular gas as $Q_{\rm Primary}(\vec{x})\propto (n_{\rm H2}(\vec{x}))^{n_s}$.  The remaining fraction is distributed according to the observed azimuthally averaged surface density of supernova remnants~\cite{Case:1998}.  Mod A is a benchmark model from Ref.~\cite{Calore:2015}. {\bf Center:} $\Delta\chi^2$ for the inner Galaxy analysis as \fh is varied.  Red (blue) curves show the $\Delta\chi^2$ with (without) a GCE template included, with negative values indicating a better fit than Mod A+GCE. Inset numbers indicate the statistical preference (TS) for the inclusion of a GCE template in the fit. {\bf Right:} $\Delta\chi^2$ for the three region global $X_{\rm CO}$ fitting analysis (no GCE template is included in the global fitting). $\Delta\chi^2=0$ in this column corresponds to the the $f_{\rm H2}=0$ model, with negative values indicating an improved fit. The inner Galaxy and total-global ROIs have $1.65 \times 10^5$ and $1.89 \times 10^7$ degrees of freedom respectively.}
 \label{fig:vary_ns_fh2_Gal}
\end{figure*}

Remarkably, almost all of the best fitting global parameters are close to the best fit values in the inner Galaxy analysis when a GCE template is not included. In Figure~\ref{fig:vary_ns_fh2_Gal} we present the most important results of this {\em paper} -- the GCE spectrum and the $\Delta \chi^2$ as we discretely vary $n_s \in [1.25, 1.5, 1.75]$ (top to bottom rows) and $f_{\rm H2}$. In the {\em left} column we show the spectrum of the $\rm NFW_{\alpha=1.05}$ GCE template in the bin-by-bin inner Galaxy analysis, with red lines from light to dark corresponding to increasing \fh from 0 to 0.3.  The Canonical model ($f_{\rm H2}=.2$) is highlighted with red error bars.  The blue error bars show the GCE spectrum for reference Mod A\footnote{The low energy spectrum of the GCE using Mod A is substantially softer here than in the original Ref.~\cite{Calore:2015}.  This is due to the combined effects of switching to Pass 8 data and using 3FGL point sources.  The GCE spectrum below 1 GeV remains quite sensitive to the choice of {\tt Galprop} parameters.}. The {\em center} column shows $\Delta \chi^2$ for the inner Galaxy analysis with (red) and without (blue) a GCE template included in the fit.  Here, negative values indicate improved fit relative to Mod A+GCE, and the difference between the blue and red lines indicates the test statistic of adding the additional GCE template (24 additional degrees of freedom).  Finally, the {\em right} column of panels shows $\Delta \chi^2$ for each of the three global fit regions, as well as the their (summed) total $\Delta \chi^2$.

As \fh is increased from zero, which corresponds to the classic SNR source distribution, the high density of gas in the inner few hundred parsecs dramatically increases the cosmic-ray injection intensity near the Galactic center. The non-linearity of the Schmidt law (when $n_s>1$) implies that cosmic-ray injection rate scales steeply with the \htwo density, concentrating cosmic-rays toward dense molecular clouds. Nowhere is the impact of this more dramatically realized than in the Central Molecular Zone and Galactic bar. Cosmic-rays younger than $10^5$ yr remain quite close to their injection site, illuminating the giant molecular structures which generated them.  As the cosmic-rays age they diffuse outward ($R_{\rm diff.}(E) =2 \sqrt{D_{xx}(E) t}$) and produce $\gamma$-rays in the ambient ISM.  For electrons near the Galactic center, the magnetic fields and ISRF energy densities are sufficiently large that energy losses strongly limit the diffusion timescale, leaving behind a sharply peaked, and approximately spherical inverse Compton component.  Thus, as \fh is increased, the CR population at the Galactic center is enhanced and the GCE is strongly reduced until the diffuse emission eventually over-saturates the observed emission from the inner Galaxy for high values of \fh.

As $n_s$ is increased, cosmic-ray sources become increasingly concentrated in the most gas dense regions -- e.g. the CMZ.  For the inner Galaxy, the effect is quite similar to increasing $f_{\rm H2}$.  Thus for larger values of $n_s$, the same level of GCE reduction (and CMZ injecttion rate) is achieved by smaller $f_{\rm H2}$.  Statistically, this is evidenced by the compression of the $\Delta \chi^2$ versus \fh profile as $n_s$ becomes larger.  The global $\gamma$-ray fit improves somewhat for larger values of $n_s$, though the difference is subdominant compared with changing $f_{\rm H2}$.  Because diffusion washes out much of the peaked structures, increasing $n_s$ adds more cosmic-rays to the densest gas clouds, effectively rescaling the action of $f_{\rm H2}$.  We therefore choose $n_s=1.5$ for the remainder of this \emph{paper} and relegate further study to the future.

Without invoking an extra GCE template, the fit in the inner Galaxy analysis shows marked improvement using our star formation source model, preferring  $f_{\rm H2}\approx 0.15-0.25$ at very high significance ($\Delta \chi^2\approx 4000$) over the pure SNR distribution.  When a GCE template is added, the fit is more agnostic to changes in \fh and has a shallower profile which slightly prefers $f_{\rm H2}=0.1$.  The test statistic (TS) for the addition of the GCE template is given by the difference between the red and blue curves and our Canonical model with $n_s=1.5$ and $f_{\rm H2}\approx 0.1$ reduces the significance of the excess from TS$\approx$4000 (in the case of Mod A), down to TS$\sim$1000\footnote{We note here that the $\rm \sqrt{TS}$ cannot be interpreted straightforwardly as a significance due to the large unresolved systematic uncertainties~\cite{Calore:2015}.  In particular, no GDE model currently describes the data even remotely close to the level of Poisson noise making an interpretation of $\Delta\chi^2$ in terms of significance difficult.  A study of the (much larger) correlated systematic uncertainties along the Galactic plane can be found in ref.~\cite{Calore:2015}, and is used at several points below.} These two results indicate the strong degeneracy between the GCE template and GDE models containing a cosmic-ray emitting CMZ. In addition, it is intriguing to note that the globally preferred values of \fh are the same as those that maximally reduce the significance of the GCE. Specifically, the TS of the GCE is reduced further to TS~=~333 when the best fit global model of $n_s=1.5$ and $f_{\rm H2}\approx 0.2$ is employed, and remains highly suppressed when larger \fh are used (as preferred by the CMZ SFR constraints).  On the other hand, the statistical significance of this component is still reasonably high, motivating us to study the residual properties of the GCE in detail in Section~\ref{sec:characterize_residuals}.

\begin{figure*}[tbp!]
  \centering
  \includegraphics[width=.95\textwidth]{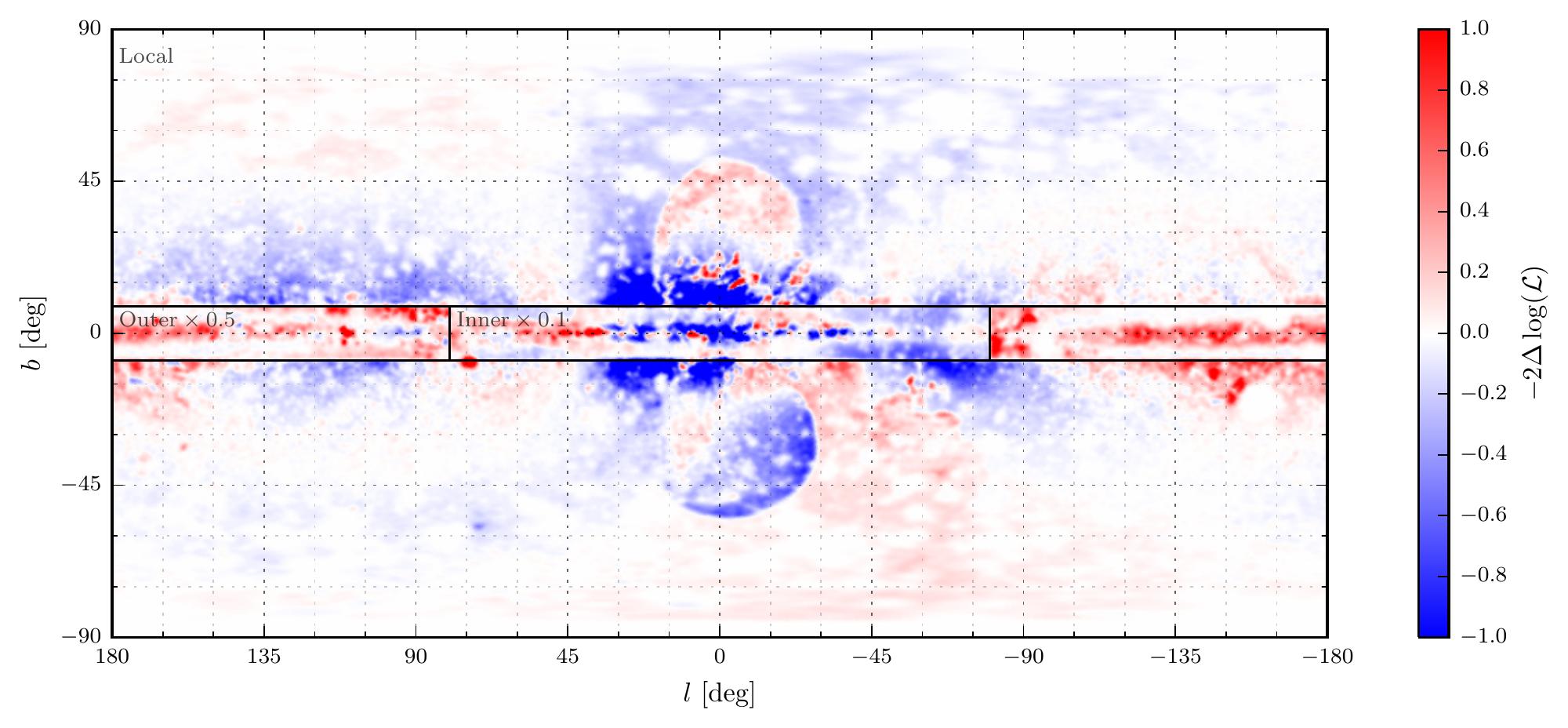}
  \caption{Pixel-by-pixel $-2\Delta \ln(\mathcal{L})$ for $f_{\rm H2}=0.2$ against the null model $f_{\rm H2}=0.0$, integrated over all energy bands for the global $\gamma$-ray analysis in the local($|b|\geq 8^\circ$), outer ($|b|<8^\circ, |l|>80^\circ$), and inner Galaxy ($|b|<8^\circ, |l|<80^\circ$) regions of interest, smoothed by a $0.5^\circ$ Gaussian kernel. Blue regions represent an improved fit compared with the axisymmetric source distributions.  The outer and inner regions have been rescaled by a factor 1/2 and 1/10, respectively and the white `holes' are due to point source masking, where the pixels have been weighted according to Eq.~(\ref{eqn:psc_weights}). Boxes indicate the edges of each global analysis ROI, and may produce discontinuities in the residuals since different model fits are imposed.} 
 \label{fig:delta_chi2}
\end{figure*}

\begin{figure}[thb]
  \centering
  \includegraphics[width=.45\textwidth]{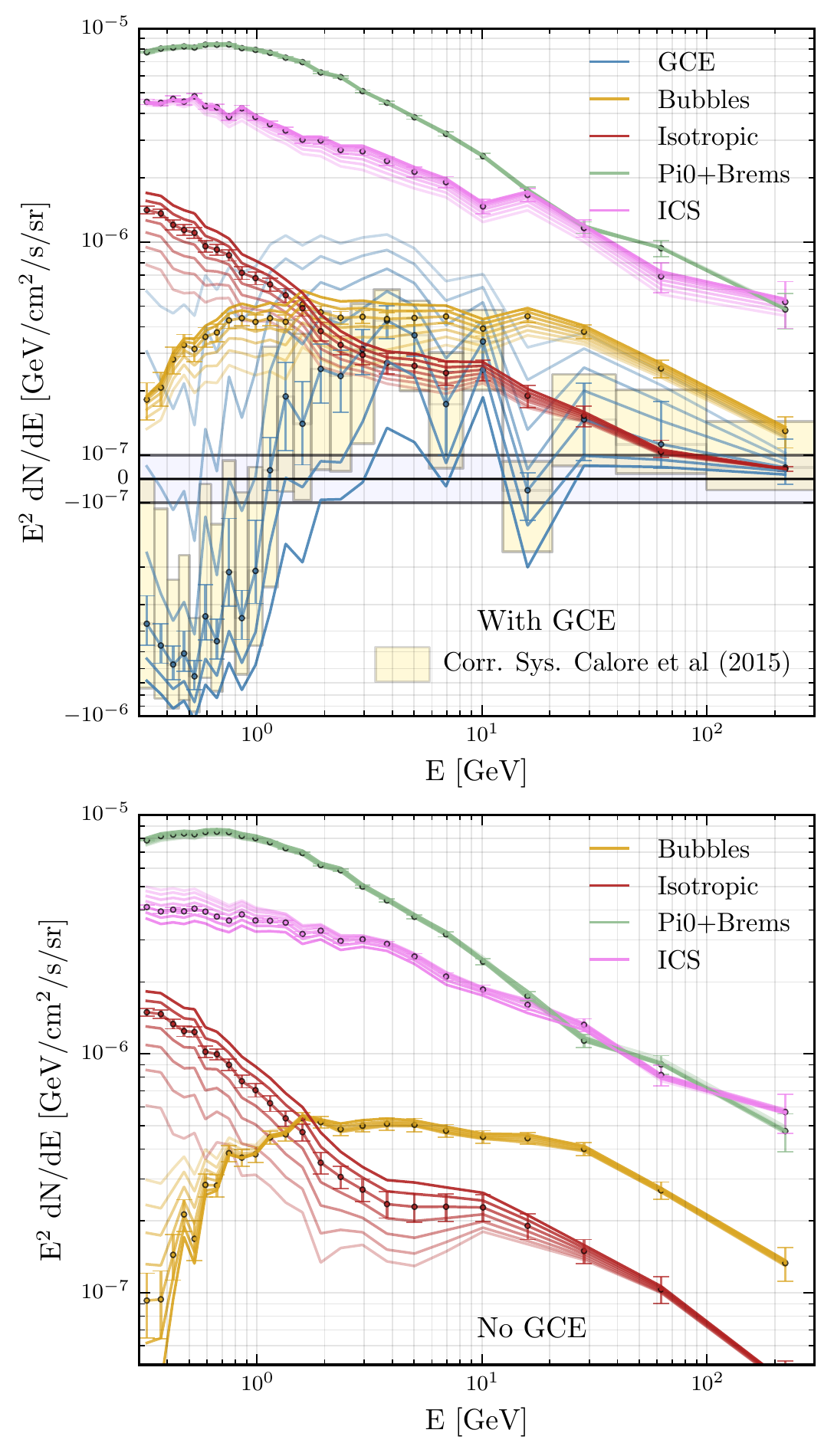}
  \caption{Inner Galaxy spectra of diffuse emission components with a GCE template (top) and without (bottom).  Curves from transparent to opaque increase \fh from 0 to 0.3 in increments of 0.05, with the \fh =~0.2 case marked by error bars.  In the top panel, absolute fluxes below $10^{-7} \rm ~ GeV/cm^2/s/sr$ have been linearized in order to show negative fit values.  The filled yellow error bars show correlated systematic uncertainties taken from ref.~\cite{Calore:2015}. We have assumed here (with some motivation as described below) that these are comparable to the systematic errors of our new GDE models.  Note that although the Fermi bubbles and isotropic spectra are allowed to float, deviations from the values determined using larger regions of interest are penalized by an externally imposed $\chi^2_{\rm ext}$, as described in Sec.~\ref{sec:global_analysis}.}
 \label{fig:spectra}
\end{figure}

Globally, our \fh models perform much better than the default SNR case (and better than Mod A, though this is not shown). One can examine $\Delta\chi^2$ for each pixel in order to determine which regions improve as \fh is increased.  This is shown in Figure~\ref{fig:delta_chi2}, where the delta-log-likelihoods for $f_{\rm H2}=0.2$ versus $f_{\rm H2}=0.0$ (null model) are presented for the three regions used in the Global analysis.  Blue regions highlight lines-of-sight where the addition cosmic-ray sources tracing the H$_2$ density provide an improved fit relative to the axisymmetric SNR model~\cite{Case:1998}.  

In the global-inner Galaxy, the redistribution of cosmic rays dramatically improves the fit for $45^\circ<l<30^\circ$. In the plane, diffuse Galactic $\gamma$-ray emission is dominated by $\pi^0$ decays following the hadronic
interactions of cosmic-ray protons with {\em molecular} hydrogen. Because the CO$\to \rm H_2$ conversion factor ($X_{\rm CO}$) has been
refit for each model, the fit improvements in this region must originate from (i) non-axisymmetric features of the
cosmic-ray injection morphology and/or (ii) an improved steady-state distribution of cosmic-rays which illuminate
the fixed {\em atomic} and {\em ionized} Hydrogen gas components.  In either case, the improved fit indicates that the new source models are resolving important cosmic-ray emitting structures toward the inner Galaxy. 

The outer Galaxy analysis produces very different outcomes, with non-zero \fh resulting in an inferior fit. However, this is likely to be a red herring, as the under-performing pixels lie above a few degrees latitude, where the thick disks of HI and HII dominate the gas density (rather than $\rm H_2$). Because HI is directly observable and the conversion from 21cm line intensity to gas density is requires only a single parameter (the hydrogen spin
temperature, which is typically treated as globally constant) the radial profile of atomic hydrogen is fixed.  This is in contrast to 
H$_2$ where the \xco conversion factor is allowed to vary.  As we have seen in Fig.~\ref{fig:CR_steady_state}, increasing \fh centrally concentrates the cosmic-rays causing the fixed HI and HII $\gamma$-ray emission to become dimmer. This leads to a worse fit which can only be compensated by \xco at low latitudes.  Given that the gas surveys are complete over these regions, it is notable that the outer Galaxy appears to have either a significant abundance of dark gas, or an increased population of cosmic-ray sources relative to observations (e.g. supernova or pulsar counts). This conclusion is consistent with previous determinations of the radial \xco profile from $\gamma$-ray data~\cite{fermi_diffuse,MS:2004}.

\begin{figure}[tb]
  \centering
  \includegraphics[width=.45\textwidth]{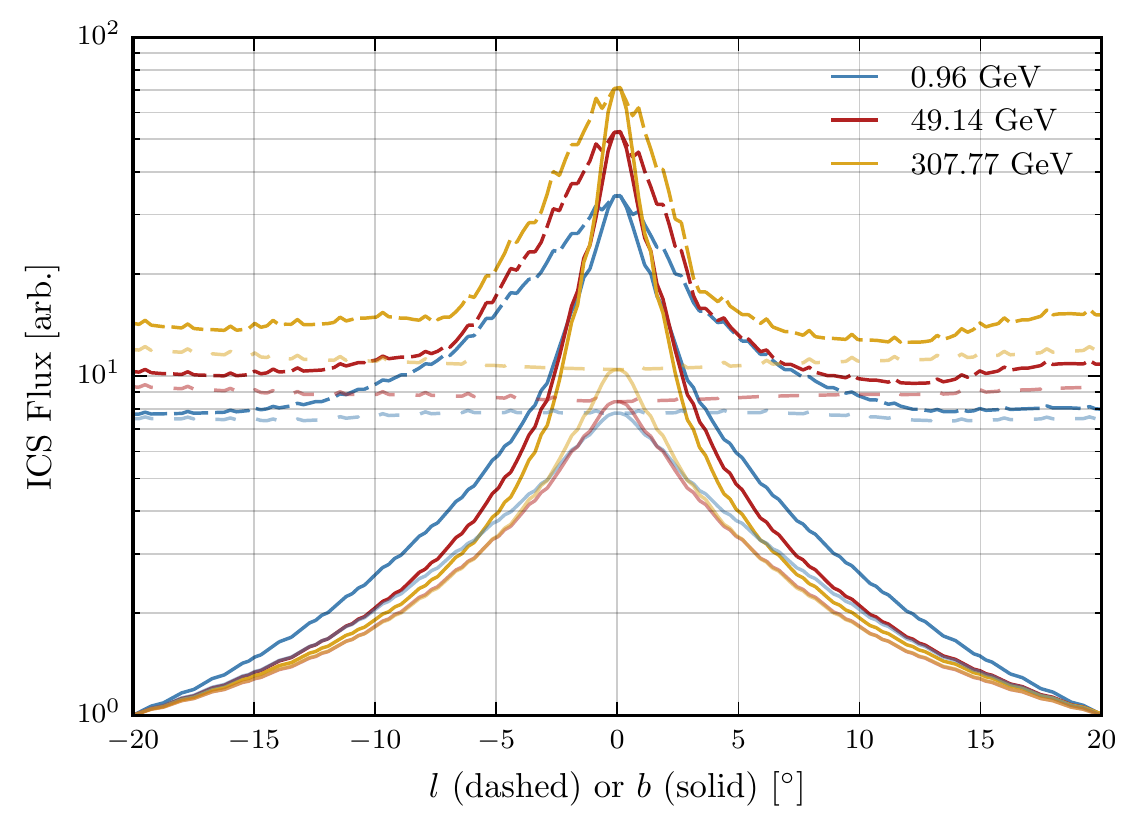}
  \caption{ICS flux as a function of longitude (dashed) and latitude (solid) along $b=0$ and $l=0$ for the cases of $f_{\rm H2}=0$ (transparent lines) and $f_{\rm H2}=0.2$ (opaque lines).  Each line has been normalized to unity at $(l,b)=(0,20^\circ)$. }
 \label{fig:ics_comparison}
\end{figure}

\begin{figure}[tb]
  \centering
  \includegraphics[width=.45\textwidth]{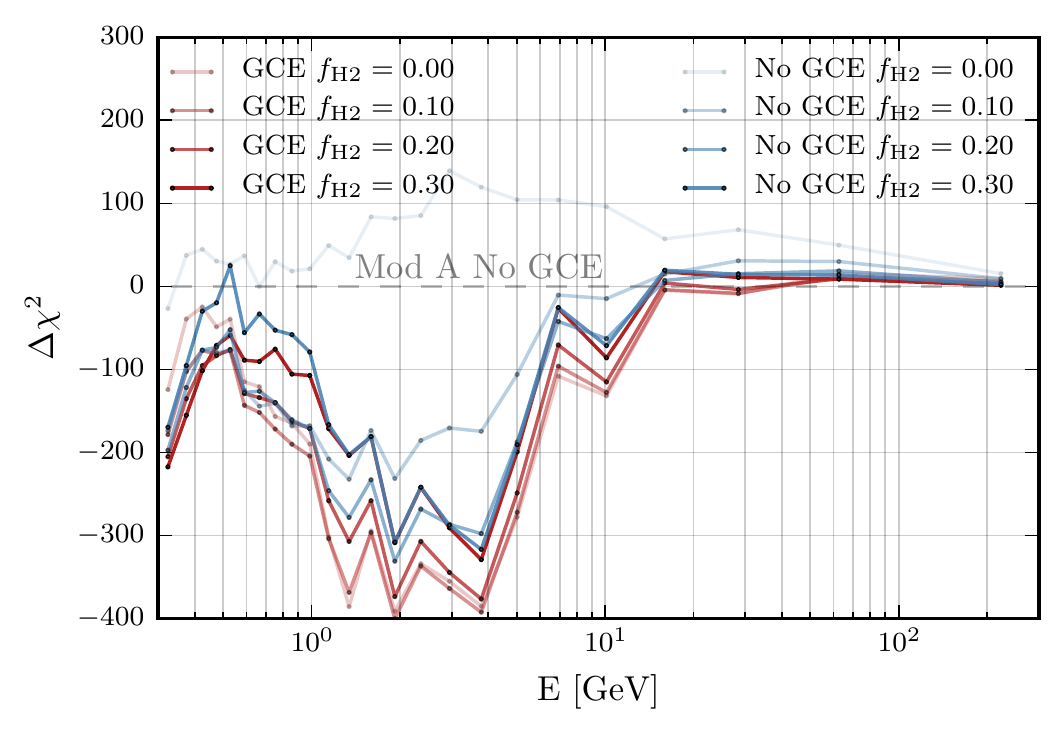}
  \caption{Inner Galaxy $\Delta\chi^2$ as a function of energy bin for representative values of $f_{\rm H2}$ with (red) and without (blue) dark matter.  The zero point is with respect to Mod A without dark matter.}
 \label{fig:TS_vs_energy}
\end{figure}

\medskip
In Figure~\ref{fig:spectra} we show the spectrum of each diffuse component (omitting the fixed point source template) in the inner Galaxy analysis as \fh is increased, with transparent to opaque lines showing $f_{\rm H2}=0\to0.3$ in increments of 0.05. In the top panel, we include a GCE template in the fit and as \fh is increased, the GCE flux is rapidly diminished at all energies, hardening at low energies and eventually becoming over-subtracted {\em if the DM template is allowed to take on negative values}.  The yellow uncertainty bands correspond to 1$\sigma$ diagonal elements of the full correlated systematic uncertainties\footnote{While these were derived using a different GDE model, we show in Sec~\ref{sec:plane_residuals} that GCE-like residuals along the Galactic plane do not change as dramatically as the Galactic center excess with increasing $f_{\rm H2}$, implying that the errors for the $f_{\rm H2}=0.2$ case should be comparable to those of Mod A.} from Ref.~\cite{Calore:2015}.  These are are most significant at energies $\lesssim1$~GeV, where the {\em Fermi's} point spread function becomes large, making it difficult to distinguish components based on morphology alone. The ICS component, Fermi bubbles, and isotropic templates gain some power across all energies while the $\pi^0$ component is largely unchanged.  

In the lower panel, we include only known astrophysical components (no GCE) in the fit. The Fermi bubbles spectrum is now very stable as a function of \fh while the isotropic component changes by a factor 2-3 below 10 GeV becoming more akin to a smooth power-law for larger $f_{\rm H2}$.  In all cases, the isotropic and Bubbles spectra are constrained by larger ROIs.  Even so, at low energies the point spread function is large, and the effective ROI of the inner Galaxy is small due to the large number of point sources in the field.  The spectrum of each component is thus much more uncertain than the statistical error bars shown, and can include contributions from mismodeled point sources, gas, or unmodeled diffuse components that are not present in other regions of the Galaxy.  Finally, while the the total ICS component is reduced, this does not imply that the ICS emission is reduced near the Galactic center. In fact, models with $f_{H2}\approx0.2$ produce a several fold enhancement within the inner few degrees compared with the pure SNR case.

In Figure~\ref{fig:ics_comparison} we show the longitude and latitude profiles of the ICS emission along $b=0$ and $l=0$ for the cases of $f_{\rm H2}=0$ (transparent lines) and $f_{\rm H2}=0.2$ (opaque lines).  Each line has been normalized to unity at $(l,b)=(0,20^\circ)$.  Most apparent is the completely flat longitudinal profile for the traditional models, highlighting the missing CMZ contribution.  As we turn on the new source distribution, a large spike appears, peaked at the Galactic center. In longitude, the old model includes a peak at $b=0$ due to the traversal of the Galactic plane, where the electron density is large throughout.  The new spike is approximately spherical (keeping in mind this figure shows the {\em total} ICS emission with minor elongation along the plane due to the partly disk-oriented injection morphology.  It is precisely this spike component which becomes highly degenerate with the observed properties of the GCE.  As will be briefly discussed later, the true ICS profile may be steeper than shown here if one includes a more realistic model of the optical and infrared radiation field morphology near the CMZ.  This ISRF structure is not present, and directly impacts the morphology and spectrum of the ICS emission.

In Figure~\ref{fig:TS_vs_energy} we show $\Delta \chi^2$ as a function of the energy for different values of $f_{\rm H2}$.  Increasing \fh greatly improves the fit without a GCE template between 1 and 20 GeV where the Galactic center excess is brightest. At lower energies, the improvement is smaller owing to the heavy PSF masking of the ROI, with only marginal improvements up to $f_{\rm H2}=0.1$, and reversing for higher values where the GCE template becomes over-subtracted.  Again, these $\Delta\chi^2$ curves are purely statistical and do not take into account the very large systematic uncertainties present below 1 GeV, where point sources in the Galactic plane can strongly influence the results, and the diffuse components become more degenerate. The greatest improvement occurs at energies near the peak of the GCE spectrum, where the number of residual photons is greatest in previous models, and the point spread function is small enough to preserve morphological details.  We see that when including dark matter, the fit is only marginally improved near the peak GCE energies, compared to $f_{\rm H2}=0.2$.  The similarity of these curves points to the strong statistical degeneracy between the astrophysical ICS emission of the Canonical model and dark-matter-like GCE template. 

\begin{figure*}[thb]
  \centering
  \includegraphics[width=\textwidth]{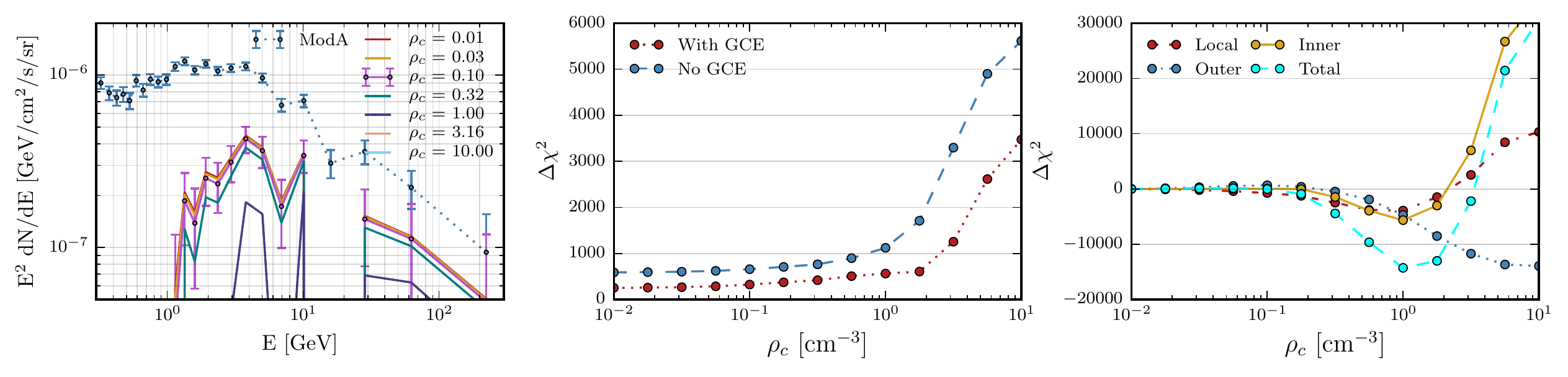}
  \caption{{\bf Left:} Flux of the GCE template, for the inner Galaxy analysis as the star formation threshold density $\rho_s$ is varied.  {\bf Center:} $\Delta\chi^2$ for the inner Galaxy analysis with and without a GCE template included in the fit.  {\bf Right:} $\Delta\chi^2$ for the global analysis.}
 \label{fig:threshold}
\end{figure*}

Our star formation model has one additional parameter: the critical density, $\rho_c$, which sets the minimum gas threshold to initiate star formation, and thus to inject cosmic-rays. In Figure~\ref{fig:threshold}, we show variations away from the default $0.1\ \rm cm^{-3}$.The impact on the GCE negligible for $\rm \rho_c\leq 0.1 \ cm^{-3}$, indicating that $\rm H_2$ densities below $\rho_c=0.1 \rm ~n_{\rm H2}/cm^{3}$ do not contribute significantly to the primary source population near the Galactic center.  At higher thresholds, lower-density diffuse gas clouds contribute less.  For a fixed value of $f_{\rm H2}$, the number of sources in dense regions is thus increased, with an effect that is similar to increasing $n_s$.  Globally, slightly higher thresholds $\rho_c\approx 1\ \rm cm^{-3}$ are preferred, larger than those typically implemented theoretically and in hydrodynamic simulations~\cite{Schaye2008}, but within range of some models~\cite{Kravtsov2003,Li2005}.  All but the largest star forming regions are below the 500 pc resolution of our simulation and 100 pc resolution of the gas distributions, making this parameter more phenomenological than physical.  Furthermore, the gas density is averaged over the lattice cell, and at sub-grid scales will contain much higher densities.  The high threshold preference in the outer Galaxy is due to the redistribution of cosmic-rays to large radius, though, as mentioned above, this region appears to be biased in the $\gamma$-ray fits.  Regardless of the specific value, the net effect on the Galactic center excess is to change the effective normalization of the CMZ region, since -- for a given \fh -- excluding the low density diffuse cosmic-ray sources assigns the more sources to the very dense GC (and to the $R>8.5$ kpc Galaxy).  We therefore consider $\rho_c$ as essentially degenerate with $f_{\rm H2}$, and do not consider further variations here.

\begin{figure*}[thb]
  \centering
  \includegraphics[width=\textwidth]{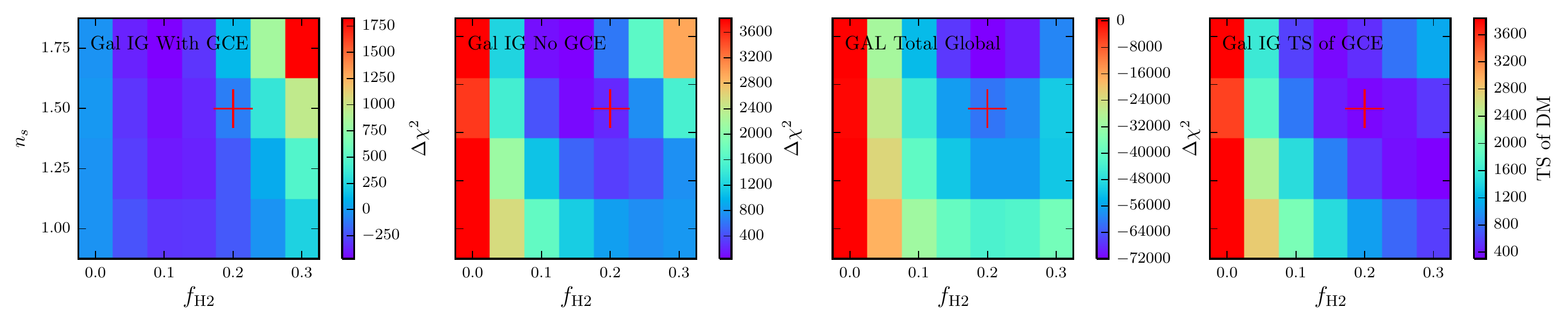}\\
  \includegraphics[width=\textwidth]{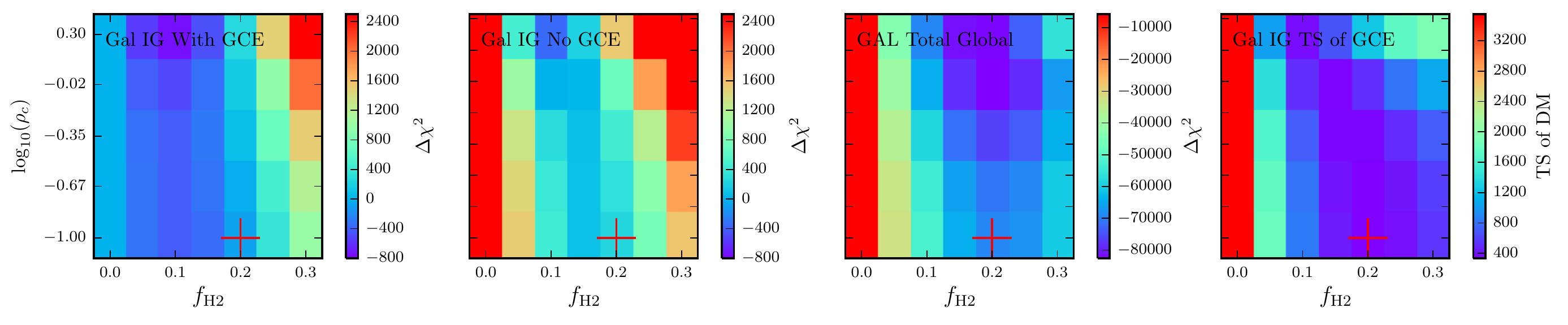}\\
  \includegraphics[width=\textwidth]{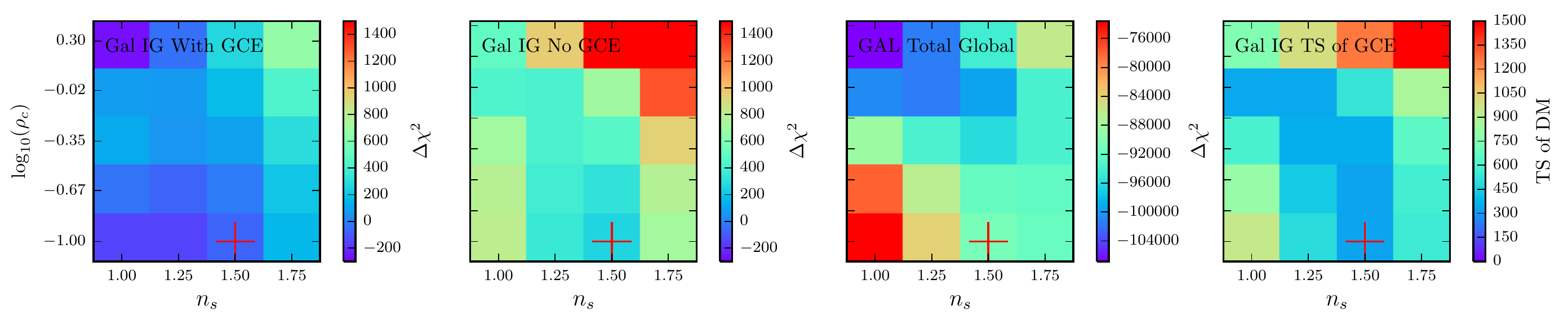}
  \caption{Fit statistics as star formation model parameters are varied.  From left to right columns we show $\Delta\chi^2$ for the inner Galaxy with and without a GCE template (first two columns), total global $\Delta\chi^2$ with respect to $f_{\rm H2}=0$ (third column), and the test statistic of the GCE template (right column). Lower (purple) values correspond to better fits except for the the rightmost column where purple regions indicate the minimal significance of an additional GCE component.  Red `+'s indicate the Canonical model.}
 \label{fig:vary_ns_fh2_plane}
\end{figure*}

In Figure~\ref{fig:vary_ns_fh2_plane} we summarize the star formation model parameter space by showing statistics for a variety of fits in three cross-sectional planes involving $f_{\rm H2}, n_s, \rho_c$.  From left to right columns we show the inner Galaxy $\Delta\chi^2$ with a GCE and without a GCE template (relative to Mod A + GCE), the total Global $\Delta\chi^2$ (relative to the Canonical model with $f_{\rm H2}=0$), and the test statistic of the GCE template in the inner Galaxy analysis. For the first three columns, lower values correspond to better fits, while in the right column, lower values correspond to lower GCE significance. 

In the first row, we show the $n_s$ versus $f_{\rm H2}$ plane. The inner Galaxy fits with and without a GCE template prefer $n_s\approx1.5-1.75$ and smaller $f_{\rm H2}\approx .1-.2$.  Globally, high $n_s$ are preferred which increases the number of sources in very dense gas regions, and decreases the fraction of cosmic-ray injection stemming from diffuse low-density sources.  Because of this, a larger $n_s$ requires a lower \fh to achieve the same level of structure. This inverse proportionality is clearly visible in all panels, and is especially prominent in the significance of the GCE template. 

In the $\rho_c$ versus \fh plane, a similar story unfolds.  For a fixed $f_{\rm H2}$, a larger star-formation threshold will enhance the number of sources in over-dense regions.  For the CMZ in particular, the gas density is well over threshold, and an increase in $\rho_c$ is nearly identical to increasing $f_{\rm H2}$.  As we discussed in Fig~\ref{fig:threshold}, the global fit improvement at $\rho_c\approx 1 \rm cm^{-3}$ is driven mostly by the biased outer Galaxy analysis.  The inner Galaxy fit with and without a GCE template is marginally improved for $(f_{\rm H2},\rho_c)\approx (0.1, 2 \rm ~cm^{-3})$, but the GCE significance remains largely indifferent for similar CMZ brightnesses.

Finally, the third row shows $\rho_c$ versus $n_s$.  The IG fits which include a GCE template prefer a large threshold and low $n_s$, showing that the parameters are not completely degenerate with each other. Similarly, the global fits marginally prefer high $n_s$ and $\rho_c$, again due to the improved outer Galaxy fit. With no GCE template in the IG, our Canonical model performs well.  Perhaps most important is that both the IG No GCE fit and the GCE template significance are roughly constant along an elliptical ridge (blue arc moving counter-clockwise from $\log_{10}(\rho_c)=0$ to $n_s=1.5$). This highlights the strong degeneracy between $n_s$ and $\rho_c$.  
\bigskip

In summary, we have shown that our star formation model parameters are highly covariant with each other, and that the full model space is conveniently approximated by a single parameter $f_{\rm H2}$ over the interesting subspace. Globally, the $\gamma$-ray data strongly prefer $f_{\rm H2}\approx 0.2-0.25$ overall, and even higher values toward the global-local and global-inner regions.  Remarkably, this parameter space is compatible with independent measures of the CMZ SNe rate (see Fig.~\ref{fig:CR_sources}).  When focusing only on the inner Galaxy ROI, fits including only the Galactic diffuse emission components very strongly prefer $f_{\rm H2}\approx 0.15-0.20$.  These models produce an ICS emission spike which is highly degenerate with the properties of the GCE.  However, when a GCE template is added, the fit is still significantly improved, though a lower value of $f_{H2}\approx0.1-0.15$ is preferred.  Below we will first study the Galactic center ROI, and will then characterize the spectrum and morphology of the residual emission in each analysis.

\subsection{Spectrum and Statistics at the Galactic Center }
\label{sec:gc}

\begin{figure*}[thb]
  \centering
  \includegraphics[width=1.00\textwidth]{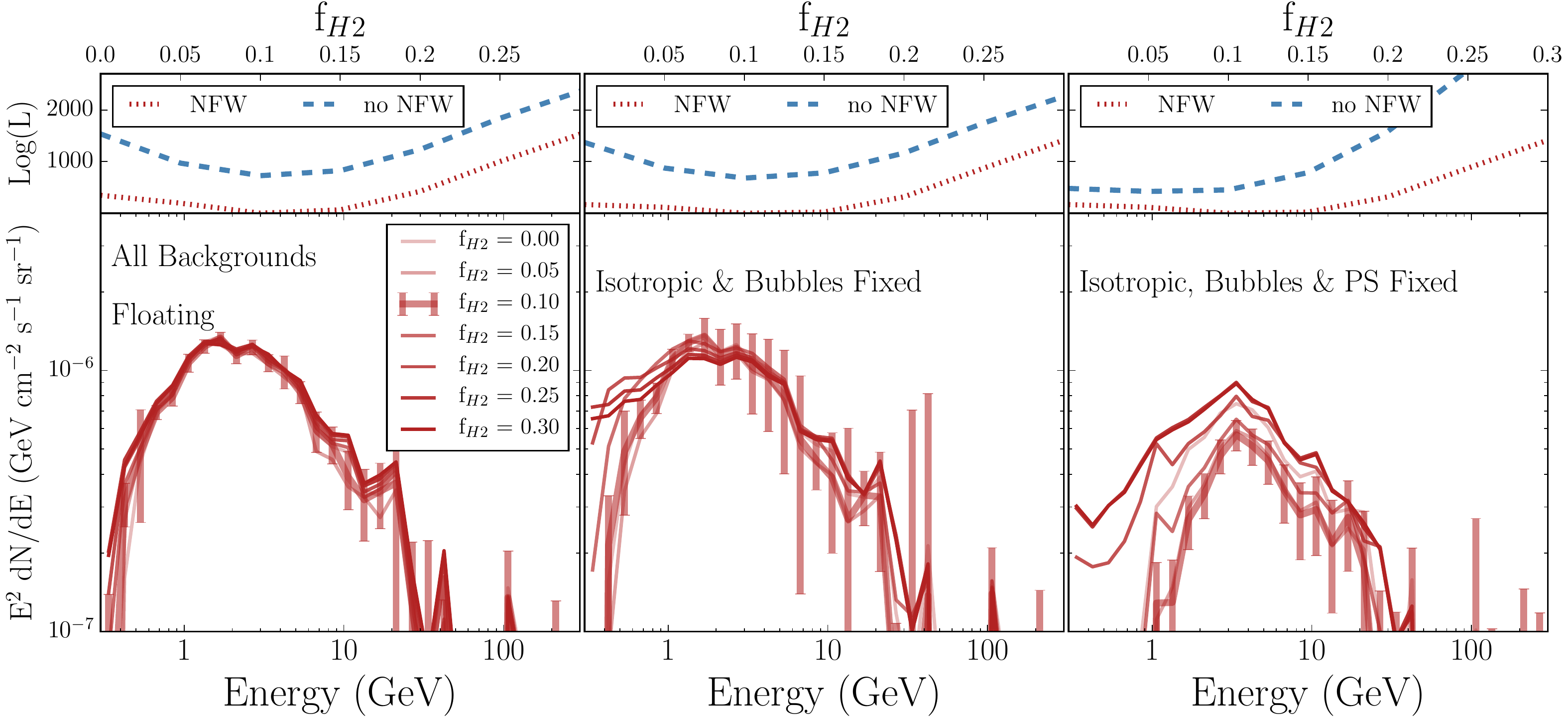}
\caption{The log-likelihood fit (top) and best fit GCE spectrum (bottom) for values of $f_{\rm H2}$~=~0.0 -- 0.3, in models where all backgrounds are allowed to float independently in each energy bin (left), the isotropic and bubbles templates are fixed to their putative value in full sky fits to the data (center), and the isotropic, bubbles, and 3FGL point source templates are fixed to their nominal values (right). In nearly all cases a value of f$_{H2}$~=~0.1 is preferred by the data. We note that the NFW template remains statistically significant and maintains a consistent spectrum in all cases except for models where the 3FGL point sources are fixed to their default values, a result that is expected due to the significant degeneracy between point sources near the GC and the GCE template. }
 \label{fig:gc_fits_no_mask}
\end{figure*}

\begin{figure*}[thb]
  \centering
  \includegraphics[width=1.00\textwidth]{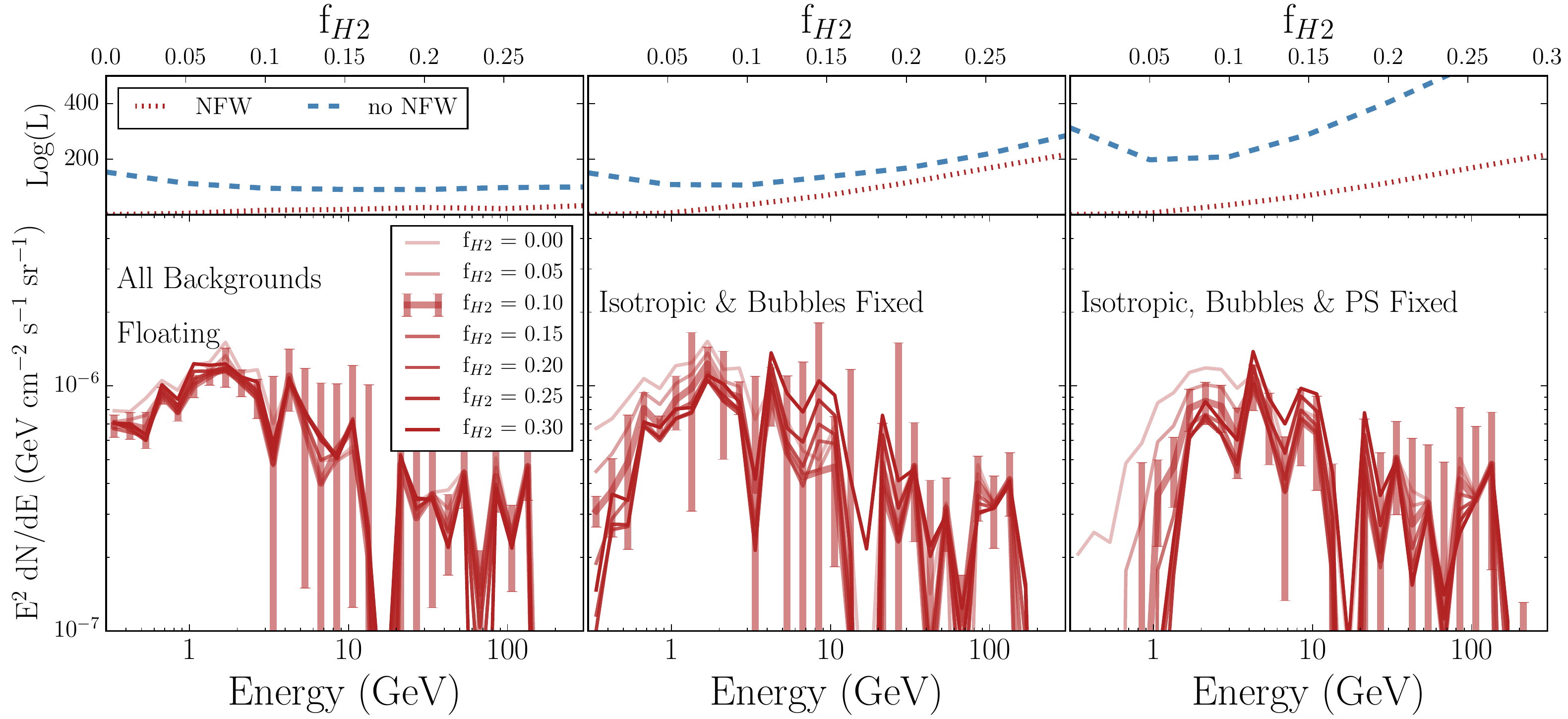}
\caption{Same as Figure~\ref{fig:gc_fits_no_mask} for an analysis which masks the Galactic plane ($|$b$|$~$<$~2$^\circ$) from the $15^\circ \times 15^\circ$ ROI surrounding the GC.}
 \label{fig:gc_fits_mask}
\end{figure*}

\begin{figure*}[thb]
  \centering
  \includegraphics[width=.8\textwidth]{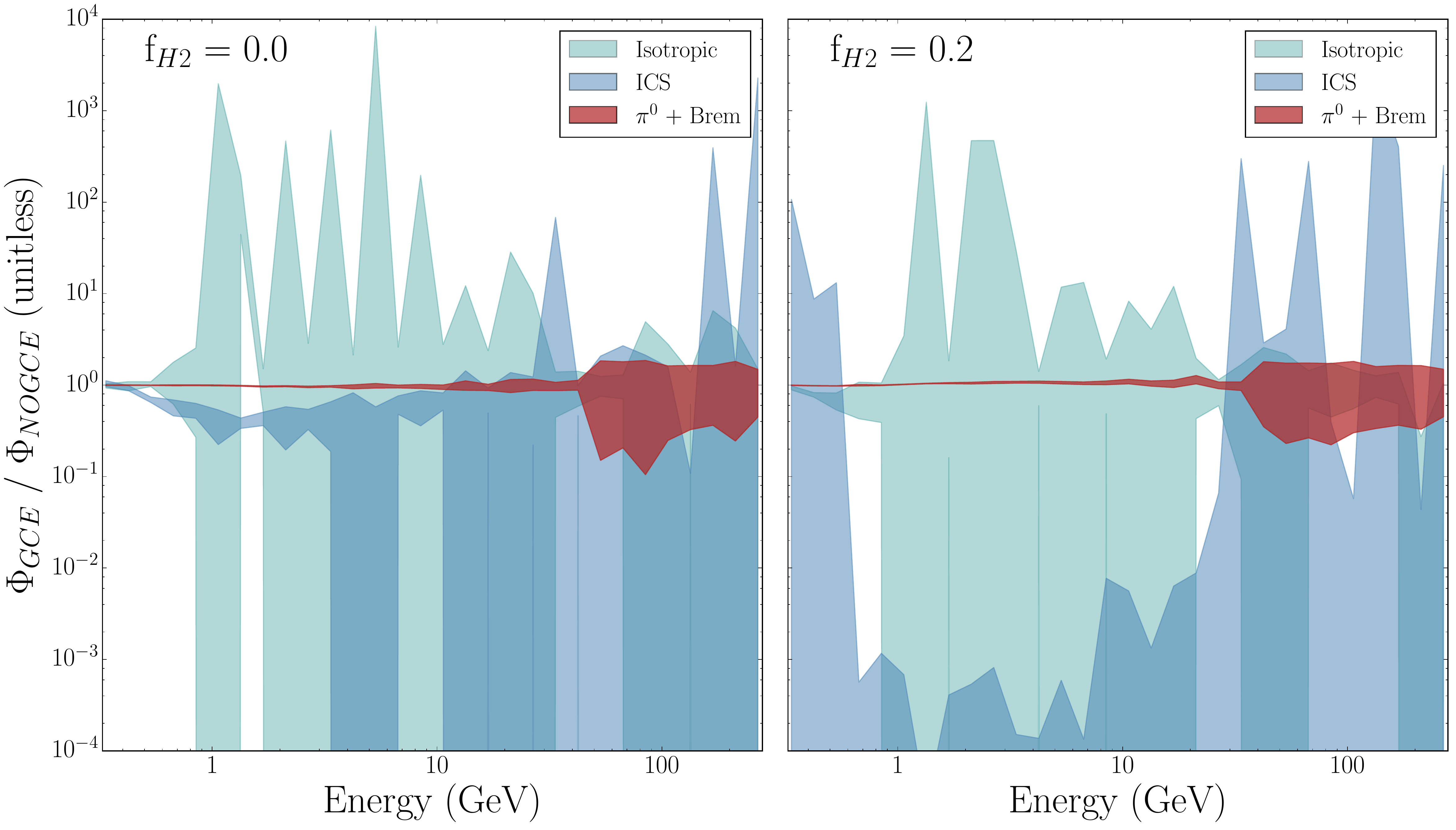}
\caption{Here we show 1$\sigma$ uncertainty bands on the relative normalizations of diffuse background components in the unmasked GC analysis for fits including and excluding a GCE template, for the case of $f_{\rm H2}$~=~0.0 (left) or $f_{\rm H2}$~=~0.2 (right). Components include the combined $\pi^0$ and bremsstrahlung template (red), inverse Compton scattering template (ICS, blue), and isotropic background template (green). The GCE template in the Galactic center analysis is highly degenerate with the ICS template, especially in models with higher values of $f_{\rm H2}$.  The isotropic template is poorly constrained over the ROI. All results are shown over the 15$^\circ\times$15$^\circ$ ROI of the GC analysis with no latitude mask applied.}
 \label{fig:astrophysical_background_gc}
\end{figure*}

In this section we present results for the $15^\circ \times 15^\circ$ region surrounding the Galactic center. In this analysis, the GCE template produces a significant fraction of the total $\gamma$-ray emission throughout the entire ROI. This contrasts with the Inner Galaxy analysis, where the astrophysical emission components are, in large part, fit to the data in regions where the GCE template provides only a marginal contribution to the total $\gamma$-ray flux.  Furthermore, when the bright Galactic plane is included in the analysis window, the analysis becomes sensitive to both emission along the plane, and to the GCE profile within 2 degrees of the Galactic center. For this reason, we perform fits over two analysis windows: one including the Galactic plane and one with the plane ($|b|<2^\circ$) masked. 

In Figure~\ref{fig:gc_fits_no_mask} we show the log-likelihood preference for the GCE template as well as the best-fitting NFW spectrum for various choices of $f_{\rm H2}$. We note two important conclusions: (1) the normalization and spectrum of the NFW template remain robust to changes in $f_{\rm H2}$, maintaining a total intensity that varies by less than 10\% in the 1-10~GeV energy range for all astrophysical diffuse emission models, (2) The value of $f_{\rm H2}$~=~0.1 is preferred for our standard analysis for both fits that do, or do not, include a GCE component. This result is somewhat lower than in the Inner Galaxy analysis in the case that no GCE source is present, but is consistent in the case that the GCE component remains in the analysis.  

We also show the resulting spectra and normalizations of the GCE template in models where the isotropic emission component and bubbles emission component, as well as all 3FGL sources, are fixed to their standard values from analyses of larger ROIs. We find that the spectrum and normalization of the GCE template remain robust when the isotropic and bubbles emission templates are fixed, showing the lack of degeneracy between these diffuse emission models and the GCE component in the Galactic Center ROI. However, the emission in the GCE component decreases significantly when 3FGL sources are fit to their nominal values. This is not unexpected, as there are several bright sources within $\sim$1$^\circ$ of the Galactic center that are highly degenerate with the addition of an NFW template~\cite{Daylan:2014rsa}. In interpretations where the GCE is a real emission component, this degeneracy is easily explained as a mismodeling of 3FGL point sources due to a miscalibration of the background diffuse emission.

However, one might worry that the robustness of the GCE in the Galactic center analysis stems from its large fractional intensity in the inner few degrees surrounding the GC. If the GCE template is highly favored close to the Galactic center, it may remain bright in a Galactic center analysis even if it provides a poorer fit to the $\gamma$-ray emission in regions several degrees from the GC. To investigate this possibility, we modify the GC analysis in order to mask regions of the sky with $|b|<2^\circ$, identical to the mask employed in the IG analysis. In Figure~\ref{fig:gc_fits_mask} we show the resulting normalization and spectrum of the NFW profile for all choices of $f_{\rm H2}$. 

We note three immediate results: (1) the best fit value of \fh is 0.0 in scenarios where the GCE template is included. However, this result is not particularly statistically significant, and a value f$_{H2}$~=~0.1 is disfavored at only TS$\sim$31. (2) The best fit spectrum of the GCE component remains similar to fits over the full Galactic Center ROI, albeit with the addition of some low-energy emission that may be due to leakage from the masked plane region, and (3) the statistical significance of the GCE component is, however, substantially reduced, from TS$\sim$1450 to TS$\sim$160. While some reduction in the TS is expected from the smaller ROI of the masked analysis, we note that cutting the region $|b|<2^\circ$ removes only 61\% of the photons above 1~GeV. Since TS is, roughly, a photon counting statistic, we would expect a similar reduction in the TS, compared to the observed 90\%. Instead, this result indicates that there is a greater degeneracy between the astrophysical diffuse emission and the GCE in the region $|b|<2^\circ$.

As mentioned above, the GCE flux is comparable to that of the astrophysical emission components over the small ROI.  Thus, in the Galactic center analysis, the addition of a GCE template is likely to significantly alter the flux of astrophysical emission components, compared to models of the Inner Galaxy. In Figure~\ref{fig:astrophysical_background_gc} we show 1$\sigma$ statistical uncertainties on relative normalization of the astrophysical diffuse background components after the addition of a GCE template in the unmasked GC analysis. The large reduction in the ICS normalization after the GCE is added shows that the ICS emission (in particular) is highly degenerate with the properties of the excess in the case where $f_{\rm H2}=0.2$. For visual clarity, we do not show the relative normalization of the bubbles component, but note that the flux uncertainty of the Fermi bubbles component is significantly larger than its flux in nearly all energy bins, implying that the component is unimportant for fits in the Galactic center ROI.  This is reasonable considering that the Fermi Bubbles template is has uniform brightness and covers almost the full GC ROI.  Intriguingly, this figure depicts the Galactic center analog to the decreasing intensity of the GCE component when \fh is increased in the Inner Galaxy analysis. In analyses of the Galactic center ROI, the degeneracy between the GCE component and ICS component statistically favors the fit from the GCE. Turning on a GCE component thus significantly decreases the emission stemming from the ICS component, producing a spectral dip mimicking the GCE emission.

\subsection{Characterizing Residual Emission}
\label{sec:characterize_residuals}
In this section we study the emission morphology and spectrum of the GCE component as determined by the Inner Galaxy and Galactic center analyses. We will first examine IG residuals as a function of $f_{\rm H2}$ (\ref{sec:residuals}), followed by a comparison of the GCE with residuals along the Galactic plane (Sec.~\ref{sec:plane_residuals}). Next we will determine radial profile derived by splitting the GCE template into annuli and determining the best fit (Sec.~\ref{sec:radial_profiles}. Then we will simultaneously vary the ellipticity and inner slope of the GCE template to determine the best fit morphology for the inner Galaxy (Sec.~\ref{sec:slope_ellipticity_IG}), and will test the energy dependence of the best-fit morphology.  Finally, we perform morphological scans on the Galactic center ROI (Sec.~\ref{sec:slope_ellipticity_GC}).

\subsubsection{Raw Residuals}
\label{sec:residuals}
\begin{figure}[tb]
  \centering
  \includegraphics[width=.45\textwidth]{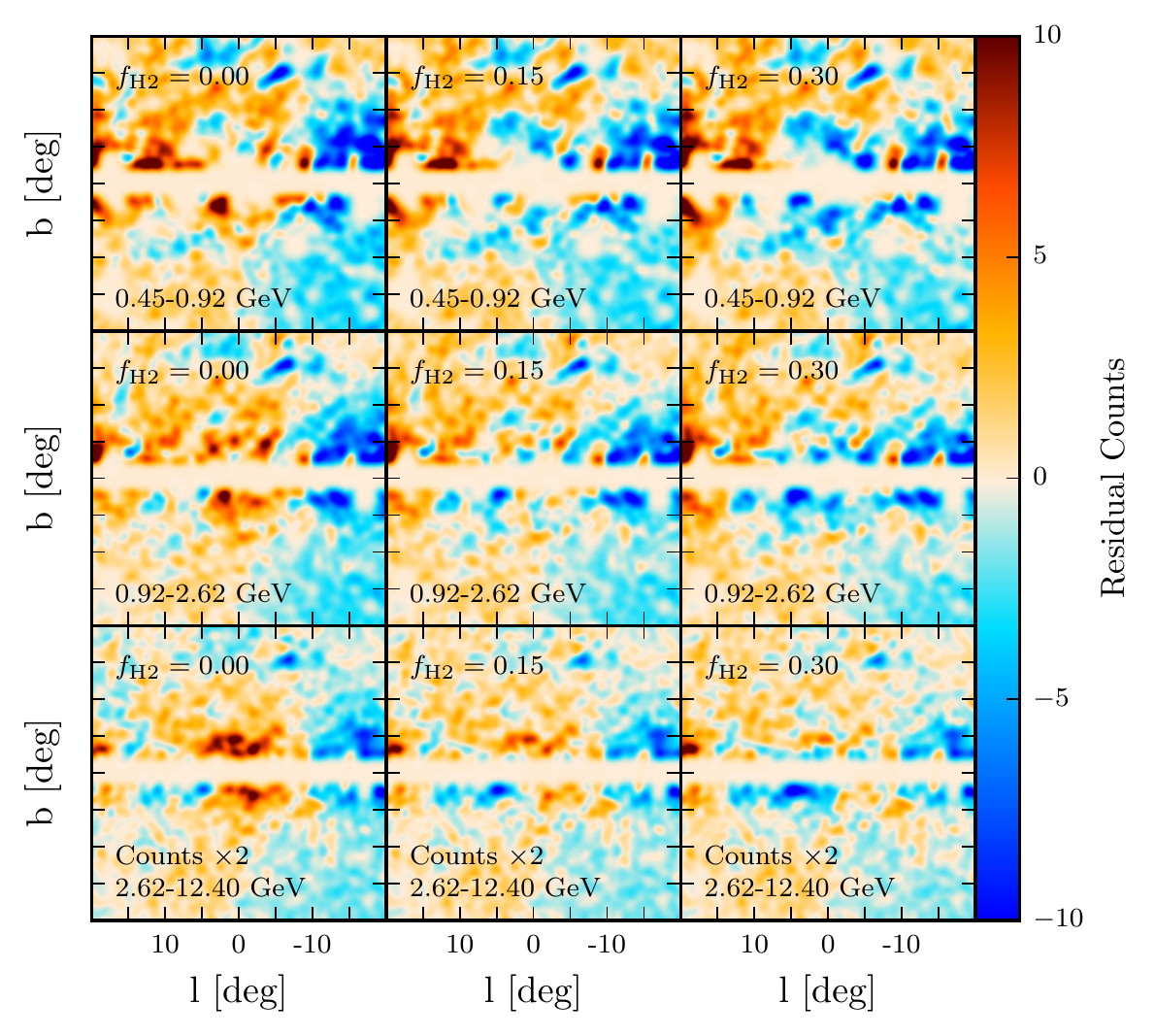}
  \caption{Residual emission maps as \fh is increased for the inner Galaxy analysis with no GCE template included in the fit. Red indicates under-subtracted regions while blue indicates regions where the diffuse model is overly bright.  All maps have been multiplied by the Galactic plane mask, weighted according to the 3FGL point-source mask defined by Eq.~(\ref{eqn:psc_weights}), and subsequently smoothed by a Gaussian kernel of $\sigma=0.5^\circ$.}
 \label{fig:residuals}
\end{figure}

In Figure~\ref{fig:residuals} we show residuals for the inner Galaxy analysis with no GCE template for $f_{\rm H2}=0, 0.15, \rm and\ 0.3$ (left to right columns), integrated over low, middle, and high energy bands (top to bottom rows).  The residuals have been multiplied by the weighted point source mask and smoothed by a Gaussian kernel with $\sigma\approx0.5^\circ$.   Visually, it is easy to see the disappearing excess in each energy band as \fh is increased, eventually leading to over-subtracted (blue) regions for $f_{\rm H2}=0.3$.  For both models, the positive residual at $l\approx10^\circ-20^\circ$ becomes brighter for large $f_{\rm H2}$.  As noted by Ref.~\cite{Calore:2015}, this is connected with the Aquila Rift \htwo star forming region which lies within a few hundred parsecs of the Earth. Unlike the GCE, the Aquila Rift spectrum is a smooth and soft power law, consistent with star forming regions.  The residual emission associated with the Aquila Rift region falls off rapidly at higher energies.  We have tried additional templates for the Aquila Rift region, where we use the PEB H2 model and sliced out the nearest 500 pc for $l>15^\circ$ (which includes the full AQ region).  This does substantially improve the fit over the positive longitude edge of the ROI, but does not impact the flux or spectrum of the GCE.    Because the inclusion of this template in fits also reduces convergence of the optimizer in many cases, we do not include it in further analyses, however, we note the utility of the PEB model~\cite{PEB} when generating templates for individual molecular or atomic hydrogen structures which need to be isolated along the line-of-sight.

As the diffuse emission model changes, so does inferred spectrum and normalization of point sources in the field.  Although these point sources should be refit for each new diffuse model, the huge number of additional parameters make this optimization impractical for the inner Galaxy ROI.  Still, $\gamma$-ray skymaps for different \fh values vary significantly, and it is inevitable that new point sources arise and that 3FGL sources become mismodeled.  To test the possibility that new point sources or leakage are significant, we compute the angular power spectra of residuals for $f_{\rm H2}=0$ and $f_{\rm H2}=0.3$ and examine the ratio of coefficients.  We find that most of the new spectral power is picked only up at low angular frequencies while the small wavelength residual power (those below the scale of the PSF) are not changed at a statistically significant level.  This provides support that the point source leakage as the diffuse model is changed is not important.  We also find only very weak sensitivity to the photon PSF class for the inner Galaxy analysis. Further photon subselections are not explored further here, although we have verified that the IG and GC results presented below remain robust when using the PSF3 events class which contains 25\% of the total photons with the best angular resolution.

\subsubsection{Galactic Plane Residuals}
\label{sec:plane_residuals}

\begin{figure}[thb]
  \centering
  \includegraphics[width=.45\textwidth]{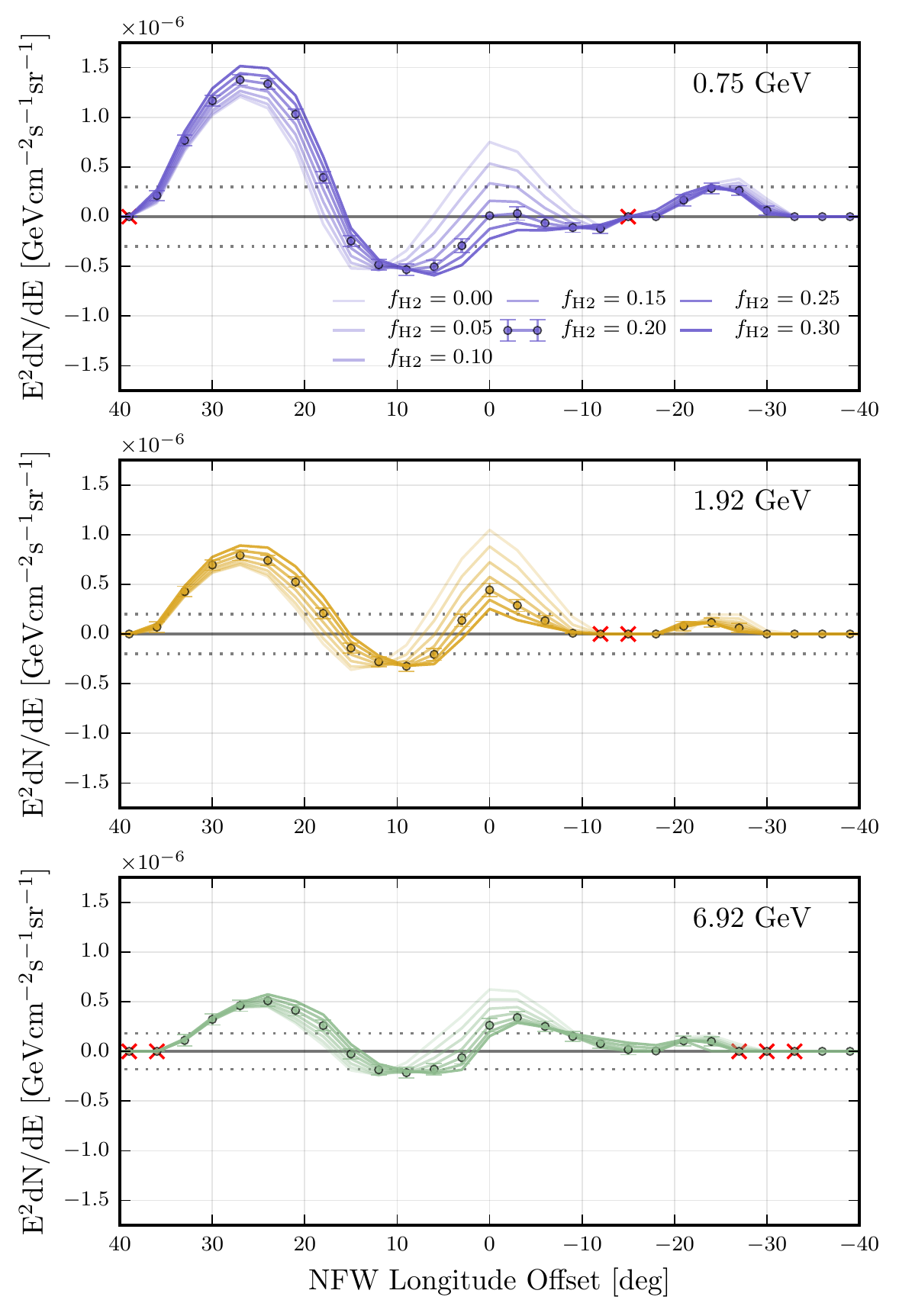}
\caption{Best fit flux for a window centered NFW$_{\alpha=1.05}$ template as the inner Galaxy analysis is transposed along the Galactic plane. Curves from light to dark increase \fh from 0 to 0.3 in increments of 0.05, with $f_{\rm H2}=0.2$ case marked with error bars.  The dotted lines are the 1$\sigma$ (highly-correlated in energy) systematic uncertainties for residuals along the Galactic plane taken from Ref.~\cite{Calore:2015}.  Red `$\times$'s mark non-convergent fits.}
 \label{fig:longitude}
\end{figure}

In order to compare the GCE intensity against residuals found along the Galactic plane, we follow the procedure of Ref.~\cite{Calore:2015}, transposing our entire inner Galaxy analysis in longitude, with the NFW$_\alpha={1.05}$ GCE template centered in each offset ROI for $|l|<40^\circ$.  We do not, however, perform a full systematic study of uncertainties as done in Ref.~\cite{Calore:2015}.  If the plane residuals are not dramatically changed by our new source models, the systematic error bars derived in Ref.~\cite{Calore:2015} should still provide a reasonable estimate of the uncertainties here.

In Figure~\ref{fig:longitude} we show the flux of the transposed GCE template at 750 MeV, 1.9 GeV, and 6.9 GeV as \fh is increased.  Error bars highlight the Canonical model.  The dotted lines (symmetric about zero) are the 1$\sigma$ systematic error bands from Ref.~\cite{Calore:2015}'s principle component analysis of Galactic plane residuals. As \fh is increased we see that the Galactic center excess is reduced well below the level of the Aquila Rift star forming region at $l\approx 25^\circ$, and is comparable to the projected molecular ring at $l\approx -25^\circ$.  The Canonical model is near or below the $1\sigma$ systematics in each case.  The residuals along the plane do increase with increasing $f_{\rm H2}$, but at a 10-20\% level, indicating that $f_{\rm H2}>0$ dominantly impacts the Galactic center excess while remaining a compatible with other regions along the Plane.  It is intriguing that increasing \fh enhances the Aquila Rift region ($l=25^\circ$) given that \htwo rich regions should be made brighter, and would seemingly reduce positive residuals which have their origins in dense \htwo star forming regions. However, it is difficult to ascertain the exact cause in a template analysis since the morphology depends on emission along the full line of sight.  Our new source model does improve the small residuals near $l=\pm 15^\circ$.  The spectrum of the non-GCE residuals is essentially unchanged, following a soft power-law consistent with star-forming regions.

\subsubsection{Radial Profiles}
\label{sec:radial_profiles}

\begin{figure*}[tb]
  \centering
  \includegraphics[width=\textwidth]{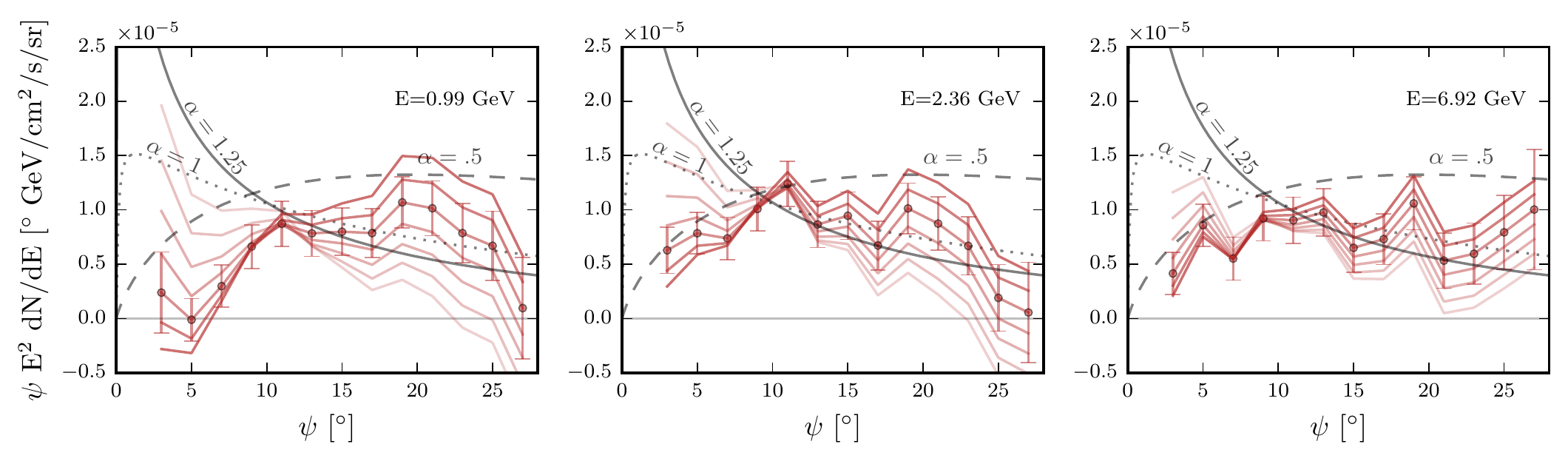}
  \caption{Radial flux profile of the NFW$_{\alpha=1.05}$ annuli at three energies representative of the Galactic center excess.  Curves from light to dark increase \fh from 0 to 0.3 in increments of 0.05, with the Canonical $f_{\rm H2}=0.2$ model indicated by error bars.  We also show arbitrarily normalized projected NFW flux profiles for inner slopes $\alpha\in \{0.5, 1, 1.25\}$.  Note that (i) the inner slope of the GCE template is fixed to $\alpha=1.05$ before subdivision into annuli, which may slightly bias results, and (ii) that the vertical axes have been rescaled by a factor $10^{-5}$.}
 \label{fig:radial}
\end{figure*}

In addition to spectral changes to the GCE, the morphology of the Galactic center excess is also sensitive to $f_{\rm H2}$. In the case of $f_{\rm H2}=0$, the residual is approximately spherical with a radial profile consistent with a standard NFW profile (though a slight adiabatic contraction to $\alpha$~=~1.05 is statistically preferred). However, as $f_{\rm H2}$ is increased, we observe the radial profile to become much shallower and we will see that the preferred ellipticity becomes {\em energy dependent} and non-spherical.  As an initial test of these distortions we begin with the inner Galaxy analysis and split the $\rm NFW_{\alpha=1.05}$ template into $2^\circ$ wide annuli, providing both the GCE intensity and spectrum as a function of radius.

In Figure~\ref{fig:radial} we show the flux as a function of the projected angle ($\psi$) from the Galactic center  at 1, 2.36, and 6.92 GeV.  Also shown, are the projected NFW profiles using inner slopes $\alpha=0.5, 1,~ \rm and~1.25$.
As \fh is increased, we observe that the emission morphology is significantly flattened at all energies. Not only is the GCE suppressed at small radii, but it is also enhanced at large radii.  This effect is most dramatic for ICS photons with $E_\gamma \lesssim 1$ GeV.  At these low energies, the electron energy loss timescale is much longer, and the electrons diffuse farther away from the CMZ. Eventually, the diffuse emission becomes too bright and saturates $\psi\lesssim 5^\circ$. This suppresses the entire ICS template, including the high latitude ICS from the disk. At large radii, the GCE template brightens to compensate.   At higher energies, larger \fh would further reduce the excess, but more efficient transport is also needed so that electrons above 30 GeV can propagate to larger radii over the same energy loss time-scale.  In Appendix~\ref{sec:calore_regions}, we show the GCE spectrum over the 10 IG regions defined by Calore et al~\cite{Calore:2015}, in which similar conclusions can be drawn.

Most importantly, the remaining excess is too flat to match any non-cored NFW emission profile. %That is to say, the presence of a bright CMZ in the model renders the remaining residual incompatible with dark matter in the inner Galaxy.  
As shown in Fig~\ref{fig:CR_sources}, models with $f_{\rm H2}<0.3$  still under-predict the supernovae rate in the CMZ, and the results shown here may be conservative.  On the other hand, low \fh models with a GCE component are still statistically preferred, and one must make a choice about which prior weights to apply on the value of $f_{\rm H2}$.

\label{sec:slope_ellipticity_IG}
\begin{figure*}[tb]
    \centering
    \includegraphics[width=.75\textwidth]{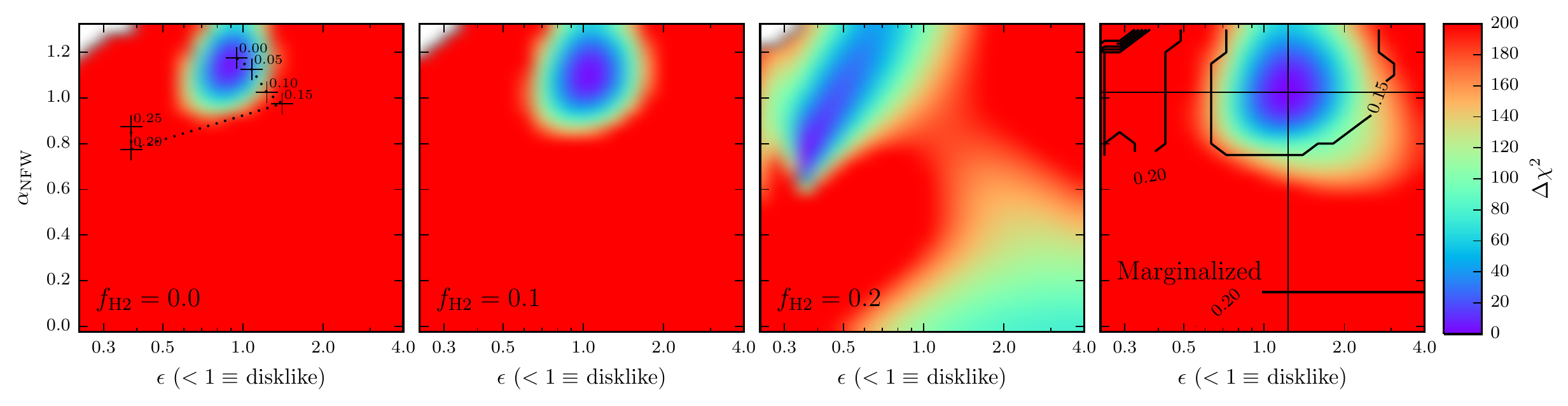} \\
  \includegraphics[width=.75\textwidth]{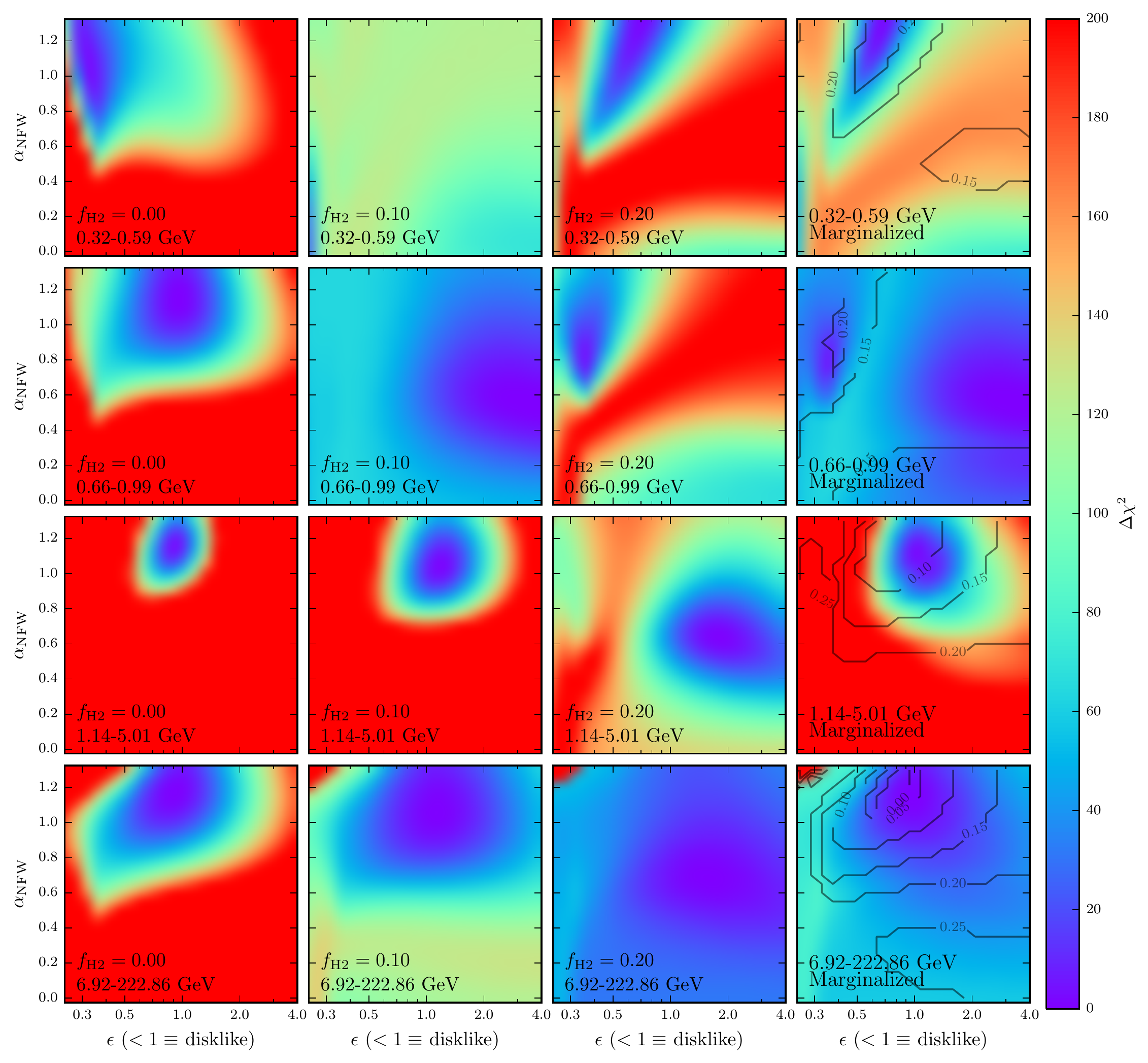}   
  \caption{{\bf Top:} Preferred morphology (ellipticity $\epsilon$ and inner-slope $\alpha_{\rm NFW}$) of the GCE template for increasing values of $f_{\rm H2}$.  The left panel shows the $f_{\rm H2}=0$ case with black `+' markers indicating the best fitting morphology for each $f_{\rm H2}$ sampled. The center two panels show the best-fitting inner Galaxy case with a GCE template and $f_{\rm H2}=0.1$, as well as the $f_{\rm H2}=0.2$ preferred by the global ROI.  The right panel shows the preferred GCE morphology after marginalizing over $f_{\rm H2}$ -- i.e. always choosing the value of $f_{\rm H2}$ which minimizes the $\chi^2$.  Here contours indicate the best fitting value of $f_{\rm H2}$, with the overall best fitting case corresponding to $f_{\rm H2}=0.1$. {\bf Bottom:} Same left to right columns as above, but divided into four energy bins (top to bottom) showing the energy dependence of the morphology as \fh is increased.  We note that for $f_{\rm H2}\gtrsim0.2$, the low energy GCE spectrum becomes negative and disk-aligned, indicating that the galactic diffuse emission model is too bright along the disk near the GC.  At higher energies, the GCE template prefers to extend out of the disk.}
 \label{fig:morphology}
\end{figure*}

\subsubsection{Inner Galaxy Slope and Ellipticity}

We have shown above that the intensity and radial profile of the residual emission are highly sensitive to $f_{\rm H2}$, with larger values resulting in a pronounced flattening.  The properties of the excess must now be evaluated in terms of the new preferred morphology. We therefore perform a new scan in the parameter space of inner slope $\alpha_{\rm NFW}$ vs ellipticity $\epsilon$ for several values of $f_{\rm H2}$.

\begin{figure}[thb]
    \centering
    \includegraphics[width=.45\textwidth]{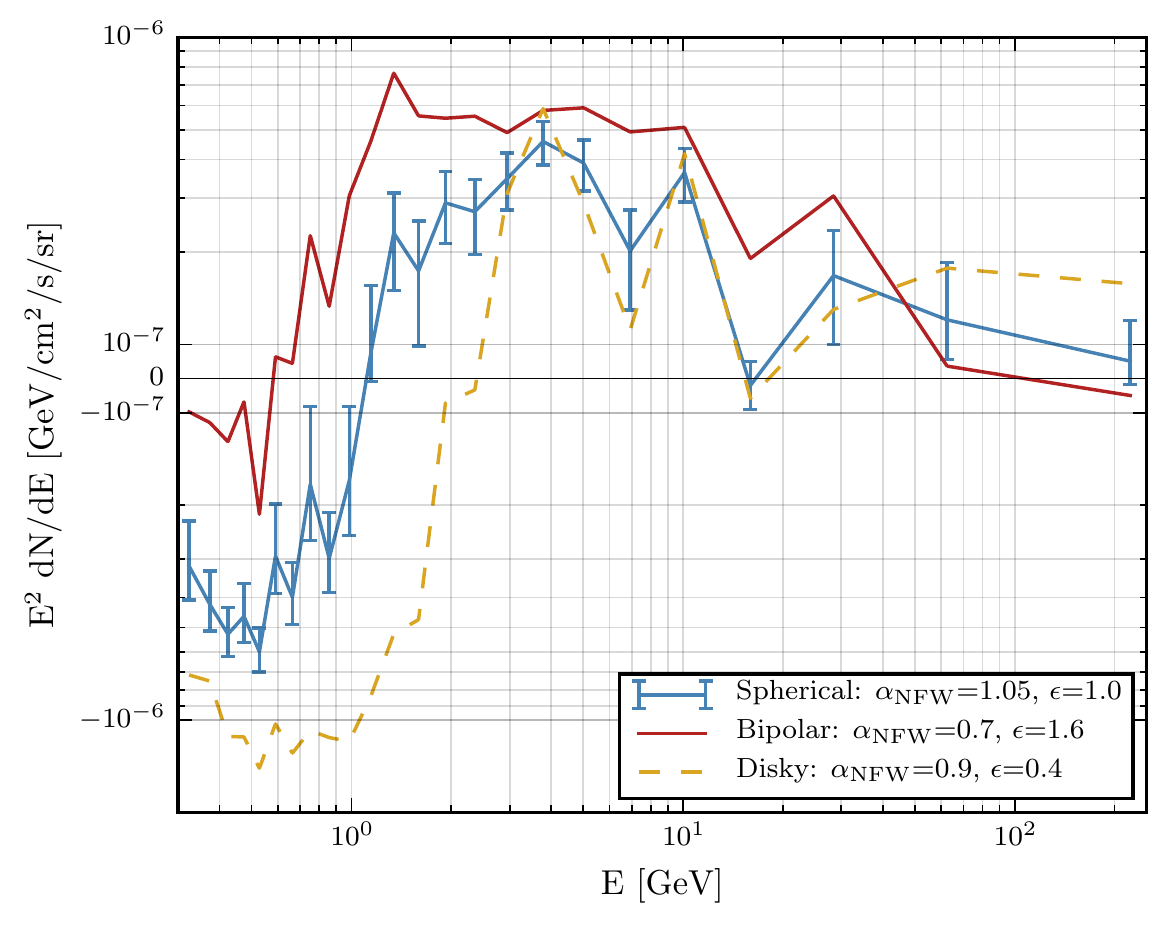}
  \caption{Spectra of the GCE template in the inner Galaxy analysis as the ellipticity $\epsilon$ and inner-slope $\alpha_{\rm NFW}$ of the GCE template is changed around our spherically symmetric Canonical $f_{\rm H2}=0.2$ model. Disk-like models are favored at low energies where they can subtract the over-brightened disk, while higher energies favor a GCE template which is both flattened and elongated perpendicular to the disk.}
 \label{fig:spectrum_after_skew}
\end{figure}

In the top panel of Figure~\ref{fig:morphology}, we present the IG $\Delta\chi^2$ in the $\alpha_{\rm NFW}$ vs $\epsilon$ plane for three values of $f_{\rm H2}$.  The left panel contains the standard case of $f_{\rm H2}=0$ which shows a highly spherical profile and a steep inner slope $\alpha_{\rm NFW}\approx 1.15$. In black '+' markers, we show the evolution of the best fitting profile as $f_{\rm H2}$ increases.  As observed in the morphological tests above, we find the profile parametrically becomes elongated and flattened as $f_{\rm H2}$ increases, up to $f_{\rm H2}=0.15$. In the second column we show the case of $f_{\rm H2}=0.1$ which is the best fitting case for the Inner Galaxy with a GCE template included. Interestingly, in this case, we find that a peaked, and roughly spherically symmetric profile is still preferred overall. The third panel shows our Canonical model, which is the preferred fit in both the full sky data, as well as the Inner Galaxy when no GCE template is included. Here, two islands form which are either highly disk oriented or prefer an ellipticity extending out of the disk.  Here the energy dependence of the GCE is manifest, and it is clear that we can not rely on an energy averaged view alone.  If the CMZ star formation rate and global $\gamma$-ray analysis are taken as a strong priors toward $f_{\rm H2}\geq0.2$ in the inner Galaxy, then a dark-matter-like profile ($\alpha_{\rm NFW}=1$, $\epsilon=1$) is ruled out by $\Delta \chi^2\approx133$, with larger values of \fh even more strongly disfavoring a dark matter interpretation. 

In the right panel, we marginalize\footnote{Typically, marginalizing implies integrating out the nuisance parameter.  Here the likelihood function is usually quite sharp in $f_{\rm H2}$ for a given choice of morphology and it suffices to just select the best fitting value.} over $f_{\rm H2}$ by choosing the best fitting case for each ($\alpha_{\rm NFW}$, $\epsilon$), and represent this best fitting $f_{\rm H2}$ via overlaid contours. Intriguingly, the best fits to the inner Galaxy (with a value of \fh~=~0.1) remains consistent with the standard assumptions for an NFW profile motivated by dark matter annihilation ($\alpha$~=~1.0, Axis Ratio~=~1.0) at the level $\Delta \chi^2=34$. For dark matter interpretations of the GCE, this remains the most important result of the present work. Using vastly improved and more realistic diffuse emission models for $\gamma$-ray generation near the GC, models with a GCE component motivated by dark matter annihilation remain compatible with the data.

More important than the energy {\em averaged} morphology is to examine the energy dependence as $f_{\rm H2}$ is increased (left to right).  In the bottom pane of Figure~\ref{fig:morphology}, we show the preferred morphology split into four bins of increasing energy (top to bottom columns).  For the $f_{\rm H2}=0$ models which have no CMZ, the morphology is spherical except for the lowest energy bin where a disky profile is favored by $\Delta \chi^2\approx 80$. Before adding CMZ cosmic-rays, the energy independent GCE mophology provides an indication that -- for a cosmic-ray interpretation of the GCE to succeed -- CR transport near the Galactic center must be dominated by energy {\em independent} mechanisms such as advection, rather than energy {\em dependent} diffusion.

For $f_{\rm H2}=0.1$, the GCE flux below 1 GeV is near zero, and the instrumental PSF is large.  This results in only weak low-energy constraints on the morphology, with a slight preference toward very flat $\alpha_{\rm NFW}\lesssim 0.6$ profiles elongated out of the disk. Above 1 GeV, the profile remains spherical and steep. These models remain compatible with a dark matter interpretation, but still dramatically underestimate the CMZ injection rate.

For $f_{\rm H2}=0.2$, the CMZ injection rate is somewhat low, and the morphology is already becoming highly energy dependent. Below 1 GeV, the GCE template normalization is {\em negative}.  Thus, the preferred morphology is not reflecting a residual, but rather parts of the GDE model which are over-saturated. In this case, the disk-aligned GCE morphology clearly implies that $\gamma$-ray emission near the disk is too bright in our Canonical model.  Above 1 GeV, where the residual still remains positive and fairly bright, the GCE morphology strongly prefers a flat ($\alpha_{\rm NFW}=0.6$) and highly elliptical ($\epsilon=2$) GCE morphology out of the disk. This trend continues for larger values of $f_{\rm H2}$.  The dual preference for a negative, steep, and disky profile versus a positive, flat, and perpendicular profile is also to be expected if the true GCE is somewhat bipolar. Similar morphological features have recently been noted in Ref.~\cite{2016arXiv160206764Y}, and would be prevelent if a bipolar Galactic center wind is present which would both clear out low-energy electrons from the disk and elongate the ICS model vertically.  In the marginalized column, large, disk oriented \fh is preferred at below 1 GeV, while the peaked emission remains spherical above 1 GeV.

As the morphology of the GCE template is adjusted to the preferred morphologies, the spectrum of the excess must be re-evaluated.  In Figure~\ref{fig:spectrum_after_skew} we show the GCE spectrum for our Canonical background model for a variety of different morphologies. In solid blue we show the typical spherical profile. In dashed-yellow we see that disky GCE profiles become even more negative at low energies. For these `disky' models, this improves the fit substantially by allowing the normalization of ICS to increase $\sim 20\%$ below 1 GeV, but requires an unphysical subtraction of the central disk by the negative GCE template.  As we flatten the profile and elongate the GCE template vertically out of the plane, the low energy GCE flux moves back to zero, with less GCE flux gained above 1 GeV.  Overall, we find that low energies in particular, are extremely sensitive to the GCE template morphology.  This seems to strongly disfavor a dark matter interpretations of the GCE for models with realistic cosmic-ray injection rates.

\subsubsection{Galactic Center Slope and Ellipticity}
\label{sec:slope_ellipticity_GC} 

\begin{figure*}[thb]
    \centering
    \includegraphics[width=.8\textwidth]{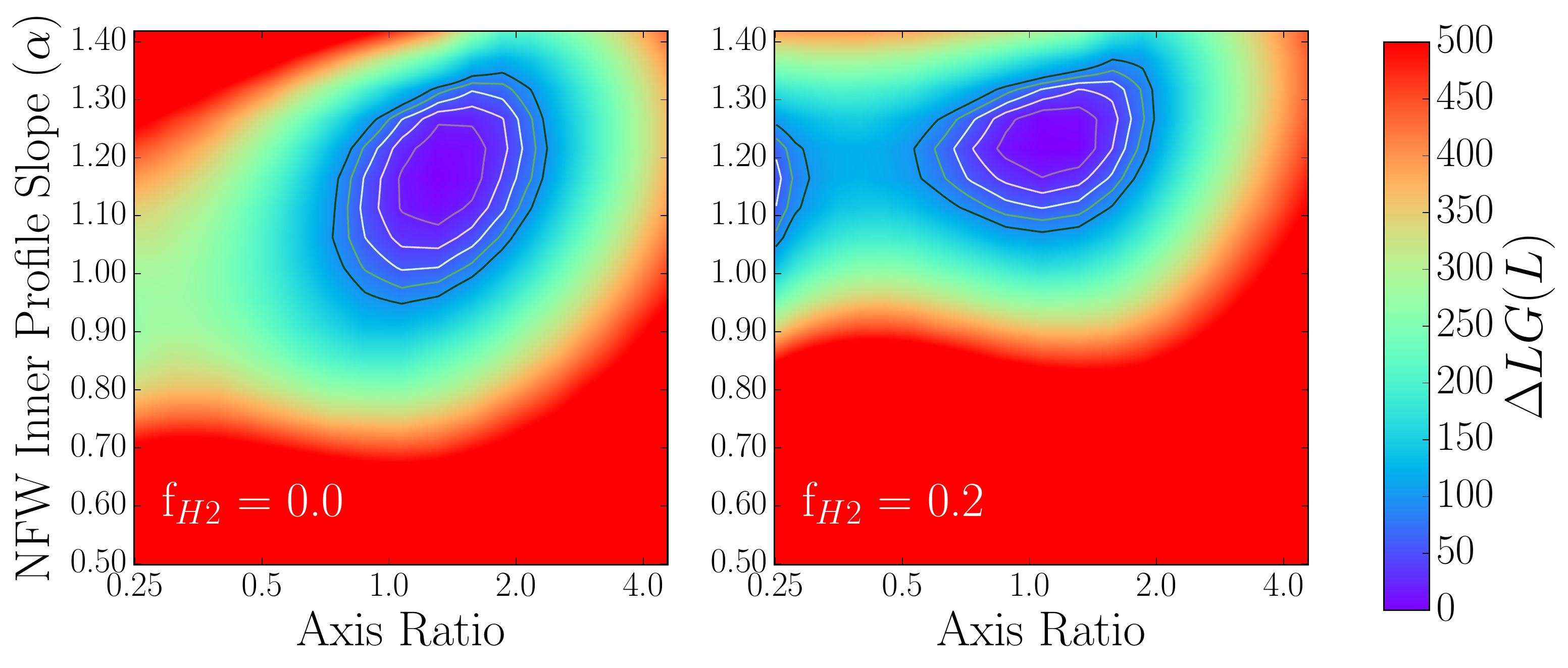}
  \caption{The log-likelihood fit of our model to data in the Galactic center analysis, as a function of the inner slope of the NFW density profile for the GCE component ($\alpha$) and the axis ratio for extension parallel to ($<$1) or perpendicular to ($>$1) the Galactic plane. Contours represent rings of $\Delta$LG($\mathcal{L}$)~=~20. In the case of $f_{\rm H2}$~=~0.0, we find that typical values ($\alpha$~$\sim$~1.0 and an Axis Ratio of approximately unity) are favored. In the case of $f_{\rm H2}$~=~0.2, this still holds, although there is some evidence for an emission component strongly elongated parallel to the Galactic plane.}
 \label{fig:gc_alpha_ellipticity_nomask}
\end{figure*}

We attempt a similar exercise in the Galactic center, distorting both the inner profile slope and the axis ratio of the NFW template which produces the GCE emission component.  We note that these simulations are conducted at a slightly lower angular resolution of 0.1$^\circ$, and we have checked that this only negligibly affects our results. In Figure~\ref{fig:gc_alpha_ellipticity_nomask} we find that ratios near the nominal value for dark matter motivated interpretations of the GCE are preferred, with a best fit $\alpha$~$\sim$~1.15 and an axis ratio of unity. When $f_{\rm H2}$ is increased to 0.2, this best fit value remains robust, although there is also some preference for a new $\gamma$-ray emission component which is strongly stretched along the Galactic plane. These results closely mirror our analysis of the Inner Galaxy ROI, and indicate that the morphology of the GCE very near the Galactic Center is not degenerate with the injection of cosmic-rays tracing the local H$_2$ density in regions very near the Galactic center.

\begin{figure*}[thb]
    \centering
    \includegraphics[width=.8\textwidth]{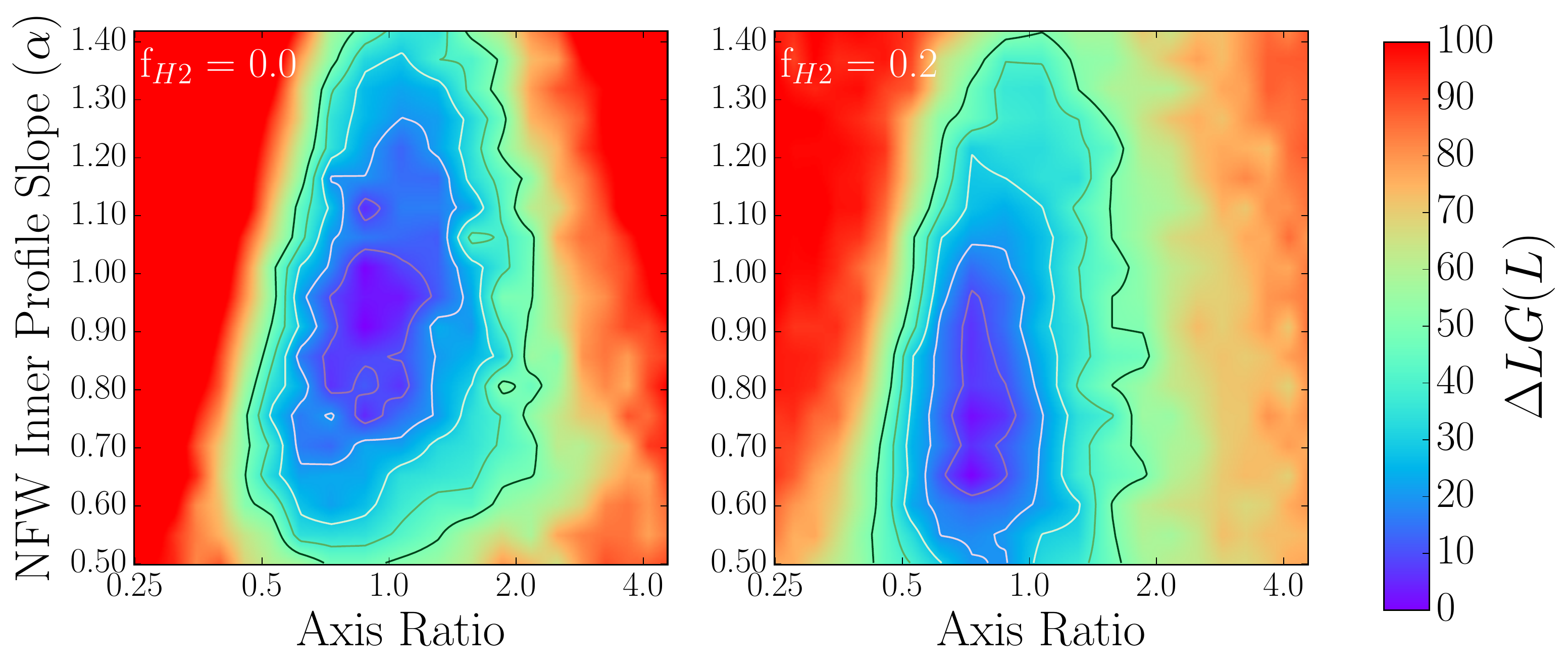}
  \caption{Same as Figure~\ref{fig:gc_alpha_ellipticity_nomask}, for a Galactic center model where the region $|b|<$2$^\circ$ is masked from the analysis. Contours now represent changes of  $\Delta\log(\mathcal{L})=10$. In the case $f_{\rm H2}$~=~0.0, we find that the resulting emission profile is still roughly consistent with dark matter predictions, although profiles that are stretched parallel to the Galactic plane, and which are slightly cored near the Galactic center provide statistically better fits to the data. In the case of $f_{\rm H2}$~=~0.2, the best fit profile becomes cored near the Galactic center, and prefers elongation parallel to the Galactic plane.}
 \label{fig:gc_alpha_ellipticity_mask2deg}
\end{figure*}

As noted in Section~\ref{sec:gc}, the strong preference for spherical symmetry and a $\alpha$~=~1.0 NFW profile in the Galactic center analysis may stem solely from the very high preference for this $\gamma$-ray morphology in the inner degree or two surrounding the GC. Thus, in Figure~\ref{fig:gc_alpha_ellipticity_mask2deg} we repeat the above exercise, but mask out regions with $|b|$~$<$~2$^\circ$. In this case, we find a small (though statistically significant) preference for slightly flatter NFW emission profiles ($\gamma$~=~0.8---1.0), and for some alignment parallel to the Galactic plane (Axis ratio 0.6---1.2). These results become more pronounced as f$_{H2}$ is increased from 0.0 to 0.2. However, we note that the statistical significance of these results has also decreased greatly after masking the Galactic plane, and standard values ($\gamma$~=~1.0, Axis Ratio = 1) are only in tension with the best fit results at the level of $\Delta \chi^2=-2\Delta$LG($L$)~$\sim$~20. While the origin of this new emission morphology is unknown, it may correlate with either the treatment of point sources very close to $|b|$~=~2$^\circ$ (regions which are masked in the Inner Galaxy analysis), or the modeling of the bubbles component, which is highly uncertain near the Galactic plane.

\bigskip

In summary of the residual analysis, we have shown that as \fh is increased, the GCE becomes suppressed in the inner Galaxy. Transposing our analysis along the Galactic plane shows that the GCE template flux which remains is reduced well below the level of nearby Galactic plane residuals, albeit with a distinct peaked spectrum.  This level of reduced residuals occurs uniquely at the Galactic center. Splitting the GCE template into annuli reveals that the radial profile of the GCE strongly flattens as \fh is enhanced, with the inner 5$^\circ$ becoming oversubracted below 1 GeV, and indicating an overabundance of electrons below $\sim$30 GeV in the Canonical model. We then scanned the ellipticity and inner slope of the GCE template, finding that the preferred emission morphology becomes highly energy dependent for increasing \fh but remaining compatible with a dark matter interpretation for $f_{\rm H2}\leq0.1$.  The spectrum and flux of the GCE were then shown to depend sensitively on the chosen GCE template morphology.  For the Galactic center analysis, a bright, spherical, and highly peaked feature remains in the inner $2^\circ$ surrounding the GC, regardless of the value of $f_{\rm H2}$, indicating that the current models do not provide an explanation for these extreme inner regions. When masking the Galactic plane, the inner profile flattens considerably and becomes disk aligned, but remains compatible with a dark matter interpretation at the level of $\Delta \chi^2\approx 50$.

\subsection{Sensitivity to Diffusion Model Parameters}
\label{sec:global_sensitivity}

%\begin{turnpage}
\begin{figure*}
  \centering
 \includegraphics[width=\textwidth]{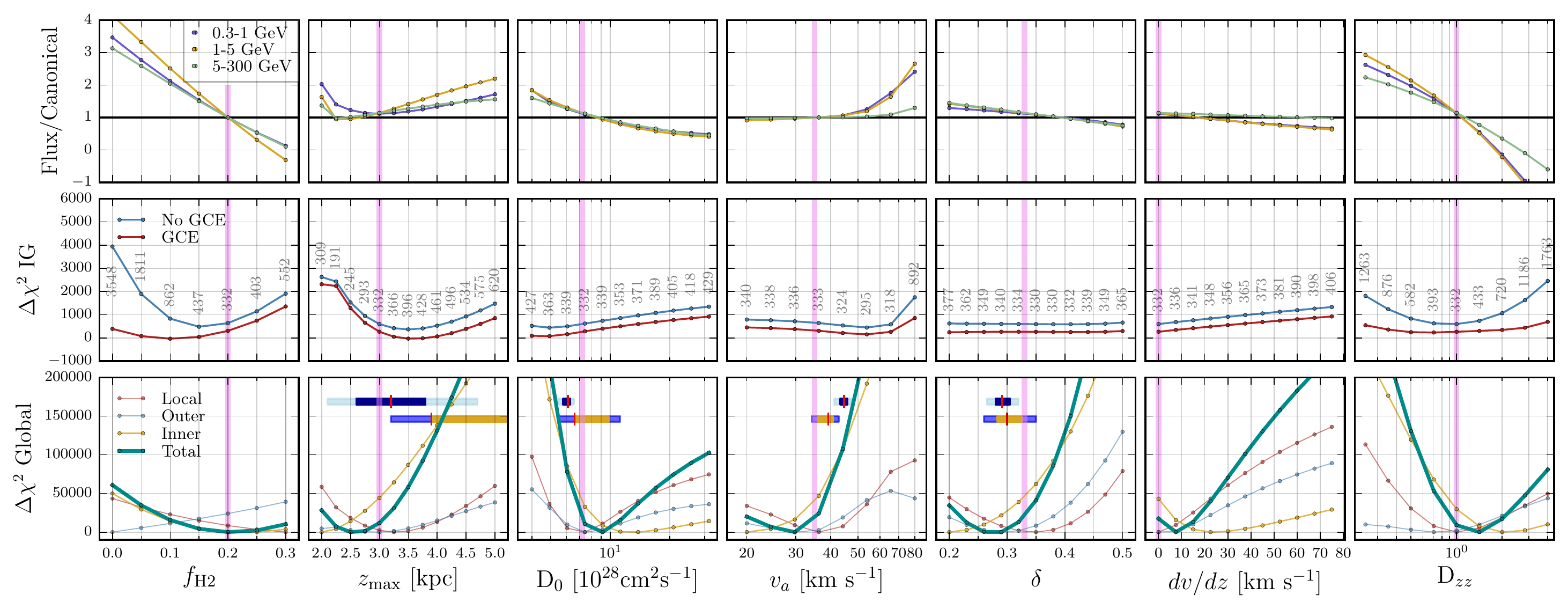}
  \includegraphics[width=\textwidth]{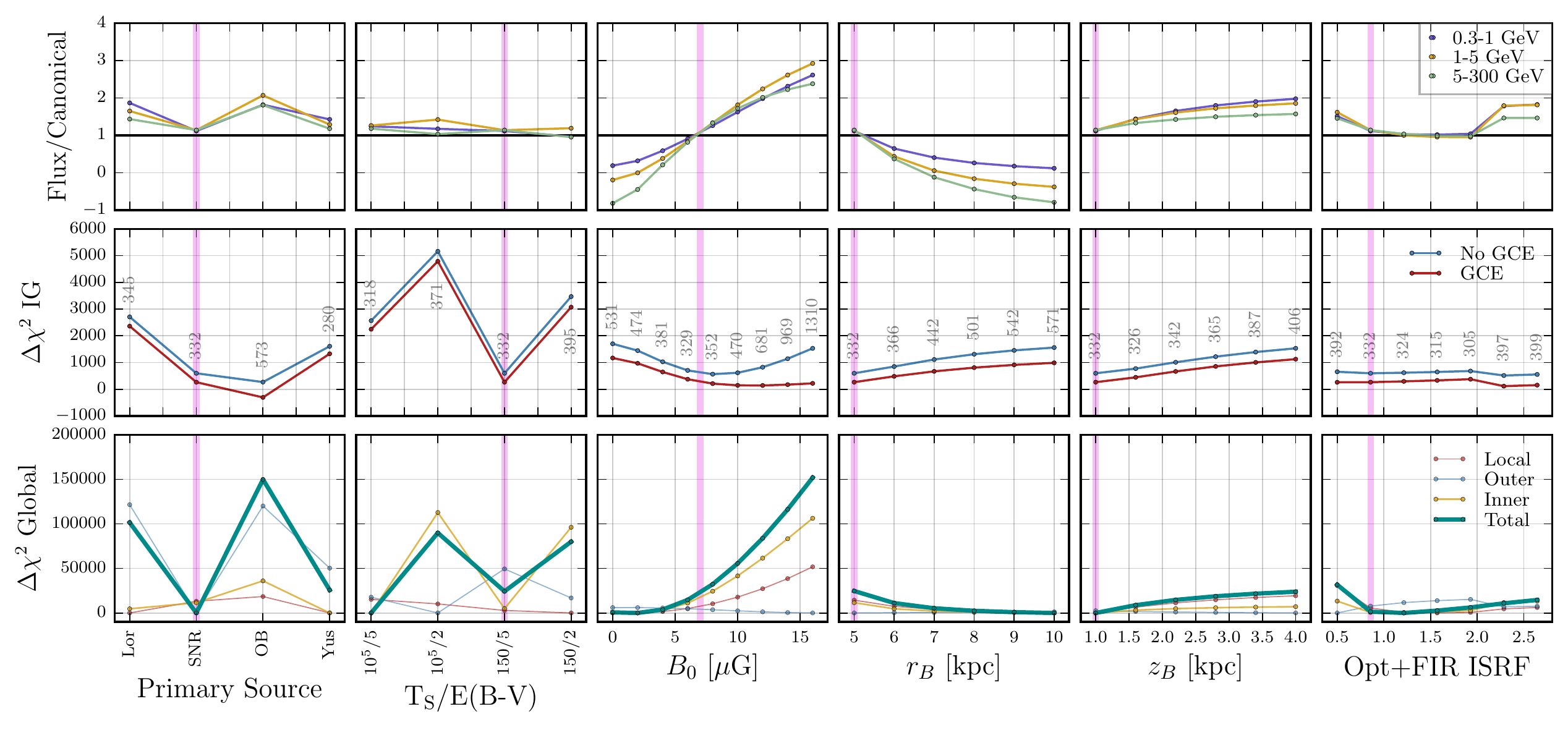}
  \caption{In two sets of panels, we show (top, middle, bottom) the GCE spectral variations, Inner Galaxy $\Delta\chi^2$, and Global $\Delta \chi^2$ as we vary global diffusion parameters around our Canonical $f_{\rm H2}$ model.  The Canonical model parameter choice is shown in each case by a vertical pink line. {\bf Top row}: The flux ratio in low/mid/high energy bands (0.3-1, 1-5, and 5-300 GeV)  of the Galactic center excess spectrum relative to the GCE spectrum obtained using the Canonical model.  Because the low-energy band (purple) is negative in the Canonical model, we reverse the slope and vertically offset the line (i.e. plot 2-Flux/Canonical) ensuring that decreasing values always indicate lower flux.  {\bf Middle Row:} Inner Galaxy $\Delta \chi^2$ with and without a GCE template.  The test-statistic of the additional GCE template is indicated for each model, noting that maximal degeneracy between the GCE and the GDE model occurs when these lines are at closest approach. {\bf Bottom Row:} Total and region-by-region global $\Delta \chi^2$ relative to the minimum over the parameter range. For $z_{\rm max}, D_0, v_a,\ \rm and\ \delta$, we also show 1-dimensional posteriors from two Global Bayesian analysis of measurements of the local cosmic-ray spectra. Blue/gold shaded bands show 68/95\% posterior ranges from Ref.~\cite{Trotta2010}, and dark/light-blue from Ref.~\cite{2015JCAP...09..049J}. Best fit parameters are indicated in each case by red lines.  See also footnote~\ref{fnt:D0}.}
\label{fig:global_grid}
\end{figure*}
%\end{turnpage}

Globally and in the inner Galaxy, the new source distribution represents a genuine quantitative improvement compared with the azimuthally symmetric case, with a $\Delta\chi^2$ comparable to that of changing between the diffusion parameters, gas distributions, or source distributions over the model space of Refs.~\cite{fermi_diffuse,Calore:2015}.  In this Section we show how the inner Galaxy GCE spectrum, inner Galaxy $\Delta \chi^2$, and Global $\Delta\chi^2$ change depending on the parameters of the {\tt Galprop} model.  Galprop's potential parameter space is large, and long computation times prohibit a full multi-dimensional exploration of the models.  Instead, we simply vary our Canonical model along each direction of the parameter space individually.  Our aim is to show (i) that the globally preferred parameter space also maximally reduces the Galactic center excess, (ii) that adding our new source distribution improves the global and inner Galaxy fits by an amount comparable to the changing most of Galprop's major parameters (within reasonable ranges inferred by local cosmic-ray measurements), and (iii) to explore how the GCE spectrum is impacted by changing global diffusion conditions.

Below we group the discussion into related parameters which include the standard diffusion parameters ($D_0$, $z_{\rm max}$, $v_a$, $\delta$, $dv/dz$) as well as source and gas distributions, ISRF variations, and magnetic field properties.  We also show the impact of global anisotropic diffusion perpendicular to the plane. In most cases the results are similar to the findings of Ref.~\cite{Calore:2015}, where the diffusion parameters do not strongly change the GCE spectrum except occasionally at low energies. However, several parameters reshape the {\em new} CMZ cosmic-ray population, and the morphology of the resulting $\gamma$-ray emission depends on the diffusion conditions and energy losses at the Galactic center.  Here we limit our discussion to changing diffusion parameters {\em globally}, emphasizing that these changes effect both the CMZ electron cloud {\em and} the Galactic foreground emission. 

In Figure~\ref{fig:global_grid} we present a large grid of three-panel sets for each parameter.  The top row shows the ratio of the GCE spectrum to the Canonical model in three energy bands (0.3-1, 1-5, and 5-300 GeV), derived by averaging the individual bins (weighted by their inverse variance).  Because the low energy band is negative in the Canonical model, we have plotted this 0.3-1 GeV line as 2-flux/Canonical.  The sign-flip ensures that all decreasing trends correspond to lower (possibly negative) flux, and the offset allows for comparison with the mid/high energy bands so that one can observe the spectral reshaping. However, this also shifts the line of zero flux to +2 for the low-energy band.  In the second row, we show the $\Delta\chi^2$ for the inner Galaxy analysis, with and without a GCE template.  In the bottom row, we show the global $\Delta\chi^2$ for each region relative to the best fit.  For several of the standard diffusion parameters ($z_{\rm max}, D_0, v_a,\ \rm and\ \delta$), we plot 1-dimensional Bayesian posteriors obtained by fitting {\tt Galprop} models against a variety of nuclear cosmic-ray spectra~\cite{Trotta2010}, and more recently, AMS-02 measurements of B/C and protons~\cite{2015JCAP...09..049J}).  

Before iterating through our parameters individually, several notes are in order regarding the global $\Delta \chi^2$ and posteriors. First, the spectrum in our global fits is fixed to the {\tt Galprop} output, and we have not marginalized over injection spectra, implying that our global $\Delta \chi^2$ could be slightly biased by an improved spectral fit rather than morphological fit for some parameters (this is not the case for $f_{\rm H2}$).   Second, the cosmic-ray posteriors shown are based on the cosmic-ray spectrum in the Solar System and may not reflect populations throughout the Galaxy, where e.g. turbulence, magnetic fields, or injection spectra differ from the local ISM. Third, the results here are 1-dimensional, while many cosmic-ray parameters are notoriously degenerate.  This is particularly true when using primary-to-secondary ratios, where $D_0/z_{\rm max}$ are nearly perfectly correlated\footnote{\label{fnt:D0}Ref.~\cite{2015JCAP...09..049J} breaks some of this degeneracy using the proton spectrum observed by AMS-02, loosely constraining $z_{\rm max}$, but tightly constraining $D_0/z_{\rm max}$.  In the $D_0$ panel of Figure~\ref{fig:global_grid} we show the posterior based on our chosen $z_{\rm max}=3$~kpc, though the marginalized posterior would be much wider.}.\\

\noindent {\bf $\mathbf{f_{\rm H2}}$:} For comparison, we recast the $f_{\rm H2}$ results from Section~\ref{sec:SFR_tuning}.  Globally, the improvement going from $f_{\rm H2}=0$ to $f_{\rm H2}=0.2$ is very significant, particularly if one added additional sources to the outer Galaxy.  Relative to the global $\Delta \chi^2$ of other diffusion parameters explored here, the \fh profile is of the same order of improved $\Delta \chi^2$, noting that for most of the parameters, the shown range is much larger than the range allowed by cosmic-ray observations.  In the IG ROI, the impact of increasing \fh on the GCE template significance is matched by no other parameter or obvious combination of parameters.  Similarly, the fit quality without a GCE improves more than any other parameter, other than the Gas distributions (which provide an orthogonal improvement).  Spectrally, increasing \fh decreases the GCE steadily at all energies, and more than any other single parameter.  

\medskip
\noindent {\bf $\mathbf {z_{\rm max}}$ and $\mathbf{D_0}$:}  Using only cosmic-ray primary to secondary ratios, the ratio of these parameters is typically constant. Here however, it appears that the global $\gamma$-ray data strongly breaks this degeneracy (the $\Delta \chi^2$ are not linearly proportional) and is directly sensitive to the distribution of cosmic-rays out of the plane.  The inner and outer Galaxy in particular prefer thin diffusion halos which quickly leak out cosmic-rays.  Given that we have hugely increased the cosmic-ray density in the inner Galaxy, this might be expected, and we observe strong preferences toward efficient transport out of the plane with each of the related parameters ($D_0$, $dv/dz$, and $D_{zz}$).   The local ring includes high latitudes ($|b|>8^\circ$), and is provides a relatively clean gauge of the local diffusion halo size, preferring $z_{\rm max}$ between 2.75 and 4 kpc.  

Importantly, the spectrum of the GCE is strongly reshaped by changes in $z_{\rm max}$, with very thin halos enhancing the low energy GCE spectrum (which was previously oversubtracted) and suppressing the other bands.  Models with $z_{\rm max}=2.25$~kpc reduce the GCE significance to a mere TS=191 (adding 24 d.o.f. to the fit) over the entire inner Galaxy ROI, and reduce the GCE flux far below other Galactic plane residuals at all energies (recall that for the low-energy band, a GCE flux of zero corresponds to flux/canonical=+2 due to our offset).  This is the only cosmic-ray propagation parameter which reshapes the GCE spectrum in this way. Above a few tens of GeV, the strong energy losses at the GC mostly confine the cosmic-ray electrons below 2 kpc and thinning the diffusion halo has little impact.  At lower energies, however, the electrons can reach the boundary of the diffusion halo and escape freely.  This reduces the low-energy electron and proton populations which are otherwise confined, and correspondingly reduces the ICS emission below 1 GeV. (This was over-subtracted in the baseline Canonical model).  It is not clear why the thin halo models are disfavored by the IG ROI by $\Delta \chi^2\approx 2000$, though this is likely related to the foreground profile, noting that the global-local ROI favors thicker halos. 

Enhancing the isotropic diffusion constant $D_0$ simply broadens the width of the ICS profile as electrons at all energies diffuse to larger radii before losing their energy. This has relatively little impact on the statistics of the IG.  The GCE spectrum at all energies is reduced as the ICS spike becomes wider.  At low energies however, the ICS becomes even more oversubtracted, making large $D_0$ disfavoured in the IG.

\medskip
\noindent {\bf \alf~Velocity $\mathbf {v_a}$:} Diffusive reacceleration of cosmic-rays is quadratic in the \alf~velocity (cf. Eq~(\ref{eqn:reacceleration})).  By fractional of their kinetic energy gained, low energy particles are most strongly accelerated, and the particle spectrum is hardened. For our chosen injection spectral index, fairly typical values between 30-50 km/s are preferred globally, in line with cosmic-ray data. The GCE is not strongly effected until very high values $v_a>60$ km/s are reached at which point the low energy electrons produce harder ICS emission and do not regain some of the energy lost to synchrotron and IC cooling.  The dimmer low-energy ICS enhances the GCE spectrum in the 0.3-1 GeV band (good), but also in the 1-5 GeV band (bad), while leaving the high energy ICS unchanged.  Diffusive reacceleration in the context of leptonic burst models for the GCE are discussed further in Ref.~\cite{Cholis:2015}, where very large values were required to preserve the hard electron spectrum far from the GC in the presence of strong energy losses.

\medskip
\noindent{ $\mathbf {\delta}$:} The energy scaling of the diffusion constant is globally quite important, as it shapes the energy dependence of both the cosmic-ray residence time and the diffusive smoothing scale that smears our source distribution. Overall, we find good agreement with the previous cosmic-ray studies, strongly disfavoring $\delta \gtrsim 0.4$.  The inner Galaxy prefers very low values of $\delta$ which enhances CR diffusion at low energies. The GCE significance is basically unaffected, while the GCE spectrum decreases for larger $\delta$ as low energy cosmic rays are more confined and high energy cosmic-rays have enhanced diffusion. Over the small energy range of interest to the GCE, this effect is weak relative to other parameters.

\medskip
\noindent{\bf Convection Gradient $\mathbf{dv/dz}$:} The convection gradient is globally preferred to be zero for the local-global and outer Galaxy, while higher values from 20-40 km/s are preferred toward the inner Galaxy.  Once again, we see a preference for enhanced low-energy cosmic-ray evacuation from the inner Galaxy.  This is not unreasonable given the higher star formation rate of the inner Galaxy which generate the strong Galactic winds.  The GCE statistics and spectrum are not strongly affected.

\medskip
\noindent{\bf Anisotropic Diffusion $\mathbf{D_{zz}}$:}  Here we set the vertical diffusion coefficient equal to $D_{zz}\times D_{0}$, so that diffusion out of the plane is enhanced for $D_{zz}>1$.  In the local and outer Galaxy, the data prefer highly isotropic diffusion ($D_{zz}\approx 1$).  In the global-inner Galaxy, there is a strong preference for $D_{zz}=$2-3, driven by the large population of central cosmic-rays.  Quasi-linear theory predicts that diffusion should be enhanced along ordered magnetic field lines such as the strong poloidal fields near the Galactic center~\cite{Jansson2012}.  In the IG ROI, the statistical significance of the GCE template is minimized for $D_{zz}=1$.  Above and below unity, the electron cloud of the CMZ becomes elliptically skewed with a major axis stretched perpendicular or along the plane.  This causes the relatively spherical GCE template to regain significance. The GCE spectrum is unilaterally suppressed by enhancing $D_{zz}$. Similar to increases in $f_{H2}$, this can eliminate the mid/high energy excess, but the low energy ICS remains much to bright, forcing the GCE to be negative.  Note that our modelling here concerns only globally anisotropic diffusion and we have not studied the impact of anisotropic diffusion at the GC alone.  

\medskip
\noindent{\bf Primary Sources:}  The primary source distribution used for the axisymmetric (1-$f_{\rm H2}$) fraction of sources is globally extremely important for the outer Galaxy, with the SNR model providing the best fit there and overall.  In Figure~\ref{fig:CR_sources}, we showed that of all the distributions, the SNR CB98 contains the most sources outside the solar circle, and better illuminates the outer Galaxy.  The significance of this is very apparent here, with Yusifov models providing the second largest outer Galaxy source population.  Toward the inner and local Galaxy the SNR distribution is only weakly disfavored with the Yusifov and Lorimer pulsar based models providing slightly better fits as they concentrate cosmic-rays toward the inner Galaxy.  Tracers based on OB stars are strongly disfavored in all regions as it does not populate either the inner or outer Galaxy.  For the IG results, we see that models containing no sources in the Galactic center (OB) {\em increases} the GCE significance, while models with the largest number of central sources (Yus) {\em minimizes} the GCE significance.  This underlines Sec.~\ref{sec:sources}, which stated that existing source models systematically underestimate the population of cosmic-rays at the Galactic center.  The spectrum is somewhat sensitive to the source distribution, with the softest GCE occurring for  pulsar models and the hardest GCE arising from our chosen SNR models. 

\medskip
\noindent{\bf Gas Distributions:}  We show all combinations of the assumed hydrogen spin temperature $\rm T_s\in [100, 10^5]$~K and the reddening cut for dust corrections E(B-V)$\in [2,5]$ mag. These distributions are clearly  globally important with a strong preference given to models with maximum reddening E(B-V)=5 mag, which allows more gas to be assigned to regions of high-extinction -- i.e. the inner Galaxy. The outer Galaxy also prefers large hydrogen spin temperatures, which populate HI's triplet state and assigns a higher HI number density to a given 21 cm line temperature.  While the total fit quality of the IG ROI is highly sensitive to the the gas distribution, the GCE significance and spectrum are essentially unaffected.

\medskip
\noindent{\bf Magnetic Fields:}  Our magnetic field model is simply an exponential with scale radius $r_B$, height $z_B$, and random field intensity $B_0$. In the global-inner and local Galaxy, low field intensities are strongly preferred, perhaps due to the reduced confinement of cosmic-ray electrons to the plane where the HI correlated $\pi^0$ emission is already very bright.  In the outer Galaxy, the situation is reversed, corroborating this interpretation given that our models are under-luminous in $\gamma$-rays across the outer Galaxy.  The global data is rather agnostic to the magnetic field shape only mildly preferring large scale radii and small scale heights.  Such small values of $z_b$ are not totally inconsistent with more modern models~\cite{Jansson2012}, which contain a thin disk plus a toroidal halo with thin/thick scale heights of 0.4/4-6 kpc and radii of 10-15 kpc. $B_0$ varies in these models between 1-5$\mu G$ in these models.  The GCE spectrum is very sensitive to the magnetic field shape and intensity which dictate the shortest energy loss time-scale. The scaling radius is much larger than the CMZ so that $r_B$ has little impact on GC electrons.  However, $r_B$ does control the energy loss timescale of the foreground disk electrons which are important near the plane. The scale height $z_B$ will impact {\em both} electrons at the GC and in the foreground. Larger values of any of these three both parameters shrink the the effective foreground ICS scale height.  At the GC, $B_0$ and $z_B$ shape the central CRe population.  Larger values of either lead to stronger confinement toward the GC, and the GCE intensity becomes larger.  We also note that the strong CMZ magnetic fields~\cite{Crocker:2010xc} are not present in our models here, though they are likely to play an important role very near the GC. 

\medskip
\noindent{\bf ISRF Intensity:}  Here we globally vary the strength of the optical and FIR components of the interstellar radiation field relative to the CMB, with a value 1 corresponding to the default {\tt Galprop} model.  Globally, the {\tt Galprop} model appears to fit well.  In the inner Galaxy, results are not significantly changed until very high values begin to further confine the CMZ electrons.  The foregrounds are also effected, and like the magnetic fields, it is difficult to disentangle the two effects using simple template regression here.

\subsection{Injection Spectrum and Enhanced Transport in the CMZ}
\label{sec:winds}
Although the GCE significance is reduced from TS=3800 to 333 by the Canonical model, we have seen that the ICS emission below 1 GeV is too bright and too broadly distributed, such that it must be compensated at high latitudes by increased  isotropic template emission.  This results from an over-abundance of $E\lesssim$30 GeV electrons. {\em If our goal is to produce a self-consistent cosmic-ray model compatible with the GCE then we must reduce the central population of low-energy electrons via either (i) a hardened electron injection spectrum for the CMZ, or (ii) enhanced transport at low energies.}  For protons, saturated $\pi^0$ emission toward the inner Galaxy forces \xco to be unphysically low when realistic CMZ injection rates are imposed, and additional elements are needed to remove low energy protons from the otherwise calorimetric environment (See App.~\ref{sec:X_CO}).

The {\em global} electron spectrum and abundance are not critical to the GCE spectrum due to the freely floating ICS in each energy bin.  However, the ICS template morphology toward the GC does depend on the ratio of cosmic-ray electrons in the CMZ versus the Galactic disk.  A hardening of the CMZ injection spectrum should dim the low-energy ICS spike.  Such an injection spectrum is motivated by the hadronic $\gamma$-ray spectrum of the GC ridge as measured by HESS~\cite{2006Natur.439..695A} and VERITAS~\cite{VERITAS_ridge}, which have TeV photon spectral indices $\Gamma=$2.05-2.29 that are substantially harder than the typical Galactic $\gamma$-rays.  

We test these models by hardening the injection spectrum of both protons and electrons (above $\mathcal{R}>11.5$ and 2 GV respectively) at the CMZ $(r_{\rm 2D}< 300$ pc) by $\delta \alpha$.  Even for $0<\delta \alpha<1$, the resulting GCE spectrum and significance are negligibly effected.  Apparently, the central photon morphology drives the GCE and ICS template normalizations much more than the disk emission, such that the entire ICS template is renormalized.  This results in little change in the GCE properties from spectral hardening of the CMZ alone. These combined factors leave one obvious option for reconciling the Galactic center excess with a steady state cosmic-ray population: enhanced transport at the GC. 

Enhanced cosmic-ray transport comes in several flavors, but those explored in Sec.~\ref{sec:global_sensitivity} each come with caveats when attempting to explain the full GCE spectrum. While a very thin diffusion halo in the Galactic center can help reduce the unphysically-negative low-energy GCE, the overall inner Galaxy fit becomes significantly worse.  Alternatively one can enhance diffusion out of the plane ($D_{zz}>1$) or add a large vertical convection gradient $dv/dz$.  However, these stretch the CMZ electron cloud vertically and the corresponding ICS profile becomes less spherical, resulting in a larger GCE significance.  One can also increase the isotropic diffusion constant $D_0$, but this simply broadens the width of the ICS spike, decreasing the mid/high energy GCE, but further exacerbating the low-energy problem. Decreasing $\delta$ reduces the energy dependence of diffusion, but does not have much effect over the small energy range of interest here.  Altering the ISRF energy densities only impact high-energy electrons by reshaping their effective diffusion radius while changing the magnetic field model cannot correctly reshape the spectrum.  Large \alf~velocities strongly increase diffusive reacceleration and harden the low energy particle spectrum. And although this this produces less low energy ICS without effecting the high energy band, it cannot sufficiently suppress the GCE peak between 1-5 GeV.

No obvious combination of the above seems to solve our problem, leaving one potential cosmic-ray transport solution: high velocity winds emanating from the Galactic center region.  In order to compete with diffusion, the wind velocity must be at least several times 100 km/s ($10^3$ km/s $\equiv 1$~kpc/Myr).  In the advection dominated regime, particle transport is energy-independent. With an appreciable mixture of diffusion, the winds dominate low energy transport where the diffusion rate is lowest, and diffusive transport dominates at high energies, where large particle rigidities lead to faster propagation.  In addition to the transport rate enhancement and energy independence, the radial profile of a constant velocity advective wind is geometrically fixed to $r^{-2}$, whereas diffusion from a point-like stationary injection source results in a shallower $\sim$1/r profile over the inner kiloparsec.  These features make strong winds a natural solution to our problem.

Winds at the Galactic center are driven by intense star formation occurring throughout the CMZ, and especially from the dense stellar clusters of the inner 10 pc. With diffusion alone, the GC is highly calorimetric.  On the other hand, multiwavelength observations indicate less that more than 95\% of the non-thermal injected power must be advected from the system, despite the extreme gas densities and high magnetic fields~\cite{Crocker2010}. A detailed account of GC winds can be found in Ref.~\cite{Crocker2010} and references therein, but is briefly reviewed here.  Perhaps most significant are observations of the `GC lobe', a rising $1^\circ$ tall and $\lesssim 0.5^\circ$ radial shell of 10 GHz radio continuum emission~\cite{1984Natur.310..568S} with associated mid-infrared filaments~\cite{2003ApJ...582..246B}, X-ray shells~\cite{2000ApJ...540..224S}, and optical and radio recombination lines which point to nested shells of ionized gas, synchrotron emission, and dust entrained in the outflow whose pressure and energetics are consistent with star formation or nuclear activity from the central 10 pc of the Galaxy~\cite{1992ApJ...397L..39C,2003ApJ...582..246B,2011MNRAS.411L..11C}.  More recent radio observations combined with multiwavelength modeling~\cite{2016ApJ...817..171Z} have confirmed these features, finding additional X-ray counterparts and associations with the circum-nuclear disk.  In addition, from the perspective of extragalactic star-forming galaxies the SFR within the CMZ is expected to drive powerful outflows~\cite{Crocker2010}. 

We consider here the addition of such a wind, modelled as a purely radial outflow with constant velocity $v_{\rm wind}$ within $r_{\rm 3D}\lesssim 2$ kpc of the Galactic center, which is assumed to stall and vanish beyond this.  Explicitly, we describe the wind in terms of a Fermi-Dirac distribution with a boundary width of 200 pc. A stall zone at 2 kpc is likely conservatively small based on recent modeling~\cite{2009A&A...501..411R}, and, in the vertical direction, lies outside of our inner Galaxy ROI.  

\begin{eqnarray}
\vec{V}_{\rm wind}(r) =\frac{v_{\rm wind} \hat{r}_{\rm 3D}}{e^{(r_{\rm 3D}-2\rm~kpc)/0.2~kpc}+1}
\end{eqnarray}
where we vary the value $v_{\rm wind}$ between $0-2000~\rm km~ s^{-1}$.

\begin{figure*}[thb]
  \centering
  \includegraphics[width=\textwidth]{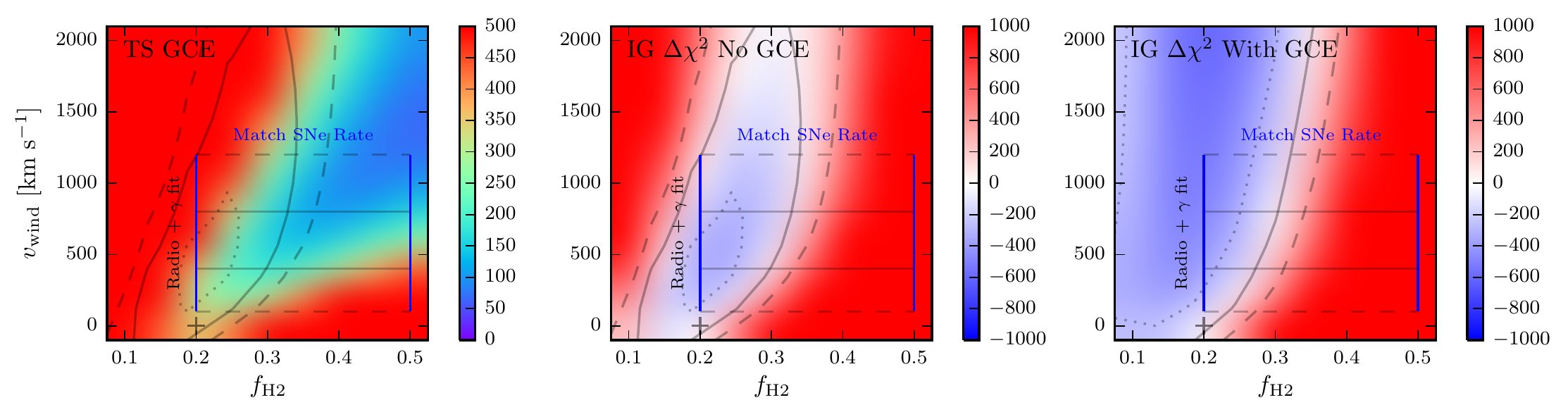}
\caption{Statistics for the Inner Galaxy when varying \fh and $v_{\rm wind}$.  {\bf Left:} Test Statistic of the GCE template, with contours showing $\Delta\chi^2\ \in \{-300,0,300\}$ (dot, solid, dashed) for the `No GCE' Inner Galaxy fits. {\bf Center:} $\Delta \chi^2$ for IG fits with No GCE template and the same contour levels, relative to the Canonical model.  {\bf Right:} Same as center, but for fits including a GCE template.  In all panels, the box is comprised of `bounds' and `most probable' wind velocities from Ref.~\cite{Crocker2010} on the vertical axis, and models which match the CMZ injection rates (See Sec.~\ref{sec:sources}) on the horizontal.}
 \label{fig:vary_winds}
\end{figure*}

Although the wind is expected to be bipolar we do not explicitly model an opening angle here, leaving such studies to future work. In modeling the GC wind, we also increase the simulation's planar resolution to $dx=dy=250$ pc and set $dz=100$ pc.

In Figure~\ref{fig:vary_winds}, we show the statistics of the IG fit at the Galactic center as we simultaneously vary $f_{\rm H2}$ and $v_{\rm wind}$.  Specifically we show the TS of the GCE template (left panel), the $\Delta \chi^2$ for fits with no GCE template, and the $\Delta \chi^2$ for fits with a GCE template.  The inset box shows the range of wind velocities compatible with radio observations of the GC Lobe and TeV $\gamma$-ray observations of the HESS region~\cite{Crocker2010}. Alternative modelling assumptions~\cite{Yoast-Hull2014} suggest potentially higher wind speeds depending on the magnetic field strength of the CMZ.  Horizontal bands highlight values of $f_{\rm H2}$ which approximately match the observed SNe rates at the Galactic center assuming typical SNR cosmic-ray acceleration efficiencies (See Sec.~\ref{sec:sources}).

In the left panel, we see clearly that a simultaneous increase of the injection rate and wind speed leads to strong reduction of the GCE significance, with reasonable parameters (from multi-wavelength data) reducing the entire GCE template to TS$\approx$100. On the other hand, models which reduce the GCE significance the most are not strongly preferred by the inner Galaxy fits, and the GCE significance can only be reduced to TS$\approx250$ while also providing an improved IG fit.  As the wind velocity increases, the low-energy electron cloud becomes both dimmer in ICS and more sharply peaked, with electrons above 50 GeV remaining unaffected by the wind.  This sharply reduces the negative low-energy residuals picked up by the GCE template in windless models. 

Examining the center panel we see that the addition of a GC wind not only reduces the GCE significance, but can also improves the inner Galaxy fit by up to $\Delta \chi^2=-375$ over the Canonical zero-wind model.  Although these are not fully overlapping parameter spaces, a mutual reduction of the GCE and improved fit in the inner Galaxy is achieved over a substantial portion of the parameter space.  Importantly, the preferred parameter space overlaps well with multi-wavelength expectations and statistically excludes a purely diffusive transport at $\Delta \chi^2\approx350$ for only two additional parameters\footnote{It is important to note that this is a purely statistical significance and that systematic uncertainties are much larger, in particular due to the magnetic-field and ISRF uncertainties from the CMZ, and the transport parameters discussed in Sec.~\ref{sec:global_sensitivity}.}.  The right panel shows that even with the addition of the GCE template, a non-zero wind velocity is still favored, and is highly degenerate with the GCE template up to 2000 km/s provided that $f_{\rm H2}\lesssim 0.3$. As we show in Appendix~\ref{sec:X_CO}, these models also provide more reasonable values for the inner Galaxy \xco conversion factor as they remove low energy protons which otherwise saturate the gas-correlated $\gamma$-ray emission.

\begin{figure}[tb]
  \centering
  \includegraphics[width=.45\textwidth]{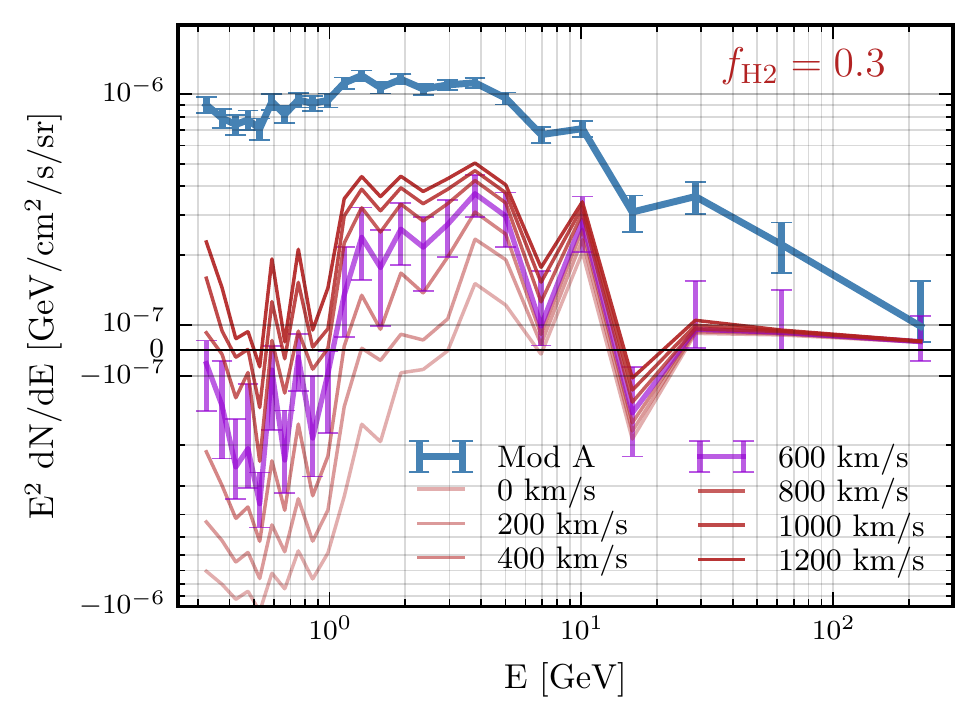}
\caption{Spectrum of the Galactic center excess as the wind velocity $v_{\rm wind}$ is varied for $f_{\rm H2}=0.3$ which well reproduces the observed SNe rate of the CMZ. Best fitting models without a GCE prefer wind velocities from 500-1000 km/s for $f_{\rm H2}=0.3$.  In all cases the GCE is reduced very far below the GCE spectrum of Mod A.}
 \label{fig:vary_winds_spectrum}
\end{figure}

In Figure~\ref{fig:vary_winds_spectrum} we show the GCE spectrum as the wind velocity is increased for models which correctly reproduce the CMZ supernovae rate ($f_{\rm H2}=0.3$).  The addition of winds strongly reduces the low-energy electron and proton populations at the GC which in turn (i) more sharply peaks the associated ICS emission, and (ii) reduces $\pi^0$ emission associated with the thick atomic and ionized hydrogen disks (which cannot be compensated by adjusting $X_{\rm CO}$). This strongly reduces the negative normalization of the GCE template while only weakly impacting emission above 5 GeV. Thus the spectrum and morphology of cosmic-rays at the GC are reshaped in precisely the way needed to further reduce the Galactic center excess. {\tt Galprop} allows us to directly examining the electron spectrum at the Galactic center and reveals that the wind induces a broad break in the steady-state spectrum between 10-50 GeV (depending on the wind velocity).  Assuming a CMZ magnetic field strength $B\gtrsim 100\ \mu G$, this break energy is nearly equivalent to that  inferred from the 1 GHz spectral break in the Galactic ridge synchrotron spectrum~\cite{Crocker2010}.

\begin{figure*}[thb]
    \centering
    \includegraphics[width=.8\textwidth]{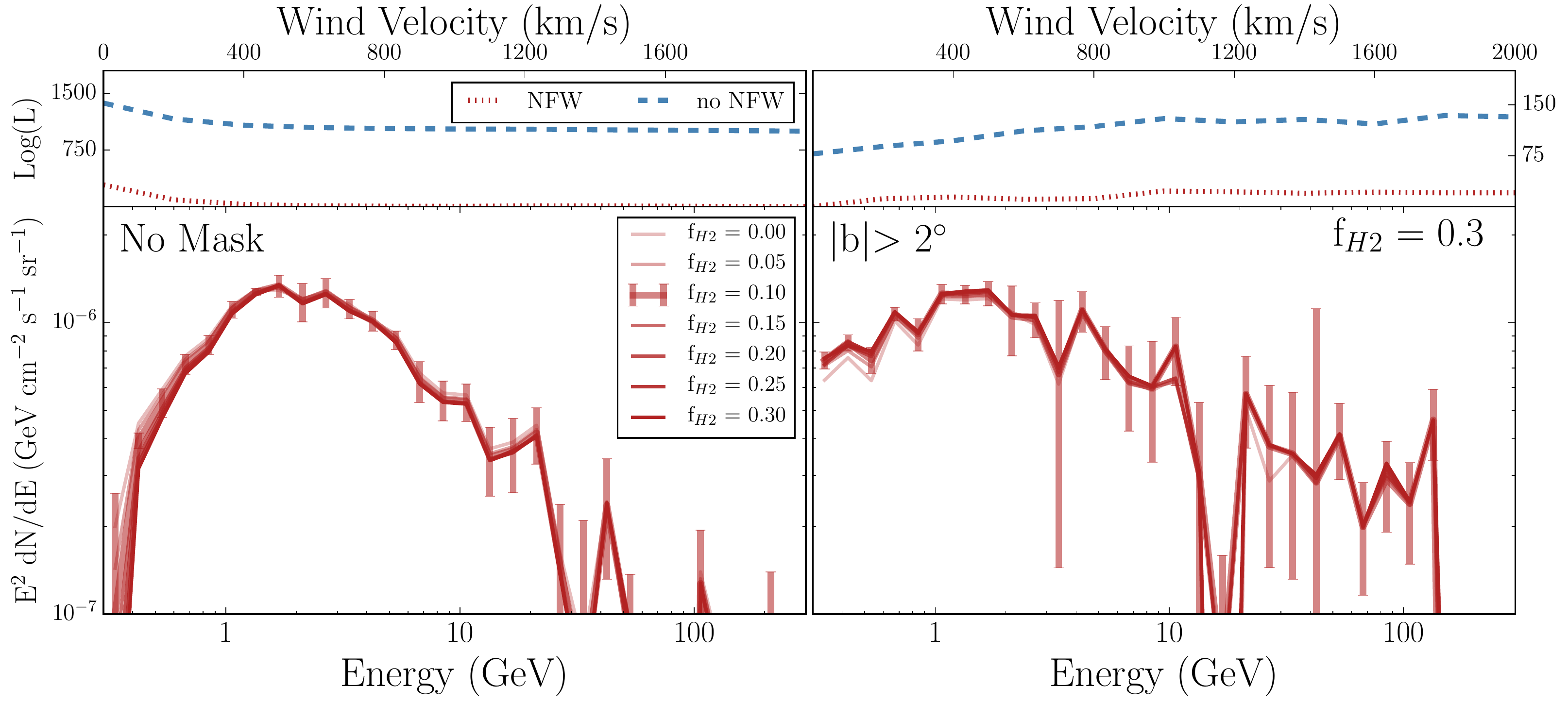}
  \caption{Same as Figure~\ref{fig:vary_winds_spectrum}, for a Galactic center analysis in cases where the full 15$^\circ\times$15$^\circ$ ROI is analyzed, as well as an analysis where the region $|b|<2^\circ$ is masked. Note the different scales for the log-likelihood fits to the data in each analysis. In all cases, the excess remains relatively bright, similar to the default results shown in Figures~\ref{fig:gc_fits_no_mask} and ~\ref{fig:gc_fits_mask}.}
 \label{fig:gc_winds_spectrum}
\end{figure*}

In Figure~\ref{fig:gc_winds_spectrum} we show results for the same diffuse emission models applied in our Galactic center analysis, utilizing both the full 15$^\circ$$\times$15$^\circ$ ROI as well as an analysis which masks the region $|b|$$<$2$^\circ$ from the Galactic plane. In the analysis of the full-sky ROI, we find that, similar to the analysis of the IG region, models with non-zero wind velocities are statistically preferred in the data. Specifically, we find the data to be best fit by a wind velocity of 2000 km/s both in models that include, or ignore the GCE component. However, we note that wind velocities of 600 km/s are disfavored by only a $\Delta$LG(L) of 10 (48) in the case that the GCE component is included (not included) in the model. A model with no winds, however, is disfavored by a $\Delta$LG(L) of 288 (372). In an analysis that masks the region $|b|<$2$^\circ$, models with no winds are slightly preferred by the data, but at a level of only $\Delta$LG(L) of 11 (33) compared to a model with 600 km/s winds.

Unlike for models with non-zero winds in the Inner Galaxy analysis, the spectrum and intensity of the GCE component appears to be unaffected by the presence of strong GC winds, similar to our results for models with varying values of \fh in the Galactic Center ROI. The effect of the addition of the GCE template in the Galactic Center ROI is to significantly suppress the normalization of the ICS emission template. 

%\medskip
The results above show that for any realistic value of $f_{\rm H2}$, {\em Fermi GeV data toward the inner Galaxy strongly favor the presence of a Galactic center wind based on the morphology of the low energy ICS emission}. Furthermore, the presence of this wind increases the degeneracy between the peaked central ICS emission and templates for a GCE component. In the Galactic Center ROI we also see some evidence for Galactic center winds in the overall fit to the $\gamma$-ray data, but do not find any degeneracy between the strength of the GC winds and the normalization and spectrum of the GCE component. In Appendix~\ref{sec:X_CO} we provide additional evidence for Galactic center winds based on the over saturation of $E\sim1$~GeV $\pi^0$ emission in the central 2 kpc of the Galaxy.

\section{Discussion}
\label{sec:discussion}

In this \emph{paper} we have presented novel, physically-motivated, and significantly improved models for Galactic diffuse $\gamma$-ray emission in the Milky Way, focusing on understanding the diffuse sources of $\sim$GeV $\gamma$-ray emission in the complex Galactic center region.  In contradiction with multi-wavelength observations, previous models of the Galaxy's diffuse Galactic emission have neglected cosmic-rays from the Central Molecular Zone, which is known to harbor a significant fraction of the Milky Way's supernova power (and thus cosmic-ray injection).  We have rigorously examined the robustness, spectrum, and morphology of the GeV Galactic Center excess in the presence of this new CMZ associated $\gamma$-ray emission. Our primary results can be condensed into three key findings:

\begin{enumerate}
\item There exists a clear degeneracy between the intensity, spectrum, and morphology of the GCE and the ICS emission following from realistic cosmic-ray injection near the Galactic center. 

\item Models with \fh=~0.2---0.25 both reproduce the correct CMZ injection rate and provide the best fit to the $\gamma$-ray data in regions far from the Galactic center (i.e. the full sky, and the inner Galaxy fits with no GCE template). When these models are employed in the Inner Galaxy analysis, they substantially decrease the intensity and significance of the Galactic center excess, and also distort its morphology and spectrum. When fitting these models in the Galactic Center ROI, the spectrum and intensity of the GCE remain high, but the statistical significance of the excess decreases drastically if the region $|b|<$2$^\circ$ is masked.

\item When including a GCE template, fits in the Inner Galaxy and Galactic center analyses statistically prefer $f_{\rm H2}\approx 0.1-0.15$. For these models, the intensity of the GCE component generally decreases by $\sim$30\%, while the morphology and spectrum remain consistent with previous results -- i.e. compatible with a dark matter interpretation. 
\end{enumerate}

These results can thus be interpreted in two ways. If global $\gamma$-ray fits and the multi-wavelength evidence for large cosmic-ray injection in the CMZ are accepted as priors on the value of $f_{\rm H2}$, then the low-intensity, hard spectrum residual that remains may be potentially viewed as a systematic issue of uncertain origin. On the other hand, if these priors are taken to be weak against the statistical preference for lower \fh in the narrower Inner Galaxy and Galactic center ROIs, then the resilience of the GCE in light of our new models reaffirms previous determinations of the GCE properties.  Large and presently unknown systematic uncertainties from both analysis procedures and from GDE modeling make it difficult to objectively balance the weights of the high $f_{\rm H2}$ priors against the low $f_{\rm H2}$ IG+GC likelihood function. Ultimately, the properties of the GCE depend sensitively on these assumptions and its remains an open question.

In addition to the results above, we find that GDE models with appreciable CMZ injection rates predict overly bright $\gamma$-ray emission below $\sim$1 GeV near the plane.  This strongly impacts the spectrum of the low-energy GCE, and the inferred \xco conversion factor toward the inner Galaxy. We discuss multiple possible systematics that could produce this signal, and show that the majority of (global) diffusion parameters can not alleviate these issues. 

Motivated by substantial multi-wavelength evidence, we implement a radially outflowing wind at the Galactic center.  For wind speeds $\gtrsim 500$ km/s, our models are improved simultaneously in three separate ways:  (i) low-energy cosmic rays are advected from the region so that the low-energy GCE spectrum is no longer compensating an oversaturated GDE model, allowing for more realistic values of \xco in the inner Galaxy; (ii) Inner Galaxy and Galactic Center fits (both with and without a GCE template) are substantially improved; (iii)  the best fit value of \fh in the Inner Galaxy ROI is increased, bringing the preferred model space into better agreement with the full-sky $\gamma$-ray data and with the observed CMZ supernovae rate.  Furthermore, when both \fh and the wind velocity are large, the GCE can be even further suppressed.  However, we again find that the models which best fit the $\gamma$-ray data in both the Galactic Center and Inner Galaxy ROIs include a significant GCE component with a spectrum and intensity consistent with previous analyses.

There remain multiple avenues worth of future exploration, and additional modeling improvements are necessary to fully understand diffuse $\gamma$-ray emission from regions near the Galactic center.  At present, our models employ ISRFs and magnetic field energy densities that are not consistent with multiwavelength observations of the CMZ. Specifically, radio surveys find that 50---200$\mu$G magnetic fields permeate the CMZ~\citep{Crocker:2010xc, Crocker:2011}, while stellar populations spanning the inner few degrees~\cite{0004-637X-702-1-178,2015PASJ...67..123Y} and the dense nuclear star clusters surrounding Sgr A* indicate ISRF energy densities in the inner 10~pc which are several orders of magnitude higher than current {\tt Galprop} models~\citep{2010RvMP...82.3121G, 2013ApJ...764..154D, 2013ApJ...764..155L}. A realistic model of diffuse $\gamma$-ray emission will need to investigate both: 

(i) the short electron cooling timescales indicated by these observations, as well as 

(ii) utilize a detailed model of the stellar (and dust-reprocessed FIR) ISRF in order to determine the precise morphology of ICS $\gamma$-rays in the inner few degrees surrounding the GC.  

Finally, gas maps for both $\gamma$-ray generation, and for our our H$_2$ based source model, may be improved using more recent molecular line surveys of the CMZ, such as MOPRA~\cite{2012MNRAS.426.1972P}, which samples multiple organic species in order to better probe the variety of gas phases that are not well traced by CO alone.  Alongside the improved $\gamma$-ray predictions (which also include the TeV regime), more precise models can be better constrained by cosmic-ray calorimetry and synchrotron emission modeling.

\section{Conclusions}
\label{sec:conclusions}
The conclusive determination of the existence of genuinely {\em excess} diffuse emission from the inner regions of the Galaxy depends crucially on the use of {\em reliable} and {\em physically motivated} models for the Galactic diffuse emission. It is clear that in studies thus far available a critically important population of cosmic rays in the region has been neglected, with potentially dramatic implications for the determination of both the existence and the properties of any excess emission. 

In a recent short note, Ref.~\cite{paper_one}, we proposed a physically well-motivated model for the three-dimensional Galactic cosmic-ray injection source distribution, based on a fraction \fh of such sources being associated with relatively recent injection in star-forming regions. We employed observational data on the density of H$_2$ alongside a simple prescription for star-formation efficiency, and showed that the resulting Galactic diffuse emission models with \fh$\sim0.1-0.3$ are {\em statistically strongly favored} by the $\gamma$-ray data over models with \fh$=0$.

The present extensive study explores in detail the implications of our original finding \cite{paper_one} on models for the diffuse Galactic emission, and the resulting impact on the existence and properties of any Galactic center ``excess''. We found that the new cosmic ray population is consistent with the CMZ supernova rate (as determined by multi-wavelength observations) and produces a significant $\gamma$-ray emission that is degenerate with many properties of the $\gamma$-ray excess. The choice of the region of interest impacts the details of this degeneracy. Specifically, without employing a GC excess template there is a strong preference for \fh$\sim0.1$ in the (narrower) Galactic Center analysis, and for \fh$\sim0.2$ in the (broader) Inner Galaxy and in the full-sky analyses. Including a GC excess template, but utilizing a value \fh$\sim0.2$ as a {\em prior} in the Inner Galaxy analysis, the GC excess is strongly affected, while the Galactic Center analysis still favors the existence of a bright residual emission. {\em In all cases, we find a clear statistical preference for the existence of a Galactic Center Excess residual emission template, but with an intensity, spectrum, and morphology which are sensitive to $f_{\rm H2}$}. 

An additional important finding of the present study is that low-energy ($E\lesssim$20 GeV) protons and electrons are over-produced in the GC region, as determined by the negative $\gamma$-ray residuals below 1 GeV. We explored possible solutions to this issue, and argue that the most plausible solution is the addition of significant advective winds which remove the excess electron and proton populations from the central Galactic regions. We showed that the existence of such winds is physically well-motivated. The inclusion of strong Galactocentric winds significantly improves our fit to the $\gamma$-ray data both in models that include (exclude), a GCE component, at the level of $\Delta \chi^2$=605 (375) in an Inner Galaxy type analysis. For models which include a GCE component, the addition of strong Galactic winds removes the negative residuals typically absorbed by the GCE template, producing a GCE $\gamma$-ray spectrum which is physically realistic over the full energy range.

We believe that the models for the diffuse Galactic emission discussed here present a clear step forward in the current state of the art in this field. Additionally, it is clear that a solid determination of the properties of a Galactic center excess, and its association with new physical phenomena such as dark matter annihilation, hinges on further improving our modeling of the Central Molecular Zone. At present, a precision determination of diffuse emission in the Galaxy remains the most significant barrier to using high-energy gamma rays as a probe of new physics at the Galactic center.

\acknowledgments
We thank Christoph Weniger, Mark Krumholz, and Andy Strong for their very fruitful discussions as well as Gudlaugur J\'{o}hannesson, Hiroyuki Nakanishi, and Martin Pohl for their discussions and access to various datasets used in this analysis. The simulations for this research were carried out on the UCSC supercomputer Hyades, which is supported by National Science Foundation (award number AST-1229745) and UCSC. This work also made use of computing resources and support provided by the Research Computing Center at the University of Chicago. EC is supported by a NASA Graduate Research Fellowship under NASA NESSF Grant No. NNX13AO63H. TL is supported by the National Aeronautics and Space Administration through Einstein Postdoctoral Fellowship Award Number PF3-140110. SP is partly supported by the US Department of Energy, Contract DE-SC0010107-001.

\bibliography{biblio}

\begin{thebibliography}{134}%
\makeatletter
\providecommand \@ifxundefined [1]{%
 \@ifx{#1\undefined}
}%
\providecommand \@ifnum [1]{%
 \ifnum #1\expandafter \@firstoftwo
 \else \expandafter \@secondoftwo
 \fi
}%
\providecommand \@ifx [1]{%
 \ifx #1\expandafter \@firstoftwo
 \else \expandafter \@secondoftwo
 \fi
}%
\providecommand \natexlab [1]{#1}%
\providecommand \enquote  [1]{``#1''}%
\providecommand \bibnamefont  [1]{#1}%
\providecommand \bibfnamefont [1]{#1}%
\providecommand \citenamefont [1]{#1}%
\providecommand \href@noop [0]{\@secondoftwo}%
\providecommand \href [0]{\begingroup \@sanitize@url \@href}%
\providecommand \@href[1]{\@@startlink{#1}\@@href}%
\providecommand \@@href[1]{\endgroup#1\@@endlink}%
\providecommand \@sanitize@url [0]{\catcode `\\12\catcode `\$12\catcode
  `\&12\catcode `\#12\catcode `\^12\catcode `\_12\catcode `\%12\relax}%
\providecommand \@@startlink[1]{}%
\providecommand \@@endlink[0]{}%
\providecommand \url  [0]{\begingroup\@sanitize@url \@url }%
\providecommand \@url [1]{\endgroup\@href {#1}{\urlprefix }}%
\providecommand \urlprefix  [0]{URL }%
\providecommand \Eprint [0]{\href }%
\providecommand \doibase [0]{http://dx.doi.org/}%
\providecommand \selectlanguage [0]{\@gobble}%
\providecommand \bibinfo  [0]{\@secondoftwo}%
\providecommand \bibfield  [0]{\@secondoftwo}%
\providecommand \translation [1]{[#1]}%
\providecommand \BibitemOpen [0]{}%
\providecommand \bibitemStop [0]{}%
\providecommand \bibitemNoStop [0]{.\EOS\space}%
\providecommand \EOS [0]{\spacefactor3000\relax}%
\providecommand \BibitemShut  [1]{\csname bibitem#1\endcsname}%
\let\auto@bib@innerbib\@empty
%</preamble>
\bibitem [{\citenamefont {Goodenough}\ and\ \citenamefont
  {Hooper}(2009)}]{Goodenough:2009gk}%
  \BibitemOpen
  \bibfield  {author} {\bibinfo {author} {\bibfnamefont {L.}~\bibnamefont
  {Goodenough}}\ and\ \bibinfo {author} {\bibfnamefont {D.}~\bibnamefont
  {Hooper}},\ }\href@noop {} {\  (\bibinfo {year} {2009})},\ \Eprint
  {http://arxiv.org/abs/0910.2998} {arXiv:0910.2998 [hep-ph]} \BibitemShut
  {NoStop}%
%%CITATION = ARXIV:0910.2998;%%
\bibitem [{\citenamefont {Hooper}\ and\ \citenamefont
  {Goodenough}(2011)}]{Hooper:2010mq}%
  \BibitemOpen
  \bibfield  {author} {\bibinfo {author} {\bibfnamefont {D.}~\bibnamefont
  {Hooper}}\ and\ \bibinfo {author} {\bibfnamefont {L.}~\bibnamefont
  {Goodenough}},\ }\href {\doibase 10.1016/j.physletb.2011.02.029} {\bibfield
  {journal} {\bibinfo  {journal} {Phys.Lett.}\ }\textbf {\bibinfo {volume}
  {B697}},\ \bibinfo {pages} {412} (\bibinfo {year} {2011})},\ \Eprint
  {http://arxiv.org/abs/1010.2752} {arXiv:1010.2752 [hep-ph]} \BibitemShut
  {NoStop}%
%%CITATION = ARXIV:1010.2752;%%
\bibitem [{\citenamefont {Hooper}\ and\ \citenamefont
  {Linden}(2011)}]{Hooper:2011ti}%
  \BibitemOpen
  \bibfield  {author} {\bibinfo {author} {\bibfnamefont {D.}~\bibnamefont
  {Hooper}}\ and\ \bibinfo {author} {\bibfnamefont {T.}~\bibnamefont
  {Linden}},\ }\href {\doibase 10.1103/PhysRevD.84.123005} {\bibfield
  {journal} {\bibinfo  {journal} {Phys.Rev.}\ }\textbf {\bibinfo {volume}
  {D84}},\ \bibinfo {pages} {123005} (\bibinfo {year} {2011})},\ \Eprint
  {http://arxiv.org/abs/1110.0006} {arXiv:1110.0006 [astro-ph.HE]} \BibitemShut
  {NoStop}%
%%CITATION = ARXIV:1110.0006;%%
\bibitem [{\citenamefont {Abazajian}\ and\ \citenamefont
  {Kaplinghat}(2012)}]{Abazajian:2012pn}%
  \BibitemOpen
  \bibfield  {author} {\bibinfo {author} {\bibfnamefont {K.~N.}\ \bibnamefont
  {Abazajian}}\ and\ \bibinfo {author} {\bibfnamefont {M.}~\bibnamefont
  {Kaplinghat}},\ }\href {\doibase 10.1103/PhysRevD.86.083511} {\bibfield
  {journal} {\bibinfo  {journal} {Phys.Rev.}\ }\textbf {\bibinfo {volume}
  {D86}},\ \bibinfo {pages} {083511} (\bibinfo {year} {2012})},\ \Eprint
  {http://arxiv.org/abs/1207.6047} {arXiv:1207.6047 [astro-ph.HE]} \BibitemShut
  {NoStop}%
%%CITATION = ARXIV:1207.6047;%%
\bibitem [{\citenamefont {{Gordon}}\ and\ \citenamefont
  {{Mac{\'{\i}}as}}(2013)}]{gordon_macias:2013}%
  \BibitemOpen
  \bibfield  {author} {\bibinfo {author} {\bibfnamefont {C.}~\bibnamefont
  {{Gordon}}}\ and\ \bibinfo {author} {\bibfnamefont {O.}~\bibnamefont
  {{Mac{\'{\i}}as}}},\ }\href {\doibase 10.1103/PhysRevD.88.083521} {\bibfield
  {journal} {\bibinfo  {journal} {\prd}\ }\textbf {\bibinfo {volume} {88}},\
  \bibinfo {eid} {083521} (\bibinfo {year} {2013})},\ \Eprint
  {http://arxiv.org/abs/1306.5725} {arXiv:1306.5725 [astro-ph.HE]} \BibitemShut
  {NoStop}%
\bibitem [{\citenamefont {Hooper}\ and\ \citenamefont
  {Slatyer}(2013)}]{Hooper:2013rwa}%
  \BibitemOpen
  \bibfield  {author} {\bibinfo {author} {\bibfnamefont {D.}~\bibnamefont
  {Hooper}}\ and\ \bibinfo {author} {\bibfnamefont {T.~R.}\ \bibnamefont
  {Slatyer}},\ }\href {\doibase 10.1016/j.dark.2013.06.003} {\bibfield
  {journal} {\bibinfo  {journal} {Phys. Dark Univ.}\ }\textbf {\bibinfo
  {volume} {2}},\ \bibinfo {pages} {118} (\bibinfo {year} {2013})},\ \Eprint
  {http://arxiv.org/abs/1302.6589} {arXiv:1302.6589 [astro-ph.HE]} \BibitemShut
  {NoStop}%
%%CITATION = ARXIV:1302.6589;%%
\bibitem [{\citenamefont {{Abazajian}}\ \emph {et~al.}(2014)\citenamefont
  {{Abazajian}}, \citenamefont {{Canac}}, \citenamefont {{Horiuchi}},\ and\
  \citenamefont {{Kaplinghat}}}]{Abazajian:2014fta}%
  \BibitemOpen
  \bibfield  {author} {\bibinfo {author} {\bibfnamefont {K.~N.}\ \bibnamefont
  {{Abazajian}}}, \bibinfo {author} {\bibfnamefont {N.}~\bibnamefont
  {{Canac}}}, \bibinfo {author} {\bibfnamefont {S.}~\bibnamefont {{Horiuchi}}},
  \ and\ \bibinfo {author} {\bibfnamefont {M.}~\bibnamefont {{Kaplinghat}}},\
  }\href@noop {} {\bibfield  {journal} {\bibinfo  {journal} {ArXiv e-prints}\ }
  (\bibinfo {year} {2014})},\ \Eprint {http://arxiv.org/abs/1402.4090}
  {arXiv:1402.4090 [astro-ph.HE]} \BibitemShut {NoStop}%
\bibitem [{\citenamefont {Daylan}\ \emph {et~al.}(2014)\citenamefont {Daylan},
  \citenamefont {Finkbeiner}, \citenamefont {Hooper}, \citenamefont {Linden},
  \citenamefont {Portillo}, \citenamefont {Rodd},\ and\ \citenamefont
  {Slatyer}}]{Daylan:2014rsa}%
  \BibitemOpen
  \bibfield  {author} {\bibinfo {author} {\bibfnamefont {T.}~\bibnamefont
  {Daylan}}, \bibinfo {author} {\bibfnamefont {D.~P.}\ \bibnamefont
  {Finkbeiner}}, \bibinfo {author} {\bibfnamefont {D.}~\bibnamefont {Hooper}},
  \bibinfo {author} {\bibfnamefont {T.}~\bibnamefont {Linden}}, \bibinfo
  {author} {\bibfnamefont {S.~K.~N.}\ \bibnamefont {Portillo}}, \bibinfo
  {author} {\bibfnamefont {N.~L.}\ \bibnamefont {Rodd}}, \ and\ \bibinfo
  {author} {\bibfnamefont {T.~R.}\ \bibnamefont {Slatyer}},\ }\href@noop {} {\
  (\bibinfo {year} {2014})},\ \Eprint {http://arxiv.org/abs/1402.6703}
  {arXiv:1402.6703 [astro-ph.HE]} \BibitemShut {NoStop}%
%%CITATION = ARXIV:1402.6703;%%
\bibitem [{\citenamefont {Zhou}\ \emph {et~al.}(2015)\citenamefont {Zhou},
  \citenamefont {Liang}, \citenamefont {Huang}, \citenamefont {Li},
  \citenamefont {Fan}, \citenamefont {Feng},\ and\ \citenamefont
  {Chang}}]{Zhou:2014lva}%
  \BibitemOpen
  \bibfield  {author} {\bibinfo {author} {\bibfnamefont {B.}~\bibnamefont
  {Zhou}}, \bibinfo {author} {\bibfnamefont {Y.-F.}\ \bibnamefont {Liang}},
  \bibinfo {author} {\bibfnamefont {X.}~\bibnamefont {Huang}}, \bibinfo
  {author} {\bibfnamefont {X.}~\bibnamefont {Li}}, \bibinfo {author}
  {\bibfnamefont {Y.-Z.}\ \bibnamefont {Fan}}, \bibinfo {author} {\bibfnamefont
  {L.}~\bibnamefont {Feng}}, \ and\ \bibinfo {author} {\bibfnamefont
  {J.}~\bibnamefont {Chang}},\ }\href {\doibase 10.1103/PhysRevD.91.123010}
  {\bibfield  {journal} {\bibinfo  {journal} {Phys. Rev.}\ }\textbf {\bibinfo
  {volume} {D91}},\ \bibinfo {pages} {123010} (\bibinfo {year} {2015})},\
  \Eprint {http://arxiv.org/abs/1406.6948} {arXiv:1406.6948 [astro-ph.HE]}
  \BibitemShut {NoStop}%
%%CITATION = ARXIV:1406.6948;%%
\bibitem [{\citenamefont {{Calore}}\ \emph {et~al.}(2015)\citenamefont
  {{Calore}}, \citenamefont {{Cholis}},\ and\ \citenamefont
  {{Weniger}}}]{Calore:2015}%
  \BibitemOpen
  \bibfield  {author} {\bibinfo {author} {\bibfnamefont {F.}~\bibnamefont
  {{Calore}}}, \bibinfo {author} {\bibfnamefont {I.}~\bibnamefont {{Cholis}}},
  \ and\ \bibinfo {author} {\bibfnamefont {C.}~\bibnamefont {{Weniger}}},\
  }\href {\doibase 10.1088/1475-7516/2015/03/038} {\bibfield  {journal}
  {\bibinfo  {journal} {\jcap}\ }\textbf {\bibinfo {volume} {3}},\ \bibinfo
  {eid} {038} (\bibinfo {year} {2015})},\ \Eprint
  {http://arxiv.org/abs/1409.0042} {arXiv:1409.0042} \BibitemShut {NoStop}%
\bibitem [{\citenamefont {Ajello}\ \emph {et~al.}(2015)\citenamefont {Ajello}
  \emph {et~al.}}]{TheFermi-LAT:2015kwa}%
  \BibitemOpen
  \bibfield  {author} {\bibinfo {author} {\bibfnamefont {M.}~\bibnamefont
  {Ajello}} \emph {et~al.} (\bibinfo {collaboration} {Fermi-LAT}),\ }\href@noop
  {} {\  (\bibinfo {year} {2015})},\ \Eprint {http://arxiv.org/abs/1511.02938}
  {arXiv:1511.02938 [astro-ph.HE]} \BibitemShut {NoStop}%
%%CITATION = ARXIV:1511.02938;%%
\bibitem [{\citenamefont {Berlin}\ \emph {et~al.}(2014)\citenamefont {Berlin},
  \citenamefont {Hooper},\ and\ \citenamefont {McDermott}}]{Berlin:2014tja}%
  \BibitemOpen
  \bibfield  {author} {\bibinfo {author} {\bibfnamefont {A.}~\bibnamefont
  {Berlin}}, \bibinfo {author} {\bibfnamefont {D.}~\bibnamefont {Hooper}}, \
  and\ \bibinfo {author} {\bibfnamefont {S.~D.}\ \bibnamefont {McDermott}},\
  }\href {\doibase 10.1103/PhysRevD.89.115022} {\bibfield  {journal} {\bibinfo
  {journal} {Phys. Rev.}\ }\textbf {\bibinfo {volume} {D89}},\ \bibinfo {pages}
  {115022} (\bibinfo {year} {2014})},\ \Eprint {http://arxiv.org/abs/1404.0022}
  {arXiv:1404.0022 [hep-ph]} \BibitemShut {NoStop}%
%%CITATION = ARXIV:1404.0022;%%
\bibitem [{\citenamefont {Agrawal}\ \emph {et~al.}(2014)\citenamefont
  {Agrawal}, \citenamefont {Batell}, \citenamefont {Hooper},\ and\
  \citenamefont {Lin}}]{Agrawal:2014una}%
  \BibitemOpen
  \bibfield  {author} {\bibinfo {author} {\bibfnamefont {P.}~\bibnamefont
  {Agrawal}}, \bibinfo {author} {\bibfnamefont {B.}~\bibnamefont {Batell}},
  \bibinfo {author} {\bibfnamefont {D.}~\bibnamefont {Hooper}}, \ and\ \bibinfo
  {author} {\bibfnamefont {T.}~\bibnamefont {Lin}},\ }\href {\doibase
  10.1103/PhysRevD.90.063512} {\bibfield  {journal} {\bibinfo  {journal} {Phys.
  Rev.}\ }\textbf {\bibinfo {volume} {D90}},\ \bibinfo {pages} {063512}
  (\bibinfo {year} {2014})},\ \Eprint {http://arxiv.org/abs/1404.1373}
  {arXiv:1404.1373 [hep-ph]} \BibitemShut {NoStop}%
%%CITATION = ARXIV:1404.1373;%%
\bibitem [{\citenamefont {Alves}\ \emph {et~al.}(2014)\citenamefont {Alves},
  \citenamefont {Profumo}, \citenamefont {Queiroz},\ and\ \citenamefont
  {Shepherd}}]{Alves:2014yha}%
  \BibitemOpen
  \bibfield  {author} {\bibinfo {author} {\bibfnamefont {A.}~\bibnamefont
  {Alves}}, \bibinfo {author} {\bibfnamefont {S.}~\bibnamefont {Profumo}},
  \bibinfo {author} {\bibfnamefont {F.~S.}\ \bibnamefont {Queiroz}}, \ and\
  \bibinfo {author} {\bibfnamefont {W.}~\bibnamefont {Shepherd}},\ }\href
  {\doibase 10.1103/PhysRevD.90.115003} {\bibfield  {journal} {\bibinfo
  {journal} {Phys. Rev.}\ }\textbf {\bibinfo {volume} {D90}},\ \bibinfo {pages}
  {115003} (\bibinfo {year} {2014})},\ \Eprint {http://arxiv.org/abs/1403.5027}
  {arXiv:1403.5027 [hep-ph]} \BibitemShut {NoStop}%
%%CITATION = ARXIV:1403.5027;%%
\bibitem [{\citenamefont {Abdullah}\ \emph {et~al.}(2014)\citenamefont
  {Abdullah}, \citenamefont {DiFranzo}, \citenamefont {Rajaraman},
  \citenamefont {Tait}, \citenamefont {Tanedo},\ and\ \citenamefont
  {Wijangco}}]{Abdullah:2014lla}%
  \BibitemOpen
  \bibfield  {author} {\bibinfo {author} {\bibfnamefont {M.}~\bibnamefont
  {Abdullah}}, \bibinfo {author} {\bibfnamefont {A.}~\bibnamefont {DiFranzo}},
  \bibinfo {author} {\bibfnamefont {A.}~\bibnamefont {Rajaraman}}, \bibinfo
  {author} {\bibfnamefont {T.~M.~P.}\ \bibnamefont {Tait}}, \bibinfo {author}
  {\bibfnamefont {P.}~\bibnamefont {Tanedo}}, \ and\ \bibinfo {author}
  {\bibfnamefont {A.~M.}\ \bibnamefont {Wijangco}},\ }\href {\doibase
  10.1103/PhysRevD.90.035004} {\bibfield  {journal} {\bibinfo  {journal} {Phys.
  Rev.}\ }\textbf {\bibinfo {volume} {D90}},\ \bibinfo {pages} {035004}
  (\bibinfo {year} {2014})},\ \Eprint {http://arxiv.org/abs/1404.6528}
  {arXiv:1404.6528 [hep-ph]} \BibitemShut {NoStop}%
%%CITATION = ARXIV:1404.6528;%%
\bibitem [{\citenamefont {Ipek}\ \emph {et~al.}(2014)\citenamefont {Ipek},
  \citenamefont {McKeen},\ and\ \citenamefont {Nelson}}]{Ipek:2014gua}%
  \BibitemOpen
  \bibfield  {author} {\bibinfo {author} {\bibfnamefont {S.}~\bibnamefont
  {Ipek}}, \bibinfo {author} {\bibfnamefont {D.}~\bibnamefont {McKeen}}, \ and\
  \bibinfo {author} {\bibfnamefont {A.~E.}\ \bibnamefont {Nelson}},\ }\href
  {\doibase 10.1103/PhysRevD.90.055021} {\bibfield  {journal} {\bibinfo
  {journal} {Phys. Rev.}\ }\textbf {\bibinfo {volume} {D90}},\ \bibinfo {pages}
  {055021} (\bibinfo {year} {2014})},\ \Eprint {http://arxiv.org/abs/1404.3716}
  {arXiv:1404.3716 [hep-ph]} \BibitemShut {NoStop}%
%%CITATION = ARXIV:1404.3716;%%
\bibitem [{\citenamefont {{Navarro}}\ \emph {et~al.}(1996)\citenamefont
  {{Navarro}}, \citenamefont {{Frenk}},\ and\ \citenamefont
  {{White}}}]{NFW:1996}%
  \BibitemOpen
  \bibfield  {author} {\bibinfo {author} {\bibfnamefont {J.~F.}\ \bibnamefont
  {{Navarro}}}, \bibinfo {author} {\bibfnamefont {C.~S.}\ \bibnamefont
  {{Frenk}}}, \ and\ \bibinfo {author} {\bibfnamefont {S.~D.~M.}\ \bibnamefont
  {{White}}},\ }\href {\doibase 10.1086/177173} {\bibfield  {journal} {\bibinfo
   {journal} {\apj}\ }\textbf {\bibinfo {volume} {462}},\ \bibinfo {pages}
  {563} (\bibinfo {year} {1996})},\ \Eprint
  {http://arxiv.org/abs/astro-ph/9508025} {astro-ph/9508025} \BibitemShut
  {NoStop}%
\bibitem [{\citenamefont {Calore}\ \emph {et~al.}(2015)\citenamefont {Calore},
  \citenamefont {Cholis}, \citenamefont {McCabe},\ and\ \citenamefont
  {Weniger}}]{Calore:2014nla}%
  \BibitemOpen
  \bibfield  {author} {\bibinfo {author} {\bibfnamefont {F.}~\bibnamefont
  {Calore}}, \bibinfo {author} {\bibfnamefont {I.}~\bibnamefont {Cholis}},
  \bibinfo {author} {\bibfnamefont {C.}~\bibnamefont {McCabe}}, \ and\ \bibinfo
  {author} {\bibfnamefont {C.}~\bibnamefont {Weniger}},\ }\href {\doibase
  10.1103/PhysRevD.91.063003} {\bibfield  {journal} {\bibinfo  {journal} {Phys.
  Rev.}\ }\textbf {\bibinfo {volume} {D91}},\ \bibinfo {pages} {063003}
  (\bibinfo {year} {2015})},\ \Eprint {http://arxiv.org/abs/1411.4647}
  {arXiv:1411.4647 [hep-ph]} \BibitemShut {NoStop}%
%%CITATION = ARXIV:1411.4647;%%
\bibitem [{\citenamefont {{Abazajian}}(2011)}]{2011JCAP...03..010A}%
  \BibitemOpen
  \bibfield  {author} {\bibinfo {author} {\bibfnamefont {K.~N.}\ \bibnamefont
  {{Abazajian}}},\ }\href {\doibase 10.1088/1475-7516/2011/03/010} {\bibfield
  {journal} {\bibinfo  {journal} {JCAP}\ }\textbf {\bibinfo {volume} {3}},\
  \bibinfo {eid} {010} (\bibinfo {year} {2011})},\ \Eprint
  {http://arxiv.org/abs/1011.4275} {arXiv:1011.4275 [astro-ph.HE]} \BibitemShut
  {NoStop}%
\bibitem [{\citenamefont {Yuan}\ and\ \citenamefont
  {Ioka}(2015)}]{Yuan:2014yda}%
  \BibitemOpen
  \bibfield  {author} {\bibinfo {author} {\bibfnamefont {Q.}~\bibnamefont
  {Yuan}}\ and\ \bibinfo {author} {\bibfnamefont {K.}~\bibnamefont {Ioka}},\
  }\href {\doibase 10.1088/0004-637X/802/2/124} {\bibfield  {journal} {\bibinfo
   {journal} {Astrophys. J.}\ }\textbf {\bibinfo {volume} {802}},\ \bibinfo
  {pages} {124} (\bibinfo {year} {2015})},\ \Eprint
  {http://arxiv.org/abs/1411.4363} {arXiv:1411.4363 [astro-ph.HE]} \BibitemShut
  {NoStop}%
%%CITATION = ARXIV:1411.4363;%%
\bibitem [{\citenamefont {Petrović}\ \emph {et~al.}(2015)\citenamefont
  {Petrović}, \citenamefont {Serpico},\ and\ \citenamefont
  {Zaharijas}}]{Petrovic:2014xra}%
  \BibitemOpen
  \bibfield  {author} {\bibinfo {author} {\bibfnamefont {J.}~\bibnamefont
  {Petrović}}, \bibinfo {author} {\bibfnamefont {P.~D.}\ \bibnamefont
  {Serpico}}, \ and\ \bibinfo {author} {\bibfnamefont {G.}~\bibnamefont
  {Zaharijas}},\ }\href {\doibase 10.1088/1475-7516/2015/02/023} {\bibfield
  {journal} {\bibinfo  {journal} {JCAP}\ }\textbf {\bibinfo {volume} {1502}},\
  \bibinfo {pages} {023} (\bibinfo {year} {2015})},\ \Eprint
  {http://arxiv.org/abs/1411.2980} {arXiv:1411.2980 [astro-ph.HE]} \BibitemShut
  {NoStop}%
%%CITATION = ARXIV:1411.2980;%%
\bibitem [{\citenamefont {O'Leary}\ \emph {et~al.}(2015)\citenamefont
  {O'Leary}, \citenamefont {Kistler}, \citenamefont {Kerr},\ and\ \citenamefont
  {Dexter}}]{O'Leary:2015gfa}%
  \BibitemOpen
  \bibfield  {author} {\bibinfo {author} {\bibfnamefont {R.~M.}\ \bibnamefont
  {O'Leary}}, \bibinfo {author} {\bibfnamefont {M.~D.}\ \bibnamefont
  {Kistler}}, \bibinfo {author} {\bibfnamefont {M.}~\bibnamefont {Kerr}}, \
  and\ \bibinfo {author} {\bibfnamefont {J.}~\bibnamefont {Dexter}},\
  }\href@noop {} {\  (\bibinfo {year} {2015})},\ \Eprint
  {http://arxiv.org/abs/1504.02477} {arXiv:1504.02477 [astro-ph.HE]}
  \BibitemShut {NoStop}%
%%CITATION = ARXIV:1504.02477;%%
\bibitem [{\citenamefont {{Carlson}}\ and\ \citenamefont
  {{Profumo}}(2014)}]{Carlson:2014}%
  \BibitemOpen
  \bibfield  {author} {\bibinfo {author} {\bibfnamefont {E.}~\bibnamefont
  {{Carlson}}}\ and\ \bibinfo {author} {\bibfnamefont {S.}~\bibnamefont
  {{Profumo}}},\ }\href {\doibase 10.1103/PhysRevD.90.023015} {\bibfield
  {journal} {\bibinfo  {journal} {\prd}\ }\textbf {\bibinfo {volume} {90}},\
  \bibinfo {eid} {023015} (\bibinfo {year} {2014})},\ \Eprint
  {http://arxiv.org/abs/1405.7685} {arXiv:1405.7685 [astro-ph.HE]} \BibitemShut
  {NoStop}%
\bibitem [{\citenamefont {{Petrovi{\'c}}}\ \emph {et~al.}(2014)\citenamefont
  {{Petrovi{\'c}}}, \citenamefont {{Dario Serpico}},\ and\ \citenamefont
  {{Zaharija{\v s}}}}]{Petrovic:2014}%
  \BibitemOpen
  \bibfield  {author} {\bibinfo {author} {\bibfnamefont {J.}~\bibnamefont
  {{Petrovi{\'c}}}}, \bibinfo {author} {\bibfnamefont {P.}~\bibnamefont {{Dario
  Serpico}}}, \ and\ \bibinfo {author} {\bibfnamefont {G.}~\bibnamefont
  {{Zaharija{\v s}}}},\ }\href {\doibase 10.1088/1475-7516/2014/10/052}
  {\bibfield  {journal} {\bibinfo  {journal} {\jcap}\ }\textbf {\bibinfo
  {volume} {10}},\ \bibinfo {eid} {052} (\bibinfo {year} {2014})},\ \Eprint
  {http://arxiv.org/abs/1405.7928} {arXiv:1405.7928 [astro-ph.HE]} \BibitemShut
  {NoStop}%
\bibitem [{\citenamefont {Cholis}\ \emph {et~al.}()\citenamefont {Cholis},
  \citenamefont {Evoli}, \citenamefont {Calore}, \citenamefont {Linden},
  \citenamefont {Weniger},\ and\ \citenamefont {Hooper}}]{Cholis:2015}%
  \BibitemOpen
  \bibfield  {author} {\bibinfo {author} {\bibfnamefont {I.}~\bibnamefont
  {Cholis}}, \bibinfo {author} {\bibfnamefont {C.}~\bibnamefont {Evoli}},
  \bibinfo {author} {\bibfnamefont {F.}~\bibnamefont {Calore}}, \bibinfo
  {author} {\bibfnamefont {T.}~\bibnamefont {Linden}}, \bibinfo {author}
  {\bibfnamefont {C.}~\bibnamefont {Weniger}}, \ and\ \bibinfo {author}
  {\bibfnamefont {D.}~\bibnamefont {Hooper}},\ }\href
  {http://arxiv.org/pdf/1506.05119v1.pdf http://arxiv.org/abs/1506.05119} {\ ,\
  \bibinfo {pages} {28}}\Eprint {http://arxiv.org/abs/1506.05119}
  {arXiv:1506.05119} \BibitemShut {NoStop}%
\bibitem [{\citenamefont {{Hooper}}\ \emph {et~al.}(2013)\citenamefont
  {{Hooper}}, \citenamefont {{Cholis}}, \citenamefont {{Linden}}, \citenamefont
  {{Siegal-Gaskins}},\ and\ \citenamefont {{Slatyer}}}]{Hooper:2013nhl}%
  \BibitemOpen
  \bibfield  {author} {\bibinfo {author} {\bibfnamefont {D.}~\bibnamefont
  {{Hooper}}}, \bibinfo {author} {\bibfnamefont {I.}~\bibnamefont {{Cholis}}},
  \bibinfo {author} {\bibfnamefont {T.}~\bibnamefont {{Linden}}}, \bibinfo
  {author} {\bibfnamefont {J.~M.}\ \bibnamefont {{Siegal-Gaskins}}}, \ and\
  \bibinfo {author} {\bibfnamefont {T.~R.}\ \bibnamefont {{Slatyer}}},\ }\href
  {\doibase 10.1103/PhysRevD.88.083009} {\bibfield  {journal} {\bibinfo
  {journal} {\prd}\ }\textbf {\bibinfo {volume} {88}},\ \bibinfo {eid} {083009}
  (\bibinfo {year} {2013})},\ \Eprint {http://arxiv.org/abs/1305.0830}
  {arXiv:1305.0830 [astro-ph.HE]} \BibitemShut {NoStop}%
\bibitem [{\citenamefont {Cholis}\ \emph {et~al.}(2015)\citenamefont {Cholis},
  \citenamefont {Hooper},\ and\ \citenamefont {Linden}}]{Cholis:2014lta}%
  \BibitemOpen
  \bibfield  {author} {\bibinfo {author} {\bibfnamefont {I.}~\bibnamefont
  {Cholis}}, \bibinfo {author} {\bibfnamefont {D.}~\bibnamefont {Hooper}}, \
  and\ \bibinfo {author} {\bibfnamefont {T.}~\bibnamefont {Linden}},\ }\href
  {\doibase 10.1088/1475-7516/2015/06/043} {\bibfield  {journal} {\bibinfo
  {journal} {JCAP}\ }\textbf {\bibinfo {volume} {1506}},\ \bibinfo {pages}
  {043} (\bibinfo {year} {2015})},\ \Eprint {http://arxiv.org/abs/1407.5625}
  {arXiv:1407.5625 [astro-ph.HE]} \BibitemShut {NoStop}%
%%CITATION = ARXIV:1407.5625;%%
\bibitem [{\citenamefont {Lee}\ \emph {et~al.}(2015)\citenamefont {Lee},
  \citenamefont {Lisanti}, \citenamefont {Safdi}, \citenamefont {Slatyer},\
  and\ \citenamefont {Xue}}]{Lee:2015fea}%
  \BibitemOpen
  \bibfield  {author} {\bibinfo {author} {\bibfnamefont {S.~K.}\ \bibnamefont
  {Lee}}, \bibinfo {author} {\bibfnamefont {M.}~\bibnamefont {Lisanti}},
  \bibinfo {author} {\bibfnamefont {B.~R.}\ \bibnamefont {Safdi}}, \bibinfo
  {author} {\bibfnamefont {T.~R.}\ \bibnamefont {Slatyer}}, \ and\ \bibinfo
  {author} {\bibfnamefont {W.}~\bibnamefont {Xue}},\ }\href@noop {} {\
  (\bibinfo {year} {2015})},\ \Eprint {http://arxiv.org/abs/1506.05124}
  {arXiv:1506.05124 [astro-ph.HE]} \BibitemShut {NoStop}%
%%CITATION = ARXIV:1506.05124;%%
\bibitem [{\citenamefont {Bartels}\ \emph {et~al.}(2015)\citenamefont
  {Bartels}, \citenamefont {Krishnamurthy},\ and\ \citenamefont
  {Weniger}}]{Bartels:2015aea}%
  \BibitemOpen
  \bibfield  {author} {\bibinfo {author} {\bibfnamefont {R.}~\bibnamefont
  {Bartels}}, \bibinfo {author} {\bibfnamefont {S.}~\bibnamefont
  {Krishnamurthy}}, \ and\ \bibinfo {author} {\bibfnamefont {C.}~\bibnamefont
  {Weniger}},\ }\href@noop {} {\  (\bibinfo {year} {2015})},\ \Eprint
  {http://arxiv.org/abs/1506.05104} {arXiv:1506.05104 [astro-ph.HE]}
  \BibitemShut {NoStop}%
%%CITATION = ARXIV:1506.05104;%%
\bibitem [{\citenamefont {{Carlson}}\ \emph {et~al.}(2015)\citenamefont
  {{Carlson}}, \citenamefont {{Linden}},\ and\ \citenamefont
  {{Profumo}}}]{paper_one}%
  \BibitemOpen
  \bibfield  {author} {\bibinfo {author} {\bibfnamefont {E.}~\bibnamefont
  {{Carlson}}}, \bibinfo {author} {\bibfnamefont {T.}~\bibnamefont {{Linden}}},
  \ and\ \bibinfo {author} {\bibfnamefont {S.}~\bibnamefont {{Profumo}}},\
  }\href@noop {} {\bibfield  {journal} {\bibinfo  {journal} {ArXiv e-prints}\ }
  (\bibinfo {year} {2015})},\ \Eprint {http://arxiv.org/abs/1510.04698}
  {arXiv:1510.04698 [astro-ph.HE]} \BibitemShut {NoStop}%
\bibitem [{\citenamefont {{Crocker}}\ \emph
  {et~al.}(2011{\natexlab{a}})\citenamefont {{Crocker}}, \citenamefont
  {{Jones}}, \citenamefont {{Aharonian}}, \citenamefont {{Law}}, \citenamefont
  {{Melia}}, \citenamefont {{Oka}},\ and\ \citenamefont
  {{Ott}}}]{Crocker:2011}%
  \BibitemOpen
  \bibfield  {author} {\bibinfo {author} {\bibfnamefont {R.~M.}\ \bibnamefont
  {{Crocker}}}, \bibinfo {author} {\bibfnamefont {D.~I.}\ \bibnamefont
  {{Jones}}}, \bibinfo {author} {\bibfnamefont {F.}~\bibnamefont
  {{Aharonian}}}, \bibinfo {author} {\bibfnamefont {C.~J.}\ \bibnamefont
  {{Law}}}, \bibinfo {author} {\bibfnamefont {F.}~\bibnamefont {{Melia}}},
  \bibinfo {author} {\bibfnamefont {T.}~\bibnamefont {{Oka}}}, \ and\ \bibinfo
  {author} {\bibfnamefont {J.}~\bibnamefont {{Ott}}},\ }\href {\doibase
  10.1111/j.1365-2966.2010.18170.x} {\bibfield  {journal} {\bibinfo  {journal}
  {\mnras}\ }\textbf {\bibinfo {volume} {413}},\ \bibinfo {pages} {763}
  (\bibinfo {year} {2011}{\natexlab{a}})},\ \Eprint
  {http://arxiv.org/abs/1011.0206} {arXiv:1011.0206 [astro-ph.GA]} \BibitemShut
  {NoStop}%
\bibitem [{\citenamefont {{Crocker}}\ \emph
  {et~al.}(2011{\natexlab{b}})\citenamefont {{Crocker}}, \citenamefont
  {{Jones}}, \citenamefont {{Aharonian}}, \citenamefont {{Law}}, \citenamefont
  {{Melia}},\ and\ \citenamefont {{Ott}}}]{2011MNRAS.411L..11C}%
  \BibitemOpen
  \bibfield  {author} {\bibinfo {author} {\bibfnamefont {R.~M.}\ \bibnamefont
  {{Crocker}}}, \bibinfo {author} {\bibfnamefont {D.~I.}\ \bibnamefont
  {{Jones}}}, \bibinfo {author} {\bibfnamefont {F.}~\bibnamefont
  {{Aharonian}}}, \bibinfo {author} {\bibfnamefont {C.~J.}\ \bibnamefont
  {{Law}}}, \bibinfo {author} {\bibfnamefont {F.}~\bibnamefont {{Melia}}}, \
  and\ \bibinfo {author} {\bibfnamefont {J.}~\bibnamefont {{Ott}}},\ }\href
  {\doibase 10.1111/j.1745-3933.2010.00983.x} {\bibfield  {journal} {\bibinfo
  {journal} {\mnras}\ }\textbf {\bibinfo {volume} {411}},\ \bibinfo {pages}
  {L11} (\bibinfo {year} {2011}{\natexlab{b}})},\ \Eprint
  {http://arxiv.org/abs/1009.4340} {arXiv:1009.4340} \BibitemShut {NoStop}%
\bibitem [{\citenamefont {{Sofue}}\ and\ \citenamefont
  {{Handa}}(1984)}]{1984Natur.310..568S}%
  \BibitemOpen
  \bibfield  {author} {\bibinfo {author} {\bibfnamefont {Y.}~\bibnamefont
  {{Sofue}}}\ and\ \bibinfo {author} {\bibfnamefont {T.}~\bibnamefont
  {{Handa}}},\ }\href {\doibase 10.1038/310568a0} {\bibfield  {journal}
  {\bibinfo  {journal} {\nat}\ }\textbf {\bibinfo {volume} {310}},\ \bibinfo
  {pages} {568} (\bibinfo {year} {1984})}\BibitemShut {NoStop}%
\bibitem [{\citenamefont {{Chevalier}}(1992)}]{1992ApJ...397L..39C}%
  \BibitemOpen
  \bibfield  {author} {\bibinfo {author} {\bibfnamefont {R.~A.}\ \bibnamefont
  {{Chevalier}}},\ }\href {\doibase 10.1086/186539} {\bibfield  {journal}
  {\bibinfo  {journal} {\apjl}\ }\textbf {\bibinfo {volume} {397}},\ \bibinfo
  {pages} {L39} (\bibinfo {year} {1992})}\BibitemShut {NoStop}%
\bibitem [{\citenamefont {{Bland-Hawthorn}}\ and\ \citenamefont
  {{Cohen}}(2003)}]{2003ApJ...582..246B}%
  \BibitemOpen
  \bibfield  {author} {\bibinfo {author} {\bibfnamefont {J.}~\bibnamefont
  {{Bland-Hawthorn}}}\ and\ \bibinfo {author} {\bibfnamefont {M.}~\bibnamefont
  {{Cohen}}},\ }\href {\doibase 10.1086/344573} {\bibfield  {journal} {\bibinfo
   {journal} {\apj}\ }\textbf {\bibinfo {volume} {582}},\ \bibinfo {pages}
  {246} (\bibinfo {year} {2003})},\ \Eprint
  {http://arxiv.org/abs/astro-ph/0208553} {astro-ph/0208553} \BibitemShut
  {NoStop}%
\bibitem [{\citenamefont {{Sofue}}(2000)}]{2000ApJ...540..224S}%
  \BibitemOpen
  \bibfield  {author} {\bibinfo {author} {\bibfnamefont {Y.}~\bibnamefont
  {{Sofue}}},\ }\href {\doibase 10.1086/309297} {\bibfield  {journal} {\bibinfo
   {journal} {\apj}\ }\textbf {\bibinfo {volume} {540}},\ \bibinfo {pages}
  {224} (\bibinfo {year} {2000})},\ \Eprint
  {http://arxiv.org/abs/astro-ph/9912528} {astro-ph/9912528} \BibitemShut
  {NoStop}%
\bibitem [{\citenamefont {{Zhao}}\ \emph {et~al.}(2016)\citenamefont {{Zhao}},
  \citenamefont {{Morris}},\ and\ \citenamefont
  {{Goss}}}]{2016ApJ...817..171Z}%
  \BibitemOpen
  \bibfield  {author} {\bibinfo {author} {\bibfnamefont {J.-H.}\ \bibnamefont
  {{Zhao}}}, \bibinfo {author} {\bibfnamefont {M.~R.}\ \bibnamefont
  {{Morris}}}, \ and\ \bibinfo {author} {\bibfnamefont {W.~M.}\ \bibnamefont
  {{Goss}}},\ }\href {\doibase 10.3847/0004-637X/817/2/171} {\bibfield
  {journal} {\bibinfo  {journal} {\apj}\ }\textbf {\bibinfo {volume} {817}},\
  \bibinfo {eid} {171} (\bibinfo {year} {2016})},\ \Eprint
  {http://arxiv.org/abs/1512.06279} {arXiv:1512.06279} \BibitemShut {NoStop}%
\bibitem [{\citenamefont {Yoast-Hull}\ \emph {et~al.}(2014)\citenamefont
  {Yoast-Hull}, \citenamefont {Gallagher},\ and\ \citenamefont
  {Zweibel}}]{Yoast-Hull2014}%
  \BibitemOpen
  \bibfield  {author} {\bibinfo {author} {\bibfnamefont {T.~M.}\ \bibnamefont
  {Yoast-Hull}}, \bibinfo {author} {\bibfnamefont {J.~S.}\ \bibnamefont
  {Gallagher}}, \ and\ \bibinfo {author} {\bibfnamefont {E.~G.}\ \bibnamefont
  {Zweibel}},\ }\href {http://arxiv.org/pdf/1405.7059.pdf
  http://arxiv.org/abs/1405.7059} {\ \textbf {\bibinfo {volume} {i}},\ \bibinfo
  {pages} {9} (\bibinfo {year} {2014})},\ \Eprint
  {http://arxiv.org/abs/1405.7059} {arXiv:1405.7059} \BibitemShut {NoStop}%
\bibitem [{\citenamefont {Gaggero}\ \emph {et~al.}(2015)\citenamefont
  {Gaggero}, \citenamefont {Taoso}, \citenamefont {Urbano}, \citenamefont
  {Valli},\ and\ \citenamefont {Ullio}}]{Gaggero2015}%
  \BibitemOpen
  \bibfield  {author} {\bibinfo {author} {\bibfnamefont {D.}~\bibnamefont
  {Gaggero}}, \bibinfo {author} {\bibfnamefont {M.}~\bibnamefont {Taoso}},
  \bibinfo {author} {\bibfnamefont {A.}~\bibnamefont {Urbano}}, \bibinfo
  {author} {\bibfnamefont {M.}~\bibnamefont {Valli}}, \ and\ \bibinfo {author}
  {\bibfnamefont {P.}~\bibnamefont {Ullio}},\ }\href
  {http://arxiv.org/abs/1507.06129} {\  (\bibinfo {year} {2015})},\ \Eprint
  {http://arxiv.org/abs/1507.06129} {arXiv:1507.06129} \BibitemShut {NoStop}%
\bibitem [{gal(2015{\natexlab{a}})}]{galprop_sourceforge}%
  \BibitemOpen
  \href {http://sourceforge.net/projects/galprop/} {\enquote {\bibinfo {title}
  {http://sourceforge.net/projects/galprop/},}\ } (\bibinfo {year}
  {2015}{\natexlab{a}})\BibitemShut {NoStop}%
\bibitem [{gal(2015{\natexlab{b}})}]{galprop_stanford}%
  \BibitemOpen
  \href {http://galprop.stanford.edu/} {\enquote {\bibinfo {title}
  {http://galprop.stanford.edu/},}\ } (\bibinfo {year}
  {2015}{\natexlab{b}})\BibitemShut {NoStop}%
\bibitem [{\citenamefont {{Moskalenko}}\ and\ \citenamefont
  {{Strong}}(1998)}]{galprop0}%
  \BibitemOpen
  \bibfield  {author} {\bibinfo {author} {\bibfnamefont {I.~V.}\ \bibnamefont
  {{Moskalenko}}}\ and\ \bibinfo {author} {\bibfnamefont {A.~W.}\ \bibnamefont
  {{Strong}}},\ }\href {\doibase 10.1086/305152} {\bibfield  {journal}
  {\bibinfo  {journal} {\apj}\ }\textbf {\bibinfo {volume} {493}},\ \bibinfo
  {pages} {694} (\bibinfo {year} {1998})},\ \Eprint
  {http://arxiv.org/abs/astro-ph/9710124} {astro-ph/9710124} \BibitemShut
  {NoStop}%
\bibitem [{\citenamefont {{Strong}}\ \emph {et~al.}(2000)\citenamefont
  {{Strong}}, \citenamefont {{Moskalenko}},\ and\ \citenamefont
  {{Reimer}}}]{galprop1}%
  \BibitemOpen
  \bibfield  {author} {\bibinfo {author} {\bibfnamefont {A.~W.}\ \bibnamefont
  {{Strong}}}, \bibinfo {author} {\bibfnamefont {I.~V.}\ \bibnamefont
  {{Moskalenko}}}, \ and\ \bibinfo {author} {\bibfnamefont {O.}~\bibnamefont
  {{Reimer}}},\ }\href {\doibase 10.1086/309038} {\bibfield  {journal}
  {\bibinfo  {journal} {\apj}\ }\textbf {\bibinfo {volume} {537}},\ \bibinfo
  {pages} {763} (\bibinfo {year} {2000})},\ \Eprint
  {http://arxiv.org/abs/astro-ph/9811296} {astro-ph/9811296} \BibitemShut
  {NoStop}%
\bibitem [{\citenamefont {{Moskalenko}}\ \emph {et~al.}(1998)\citenamefont
  {{Moskalenko}}, \citenamefont {{Strong}},\ and\ \citenamefont
  {{Reimer}}}]{galprop2}%
  \BibitemOpen
  \bibfield  {author} {\bibinfo {author} {\bibfnamefont {I.~V.}\ \bibnamefont
  {{Moskalenko}}}, \bibinfo {author} {\bibfnamefont {A.~W.}\ \bibnamefont
  {{Strong}}}, \ and\ \bibinfo {author} {\bibfnamefont {O.}~\bibnamefont
  {{Reimer}}},\ }\href@noop {} {\bibfield  {journal} {\bibinfo  {journal}
  {\aap}\ }\textbf {\bibinfo {volume} {338}},\ \bibinfo {pages} {L75} (\bibinfo
  {year} {1998})},\ \Eprint {http://arxiv.org/abs/astro-ph/9808084}
  {astro-ph/9808084} \BibitemShut {NoStop}%
\bibitem [{\citenamefont {Morris}(2007)}]{Morris:2007jk}%
  \BibitemOpen
  \bibfield  {author} {\bibinfo {author} {\bibfnamefont {M.}~\bibnamefont
  {Morris}},\ }in\ \href@noop {} {\emph {\bibinfo {booktitle} {{Galactic Center
  Workshop 2006 (GC 06): From the Center of the Milky Way to Nearby Low
  Luminosity Galactic Nuclei Bad Honnef, Germany, April 18-22, 2006}}}}\
  (\bibinfo {year} {2007})\ \Eprint {http://arxiv.org/abs/astro-ph/0701050}
  {arXiv:astro-ph/0701050 [astro-ph]} \BibitemShut {NoStop}%
%%CITATION = ASTRO-PH/0701050;%%
\bibitem [{\citenamefont {{Seo}}\ and\ \citenamefont
  {{Ptuskin}}(1994)}]{1994ApJ...431..705S}%
  \BibitemOpen
  \bibfield  {author} {\bibinfo {author} {\bibfnamefont {E.~S.}\ \bibnamefont
  {{Seo}}}\ and\ \bibinfo {author} {\bibfnamefont {V.~S.}\ \bibnamefont
  {{Ptuskin}}},\ }\href {\doibase 10.1086/174520} {\bibfield  {journal}
  {\bibinfo  {journal} {\apj}\ }\textbf {\bibinfo {volume} {431}},\ \bibinfo
  {pages} {705} (\bibinfo {year} {1994})}\BibitemShut {NoStop}%
\bibitem [{\citenamefont {Crocker}\ \emph {et~al.}(2010)\citenamefont
  {Crocker}, \citenamefont {Jones}, \citenamefont {Melia}, \citenamefont
  {Ott},\ and\ \citenamefont {Protheroe}}]{Crocker:2010xc}%
  \BibitemOpen
  \bibfield  {author} {\bibinfo {author} {\bibfnamefont {R.~M.}\ \bibnamefont
  {Crocker}}, \bibinfo {author} {\bibfnamefont {D.}~\bibnamefont {Jones}},
  \bibinfo {author} {\bibfnamefont {F.}~\bibnamefont {Melia}}, \bibinfo
  {author} {\bibfnamefont {J.}~\bibnamefont {Ott}}, \ and\ \bibinfo {author}
  {\bibfnamefont {R.~J.}\ \bibnamefont {Protheroe}},\ }\href {\doibase
  10.1038/nature08635} {\bibfield  {journal} {\bibinfo  {journal} {Nature}\
  }\textbf {\bibinfo {volume} {468}},\ \bibinfo {pages} {65} (\bibinfo {year}
  {2010})},\ \Eprint {http://arxiv.org/abs/1001.1275} {arXiv:1001.1275
  [astro-ph.GA]} \BibitemShut {NoStop}%
%%CITATION = ARXIV:1001.1275;%%
\bibitem [{\citenamefont {Jansson}\ and\ \citenamefont
  {Farrar}(2012)}]{Jansson2012}%
  \BibitemOpen
  \bibfield  {author} {\bibinfo {author} {\bibfnamefont {R.}~\bibnamefont
  {Jansson}}\ and\ \bibinfo {author} {\bibfnamefont {G.~R.}\ \bibnamefont
  {Farrar}},\ }\href {\doibase 10.1088/2041-8205/761/1/L11} {\bibfield
  {journal} {\bibinfo  {journal} {The Astrophysical Journal}\ }\textbf
  {\bibinfo {volume} {761}},\ \bibinfo {pages} {L11} (\bibinfo {year}
  {2012})},\ \Eprint {http://arxiv.org/abs/1210.7820} {arXiv:1210.7820}
  \BibitemShut {NoStop}%
\bibitem [{\citenamefont {{Strong}}(2015)}]{2015arXiv150705020S}%
  \BibitemOpen
  \bibfield  {author} {\bibinfo {author} {\bibfnamefont {A.~W.}\ \bibnamefont
  {{Strong}}},\ }\href@noop {} {\bibfield  {journal} {\bibinfo  {journal}
  {ArXiv e-prints}\ } (\bibinfo {year} {2015})},\ \Eprint
  {http://arxiv.org/abs/1507.05020} {arXiv:1507.05020 [astro-ph.HE]}
  \BibitemShut {NoStop}%
\bibitem [{\citenamefont {{Ackermann}}\ \emph {et~al.}(2015)\citenamefont
  {{Ackermann}}, \citenamefont {{Ajello}}, \citenamefont {{Albert}},
  \citenamefont {{Atwood}}, \citenamefont {{Baldini}}, \citenamefont
  {{Ballet}}, \citenamefont {{Barbiellini}}, \citenamefont {{Bastieri}},\ and\
  \citenamefont {{et al.}}}]{isotropic}%
  \BibitemOpen
  \bibfield  {author} {\bibinfo {author} {\bibfnamefont {M.}~\bibnamefont
  {{Ackermann}}}, \bibinfo {author} {\bibfnamefont {M.}~\bibnamefont
  {{Ajello}}}, \bibinfo {author} {\bibfnamefont {A.}~\bibnamefont {{Albert}}},
  \bibinfo {author} {\bibfnamefont {W.~B.}\ \bibnamefont {{Atwood}}}, \bibinfo
  {author} {\bibfnamefont {L.}~\bibnamefont {{Baldini}}}, \bibinfo {author}
  {\bibfnamefont {J.}~\bibnamefont {{Ballet}}}, \bibinfo {author}
  {\bibfnamefont {G.}~\bibnamefont {{Barbiellini}}}, \bibinfo {author}
  {\bibfnamefont {D.}~\bibnamefont {{Bastieri}}}, \ and\ \bibinfo {author}
  {\bibnamefont {{et al.}}},\ }\href {\doibase 10.1088/0004-637X/799/1/86}
  {\bibfield  {journal} {\bibinfo  {journal} {\apj}\ }\textbf {\bibinfo
  {volume} {799}},\ \bibinfo {eid} {86} (\bibinfo {year} {2015})},\ \Eprint
  {http://arxiv.org/abs/1410.3696} {arXiv:1410.3696 [astro-ph.HE]} \BibitemShut
  {NoStop}%
\bibitem [{\citenamefont {{Moskalenko}}\ \emph {et~al.}(2006)\citenamefont
  {{Moskalenko}}, \citenamefont {{Porter}},\ and\ \citenamefont
  {{Strong}}}]{2006ApJ...640L.155M}%
  \BibitemOpen
  \bibfield  {author} {\bibinfo {author} {\bibfnamefont {I.~V.}\ \bibnamefont
  {{Moskalenko}}}, \bibinfo {author} {\bibfnamefont {T.~A.}\ \bibnamefont
  {{Porter}}}, \ and\ \bibinfo {author} {\bibfnamefont {A.~W.}\ \bibnamefont
  {{Strong}}},\ }\href {\doibase 10.1086/503524} {\bibfield  {journal}
  {\bibinfo  {journal} {\apjl}\ }\textbf {\bibinfo {volume} {640}},\ \bibinfo
  {pages} {L155} (\bibinfo {year} {2006})},\ \Eprint
  {http://arxiv.org/abs/astro-ph/0511149} {astro-ph/0511149} \BibitemShut
  {NoStop}%
\bibitem [{\citenamefont {{Porter}}\ \emph {et~al.}(2008)\citenamefont
  {{Porter}}, \citenamefont {{Moskalenko}}, \citenamefont {{Strong}},
  \citenamefont {{Orlando}},\ and\ \citenamefont
  {{Bouchet}}}]{2008ApJ...682..400P}%
  \BibitemOpen
  \bibfield  {author} {\bibinfo {author} {\bibfnamefont {T.~A.}\ \bibnamefont
  {{Porter}}}, \bibinfo {author} {\bibfnamefont {I.~V.}\ \bibnamefont
  {{Moskalenko}}}, \bibinfo {author} {\bibfnamefont {A.~W.}\ \bibnamefont
  {{Strong}}}, \bibinfo {author} {\bibfnamefont {E.}~\bibnamefont {{Orlando}}},
  \ and\ \bibinfo {author} {\bibfnamefont {L.}~\bibnamefont {{Bouchet}}},\
  }\href {\doibase 10.1086/589615} {\bibfield  {journal} {\bibinfo  {journal}
  {\apj}\ }\textbf {\bibinfo {volume} {682}},\ \bibinfo {pages} {400} (\bibinfo
  {year} {2008})},\ \Eprint {http://arxiv.org/abs/0804.1774} {arXiv:0804.1774}
  \BibitemShut {NoStop}%
\bibitem [{\citenamefont {Adriani}\ \emph {et~al.}(2013)\citenamefont
  {Adriani}, \citenamefont {Barbarino}, \citenamefont {Bazilevskaya},
  \citenamefont {Bellotti}, \citenamefont {Bianco}, \citenamefont {Boezio},
  \citenamefont {Bogomolov}, \citenamefont {Bongi},\ and\ \citenamefont
  {et~al.}}]{PhysRevLett.111.081102}%
  \BibitemOpen
  \bibfield  {author} {\bibinfo {author} {\bibfnamefont {O.}~\bibnamefont
  {Adriani}}, \bibinfo {author} {\bibfnamefont {G.~C.}\ \bibnamefont
  {Barbarino}}, \bibinfo {author} {\bibfnamefont {G.~A.}\ \bibnamefont
  {Bazilevskaya}}, \bibinfo {author} {\bibfnamefont {R.}~\bibnamefont
  {Bellotti}}, \bibinfo {author} {\bibfnamefont {A.}~\bibnamefont {Bianco}},
  \bibinfo {author} {\bibfnamefont {M.}~\bibnamefont {Boezio}}, \bibinfo
  {author} {\bibfnamefont {E.~A.}\ \bibnamefont {Bogomolov}}, \bibinfo {author}
  {\bibfnamefont {M.}~\bibnamefont {Bongi}}, \ and\ \bibinfo {author}
  {\bibnamefont {et~al.}},\ }\href {\doibase 10.1103/PhysRevLett.111.081102}
  {\bibfield  {journal} {\bibinfo  {journal} {Phys. Rev. Lett.}\ }\textbf
  {\bibinfo {volume} {111}},\ \bibinfo {pages} {081102} (\bibinfo {year}
  {2013})}\BibitemShut {NoStop}%
\bibitem [{\citenamefont {Aguilar}\ \emph {et~al.}(2013)\citenamefont
  {Aguilar}, \citenamefont {Alberti}, \citenamefont {Alpat}, \citenamefont
  {Alvino}, \citenamefont {Ambrosi}, \citenamefont {Andeen}, \citenamefont
  {Anderhub}, \citenamefont {Arruda},\ and\ \citenamefont
  {et~al.}}]{PhysRevLett.110.141102}%
  \BibitemOpen
  \bibfield  {author} {\bibinfo {author} {\bibfnamefont {M.}~\bibnamefont
  {Aguilar}}, \bibinfo {author} {\bibfnamefont {G.}~\bibnamefont {Alberti}},
  \bibinfo {author} {\bibfnamefont {B.}~\bibnamefont {Alpat}}, \bibinfo
  {author} {\bibfnamefont {A.}~\bibnamefont {Alvino}}, \bibinfo {author}
  {\bibfnamefont {G.}~\bibnamefont {Ambrosi}}, \bibinfo {author} {\bibfnamefont
  {K.}~\bibnamefont {Andeen}}, \bibinfo {author} {\bibfnamefont
  {H.}~\bibnamefont {Anderhub}}, \bibinfo {author} {\bibfnamefont
  {L.}~\bibnamefont {Arruda}}, \ and\ \bibinfo {author} {\bibnamefont {et~al.}}
  (\bibinfo {collaboration} {AMS Collaboration}),\ }\href {\doibase
  10.1103/PhysRevLett.110.141102} {\bibfield  {journal} {\bibinfo  {journal}
  {Phys. Rev. Lett.}\ }\textbf {\bibinfo {volume} {110}},\ \bibinfo {pages}
  {141102} (\bibinfo {year} {2013})}\BibitemShut {NoStop}%
\bibitem [{\citenamefont {{The Fermi-LAT
  Collaboration}}(2012)}]{fermi_diffuse}%
  \BibitemOpen
  \bibfield  {author} {\bibinfo {author} {\bibnamefont {{The Fermi-LAT
  Collaboration}}},\ }\href@noop {} {\bibfield  {journal} {\bibinfo  {journal}
  {ArXiv e-prints}\ } (\bibinfo {year} {2012})},\ \Eprint
  {http://arxiv.org/abs/1202.4039} {arXiv:1202.4039 [astro-ph.HE]} \BibitemShut
  {NoStop}%
\bibitem [{\citenamefont {{Kamae}}\ \emph {et~al.}(2006)\citenamefont
  {{Kamae}}, \citenamefont {{Karlsson}}, \citenamefont {{Mizuno}},
  \citenamefont {{Abe}},\ and\ \citenamefont {{Koi}}}]{2006ApJ...647..692K}%
  \BibitemOpen
  \bibfield  {author} {\bibinfo {author} {\bibfnamefont {T.}~\bibnamefont
  {{Kamae}}}, \bibinfo {author} {\bibfnamefont {N.}~\bibnamefont {{Karlsson}}},
  \bibinfo {author} {\bibfnamefont {T.}~\bibnamefont {{Mizuno}}}, \bibinfo
  {author} {\bibfnamefont {T.}~\bibnamefont {{Abe}}}, \ and\ \bibinfo {author}
  {\bibfnamefont {T.}~\bibnamefont {{Koi}}},\ }\href {\doibase 10.1086/505189}
  {\bibfield  {journal} {\bibinfo  {journal} {\apj}\ }\textbf {\bibinfo
  {volume} {647}},\ \bibinfo {pages} {692} (\bibinfo {year} {2006})},\ \Eprint
  {http://arxiv.org/abs/astro-ph/0605581} {astro-ph/0605581} \BibitemShut
  {NoStop}%
\bibitem [{\citenamefont {{Kachelrie{\ss}}}\ and\ \citenamefont
  {{Ostapchenko}}(2012)}]{Kachelriess:2012}%
  \BibitemOpen
  \bibfield  {author} {\bibinfo {author} {\bibfnamefont {M.}~\bibnamefont
  {{Kachelrie{\ss}}}}\ and\ \bibinfo {author} {\bibfnamefont {S.}~\bibnamefont
  {{Ostapchenko}}},\ }\href {\doibase 10.1103/PhysRevD.86.043004} {\bibfield
  {journal} {\bibinfo  {journal} {\prd}\ }\textbf {\bibinfo {volume} {86}},\
  \bibinfo {eid} {043004} (\bibinfo {year} {2012})},\ \Eprint
  {http://arxiv.org/abs/1206.4705} {arXiv:1206.4705 [astro-ph.HE]} \BibitemShut
  {NoStop}%
\bibitem [{\citenamefont {Dermer}\ \emph {et~al.}(2013)\citenamefont {Dermer},
  \citenamefont {Strong}, \citenamefont {Orlando}, \citenamefont {Tibaldo},\
  and\ \citenamefont {Collaboration}}]{Dermer2013}%
  \BibitemOpen
  \bibfield  {author} {\bibinfo {author} {\bibfnamefont {C.~D.}\ \bibnamefont
  {Dermer}}, \bibinfo {author} {\bibfnamefont {A.~W.}\ \bibnamefont {Strong}},
  \bibinfo {author} {\bibfnamefont {E.}~\bibnamefont {Orlando}}, \bibinfo
  {author} {\bibfnamefont {L.}~\bibnamefont {Tibaldo}}, \ and\ \bibinfo
  {author} {\bibfnamefont {f.~t.~F.}\ \bibnamefont {Collaboration}},\ }\href
  {http://arxiv.org/pdf/1307.0497v1.pdf http://arxiv.org/abs/1307.0497} {\ ,\
  \bibinfo {pages} {4} (\bibinfo {year} {2013})},\ \Eprint
  {http://arxiv.org/abs/1307.0497} {arXiv:1307.0497} \BibitemShut {NoStop}%
\bibitem [{\citenamefont {{Kalberla}}\ \emph {et~al.}(2005)\citenamefont
  {{Kalberla}}, \citenamefont {{Burton}}, \citenamefont {{Hartmann}},
  \citenamefont {{Arnal}}, \citenamefont {{Bajaja}}, \citenamefont {{Morras}},\
  and\ \citenamefont {{P{\"o}ppel}}}]{LAB}%
  \BibitemOpen
  \bibfield  {author} {\bibinfo {author} {\bibfnamefont {P.~M.~W.}\
  \bibnamefont {{Kalberla}}}, \bibinfo {author} {\bibfnamefont {W.~B.}\
  \bibnamefont {{Burton}}}, \bibinfo {author} {\bibfnamefont {D.}~\bibnamefont
  {{Hartmann}}}, \bibinfo {author} {\bibfnamefont {E.~M.}\ \bibnamefont
  {{Arnal}}}, \bibinfo {author} {\bibfnamefont {E.}~\bibnamefont {{Bajaja}}},
  \bibinfo {author} {\bibfnamefont {R.}~\bibnamefont {{Morras}}}, \ and\
  \bibinfo {author} {\bibfnamefont {W.~G.~L.}\ \bibnamefont {{P{\"o}ppel}}},\
  }\href {\doibase 10.1051/0004-6361:20041864} {\bibfield  {journal} {\bibinfo
  {journal} {\aap}\ }\textbf {\bibinfo {volume} {440}},\ \bibinfo {pages} {775}
  (\bibinfo {year} {2005})},\ \Eprint {http://arxiv.org/abs/astro-ph/0504140}
  {astro-ph/0504140} \BibitemShut {NoStop}%
\bibitem [{\citenamefont {{Dame}}\ \emph {et~al.}(2001)\citenamefont {{Dame}},
  \citenamefont {{Hartmann}},\ and\ \citenamefont {{Thaddeus}}}]{Dame:2001}%
  \BibitemOpen
  \bibfield  {author} {\bibinfo {author} {\bibfnamefont {T.~M.}\ \bibnamefont
  {{Dame}}}, \bibinfo {author} {\bibfnamefont {D.}~\bibnamefont {{Hartmann}}},
  \ and\ \bibinfo {author} {\bibfnamefont {P.}~\bibnamefont {{Thaddeus}}},\
  }\href {\doibase 10.1086/318388} {\bibfield  {journal} {\bibinfo  {journal}
  {\apj}\ }\textbf {\bibinfo {volume} {547}},\ \bibinfo {pages} {792} (\bibinfo
  {year} {2001})},\ \Eprint {http://arxiv.org/abs/astro-ph/0009217}
  {astro-ph/0009217} \BibitemShut {NoStop}%
\bibitem [{\citenamefont {{Clemens}}(1985)}]{Clemens:1985}%
  \BibitemOpen
  \bibfield  {author} {\bibinfo {author} {\bibfnamefont {D.~P.}\ \bibnamefont
  {{Clemens}}},\ }\href {\doibase 10.1086/163386} {\bibfield  {journal}
  {\bibinfo  {journal} {\apj}\ }\textbf {\bibinfo {volume} {295}},\ \bibinfo
  {pages} {422} (\bibinfo {year} {1985})}\BibitemShut {NoStop}%
\bibitem [{\citenamefont {{Pohl}}\ \emph {et~al.}(2008)\citenamefont {{Pohl}},
  \citenamefont {{Englmaier}},\ and\ \citenamefont {{Bissantz}}}]{PEB}%
  \BibitemOpen
  \bibfield  {author} {\bibinfo {author} {\bibfnamefont {M.}~\bibnamefont
  {{Pohl}}}, \bibinfo {author} {\bibfnamefont {P.}~\bibnamefont {{Englmaier}}},
  \ and\ \bibinfo {author} {\bibfnamefont {N.}~\bibnamefont {{Bissantz}}},\
  }\href {\doibase 10.1086/529004} {\bibfield  {journal} {\bibinfo  {journal}
  {\apj}\ }\textbf {\bibinfo {volume} {677}},\ \bibinfo {pages} {283} (\bibinfo
  {year} {2008})},\ \Eprint {http://arxiv.org/abs/0712.4264} {arXiv:0712.4264}
  \BibitemShut {NoStop}%
\bibitem [{\citenamefont {Bissantz}\ \emph {et~al.}(2003)\citenamefont
  {Bissantz}, \citenamefont {Englmaier},\ and\ \citenamefont
  {Gerhard}}]{MNR:MNR6358}%
  \BibitemOpen
  \bibfield  {author} {\bibinfo {author} {\bibfnamefont {N.}~\bibnamefont
  {Bissantz}}, \bibinfo {author} {\bibfnamefont {P.}~\bibnamefont {Englmaier}},
  \ and\ \bibinfo {author} {\bibfnamefont {O.}~\bibnamefont {Gerhard}},\ }\href
  {\doibase 10.1046/j.1365-8711.2003.06358.x} {\bibfield  {journal} {\bibinfo
  {journal} {Monthly Notices of the Royal Astronomical Society}\ }\textbf
  {\bibinfo {volume} {340}},\ \bibinfo {pages} {949} (\bibinfo {year}
  {2003})}\BibitemShut {NoStop}%
\bibitem [{\citenamefont {Gaggero}\ \emph {et~al.}(2014)\citenamefont
  {Gaggero}, \citenamefont {Maccione}, \citenamefont {Grasso}, \citenamefont
  {Di~Bernardo},\ and\ \citenamefont {Evoli}}]{PhysRevD.89.083007}%
  \BibitemOpen
  \bibfield  {author} {\bibinfo {author} {\bibfnamefont {D.}~\bibnamefont
  {Gaggero}}, \bibinfo {author} {\bibfnamefont {L.}~\bibnamefont {Maccione}},
  \bibinfo {author} {\bibfnamefont {D.}~\bibnamefont {Grasso}}, \bibinfo
  {author} {\bibfnamefont {G.}~\bibnamefont {Di~Bernardo}}, \ and\ \bibinfo
  {author} {\bibfnamefont {C.}~\bibnamefont {Evoli}},\ }\href {\doibase
  10.1103/PhysRevD.89.083007} {\bibfield  {journal} {\bibinfo  {journal} {Phys.
  Rev. D}\ }\textbf {\bibinfo {volume} {89}},\ \bibinfo {pages} {083007}
  (\bibinfo {year} {2014})}\BibitemShut {NoStop}%
\bibitem [{\citenamefont {{Gaggero}}\ \emph {et~al.}(2013)\citenamefont
  {{Gaggero}}, \citenamefont {{Maccione}}, \citenamefont {{Di Bernardo}},
  \citenamefont {{Evoli}},\ and\ \citenamefont
  {{Grasso}}}]{2013PhRvL.111b1102G}%
  \BibitemOpen
  \bibfield  {author} {\bibinfo {author} {\bibfnamefont {D.}~\bibnamefont
  {{Gaggero}}}, \bibinfo {author} {\bibfnamefont {L.}~\bibnamefont
  {{Maccione}}}, \bibinfo {author} {\bibfnamefont {G.}~\bibnamefont {{Di
  Bernardo}}}, \bibinfo {author} {\bibfnamefont {C.}~\bibnamefont {{Evoli}}}, \
  and\ \bibinfo {author} {\bibfnamefont {D.}~\bibnamefont {{Grasso}}},\ }\href
  {\doibase 10.1103/PhysRevLett.111.021102} {\bibfield  {journal} {\bibinfo
  {journal} {Physical Review Letters}\ }\textbf {\bibinfo {volume} {111}},\
  \bibinfo {eid} {021102} (\bibinfo {year} {2013})},\ \Eprint
  {http://arxiv.org/abs/1304.6718} {arXiv:1304.6718 [astro-ph.HE]} \BibitemShut
  {NoStop}%
\bibitem [{\citenamefont {Werner}\ \emph {et~al.}(2015)\citenamefont {Werner},
  \citenamefont {Kissmann}, \citenamefont {Strong},\ and\ \citenamefont
  {Reimer}}]{Werner201518}%
  \BibitemOpen
  \bibfield  {author} {\bibinfo {author} {\bibfnamefont {M.}~\bibnamefont
  {Werner}}, \bibinfo {author} {\bibfnamefont {R.}~\bibnamefont {Kissmann}},
  \bibinfo {author} {\bibfnamefont {A.}~\bibnamefont {Strong}}, \ and\ \bibinfo
  {author} {\bibfnamefont {O.}~\bibnamefont {Reimer}},\ }\href {\doibase
  http://dx.doi.org/10.1016/j.astropartphys.2014.10.005} {\bibfield  {journal}
  {\bibinfo  {journal} {Astroparticle Physics}\ }\textbf {\bibinfo {volume}
  {64}},\ \bibinfo {pages} {18 } (\bibinfo {year} {2015})}\BibitemShut
  {NoStop}%
\bibitem [{\citenamefont {Kissmann}\ \emph {et~al.}(2015)\citenamefont
  {Kissmann}, \citenamefont {Werner}, \citenamefont {Reimer},\ and\
  \citenamefont {Strong}}]{Kissmann201539}%
  \BibitemOpen
  \bibfield  {author} {\bibinfo {author} {\bibfnamefont {R.}~\bibnamefont
  {Kissmann}}, \bibinfo {author} {\bibfnamefont {M.}~\bibnamefont {Werner}},
  \bibinfo {author} {\bibfnamefont {O.}~\bibnamefont {Reimer}}, \ and\ \bibinfo
  {author} {\bibfnamefont {A.}~\bibnamefont {Strong}},\ }\href {\doibase
  http://dx.doi.org/10.1016/j.astropartphys.2015.04.003} {\bibfield  {journal}
  {\bibinfo  {journal} {Astroparticle Physics}\ }\textbf {\bibinfo {volume}
  {70}},\ \bibinfo {pages} {39 } (\bibinfo {year} {2015})}\BibitemShut
  {NoStop}%
\bibitem [{\citenamefont {Case}\ and\ \citenamefont
  {Bhattacharya}(1998)}]{Case:1998}%
  \BibitemOpen
  \bibfield  {author} {\bibinfo {author} {\bibfnamefont {G.~L.}\ \bibnamefont
  {Case}}\ and\ \bibinfo {author} {\bibfnamefont {D.}~\bibnamefont
  {Bhattacharya}},\ }\href {\doibase 10.1086/306089} {\bibfield  {journal}
  {\bibinfo  {journal} {The Astrophysical Journal}\ }\textbf {\bibinfo {volume}
  {504}},\ \bibinfo {pages} {761} (\bibinfo {year} {1998})},\ \Eprint
  {http://arxiv.org/abs/9807162} {arXiv:9807162 [astro-ph]} \BibitemShut
  {NoStop}%
\bibitem [{\citenamefont {{Green}}(2015)}]{Green:2015}%
  \BibitemOpen
  \bibfield  {author} {\bibinfo {author} {\bibfnamefont {D.~A.}\ \bibnamefont
  {{Green}}},\ }\href {\doibase 10.1093/mnras/stv1885} {\bibfield  {journal}
  {\bibinfo  {journal} {\mnras}\ }\textbf {\bibinfo {volume} {454}},\ \bibinfo
  {pages} {1517} (\bibinfo {year} {2015})},\ \Eprint
  {http://arxiv.org/abs/1508.02931} {arXiv:1508.02931 [astro-ph.HE]}
  \BibitemShut {NoStop}%
\bibitem [{\citenamefont {{Lorimer}}(2004)}]{Lorimer:2004}%
  \BibitemOpen
  \bibfield  {author} {\bibinfo {author} {\bibfnamefont {D.~R.}\ \bibnamefont
  {{Lorimer}}},\ }in\ \href@noop {} {\emph {\bibinfo {booktitle} {Young Neutron
  Stars and Their Environments}}},\ \bibinfo {series} {IAU Symposium}, Vol.\
  \bibinfo {volume} {218},\ \bibinfo {editor} {edited by\ \bibinfo {editor}
  {\bibfnamefont {F.}~\bibnamefont {{Camilo}}}\ and\ \bibinfo {editor}
  {\bibfnamefont {B.~M.}\ \bibnamefont {{Gaensler}}}}\ (\bibinfo {year}
  {2004})\ p.\ \bibinfo {pages} {105},\ \Eprint
  {http://arxiv.org/abs/astro-ph/0308501} {astro-ph/0308501} \BibitemShut
  {NoStop}%
\bibitem [{\citenamefont {Lorimer}\ \emph {et~al.}(2006)\citenamefont
  {Lorimer}, \citenamefont {Faulkner}, \citenamefont {Lyne}, \citenamefont
  {Manchester}, \citenamefont {Kramer}, \citenamefont {McLaughlin},
  \citenamefont {Hobbs}, \citenamefont {Possenti}, \citenamefont {Stairs},
  \citenamefont {Camilo}, \citenamefont {Burgay}, \citenamefont {D'Amico},
  \citenamefont {Corongiu},\ and\ \citenamefont {Crawford}}]{Lorimer:2006}%
  \BibitemOpen
  \bibfield  {author} {\bibinfo {author} {\bibfnamefont {D.~R.}\ \bibnamefont
  {Lorimer}}, \bibinfo {author} {\bibfnamefont {A.~J.}\ \bibnamefont
  {Faulkner}}, \bibinfo {author} {\bibfnamefont {A.~G.}\ \bibnamefont {Lyne}},
  \bibinfo {author} {\bibfnamefont {R.~N.}\ \bibnamefont {Manchester}},
  \bibinfo {author} {\bibfnamefont {M.}~\bibnamefont {Kramer}}, \bibinfo
  {author} {\bibfnamefont {M.~A.}\ \bibnamefont {McLaughlin}}, \bibinfo
  {author} {\bibfnamefont {G.}~\bibnamefont {Hobbs}}, \bibinfo {author}
  {\bibfnamefont {A.}~\bibnamefont {Possenti}}, \bibinfo {author}
  {\bibfnamefont {I.~H.}\ \bibnamefont {Stairs}}, \bibinfo {author}
  {\bibfnamefont {F.}~\bibnamefont {Camilo}}, \bibinfo {author} {\bibfnamefont
  {M.}~\bibnamefont {Burgay}}, \bibinfo {author} {\bibfnamefont
  {N.}~\bibnamefont {D'Amico}}, \bibinfo {author} {\bibfnamefont
  {A.}~\bibnamefont {Corongiu}}, \ and\ \bibinfo {author} {\bibfnamefont
  {F.}~\bibnamefont {Crawford}},\ }\href {\doibase
  10.1111/j.1365-2966.2006.10887.x} {\bibfield  {journal} {\bibinfo  {journal}
  {Monthly Notices of the Royal Astronomical Society}\ }\textbf {\bibinfo
  {volume} {372}},\ \bibinfo {pages} {777} (\bibinfo {year} {2006})},\ \Eprint
  {http://arxiv.org/abs/0607640v1} {arXiv:0607640v1 [arXiv:astro-ph]}
  \BibitemShut {NoStop}%
\bibitem [{\citenamefont {Yusifov}\ and\ \citenamefont
  {Kucuk}(2004)}]{Yusifov:2004}%
  \BibitemOpen
  \bibfield  {author} {\bibinfo {author} {\bibfnamefont {I.}~\bibnamefont
  {Yusifov}}\ and\ \bibinfo {author} {\bibfnamefont {I.}~\bibnamefont
  {Kucuk}},\ }\href {\doibase 10.1051/0004-6361:20040152} {\ ,\ \bibinfo
  {pages} {9} (\bibinfo {year} {2004})},\ \Eprint
  {http://arxiv.org/abs/0405559} {arXiv:0405559 [astro-ph]} \BibitemShut
  {NoStop}%
\bibitem [{\citenamefont {{Bronfman}}\ \emph {et~al.}(2000)\citenamefont
  {{Bronfman}}, \citenamefont {{Casassus}}, \citenamefont {{May}},\ and\
  \citenamefont {{Nyman}}}]{Bronfman:2000}%
  \BibitemOpen
  \bibfield  {author} {\bibinfo {author} {\bibfnamefont {L.}~\bibnamefont
  {{Bronfman}}}, \bibinfo {author} {\bibfnamefont {S.}~\bibnamefont
  {{Casassus}}}, \bibinfo {author} {\bibfnamefont {J.}~\bibnamefont {{May}}}, \
  and\ \bibinfo {author} {\bibfnamefont {L.-{\AA}.}\ \bibnamefont {{Nyman}}},\
  }\href@noop {} {\bibfield  {journal} {\bibinfo  {journal} {\aap}\ }\textbf
  {\bibinfo {volume} {358}},\ \bibinfo {pages} {521} (\bibinfo {year}
  {2000})},\ \Eprint {http://arxiv.org/abs/astro-ph/0006104} {astro-ph/0006104}
  \BibitemShut {NoStop}%
\bibitem [{\citenamefont {Montmerle}(1979)}]{Montmerle1979}%
  \BibitemOpen
  \bibfield  {author} {\bibinfo {author} {\bibfnamefont {T.}~\bibnamefont
  {Montmerle}},\ }\href {\doibase 10.1086/157166} {\bibfield  {journal}
  {\bibinfo  {journal} {The Astrophysical Journal}\ }\textbf {\bibinfo {volume}
  {231}},\ \bibinfo {pages} {95} (\bibinfo {year} {1979})}\BibitemShut
  {NoStop}%
\bibitem [{\citenamefont {Montmerle}(2009)}]{Montmerle2009}%
  \BibitemOpen
  \bibfield  {author} {\bibinfo {author} {\bibfnamefont {T.}~\bibnamefont
  {Montmerle}},\ }\href {http://arxiv.org/abs/0909.0222} {\ \textbf {\bibinfo
  {volume} {i}},\ \bibinfo {pages} {1} (\bibinfo {year} {2009})},\ \Eprint
  {http://arxiv.org/abs/0909.0222} {arXiv:0909.0222} \BibitemShut {NoStop}%
\bibitem [{\citenamefont {{Bally}}\ \emph {et~al.}(1987)\citenamefont
  {{Bally}}, \citenamefont {{Stark}}, \citenamefont {{Wilson}},\ and\
  \citenamefont {{Henkel}}}]{Bally:1987}%
  \BibitemOpen
  \bibfield  {author} {\bibinfo {author} {\bibfnamefont {J.}~\bibnamefont
  {{Bally}}}, \bibinfo {author} {\bibfnamefont {A.~A.}\ \bibnamefont
  {{Stark}}}, \bibinfo {author} {\bibfnamefont {R.~W.}\ \bibnamefont
  {{Wilson}}}, \ and\ \bibinfo {author} {\bibfnamefont {C.}~\bibnamefont
  {{Henkel}}},\ }\href {\doibase 10.1086/191217} {\bibfield  {journal}
  {\bibinfo  {journal} {\apjs}\ }\textbf {\bibinfo {volume} {65}},\ \bibinfo
  {pages} {13} (\bibinfo {year} {1987})}\BibitemShut {NoStop}%
\bibitem [{\citenamefont {{Ferri{\`e}re}}\ \emph {et~al.}(2007)\citenamefont
  {{Ferri{\`e}re}}, \citenamefont {{Gillard}},\ and\ \citenamefont
  {{Jean}}}]{Ferriere:2007}%
  \BibitemOpen
  \bibfield  {author} {\bibinfo {author} {\bibfnamefont {K.}~\bibnamefont
  {{Ferri{\`e}re}}}, \bibinfo {author} {\bibfnamefont {W.}~\bibnamefont
  {{Gillard}}}, \ and\ \bibinfo {author} {\bibfnamefont {P.}~\bibnamefont
  {{Jean}}},\ }\href {\doibase 10.1051/0004-6361:20066992} {\bibfield
  {journal} {\bibinfo  {journal} {Astronomy \& Astrophysics}\ }\textbf
  {\bibinfo {volume} {467}},\ \bibinfo {pages} {611} (\bibinfo {year}
  {2007})},\ \Eprint {http://arxiv.org/abs/astro-ph/0702532} {astro-ph/0702532}
  \BibitemShut {NoStop}%
\bibitem [{\citenamefont {{Figer}}\ \emph {et~al.}(2004)\citenamefont
  {{Figer}}, \citenamefont {{Rich}}, \citenamefont {{Kim}}, \citenamefont
  {{Morris}},\ and\ \citenamefont {{Serabyn}}}]{2004ApJ...601..319F}%
  \BibitemOpen
  \bibfield  {author} {\bibinfo {author} {\bibfnamefont {D.~F.}\ \bibnamefont
  {{Figer}}}, \bibinfo {author} {\bibfnamefont {R.~M.}\ \bibnamefont {{Rich}}},
  \bibinfo {author} {\bibfnamefont {S.~S.}\ \bibnamefont {{Kim}}}, \bibinfo
  {author} {\bibfnamefont {M.}~\bibnamefont {{Morris}}}, \ and\ \bibinfo
  {author} {\bibfnamefont {E.}~\bibnamefont {{Serabyn}}},\ }\href {\doibase
  10.1086/380392} {\bibfield  {journal} {\bibinfo  {journal} {\apj}\ }\textbf
  {\bibinfo {volume} {601}},\ \bibinfo {pages} {319} (\bibinfo {year}
  {2004})},\ \Eprint {http://arxiv.org/abs/astro-ph/0309757} {astro-ph/0309757}
  \BibitemShut {NoStop}%
\bibitem [{\citenamefont {Yusef-Zadeh}\ \emph
  {et~al.}(2009{\natexlab{a}})\citenamefont {Yusef-Zadeh}, \citenamefont
  {Hewitt}, \citenamefont {Arendt}, \citenamefont {Whitney}, \citenamefont
  {Rieke}, \citenamefont {Wardle}, \citenamefont {Hinz}, \citenamefont
  {Stolovy}, \citenamefont {Lang}, \citenamefont {Burton},\ and\ \citenamefont
  {Ramirez}}]{Yusef-Zadeh2009}%
  \BibitemOpen
  \bibfield  {author} {\bibinfo {author} {\bibfnamefont {F.}~\bibnamefont
  {Yusef-Zadeh}}, \bibinfo {author} {\bibfnamefont {J.~W.}\ \bibnamefont
  {Hewitt}}, \bibinfo {author} {\bibfnamefont {R.~G.}\ \bibnamefont {Arendt}},
  \bibinfo {author} {\bibfnamefont {B.}~\bibnamefont {Whitney}}, \bibinfo
  {author} {\bibfnamefont {G.}~\bibnamefont {Rieke}}, \bibinfo {author}
  {\bibfnamefont {M.}~\bibnamefont {Wardle}}, \bibinfo {author} {\bibfnamefont
  {J.~L.}\ \bibnamefont {Hinz}}, \bibinfo {author} {\bibfnamefont
  {S.}~\bibnamefont {Stolovy}}, \bibinfo {author} {\bibfnamefont {C.~C.}\
  \bibnamefont {Lang}}, \bibinfo {author} {\bibfnamefont {M.~G.}\ \bibnamefont
  {Burton}}, \ and\ \bibinfo {author} {\bibfnamefont {S.}~\bibnamefont
  {Ramirez}},\ }\href {\doibase 10.1088/0004-637X/702/1/178} {\bibfield
  {journal} {\bibinfo  {journal} {The Astrophysical Journal}\ }\textbf
  {\bibinfo {volume} {702}},\ \bibinfo {pages} {178} (\bibinfo {year}
  {2009}{\natexlab{a}})}\BibitemShut {NoStop}%
\bibitem [{\citenamefont {Immer}\ \emph {et~al.}(2012)\citenamefont {Immer},
  \citenamefont {Schuller}, \citenamefont {Omont},\ and\ \citenamefont
  {Menten}}]{Immer2012}%
  \BibitemOpen
  \bibfield  {author} {\bibinfo {author} {\bibfnamefont {K.}~\bibnamefont
  {Immer}}, \bibinfo {author} {\bibfnamefont {F.}~\bibnamefont {Schuller}},
  \bibinfo {author} {\bibfnamefont {A.}~\bibnamefont {Omont}}, \ and\ \bibinfo
  {author} {\bibfnamefont {K.~M.}\ \bibnamefont {Menten}},\ }\href {\doibase
  10.1051/0004-6361/201117857} {\bibfield  {journal} {\bibinfo  {journal}
  {Astronomy {\&} Astrophysics}\ }\textbf {\bibinfo {volume} {537}},\ \bibinfo
  {pages} {A121} (\bibinfo {year} {2012})}\BibitemShut {NoStop}%
\bibitem [{\citenamefont {Longmore}\ \emph {et~al.}(2013)\citenamefont
  {Longmore}, \citenamefont {Bally}, \citenamefont {Testi}, \citenamefont
  {Purcell}, \citenamefont {Walsh}, \citenamefont {Bressert}, \citenamefont
  {Pestalozzi}, \citenamefont {Molinari}, \citenamefont {Ott}, \citenamefont
  {Cortese}, \citenamefont {Battersby}, \citenamefont {Murray}, \citenamefont
  {Lee}, \citenamefont {Kruijssen}, \citenamefont {Schisano},\ and\
  \citenamefont {Elia}}]{Longmore2013}%
  \BibitemOpen
  \bibfield  {author} {\bibinfo {author} {\bibfnamefont {S.~N.}\ \bibnamefont
  {Longmore}}, \bibinfo {author} {\bibfnamefont {J.}~\bibnamefont {Bally}},
  \bibinfo {author} {\bibfnamefont {L.}~\bibnamefont {Testi}}, \bibinfo
  {author} {\bibfnamefont {C.~R.}\ \bibnamefont {Purcell}}, \bibinfo {author}
  {\bibfnamefont {A.~J.}\ \bibnamefont {Walsh}}, \bibinfo {author}
  {\bibfnamefont {E.}~\bibnamefont {Bressert}}, \bibinfo {author}
  {\bibfnamefont {M.}~\bibnamefont {Pestalozzi}}, \bibinfo {author}
  {\bibfnamefont {S.}~\bibnamefont {Molinari}}, \bibinfo {author}
  {\bibfnamefont {J.}~\bibnamefont {Ott}}, \bibinfo {author} {\bibfnamefont
  {L.}~\bibnamefont {Cortese}}, \bibinfo {author} {\bibfnamefont
  {C.}~\bibnamefont {Battersby}}, \bibinfo {author} {\bibfnamefont
  {N.}~\bibnamefont {Murray}}, \bibinfo {author} {\bibfnamefont
  {E.}~\bibnamefont {Lee}}, \bibinfo {author} {\bibfnamefont {J.~M.~D.}\
  \bibnamefont {Kruijssen}}, \bibinfo {author} {\bibfnamefont {E.}~\bibnamefont
  {Schisano}}, \ and\ \bibinfo {author} {\bibfnamefont {D.}~\bibnamefont
  {Elia}},\ }\href {\doibase 10.1093/mnras/sts376} {\bibfield  {journal}
  {\bibinfo  {journal} {Monthly Notices of the Royal Astronomical Society}\
  }\textbf {\bibinfo {volume} {429}},\ \bibinfo {pages} {987} (\bibinfo {year}
  {2013})},\ \Eprint {http://arxiv.org/abs/1208.4256v1} {arXiv:1208.4256v1}
  \BibitemShut {NoStop}%
\bibitem [{\citenamefont {Licquia}\ and\ \citenamefont
  {Newman}(2015)}]{Licquia2015}%
  \BibitemOpen
  \bibfield  {author} {\bibinfo {author} {\bibfnamefont {T.~C.}\ \bibnamefont
  {Licquia}}\ and\ \bibinfo {author} {\bibfnamefont {J.~A.}\ \bibnamefont
  {Newman}},\ }\href {\doibase 10.1088/0004-637X/806/1/96} {\bibfield
  {journal} {\bibinfo  {journal} {Astrophys J}\ }\textbf {\bibinfo {volume}
  {806}},\ \bibinfo {pages} {96} (\bibinfo {year} {2015})},\ \Eprint
  {http://arxiv.org/abs/1407.1078} {arXiv:1407.1078} \BibitemShut {NoStop}%
\bibitem [{\citenamefont {Rosslowe}\ and\ \citenamefont
  {Crowther}(2015)}]{Rosslowe2015}%
  \BibitemOpen
  \bibfield  {author} {\bibinfo {author} {\bibfnamefont {C.~K.}\ \bibnamefont
  {Rosslowe}}\ and\ \bibinfo {author} {\bibfnamefont {P.~a.}\ \bibnamefont
  {Crowther}},\ }\href {\doibase 10.1093/mnras/stv502} {\bibfield  {journal}
  {\bibinfo  {journal} {Monthly Notices of the Royal Astronomical Society}\
  }\textbf {\bibinfo {volume} {449}},\ \bibinfo {pages} {2436} (\bibinfo {year}
  {2015})},\ \Eprint {http://arxiv.org/abs/1412.0699} {arXiv:1412.0699}
  \BibitemShut {NoStop}%
\bibitem [{\citenamefont {{Murray}}(2011)}]{2011ApJ...729..133M}%
  \BibitemOpen
  \bibfield  {author} {\bibinfo {author} {\bibfnamefont {N.}~\bibnamefont
  {{Murray}}},\ }\href {\doibase 10.1088/0004-637X/729/2/133} {\bibfield
  {journal} {\bibinfo  {journal} {\apj}\ }\textbf {\bibinfo {volume} {729}},\
  \bibinfo {eid} {133} (\bibinfo {year} {2011})},\ \Eprint
  {http://arxiv.org/abs/1007.3270} {arXiv:1007.3270} \BibitemShut {NoStop}%
\bibitem [{\citenamefont {{Schmidt}}(1959)}]{1959ApJ...129..243S}%
  \BibitemOpen
  \bibfield  {author} {\bibinfo {author} {\bibfnamefont {M.}~\bibnamefont
  {{Schmidt}}},\ }\href {\doibase 10.1086/146614} {\bibfield  {journal}
  {\bibinfo  {journal} {\apj}\ }\textbf {\bibinfo {volume} {129}},\ \bibinfo
  {pages} {243} (\bibinfo {year} {1959})}\BibitemShut {NoStop}%
\bibitem [{\citenamefont {{Kennicutt}}(1998)}]{1998ApJ...498..541K}%
  \BibitemOpen
  \bibfield  {author} {\bibinfo {author} {\bibfnamefont {R.~C.}\ \bibnamefont
  {{Kennicutt}}, \bibfnamefont {Jr.}},\ }\href {\doibase 10.1086/305588}
  {\bibfield  {journal} {\bibinfo  {journal} {\apj}\ }\textbf {\bibinfo
  {volume} {498}},\ \bibinfo {pages} {541} (\bibinfo {year} {1998})},\ \Eprint
  {http://arxiv.org/abs/astro-ph/9712213} {astro-ph/9712213} \BibitemShut
  {NoStop}%
\bibitem [{\citenamefont {Schaye}\ and\ \citenamefont {{Dalla
  Vecchia}}(2008)}]{Schaye2008}%
  \BibitemOpen
  \bibfield  {author} {\bibinfo {author} {\bibfnamefont {J.}~\bibnamefont
  {Schaye}}\ and\ \bibinfo {author} {\bibfnamefont {C.}~\bibnamefont {{Dalla
  Vecchia}}},\ }\href {\doibase 10.1111/j.1365-2966.2007.12639.x} {\bibfield
  {journal} {\bibinfo  {journal} {Monthly Notices of the Royal Astronomical
  Society}\ }\textbf {\bibinfo {volume} {383}},\ \bibinfo {pages} {1210}
  (\bibinfo {year} {2008})},\ \Eprint {http://arxiv.org/abs/0709.0292}
  {arXiv:0709.0292} \BibitemShut {NoStop}%
\bibitem [{\citenamefont {Krumholz}\ and\ \citenamefont
  {McKee}(2005)}]{0004-637X-630-1-250}%
  \BibitemOpen
  \bibfield  {author} {\bibinfo {author} {\bibfnamefont {M.~R.}\ \bibnamefont
  {Krumholz}}\ and\ \bibinfo {author} {\bibfnamefont {C.~F.}\ \bibnamefont
  {McKee}},\ }\href {http://stacks.iop.org/0004-637X/630/i=1/a=250} {\bibfield
  {journal} {\bibinfo  {journal} {The Astrophysical Journal}\ }\textbf
  {\bibinfo {volume} {630}},\ \bibinfo {pages} {250} (\bibinfo {year}
  {2005})}\BibitemShut {NoStop}%
\bibitem [{\citenamefont {Cunningham}\ \emph {et~al.}(2011)\citenamefont
  {Cunningham}, \citenamefont {Klein}, \citenamefont {Krumholz},\ and\
  \citenamefont {McKee}}]{0004-637X-740-2-107}%
  \BibitemOpen
  \bibfield  {author} {\bibinfo {author} {\bibfnamefont {A.~J.}\ \bibnamefont
  {Cunningham}}, \bibinfo {author} {\bibfnamefont {R.~I.}\ \bibnamefont
  {Klein}}, \bibinfo {author} {\bibfnamefont {M.~R.}\ \bibnamefont {Krumholz}},
  \ and\ \bibinfo {author} {\bibfnamefont {C.~F.}\ \bibnamefont {McKee}},\
  }\href {http://stacks.iop.org/0004-637X/740/i=2/a=107} {\bibfield  {journal}
  {\bibinfo  {journal} {The Astrophysical Journal}\ }\textbf {\bibinfo {volume}
  {740}},\ \bibinfo {pages} {107} (\bibinfo {year} {2011})}\BibitemShut
  {NoStop}%
\bibitem [{\citenamefont {Brunetti}\ and\ \citenamefont
  {Codino}(2000)}]{0004-637X-528-2-789}%
  \BibitemOpen
  \bibfield  {author} {\bibinfo {author} {\bibfnamefont {M.~T.}\ \bibnamefont
  {Brunetti}}\ and\ \bibinfo {author} {\bibfnamefont {A.}~\bibnamefont
  {Codino}},\ }\href {http://stacks.iop.org/0004-637X/528/i=2/a=789} {\bibfield
   {journal} {\bibinfo  {journal} {The Astrophysical Journal}\ }\textbf
  {\bibinfo {volume} {528}},\ \bibinfo {pages} {789} (\bibinfo {year}
  {2000})}\BibitemShut {NoStop}%
\bibitem [{\citenamefont {{Codino}}(1999)}]{1999ICRC....4..314C}%
  \BibitemOpen
  \bibfield  {author} {\bibinfo {author} {\bibfnamefont {A.}~\bibnamefont
  {{Codino}}},\ }\href@noop {} {\bibfield  {journal} {\bibinfo  {journal}
  {International Cosmic Ray Conference}\ }\textbf {\bibinfo {volume} {4}},\
  \bibinfo {pages} {314} (\bibinfo {year} {1999})}\BibitemShut {NoStop}%
\bibitem [{\citenamefont {Sandstrom}\ \emph {et~al.}(2013)\citenamefont
  {Sandstrom}, \citenamefont {Leroy}, \citenamefont {Walter}, \citenamefont
  {Bolatto}, \citenamefont {Croxall} \emph {et~al.}}]{Sandstrom:2012ni}%
  \BibitemOpen
  \bibfield  {author} {\bibinfo {author} {\bibfnamefont {K.}~\bibnamefont
  {Sandstrom}}, \bibinfo {author} {\bibfnamefont {A.}~\bibnamefont {Leroy}},
  \bibinfo {author} {\bibfnamefont {F.}~\bibnamefont {Walter}}, \bibinfo
  {author} {\bibfnamefont {A.}~\bibnamefont {Bolatto}}, \bibinfo {author}
  {\bibfnamefont {K.}~\bibnamefont {Croxall}},  \emph {et~al.},\ }\href
  {\doibase 10.1088/0004-637X/777/1/5} {\bibfield  {journal} {\bibinfo
  {journal} {Astrophys.J.}\ }\textbf {\bibinfo {volume} {777}},\ \bibinfo
  {pages} {5} (\bibinfo {year} {2013})},\ \Eprint
  {http://arxiv.org/abs/1212.1208} {arXiv:1212.1208 [astro-ph.CO]} \BibitemShut
  {NoStop}%
%%CITATION = ARXIV:1212.1208;%%
\bibitem [{\citenamefont {{Strong}}\ \emph
  {et~al.}(2004{\natexlab{a}})\citenamefont {{Strong}}, \citenamefont
  {{Moskalenko}}, \citenamefont {{Reimer}}, \citenamefont {{Digel}},\ and\
  \citenamefont {{Diehl}}}]{MS:2004}%
  \BibitemOpen
  \bibfield  {author} {\bibinfo {author} {\bibfnamefont {A.~W.}\ \bibnamefont
  {{Strong}}}, \bibinfo {author} {\bibfnamefont {I.~V.}\ \bibnamefont
  {{Moskalenko}}}, \bibinfo {author} {\bibfnamefont {O.}~\bibnamefont
  {{Reimer}}}, \bibinfo {author} {\bibfnamefont {S.}~\bibnamefont {{Digel}}}, \
  and\ \bibinfo {author} {\bibfnamefont {R.}~\bibnamefont {{Diehl}}},\ }\href
  {\doibase 10.1051/0004-6361:20040172} {\bibfield  {journal} {\bibinfo
  {journal} {\aap}\ }\textbf {\bibinfo {volume} {422}},\ \bibinfo {pages} {L47}
  (\bibinfo {year} {2004}{\natexlab{a}})},\ \Eprint
  {http://arxiv.org/abs/astro-ph/0405275} {astro-ph/0405275} \BibitemShut
  {NoStop}%
\bibitem [{\citenamefont {{Heyer}}\ and\ \citenamefont
  {{Dame}}(2015)}]{2015ARA&A..53..583H}%
  \BibitemOpen
  \bibfield  {author} {\bibinfo {author} {\bibfnamefont {M.}~\bibnamefont
  {{Heyer}}}\ and\ \bibinfo {author} {\bibfnamefont {T.~M.}\ \bibnamefont
  {{Dame}}},\ }\href {\doibase 10.1146/annurev-astro-082214-122324} {\bibfield
  {journal} {\bibinfo  {journal} {\araa}\ }\textbf {\bibinfo {volume} {53}},\
  \bibinfo {pages} {583} (\bibinfo {year} {2015})}\BibitemShut {NoStop}%
\bibitem [{\citenamefont {{Bolatto}}\ \emph {et~al.}(2013)\citenamefont
  {{Bolatto}}, \citenamefont {{Wolfire}},\ and\ \citenamefont
  {{Leroy}}}]{Bolatto:2013}%
  \BibitemOpen
  \bibfield  {author} {\bibinfo {author} {\bibfnamefont {A.~D.}\ \bibnamefont
  {{Bolatto}}}, \bibinfo {author} {\bibfnamefont {M.}~\bibnamefont
  {{Wolfire}}}, \ and\ \bibinfo {author} {\bibfnamefont {A.~K.}\ \bibnamefont
  {{Leroy}}},\ }\href {\doibase 10.1146/annurev-astro-082812-140944} {\bibfield
   {journal} {\bibinfo  {journal} {\araa}\ }\textbf {\bibinfo {volume} {51}},\
  \bibinfo {pages} {207} (\bibinfo {year} {2013})},\ \Eprint
  {http://arxiv.org/abs/1301.3498} {arXiv:1301.3498} \BibitemShut {NoStop}%
\bibitem [{\citenamefont {Crocker}\ and\ \citenamefont
  {Aharonian}(2010)}]{Crocker2010}%
  \BibitemOpen
  \bibfield  {author} {\bibinfo {author} {\bibfnamefont {R.~M.}\ \bibnamefont
  {Crocker}}\ and\ \bibinfo {author} {\bibfnamefont {F.}~\bibnamefont
  {Aharonian}},\ }\href {\doibase 10.1103/PhysRevLett.106.101102} {\ ,\
  \bibinfo {pages} {4} (\bibinfo {year} {2010})},\ \Eprint
  {http://arxiv.org/abs/1008.2658} {arXiv:1008.2658} \BibitemShut {NoStop}%
\bibitem [{\citenamefont {{Bartko}}\ \emph {et~al.}(2010)\citenamefont
  {{Bartko}}, \citenamefont {{Martins}}, \citenamefont {{Trippe}},
  \citenamefont {{Fritz}}, \citenamefont {{Genzel}}, \citenamefont {{Ott}},
  \citenamefont {{Eisenhauer}}, \citenamefont {{Gillessen}}, \citenamefont
  {{Paumard}}, \citenamefont {{Alexander}}, \citenamefont {{Dodds-Eden}},
  \citenamefont {{Gerhard}}, \citenamefont {{Levin}}, \citenamefont
  {{Mascetti}}, \citenamefont {{Nayakshin}}, \citenamefont {{Perets}},
  \citenamefont {{Perrin}}, \citenamefont {{Pfuhl}}, \citenamefont {{Reid}},
  \citenamefont {{Rouan}}, \citenamefont {{Zilka}},\ and\ \citenamefont
  {{Sternberg}}}]{2010ApJ...708..834B}%
  \BibitemOpen
  \bibfield  {author} {\bibinfo {author} {\bibfnamefont {H.}~\bibnamefont
  {{Bartko}}}, \bibinfo {author} {\bibfnamefont {F.}~\bibnamefont {{Martins}}},
  \bibinfo {author} {\bibfnamefont {S.}~\bibnamefont {{Trippe}}}, \bibinfo
  {author} {\bibfnamefont {T.~K.}\ \bibnamefont {{Fritz}}}, \bibinfo {author}
  {\bibfnamefont {R.}~\bibnamefont {{Genzel}}}, \bibinfo {author}
  {\bibfnamefont {T.}~\bibnamefont {{Ott}}}, \bibinfo {author} {\bibfnamefont
  {F.}~\bibnamefont {{Eisenhauer}}}, \bibinfo {author} {\bibfnamefont
  {S.}~\bibnamefont {{Gillessen}}}, \bibinfo {author} {\bibfnamefont
  {T.}~\bibnamefont {{Paumard}}}, \bibinfo {author} {\bibfnamefont
  {T.}~\bibnamefont {{Alexander}}}, \bibinfo {author} {\bibfnamefont
  {K.}~\bibnamefont {{Dodds-Eden}}}, \bibinfo {author} {\bibfnamefont
  {O.}~\bibnamefont {{Gerhard}}}, \bibinfo {author} {\bibfnamefont
  {Y.}~\bibnamefont {{Levin}}}, \bibinfo {author} {\bibfnamefont
  {L.}~\bibnamefont {{Mascetti}}}, \bibinfo {author} {\bibfnamefont
  {S.}~\bibnamefont {{Nayakshin}}}, \bibinfo {author} {\bibfnamefont {H.~B.}\
  \bibnamefont {{Perets}}}, \bibinfo {author} {\bibfnamefont {G.}~\bibnamefont
  {{Perrin}}}, \bibinfo {author} {\bibfnamefont {O.}~\bibnamefont {{Pfuhl}}},
  \bibinfo {author} {\bibfnamefont {M.~J.}\ \bibnamefont {{Reid}}}, \bibinfo
  {author} {\bibfnamefont {D.}~\bibnamefont {{Rouan}}}, \bibinfo {author}
  {\bibfnamefont {M.}~\bibnamefont {{Zilka}}}, \ and\ \bibinfo {author}
  {\bibfnamefont {A.}~\bibnamefont {{Sternberg}}},\ }\href {\doibase
  10.1088/0004-637X/708/1/834} {\bibfield  {journal} {\bibinfo  {journal}
  {\apj}\ }\textbf {\bibinfo {volume} {708}},\ \bibinfo {pages} {834} (\bibinfo
  {year} {2010})},\ \Eprint {http://arxiv.org/abs/0908.2177} {arXiv:0908.2177}
  \BibitemShut {NoStop}%
\bibitem [{\citenamefont {{L{\"o}ckmann}}\ \emph {et~al.}(2010)\citenamefont
  {{L{\"o}ckmann}}, \citenamefont {{Baumgardt}},\ and\ \citenamefont
  {{Kroupa}}}]{2010MNRAS.402..519L}%
  \BibitemOpen
  \bibfield  {author} {\bibinfo {author} {\bibfnamefont {U.}~\bibnamefont
  {{L{\"o}ckmann}}}, \bibinfo {author} {\bibfnamefont {H.}~\bibnamefont
  {{Baumgardt}}}, \ and\ \bibinfo {author} {\bibfnamefont {P.}~\bibnamefont
  {{Kroupa}}},\ }\href {\doibase 10.1111/j.1365-2966.2009.15906.x} {\bibfield
  {journal} {\bibinfo  {journal} {\mnras}\ }\textbf {\bibinfo {volume} {402}},\
  \bibinfo {pages} {519} (\bibinfo {year} {2010})},\ \Eprint
  {http://arxiv.org/abs/0910.4960} {arXiv:0910.4960} \BibitemShut {NoStop}%
\bibitem [{\citenamefont {{Lu}}\ \emph {et~al.}(2013)\citenamefont {{Lu}},
  \citenamefont {{Do}}, \citenamefont {{Ghez}}, \citenamefont {{Morris}},
  \citenamefont {{Yelda}},\ and\ \citenamefont
  {{Matthews}}}]{2013ApJ...764..155L}%
  \BibitemOpen
  \bibfield  {author} {\bibinfo {author} {\bibfnamefont {J.~R.}\ \bibnamefont
  {{Lu}}}, \bibinfo {author} {\bibfnamefont {T.}~\bibnamefont {{Do}}}, \bibinfo
  {author} {\bibfnamefont {A.~M.}\ \bibnamefont {{Ghez}}}, \bibinfo {author}
  {\bibfnamefont {M.~R.}\ \bibnamefont {{Morris}}}, \bibinfo {author}
  {\bibfnamefont {S.}~\bibnamefont {{Yelda}}}, \ and\ \bibinfo {author}
  {\bibfnamefont {K.}~\bibnamefont {{Matthews}}},\ }\href {\doibase
  10.1088/0004-637X/764/2/155} {\bibfield  {journal} {\bibinfo  {journal}
  {\apj}\ }\textbf {\bibinfo {volume} {764}},\ \bibinfo {eid} {155} (\bibinfo
  {year} {2013})},\ \Eprint {http://arxiv.org/abs/1301.0540} {arXiv:1301.0540
  [astro-ph.SR]} \BibitemShut {NoStop}%
\bibitem [{\citenamefont {{Strong}}\ \emph
  {et~al.}(2004{\natexlab{b}})\citenamefont {{Strong}}, \citenamefont
  {{Moskalenko}}, \citenamefont {{Reimer}}, \citenamefont {{Digel}},\ and\
  \citenamefont {{Diehl}}}]{galprop_x_co}%
  \BibitemOpen
  \bibfield  {author} {\bibinfo {author} {\bibfnamefont {A.~W.}\ \bibnamefont
  {{Strong}}}, \bibinfo {author} {\bibfnamefont {I.~V.}\ \bibnamefont
  {{Moskalenko}}}, \bibinfo {author} {\bibfnamefont {O.}~\bibnamefont
  {{Reimer}}}, \bibinfo {author} {\bibfnamefont {S.}~\bibnamefont {{Digel}}}, \
  and\ \bibinfo {author} {\bibfnamefont {R.}~\bibnamefont {{Diehl}}},\ }\href
  {\doibase 10.1051/0004-6361:20040172} {\bibfield  {journal} {\bibinfo
  {journal} {A\&A}\ }\textbf {\bibinfo {volume} {422}},\ \bibinfo {pages} {L47}
  (\bibinfo {year} {2004}{\natexlab{b}})},\ \Eprint
  {http://arxiv.org/abs/astro-ph/0405275} {astro-ph/0405275} \BibitemShut
  {NoStop}%
\bibitem [{\citenamefont {Yuan}\ \emph {et~al.}(2013)\citenamefont {Yuan},
  \citenamefont {Funk}, \citenamefont {J\'{o}hannesson}, \citenamefont {Lande},
  \citenamefont {Tibaldo},\ and\ \citenamefont {Uchiyama}}]{fermi_snr1}%
  \BibitemOpen
  \bibfield  {author} {\bibinfo {author} {\bibfnamefont {Y.}~\bibnamefont
  {Yuan}}, \bibinfo {author} {\bibfnamefont {S.}~\bibnamefont {Funk}}, \bibinfo
  {author} {\bibfnamefont {G.}~\bibnamefont {J\'{o}hannesson}}, \bibinfo
  {author} {\bibfnamefont {J.}~\bibnamefont {Lande}}, \bibinfo {author}
  {\bibfnamefont {L.}~\bibnamefont {Tibaldo}}, \ and\ \bibinfo {author}
  {\bibfnamefont {Y.}~\bibnamefont {Uchiyama}},\ }\href
  {http://arxiv.org/pdf/1310.8287.pdf http://arxiv.org/abs/1310.8287} {\ ,\
  \bibinfo {pages} {18} (\bibinfo {year} {2013})},\ \Eprint
  {http://arxiv.org/abs/1310.8287} {arXiv:1310.8287} \BibitemShut {NoStop}%
\bibitem [{\citenamefont {Uchiyama}\ \emph {et~al.}(2012)\citenamefont
  {Uchiyama}, \citenamefont {Funk}, \citenamefont {Katagiri}, \citenamefont
  {Katsuta}, \citenamefont {Lemoine-Goumard}, \citenamefont {Tajima},
  \citenamefont {Tanaka},\ and\ \citenamefont {Torres}}]{fermi_snr2}%
  \BibitemOpen
  \bibfield  {author} {\bibinfo {author} {\bibfnamefont {Y.}~\bibnamefont
  {Uchiyama}}, \bibinfo {author} {\bibfnamefont {S.}~\bibnamefont {Funk}},
  \bibinfo {author} {\bibfnamefont {H.}~\bibnamefont {Katagiri}}, \bibinfo
  {author} {\bibfnamefont {J.}~\bibnamefont {Katsuta}}, \bibinfo {author}
  {\bibfnamefont {M.}~\bibnamefont {Lemoine-Goumard}}, \bibinfo {author}
  {\bibfnamefont {H.}~\bibnamefont {Tajima}}, \bibinfo {author} {\bibfnamefont
  {T.}~\bibnamefont {Tanaka}}, \ and\ \bibinfo {author} {\bibfnamefont
  {D.}~\bibnamefont {Torres}},\ }\href {http://arxiv.org/pdf/1203.3234.pdf
  http://arxiv.org/abs/1203.3234} {\bibfield  {journal} {\bibinfo  {journal}
  {arXiv preprint arXiv: \ldots}\ ,\ \bibinfo {pages} {5}} (\bibinfo {year}
  {2012})},\ \Eprint {http://arxiv.org/abs/1203.3234} {arXiv:1203.3234}
  \BibitemShut {NoStop}%
\bibitem [{\citenamefont {Castro}\ and\ \citenamefont
  {Slane}(2010)}]{fermi_snr3}%
  \BibitemOpen
  \bibfield  {author} {\bibinfo {author} {\bibfnamefont {D.}~\bibnamefont
  {Castro}}\ and\ \bibinfo {author} {\bibfnamefont {P.}~\bibnamefont {Slane}},\
  }\href {\doibase 10.1088/0004-637X/717/1/372} {\bibfield  {journal} {\bibinfo
   {journal} {The Astrophysical Journal}\ }\textbf {\bibinfo {volume} {717}},\
  \bibinfo {pages} {372} (\bibinfo {year} {2010})}\BibitemShut {NoStop}%
\bibitem [{\citenamefont {Giuliani}(2011)}]{fermi_snr4}%
  \BibitemOpen
  \bibfield  {author} {\bibinfo {author} {\bibfnamefont {A.~e.~a.}\
  \bibnamefont {Giuliani}},\ }\href {http://arxiv.org/abs/1111.4868} {\
  (\bibinfo {year} {2011})},\ \Eprint {http://arxiv.org/abs/1111.4868}
  {arXiv:1111.4868} \BibitemShut {NoStop}%
\bibitem [{\citenamefont {Dermer}\ and\ \citenamefont
  {Powale}(2012)}]{Dermer2013c}%
  \BibitemOpen
  \bibfield  {author} {\bibinfo {author} {\bibfnamefont {C.~D.}\ \bibnamefont
  {Dermer}}\ and\ \bibinfo {author} {\bibfnamefont {G.}~\bibnamefont
  {Powale}},\ }\href {http://arxiv.org/pdf/1210.8071v2.pdf
  http://arxiv.org/abs/1210.8071} {\ ,\ \bibinfo {pages} {1} (\bibinfo {year}
  {2012})},\ \Eprint {http://arxiv.org/abs/1210.8071} {arXiv:1210.8071}
  \BibitemShut {NoStop}%
\bibitem [{\citenamefont {Araya}(2014)}]{fermi_snr5}%
  \BibitemOpen
  \bibfield  {author} {\bibinfo {author} {\bibfnamefont {M.}~\bibnamefont
  {Araya}},\ }\href {http://arxiv.org/pdf/1405.4554.pdf
  http://arxiv.org/abs/1405.4554} {\ ,\ \bibinfo {pages} {15} (\bibinfo {year}
  {2014})},\ \Eprint {http://arxiv.org/abs/1405.4554} {arXiv:1405.4554}
  \BibitemShut {NoStop}%
\bibitem [{\citenamefont {{O'C Drury}}\ \emph {et~al.}(1996)\citenamefont {{O'C
  Drury}}, \citenamefont {{Duffy}},\ and\ \citenamefont
  {{Kirk}}}]{1996A&A...309.1002O}%
  \BibitemOpen
  \bibfield  {author} {\bibinfo {author} {\bibfnamefont {L.}~\bibnamefont {{O'C
  Drury}}}, \bibinfo {author} {\bibfnamefont {P.}~\bibnamefont {{Duffy}}}, \
  and\ \bibinfo {author} {\bibfnamefont {J.~G.}\ \bibnamefont {{Kirk}}},\
  }\href@noop {} {\bibfield  {journal} {\bibinfo  {journal} {\aap}\ }\textbf
  {\bibinfo {volume} {309}},\ \bibinfo {pages} {1002} (\bibinfo {year}
  {1996})},\ \Eprint {http://arxiv.org/abs/astro-ph/9510066} {astro-ph/9510066}
  \BibitemShut {NoStop}%
\bibitem [{\citenamefont {Malkov}\ \emph {et~al.}(2005)\citenamefont {Malkov},
  \citenamefont {Diamond},\ and\ \citenamefont {Sagdeev}}]{Malkov2005}%
  \BibitemOpen
  \bibfield  {author} {\bibinfo {author} {\bibfnamefont {M.~A.}\ \bibnamefont
  {Malkov}}, \bibinfo {author} {\bibfnamefont {P.~H.}\ \bibnamefont {Diamond}},
  \ and\ \bibinfo {author} {\bibfnamefont {R.~Z.}\ \bibnamefont {Sagdeev}},\
  }\href {\doibase 10.1086/430344} {\bibfield  {journal} {\bibinfo  {journal}
  {arXiv preprint astro-ph/0503403}\ ,\ \bibinfo {pages} {11}} (\bibinfo {year}
  {2005})},\ \Eprint {http://arxiv.org/abs/0503403} {arXiv:0503403 [astro-ph]}
  \BibitemShut {NoStop}%
\bibitem [{\citenamefont {Malkov}\ \emph {et~al.}(2011)\citenamefont {Malkov},
  \citenamefont {Diamond},\ and\ \citenamefont {Sagdeev}}]{Malkov2011}%
  \BibitemOpen
  \bibfield  {author} {\bibinfo {author} {\bibfnamefont {M.}~\bibnamefont
  {Malkov}}, \bibinfo {author} {\bibfnamefont {P.}~\bibnamefont {Diamond}}, \
  and\ \bibinfo {author} {\bibfnamefont {R.}~\bibnamefont {Sagdeev}},\ }\href
  {\doibase 10.1038/ncomms1195} {\bibfield  {journal} {\bibinfo  {journal}
  {Nature Communications}\ ,\ \bibinfo {pages} {1}} (\bibinfo {year} {2011})},\
  \Eprint {http://arxiv.org/abs/1004.4714} {arXiv:1004.4714} \BibitemShut
  {NoStop}%
\bibitem [{\citenamefont {Blasi}\ \emph {et~al.}(2012)\citenamefont {Blasi},
  \citenamefont {Morlino}, \citenamefont {Bandiera}, \citenamefont {Amato},\
  and\ \citenamefont {Caprioli}}]{Blasi2012}%
  \BibitemOpen
  \bibfield  {author} {\bibinfo {author} {\bibfnamefont {P.}~\bibnamefont
  {Blasi}}, \bibinfo {author} {\bibfnamefont {G.}~\bibnamefont {Morlino}},
  \bibinfo {author} {\bibfnamefont {R.}~\bibnamefont {Bandiera}}, \bibinfo
  {author} {\bibfnamefont {E.}~\bibnamefont {Amato}}, \ and\ \bibinfo {author}
  {\bibfnamefont {D.}~\bibnamefont {Caprioli}},\ }\href
  {http://arxiv.org/pdf/1202.3080v2.pdf http://arxiv.org/abs/1202.3080v2
  http://arxiv.org/abs/1202.3080} {\  (\bibinfo {year} {2012})},\ \Eprint
  {http://arxiv.org/abs/1202.3080} {arXiv:1202.3080} \BibitemShut {NoStop}%
\bibitem [{\citenamefont {For}\ and\ \citenamefont {In}(1999)}]{uchiyama2010}%
  \BibitemOpen
  \bibfield  {author} {\bibinfo {author} {\bibfnamefont {C.}~\bibnamefont
  {For}}\ and\ \bibinfo {author} {\bibfnamefont {P.}~\bibnamefont {In}},\
  }\href {http://arxiv.org/pdf/1008.1840v2.pdf} {\ ,\ \bibinfo {pages} {1}
  (\bibinfo {year} {1999})},\ \Eprint {http://arxiv.org/abs/arXiv:1008.1840v2}
  {arXiv:arXiv:1008.1840v2} \BibitemShut {NoStop}%
\bibitem [{\citenamefont {{Acero}}\ \emph {et~al.}(2015)\citenamefont
  {{Acero}}, \citenamefont {{Ackermann}}, \citenamefont {{Ajello}},
  \citenamefont {{Albert}}, \citenamefont {{Atwood}}, \citenamefont
  {{Axelsson}}, \citenamefont {{Baldini}}, \citenamefont {{Ballet}},\ and\
  \citenamefont {{et al.}}}]{3FGL}%
  \BibitemOpen
  \bibfield  {author} {\bibinfo {author} {\bibfnamefont {F.}~\bibnamefont
  {{Acero}}}, \bibinfo {author} {\bibfnamefont {M.}~\bibnamefont
  {{Ackermann}}}, \bibinfo {author} {\bibfnamefont {M.}~\bibnamefont
  {{Ajello}}}, \bibinfo {author} {\bibfnamefont {A.}~\bibnamefont {{Albert}}},
  \bibinfo {author} {\bibfnamefont {W.~B.}\ \bibnamefont {{Atwood}}}, \bibinfo
  {author} {\bibfnamefont {M.}~\bibnamefont {{Axelsson}}}, \bibinfo {author}
  {\bibfnamefont {L.}~\bibnamefont {{Baldini}}}, \bibinfo {author}
  {\bibfnamefont {J.}~\bibnamefont {{Ballet}}}, \ and\ \bibinfo {author}
  {\bibnamefont {{et al.}}},\ }\href {\doibase 10.1088/0067-0049/218/2/23}
  {\bibfield  {journal} {\bibinfo  {journal} {\apjs}\ }\textbf {\bibinfo
  {volume} {218}},\ \bibinfo {eid} {23} (\bibinfo {year} {2015})},\ \Eprint
  {http://arxiv.org/abs/1501.02003} {arXiv:1501.02003 [astro-ph.HE]}
  \BibitemShut {NoStop}%
\bibitem [{\citenamefont {Su}\ \emph {et~al.}(2010)\citenamefont {Su},
  \citenamefont {Slatyer},\ and\ \citenamefont
  {Finkbeiner}}]{0004-637X-724-2-1044}%
  \BibitemOpen
  \bibfield  {author} {\bibinfo {author} {\bibfnamefont {M.}~\bibnamefont
  {Su}}, \bibinfo {author} {\bibfnamefont {T.~R.}\ \bibnamefont {Slatyer}}, \
  and\ \bibinfo {author} {\bibfnamefont {D.~P.}\ \bibnamefont {Finkbeiner}},\
  }\href {http://stacks.iop.org/0004-637X/724/i=2/a=1044} {\bibfield  {journal}
  {\bibinfo  {journal} {The Astrophysical Journal}\ }\textbf {\bibinfo {volume}
  {724}},\ \bibinfo {pages} {1044} (\bibinfo {year} {2010})}\BibitemShut
  {NoStop}%
\bibitem [{\citenamefont {Su}()}]{bubbles_correction}%
  \BibitemOpen
  \bibfield  {author} {\bibinfo {author} {\bibfnamefont {M.}~\bibnamefont
  {Su}},\ }\href@noop {} {}\bibinfo {howpublished} {Private
  Communication}\BibitemShut {NoStop}%
\bibitem [{\citenamefont {{Ackermann}}\ \emph {et~al.}(2014)\citenamefont
  {{Ackermann}}, \citenamefont {{Albert}}, \citenamefont {{Atwood}},
  \citenamefont {{Baldini}}, \citenamefont {{Ballet}}, \citenamefont
  {{Barbiellini}}, \citenamefont {{Bastieri}}, \citenamefont {{Bellazzini}},\
  and\ \citenamefont {{et al.}}}]{2014ApJ...793...64A}%
  \BibitemOpen
  \bibfield  {author} {\bibinfo {author} {\bibfnamefont {M.}~\bibnamefont
  {{Ackermann}}}, \bibinfo {author} {\bibfnamefont {A.}~\bibnamefont
  {{Albert}}}, \bibinfo {author} {\bibfnamefont {W.~B.}\ \bibnamefont
  {{Atwood}}}, \bibinfo {author} {\bibfnamefont {L.}~\bibnamefont {{Baldini}}},
  \bibinfo {author} {\bibfnamefont {J.}~\bibnamefont {{Ballet}}}, \bibinfo
  {author} {\bibfnamefont {G.}~\bibnamefont {{Barbiellini}}}, \bibinfo {author}
  {\bibfnamefont {D.}~\bibnamefont {{Bastieri}}}, \bibinfo {author}
  {\bibfnamefont {R.}~\bibnamefont {{Bellazzini}}}, \ and\ \bibinfo {author}
  {\bibnamefont {{et al.}}},\ }\href {\doibase 10.1088/0004-637X/793/1/64}
  {\bibfield  {journal} {\bibinfo  {journal} {\apj}\ }\textbf {\bibinfo
  {volume} {793}},\ \bibinfo {eid} {64} (\bibinfo {year} {2014})},\ \Eprint
  {http://arxiv.org/abs/1407.7905} {arXiv:1407.7905 [astro-ph.HE]} \BibitemShut
  {NoStop}%
\bibitem [{\citenamefont {{G{\'o}rski}}\ \emph {et~al.}(2005)\citenamefont
  {{G{\'o}rski}}, \citenamefont {{Hivon}}, \citenamefont {{Banday}},
  \citenamefont {{Wandelt}}, \citenamefont {{Hansen}}, \citenamefont
  {{Reinecke}},\ and\ \citenamefont {{Bartelmann}}}]{2005ApJ...622..759G}%
  \BibitemOpen
  \bibfield  {author} {\bibinfo {author} {\bibfnamefont {K.~M.}\ \bibnamefont
  {{G{\'o}rski}}}, \bibinfo {author} {\bibfnamefont {E.}~\bibnamefont
  {{Hivon}}}, \bibinfo {author} {\bibfnamefont {A.~J.}\ \bibnamefont
  {{Banday}}}, \bibinfo {author} {\bibfnamefont {B.~D.}\ \bibnamefont
  {{Wandelt}}}, \bibinfo {author} {\bibfnamefont {F.~K.}\ \bibnamefont
  {{Hansen}}}, \bibinfo {author} {\bibfnamefont {M.}~\bibnamefont
  {{Reinecke}}}, \ and\ \bibinfo {author} {\bibfnamefont {M.}~\bibnamefont
  {{Bartelmann}}},\ }\href {\doibase 10.1086/427976} {\bibfield  {journal}
  {\bibinfo  {journal} {\apj}\ }\textbf {\bibinfo {volume} {622}},\ \bibinfo
  {pages} {759} (\bibinfo {year} {2005})},\ \Eprint
  {http://arxiv.org/abs/astro-ph/0409513} {astro-ph/0409513} \BibitemShut
  {NoStop}%
\bibitem [{\citenamefont {{Gaggero}}\ \emph {et~al.}(2015)\citenamefont
  {{Gaggero}}, \citenamefont {{Taoso}}, \citenamefont {{Urbano}}, \citenamefont
  {{Valli}},\ and\ \citenamefont {{Ullio}}}]{Gaggero:2015}%
  \BibitemOpen
  \bibfield  {author} {\bibinfo {author} {\bibfnamefont {D.}~\bibnamefont
  {{Gaggero}}}, \bibinfo {author} {\bibfnamefont {M.}~\bibnamefont {{Taoso}}},
  \bibinfo {author} {\bibfnamefont {A.}~\bibnamefont {{Urbano}}}, \bibinfo
  {author} {\bibfnamefont {M.}~\bibnamefont {{Valli}}}, \ and\ \bibinfo
  {author} {\bibfnamefont {P.}~\bibnamefont {{Ullio}}},\ }\href@noop {}
  {\bibfield  {journal} {\bibinfo  {journal} {ArXiv e-prints}\ } (\bibinfo
  {year} {2015})},\ \Eprint {http://arxiv.org/abs/1507.06129} {arXiv:1507.06129
  [astro-ph.HE]} \BibitemShut {NoStop}%
\bibitem [{\citenamefont {Kravtsov}(2003)}]{Kravtsov2003}%
  \BibitemOpen
  \bibfield  {author} {\bibinfo {author} {\bibfnamefont {A.~V.}\ \bibnamefont
  {Kravtsov}},\ }\href {\doibase 10.1086/376674} {\bibfield  {journal}
  {\bibinfo  {journal} {The Astrophysical Journal}\ }\textbf {\bibinfo {volume}
  {590}},\ \bibinfo {pages} {L1} (\bibinfo {year} {2003})}\BibitemShut
  {NoStop}%
\bibitem [{\citenamefont {Li}\ \emph {et~al.}(2005)\citenamefont {Li},
  \citenamefont {{Mac Low}},\ and\ \citenamefont {Klessen}}]{Li2005}%
  \BibitemOpen
  \bibfield  {author} {\bibinfo {author} {\bibfnamefont {Y.}~\bibnamefont
  {Li}}, \bibinfo {author} {\bibfnamefont {M.}~\bibnamefont {{Mac Low}}}, \
  and\ \bibinfo {author} {\bibfnamefont {R.~S.}\ \bibnamefont {Klessen}},\
  }\href {\doibase 10.1086/430205} {\bibfield  {journal} {\bibinfo  {journal}
  {The Astrophysical Journal}\ }\textbf {\bibinfo {volume} {626}},\ \bibinfo
  {pages} {823} (\bibinfo {year} {2005})}\BibitemShut {NoStop}%
\bibitem [{\citenamefont {{Yang}}\ and\ \citenamefont
  {{Aharonian}}(2016)}]{2016arXiv160206764Y}%
  \BibitemOpen
  \bibfield  {author} {\bibinfo {author} {\bibfnamefont {R.-z.}\ \bibnamefont
  {{Yang}}}\ and\ \bibinfo {author} {\bibfnamefont {F.}~\bibnamefont
  {{Aharonian}}},\ }\href@noop {} {\bibfield  {journal} {\bibinfo  {journal}
  {ArXiv e-prints}\ } (\bibinfo {year} {2016})},\ \Eprint
  {http://arxiv.org/abs/1602.06764} {arXiv:1602.06764 [astro-ph.HE]}
  \BibitemShut {NoStop}%
\bibitem [{\citenamefont {Trotta}\ \emph {et~al.}(2010)\citenamefont {Trotta},
  \citenamefont {Johannesson}, \citenamefont {Moskalenko}, \citenamefont
  {Porter}, \citenamefont {de~Austri},\ and\ \citenamefont
  {Strong}}]{Trotta2010}%
  \BibitemOpen
  \bibfield  {author} {\bibinfo {author} {\bibfnamefont {R.}~\bibnamefont
  {Trotta}}, \bibinfo {author} {\bibfnamefont {G.}~\bibnamefont {Johannesson}},
  \bibinfo {author} {\bibfnamefont {I.~V.}\ \bibnamefont {Moskalenko}},
  \bibinfo {author} {\bibfnamefont {T.~a.}\ \bibnamefont {Porter}}, \bibinfo
  {author} {\bibfnamefont {R.~R.}\ \bibnamefont {de~Austri}}, \ and\ \bibinfo
  {author} {\bibfnamefont {a.~W.}\ \bibnamefont {Strong}},\ }\href {\doibase
  10.1088/0004-637X/729/2/106} {\ \textbf {\bibinfo {volume} {106}} (\bibinfo
  {year} {2010}),\ 10.1088/0004-637X/729/2/106},\ \Eprint
  {http://arxiv.org/abs/1011.0037} {arXiv:1011.0037} \BibitemShut {NoStop}%
\bibitem [{\citenamefont {{Jin}}\ \emph {et~al.}(2015)\citenamefont {{Jin}},
  \citenamefont {{Wu}},\ and\ \citenamefont {{Zhou}}}]{2015JCAP...09..049J}%
  \BibitemOpen
  \bibfield  {author} {\bibinfo {author} {\bibfnamefont {H.-B.}\ \bibnamefont
  {{Jin}}}, \bibinfo {author} {\bibfnamefont {Y.-L.}\ \bibnamefont {{Wu}}}, \
  and\ \bibinfo {author} {\bibfnamefont {Y.-F.}\ \bibnamefont {{Zhou}}},\
  }\href {\doibase 10.1088/1475-7516/2015/09/049} {\bibfield  {journal}
  {\bibinfo  {journal} {\jcap}\ }\textbf {\bibinfo {volume} {9}},\ \bibinfo
  {eid} {049} (\bibinfo {year} {2015})},\ \Eprint
  {http://arxiv.org/abs/1410.0171} {arXiv:1410.0171 [hep-ph]} \BibitemShut
  {NoStop}%
\bibitem [{\citenamefont {{Aharonian}}\ \emph {et~al.}(2006)\citenamefont
  {{Aharonian}}, \citenamefont {{Akhperjanian}}, \citenamefont {{Bazer-Bachi}},
  \citenamefont {{Beilicke}}, \citenamefont {{Benbow}}, \citenamefont
  {{Berge}}, \citenamefont {{Bernl{\"o}hr}}, \citenamefont {{Boisson}},
  \citenamefont {{Bolz}}, \citenamefont {{Borrel}}, \citenamefont {{Braun}},
  \citenamefont {{Breitling}}, \citenamefont {{Brown}}, \citenamefont
  {{Chadwick}}, \citenamefont {{Chounet}}, \citenamefont {{Cornils}},
  \citenamefont {{Costamante}}, \citenamefont {{Degrange}}, \citenamefont
  {{Dickinson}}, \citenamefont {{Djannati-Ata{\"i}}}, \citenamefont {{Drury}},
  \citenamefont {{Dubus}}, \citenamefont {{Emmanoulopoulos}}, \citenamefont
  {{Espigat}}, \citenamefont {{Feinstein}}, \citenamefont {{Fontaine}},
  \citenamefont {{Fuchs}}, \citenamefont {{Funk}}, \citenamefont {{Gallant}},
  \citenamefont {{Giebels}}, \citenamefont {{Gillessen}}, \citenamefont
  {{Glicenstein}}, \citenamefont {{Goret}}, \citenamefont {{Hadjichristidis}},
  \citenamefont {{Hauser}}, \citenamefont {{Hauser}}, \citenamefont
  {{Heinzelmann}}, \citenamefont {{Henri}}, \citenamefont {{Hermann}},
  \citenamefont {{Hinton}}, \citenamefont {{Hofmann}}, \citenamefont
  {{Holleran}}, \citenamefont {{Horns}}, \citenamefont {{Jacholkowska}},
  \citenamefont {{de Jager}}, \citenamefont {{Kh{\'e}lifi}}, \citenamefont
  {{Klages}}, \citenamefont {{Komin}}, \citenamefont {{Konopelko}},
  \citenamefont {{Latham}}, \citenamefont {{Le Gallou}}, \citenamefont
  {{Lemi{\`e}re}}, \citenamefont {{Lemoine-Goumard}}, \citenamefont {{Leroy}},
  \citenamefont {{Lohse}}, \citenamefont {{Marcowith}}, \citenamefont
  {{Martin}}, \citenamefont {{Martineau-Huynh}}, \citenamefont {{Masterson}},
  \citenamefont {{McComb}}, \citenamefont {{de Naurois}}, \citenamefont
  {{Nolan}}, \citenamefont {{Noutsos}}, \citenamefont {{Orford}}, \citenamefont
  {{Osborne}}, \citenamefont {{Ouchrif}}, \citenamefont {{Panter}},
  \citenamefont {{Pelletier}}, \citenamefont {{Pita}}, \citenamefont
  {{P{\"u}hlhofer}}, \citenamefont {{Punch}}, \citenamefont {{Raubenheimer}},
  \citenamefont {{Raue}}, \citenamefont {{Raux}}, \citenamefont {{Rayner}},
  \citenamefont {{Reimer}}, \citenamefont {{Reimer}}, \citenamefont {{Ripken}},
  \citenamefont {{Rob}}, \citenamefont {{Rolland}}, \citenamefont {{Rowell}},
  \citenamefont {{Sahakian}}, \citenamefont {{Saug{\'e}}}, \citenamefont
  {{Schlenker}}, \citenamefont {{Schlickeiser}}, \citenamefont {{Schuster}},
  \citenamefont {{Schwanke}}, \citenamefont {{Siewert}}, \citenamefont {{Sol}},
  \citenamefont {{Spangler}}, \citenamefont {{Steenkamp}}, \citenamefont
  {{Stegmann}}, \citenamefont {{Tavernet}}, \citenamefont {{Terrier}},
  \citenamefont {{Th{\'e}oret}}, \citenamefont {{Tluczykont}}, \citenamefont
  {{van Eldik}}, \citenamefont {{Vasileiadis}}, \citenamefont {{Venter}},
  \citenamefont {{Vincent}}, \citenamefont {{V{\"o}lk}},\ and\ \citenamefont
  {{Wagner}}}]{2006Natur.439..695A}%
  \BibitemOpen
  \bibfield  {author} {\bibinfo {author} {\bibfnamefont {F.}~\bibnamefont
  {{Aharonian}}}, \bibinfo {author} {\bibfnamefont {A.~G.}\ \bibnamefont
  {{Akhperjanian}}}, \bibinfo {author} {\bibfnamefont {A.~R.}\ \bibnamefont
  {{Bazer-Bachi}}}, \bibinfo {author} {\bibfnamefont {M.}~\bibnamefont
  {{Beilicke}}}, \bibinfo {author} {\bibfnamefont {W.}~\bibnamefont
  {{Benbow}}}, \bibinfo {author} {\bibfnamefont {D.}~\bibnamefont {{Berge}}},
  \bibinfo {author} {\bibfnamefont {K.}~\bibnamefont {{Bernl{\"o}hr}}},
  \bibinfo {author} {\bibfnamefont {C.}~\bibnamefont {{Boisson}}}, \bibinfo
  {author} {\bibfnamefont {O.}~\bibnamefont {{Bolz}}}, \bibinfo {author}
  {\bibfnamefont {V.}~\bibnamefont {{Borrel}}}, \bibinfo {author}
  {\bibfnamefont {I.}~\bibnamefont {{Braun}}}, \bibinfo {author} {\bibfnamefont
  {F.}~\bibnamefont {{Breitling}}}, \bibinfo {author} {\bibfnamefont {A.~M.}\
  \bibnamefont {{Brown}}}, \bibinfo {author} {\bibfnamefont {P.~M.}\
  \bibnamefont {{Chadwick}}}, \bibinfo {author} {\bibfnamefont {L.-M.}\
  \bibnamefont {{Chounet}}}, \bibinfo {author} {\bibfnamefont {R.}~\bibnamefont
  {{Cornils}}}, \bibinfo {author} {\bibfnamefont {L.}~\bibnamefont
  {{Costamante}}}, \bibinfo {author} {\bibfnamefont {B.}~\bibnamefont
  {{Degrange}}}, \bibinfo {author} {\bibfnamefont {H.~J.}\ \bibnamefont
  {{Dickinson}}}, \bibinfo {author} {\bibfnamefont {A.}~\bibnamefont
  {{Djannati-Ata{\"i}}}}, \bibinfo {author} {\bibfnamefont {L.~O.}\
  \bibnamefont {{Drury}}}, \bibinfo {author} {\bibfnamefont {G.}~\bibnamefont
  {{Dubus}}}, \bibinfo {author} {\bibfnamefont {D.}~\bibnamefont
  {{Emmanoulopoulos}}}, \bibinfo {author} {\bibfnamefont {P.}~\bibnamefont
  {{Espigat}}}, \bibinfo {author} {\bibfnamefont {F.}~\bibnamefont
  {{Feinstein}}}, \bibinfo {author} {\bibfnamefont {G.}~\bibnamefont
  {{Fontaine}}}, \bibinfo {author} {\bibfnamefont {Y.}~\bibnamefont {{Fuchs}}},
  \bibinfo {author} {\bibfnamefont {S.}~\bibnamefont {{Funk}}}, \bibinfo
  {author} {\bibfnamefont {Y.~A.}\ \bibnamefont {{Gallant}}}, \bibinfo {author}
  {\bibfnamefont {B.}~\bibnamefont {{Giebels}}}, \bibinfo {author}
  {\bibfnamefont {S.}~\bibnamefont {{Gillessen}}}, \bibinfo {author}
  {\bibfnamefont {J.~F.}\ \bibnamefont {{Glicenstein}}}, \bibinfo {author}
  {\bibfnamefont {P.}~\bibnamefont {{Goret}}}, \bibinfo {author} {\bibfnamefont
  {C.}~\bibnamefont {{Hadjichristidis}}}, \bibinfo {author} {\bibfnamefont
  {D.}~\bibnamefont {{Hauser}}}, \bibinfo {author} {\bibfnamefont
  {M.}~\bibnamefont {{Hauser}}}, \bibinfo {author} {\bibfnamefont
  {G.}~\bibnamefont {{Heinzelmann}}}, \bibinfo {author} {\bibfnamefont
  {G.}~\bibnamefont {{Henri}}}, \bibinfo {author} {\bibfnamefont
  {G.}~\bibnamefont {{Hermann}}}, \bibinfo {author} {\bibfnamefont {J.~A.}\
  \bibnamefont {{Hinton}}}, \bibinfo {author} {\bibfnamefont {W.}~\bibnamefont
  {{Hofmann}}}, \bibinfo {author} {\bibfnamefont {M.}~\bibnamefont
  {{Holleran}}}, \bibinfo {author} {\bibfnamefont {D.}~\bibnamefont {{Horns}}},
  \bibinfo {author} {\bibfnamefont {A.}~\bibnamefont {{Jacholkowska}}},
  \bibinfo {author} {\bibfnamefont {O.~C.}\ \bibnamefont {{de Jager}}},
  \bibinfo {author} {\bibfnamefont {B.}~\bibnamefont {{Kh{\'e}lifi}}}, \bibinfo
  {author} {\bibfnamefont {S.}~\bibnamefont {{Klages}}}, \bibinfo {author}
  {\bibfnamefont {N.}~\bibnamefont {{Komin}}}, \bibinfo {author} {\bibfnamefont
  {A.}~\bibnamefont {{Konopelko}}}, \bibinfo {author} {\bibfnamefont {I.~J.}\
  \bibnamefont {{Latham}}}, \bibinfo {author} {\bibfnamefont {R.}~\bibnamefont
  {{Le Gallou}}}, \bibinfo {author} {\bibfnamefont {A.}~\bibnamefont
  {{Lemi{\`e}re}}}, \bibinfo {author} {\bibfnamefont {M.}~\bibnamefont
  {{Lemoine-Goumard}}}, \bibinfo {author} {\bibfnamefont {N.}~\bibnamefont
  {{Leroy}}}, \bibinfo {author} {\bibfnamefont {T.}~\bibnamefont {{Lohse}}},
  \bibinfo {author} {\bibfnamefont {A.}~\bibnamefont {{Marcowith}}}, \bibinfo
  {author} {\bibfnamefont {J.~M.}\ \bibnamefont {{Martin}}}, \bibinfo {author}
  {\bibfnamefont {O.}~\bibnamefont {{Martineau-Huynh}}}, \bibinfo {author}
  {\bibfnamefont {C.}~\bibnamefont {{Masterson}}}, \bibinfo {author}
  {\bibfnamefont {T.~J.~L.}\ \bibnamefont {{McComb}}}, \bibinfo {author}
  {\bibfnamefont {M.}~\bibnamefont {{de Naurois}}}, \bibinfo {author}
  {\bibfnamefont {S.~J.}\ \bibnamefont {{Nolan}}}, \bibinfo {author}
  {\bibfnamefont {A.}~\bibnamefont {{Noutsos}}}, \bibinfo {author}
  {\bibfnamefont {K.~J.}\ \bibnamefont {{Orford}}}, \bibinfo {author}
  {\bibfnamefont {J.~L.}\ \bibnamefont {{Osborne}}}, \bibinfo {author}
  {\bibfnamefont {M.}~\bibnamefont {{Ouchrif}}}, \bibinfo {author}
  {\bibfnamefont {M.}~\bibnamefont {{Panter}}}, \bibinfo {author}
  {\bibfnamefont {G.}~\bibnamefont {{Pelletier}}}, \bibinfo {author}
  {\bibfnamefont {S.}~\bibnamefont {{Pita}}}, \bibinfo {author} {\bibfnamefont
  {G.}~\bibnamefont {{P{\"u}hlhofer}}}, \bibinfo {author} {\bibfnamefont
  {M.}~\bibnamefont {{Punch}}}, \bibinfo {author} {\bibfnamefont {B.~C.}\
  \bibnamefont {{Raubenheimer}}}, \bibinfo {author} {\bibfnamefont
  {M.}~\bibnamefont {{Raue}}}, \bibinfo {author} {\bibfnamefont
  {J.}~\bibnamefont {{Raux}}}, \bibinfo {author} {\bibfnamefont {S.~M.}\
  \bibnamefont {{Rayner}}}, \bibinfo {author} {\bibfnamefont {A.}~\bibnamefont
  {{Reimer}}}, \bibinfo {author} {\bibfnamefont {O.}~\bibnamefont {{Reimer}}},
  \bibinfo {author} {\bibfnamefont {J.}~\bibnamefont {{Ripken}}}, \bibinfo
  {author} {\bibfnamefont {L.}~\bibnamefont {{Rob}}}, \bibinfo {author}
  {\bibfnamefont {L.}~\bibnamefont {{Rolland}}}, \bibinfo {author}
  {\bibfnamefont {G.}~\bibnamefont {{Rowell}}}, \bibinfo {author}
  {\bibfnamefont {V.}~\bibnamefont {{Sahakian}}}, \bibinfo {author}
  {\bibfnamefont {L.}~\bibnamefont {{Saug{\'e}}}}, \bibinfo {author}
  {\bibfnamefont {S.}~\bibnamefont {{Schlenker}}}, \bibinfo {author}
  {\bibfnamefont {R.}~\bibnamefont {{Schlickeiser}}}, \bibinfo {author}
  {\bibfnamefont {C.}~\bibnamefont {{Schuster}}}, \bibinfo {author}
  {\bibfnamefont {U.}~\bibnamefont {{Schwanke}}}, \bibinfo {author}
  {\bibfnamefont {M.}~\bibnamefont {{Siewert}}}, \bibinfo {author}
  {\bibfnamefont {H.}~\bibnamefont {{Sol}}}, \bibinfo {author} {\bibfnamefont
  {D.}~\bibnamefont {{Spangler}}}, \bibinfo {author} {\bibfnamefont
  {R.}~\bibnamefont {{Steenkamp}}}, \bibinfo {author} {\bibfnamefont
  {C.}~\bibnamefont {{Stegmann}}}, \bibinfo {author} {\bibfnamefont {J.-P.}\
  \bibnamefont {{Tavernet}}}, \bibinfo {author} {\bibfnamefont
  {R.}~\bibnamefont {{Terrier}}}, \bibinfo {author} {\bibfnamefont {C.~G.}\
  \bibnamefont {{Th{\'e}oret}}}, \bibinfo {author} {\bibfnamefont
  {M.}~\bibnamefont {{Tluczykont}}}, \bibinfo {author} {\bibfnamefont
  {C.}~\bibnamefont {{van Eldik}}}, \bibinfo {author} {\bibfnamefont
  {G.}~\bibnamefont {{Vasileiadis}}}, \bibinfo {author} {\bibfnamefont
  {C.}~\bibnamefont {{Venter}}}, \bibinfo {author} {\bibfnamefont
  {P.}~\bibnamefont {{Vincent}}}, \bibinfo {author} {\bibfnamefont {H.~J.}\
  \bibnamefont {{V{\"o}lk}}}, \ and\ \bibinfo {author} {\bibfnamefont {S.~J.}\
  \bibnamefont {{Wagner}}},\ }\href {\doibase 10.1038/nature04467} {\bibfield
  {journal} {\bibinfo  {journal} {\nat}\ }\textbf {\bibinfo {volume} {439}},\
  \bibinfo {pages} {695} (\bibinfo {year} {2006})},\ \Eprint
  {http://arxiv.org/abs/astro-ph/0603021} {astro-ph/0603021} \BibitemShut
  {NoStop}%
\bibitem [{\citenamefont {Archer}\ \emph {et~al.}(2016)\citenamefont {Archer}
  \emph {et~al.}}]{VERITAS_ridge}%
  \BibitemOpen
  \bibfield  {author} {\bibinfo {author} {\bibfnamefont {A.}~\bibnamefont
  {Archer}} \emph {et~al.},\ }\href@noop {} {\  (\bibinfo {year} {2016})},\
  \Eprint {http://arxiv.org/abs/1602.08522} {arXiv:1602.08522 [astro-ph.HE]}
  \BibitemShut {NoStop}%
%%CITATION = ARXIV:1602.08522;%%
\bibitem [{\citenamefont {{Rodr{\'{\i}}guez-Gonz{\'a}lez}}\ \emph
  {et~al.}(2009)\citenamefont {{Rodr{\'{\i}}guez-Gonz{\'a}lez}}, \citenamefont
  {{Raga}},\ and\ \citenamefont {{Cant{\'o}}}}]{2009A&A...501..411R}%
  \BibitemOpen
  \bibfield  {author} {\bibinfo {author} {\bibfnamefont {A.}~\bibnamefont
  {{Rodr{\'{\i}}guez-Gonz{\'a}lez}}}, \bibinfo {author} {\bibfnamefont {A.~C.}\
  \bibnamefont {{Raga}}}, \ and\ \bibinfo {author} {\bibfnamefont
  {J.}~\bibnamefont {{Cant{\'o}}}},\ }\href {\doibase
  10.1051/0004-6361/200911693} {\bibfield  {journal} {\bibinfo  {journal}
  {\aap}\ }\textbf {\bibinfo {volume} {501}},\ \bibinfo {pages} {411} (\bibinfo
  {year} {2009})},\ \Eprint {http://arxiv.org/abs/0905.1988} {arXiv:0905.1988}
  \BibitemShut {NoStop}%
\bibitem [{\citenamefont {Yusef-Zadeh}\ \emph
  {et~al.}(2009{\natexlab{b}})\citenamefont {Yusef-Zadeh}, \citenamefont
  {Hewitt}, \citenamefont {Arendt}, \citenamefont {Whitney}, \citenamefont
  {Rieke}, \citenamefont {Wardle}, \citenamefont {Hinz}, \citenamefont
  {Stolovy}, \citenamefont {Lang}, \citenamefont {Burton},\ and\ \citenamefont
  {Ramirez}}]{0004-637X-702-1-178}%
  \BibitemOpen
  \bibfield  {author} {\bibinfo {author} {\bibfnamefont {F.}~\bibnamefont
  {Yusef-Zadeh}}, \bibinfo {author} {\bibfnamefont {J.~W.}\ \bibnamefont
  {Hewitt}}, \bibinfo {author} {\bibfnamefont {R.~G.}\ \bibnamefont {Arendt}},
  \bibinfo {author} {\bibfnamefont {B.}~\bibnamefont {Whitney}}, \bibinfo
  {author} {\bibfnamefont {G.}~\bibnamefont {Rieke}}, \bibinfo {author}
  {\bibfnamefont {M.}~\bibnamefont {Wardle}}, \bibinfo {author} {\bibfnamefont
  {J.~L.}\ \bibnamefont {Hinz}}, \bibinfo {author} {\bibfnamefont
  {S.}~\bibnamefont {Stolovy}}, \bibinfo {author} {\bibfnamefont {C.~C.}\
  \bibnamefont {Lang}}, \bibinfo {author} {\bibfnamefont {M.~G.}\ \bibnamefont
  {Burton}}, \ and\ \bibinfo {author} {\bibfnamefont {S.}~\bibnamefont
  {Ramirez}},\ }\href {http://stacks.iop.org/0004-637X/702/i=1/a=178}
  {\bibfield  {journal} {\bibinfo  {journal} {The Astrophysical Journal}\
  }\textbf {\bibinfo {volume} {702}},\ \bibinfo {pages} {178} (\bibinfo {year}
  {2009}{\natexlab{b}})}\BibitemShut {NoStop}%
\bibitem [{\citenamefont {{Yasui}}\ \emph {et~al.}(2015)\citenamefont
  {{Yasui}}, \citenamefont {{Nishiyama}}, \citenamefont {{Yoshikawa}},
  \citenamefont {{Nagatomo}}, \citenamefont {{Uchiyama}}, \citenamefont
  {{Tsuru}}, \citenamefont {{Koyama}}, \citenamefont {{Tamura}}, \citenamefont
  {{Kwon}}, \citenamefont {{Sugitani}}, \citenamefont {{Sch{\"o}del}},\ and\
  \citenamefont {{Nagata}}}]{2015PASJ...67..123Y}%
  \BibitemOpen
  \bibfield  {author} {\bibinfo {author} {\bibfnamefont {K.}~\bibnamefont
  {{Yasui}}}, \bibinfo {author} {\bibfnamefont {S.}~\bibnamefont
  {{Nishiyama}}}, \bibinfo {author} {\bibfnamefont {T.}~\bibnamefont
  {{Yoshikawa}}}, \bibinfo {author} {\bibfnamefont {S.}~\bibnamefont
  {{Nagatomo}}}, \bibinfo {author} {\bibfnamefont {H.}~\bibnamefont
  {{Uchiyama}}}, \bibinfo {author} {\bibfnamefont {T.~G.}\ \bibnamefont
  {{Tsuru}}}, \bibinfo {author} {\bibfnamefont {K.}~\bibnamefont {{Koyama}}},
  \bibinfo {author} {\bibfnamefont {M.}~\bibnamefont {{Tamura}}}, \bibinfo
  {author} {\bibfnamefont {J.}~\bibnamefont {{Kwon}}}, \bibinfo {author}
  {\bibfnamefont {K.}~\bibnamefont {{Sugitani}}}, \bibinfo {author}
  {\bibfnamefont {R.}~\bibnamefont {{Sch{\"o}del}}}, \ and\ \bibinfo {author}
  {\bibfnamefont {T.}~\bibnamefont {{Nagata}}},\ }\href {\doibase
  10.1093/pasj/psv100} {\bibfield  {journal} {\bibinfo  {journal} {\pasj}\
  }\textbf {\bibinfo {volume} {67}},\ \bibinfo {eid} {123} (\bibinfo {year}
  {2015})},\ \Eprint {http://arxiv.org/abs/1510.06832} {arXiv:1510.06832}
  \BibitemShut {NoStop}%
\bibitem [{\citenamefont {{Genzel}}\ \emph {et~al.}(2010)\citenamefont
  {{Genzel}}, \citenamefont {{Eisenhauer}},\ and\ \citenamefont
  {{Gillessen}}}]{2010RvMP...82.3121G}%
  \BibitemOpen
  \bibfield  {author} {\bibinfo {author} {\bibfnamefont {R.}~\bibnamefont
  {{Genzel}}}, \bibinfo {author} {\bibfnamefont {F.}~\bibnamefont
  {{Eisenhauer}}}, \ and\ \bibinfo {author} {\bibfnamefont {S.}~\bibnamefont
  {{Gillessen}}},\ }\href {\doibase 10.1103/RevModPhys.82.3121} {\bibfield
  {journal} {\bibinfo  {journal} {Reviews of Modern Physics}\ }\textbf
  {\bibinfo {volume} {82}},\ \bibinfo {pages} {3121} (\bibinfo {year}
  {2010})},\ \Eprint {http://arxiv.org/abs/1006.0064} {arXiv:1006.0064}
  \BibitemShut {NoStop}%
\bibitem [{\citenamefont {{Do}}\ \emph {et~al.}(2013)\citenamefont {{Do}},
  \citenamefont {{Lu}}, \citenamefont {{Ghez}}, \citenamefont {{Morris}},
  \citenamefont {{Yelda}}, \citenamefont {{Martinez}}, \citenamefont
  {{Wright}},\ and\ \citenamefont {{Matthews}}}]{2013ApJ...764..154D}%
  \BibitemOpen
  \bibfield  {author} {\bibinfo {author} {\bibfnamefont {T.}~\bibnamefont
  {{Do}}}, \bibinfo {author} {\bibfnamefont {J.~R.}\ \bibnamefont {{Lu}}},
  \bibinfo {author} {\bibfnamefont {A.~M.}\ \bibnamefont {{Ghez}}}, \bibinfo
  {author} {\bibfnamefont {M.~R.}\ \bibnamefont {{Morris}}}, \bibinfo {author}
  {\bibfnamefont {S.}~\bibnamefont {{Yelda}}}, \bibinfo {author} {\bibfnamefont
  {G.~D.}\ \bibnamefont {{Martinez}}}, \bibinfo {author} {\bibfnamefont
  {S.~A.}\ \bibnamefont {{Wright}}}, \ and\ \bibinfo {author} {\bibfnamefont
  {K.}~\bibnamefont {{Matthews}}},\ }\href {\doibase
  10.1088/0004-637X/764/2/154} {\bibfield  {journal} {\bibinfo  {journal}
  {\apj}\ }\textbf {\bibinfo {volume} {764}},\ \bibinfo {eid} {154} (\bibinfo
  {year} {2013})},\ \Eprint {http://arxiv.org/abs/1301.0539} {arXiv:1301.0539
  [astro-ph.SR]} \BibitemShut {NoStop}%
\bibitem [{\citenamefont {{Purcell}}\ \emph {et~al.}(2012)\citenamefont
  {{Purcell}}, \citenamefont {{Longmore}}, \citenamefont {{Walsh}},
  \citenamefont {{Whiting}}, \citenamefont {{Breen}}, \citenamefont
  {{Britton}}, \citenamefont {{Brooks}}, \citenamefont {{Burton}},
  \citenamefont {{Cunningham}}, \citenamefont {{Green}}, \citenamefont
  {{Harvey-Smith}}, \citenamefont {{Hindson}}, \citenamefont {{Hoare}},
  \citenamefont {{Indermuehle}}, \citenamefont {{Jones}}, \citenamefont {{Lo}},
  \citenamefont {{Lowe}}, \citenamefont {{Phillips}}, \citenamefont
  {{Thompson}}, \citenamefont {{Urquhart}}, \citenamefont {{Voronkov}},\ and\
  \citenamefont {{White}}}]{2012MNRAS.426.1972P}%
  \BibitemOpen
  \bibfield  {author} {\bibinfo {author} {\bibfnamefont {C.~R.}\ \bibnamefont
  {{Purcell}}}, \bibinfo {author} {\bibfnamefont {S.~N.}\ \bibnamefont
  {{Longmore}}}, \bibinfo {author} {\bibfnamefont {A.~J.}\ \bibnamefont
  {{Walsh}}}, \bibinfo {author} {\bibfnamefont {M.~T.}\ \bibnamefont
  {{Whiting}}}, \bibinfo {author} {\bibfnamefont {S.~L.}\ \bibnamefont
  {{Breen}}}, \bibinfo {author} {\bibfnamefont {T.}~\bibnamefont {{Britton}}},
  \bibinfo {author} {\bibfnamefont {K.~J.}\ \bibnamefont {{Brooks}}}, \bibinfo
  {author} {\bibfnamefont {M.~G.}\ \bibnamefont {{Burton}}}, \bibinfo {author}
  {\bibfnamefont {M.~R.}\ \bibnamefont {{Cunningham}}}, \bibinfo {author}
  {\bibfnamefont {J.~A.}\ \bibnamefont {{Green}}}, \bibinfo {author}
  {\bibfnamefont {L.}~\bibnamefont {{Harvey-Smith}}}, \bibinfo {author}
  {\bibfnamefont {L.}~\bibnamefont {{Hindson}}}, \bibinfo {author}
  {\bibfnamefont {M.~G.}\ \bibnamefont {{Hoare}}}, \bibinfo {author}
  {\bibfnamefont {B.}~\bibnamefont {{Indermuehle}}}, \bibinfo {author}
  {\bibfnamefont {P.~A.}\ \bibnamefont {{Jones}}}, \bibinfo {author}
  {\bibfnamefont {N.}~\bibnamefont {{Lo}}}, \bibinfo {author} {\bibfnamefont
  {V.}~\bibnamefont {{Lowe}}}, \bibinfo {author} {\bibfnamefont {C.~J.}\
  \bibnamefont {{Phillips}}}, \bibinfo {author} {\bibfnamefont {M.~A.}\
  \bibnamefont {{Thompson}}}, \bibinfo {author} {\bibfnamefont {J.~S.}\
  \bibnamefont {{Urquhart}}}, \bibinfo {author} {\bibfnamefont {M.~A.}\
  \bibnamefont {{Voronkov}}}, \ and\ \bibinfo {author} {\bibfnamefont {G.~L.}\
  \bibnamefont {{White}}},\ }\href {\doibase 10.1111/j.1365-2966.2012.21800.x}
  {\bibfield  {journal} {\bibinfo  {journal} {\mnras}\ }\textbf {\bibinfo
  {volume} {426}},\ \bibinfo {pages} {1972} (\bibinfo {year} {2012})},\ \Eprint
  {http://arxiv.org/abs/1207.6159} {arXiv:1207.6159} \BibitemShut {NoStop}%
\bibitem [{\citenamefont {{Roman-Duval}}\ \emph {et~al.}(2009)\citenamefont
  {{Roman-Duval}}, \citenamefont {{Jackson}}, \citenamefont {{Heyer}},
  \citenamefont {{Johnson}}, \citenamefont {{Rathborne}}, \citenamefont
  {{Shah}},\ and\ \citenamefont {{Simon}}}]{2009ApJ...699.1153R}%
  \BibitemOpen
  \bibfield  {author} {\bibinfo {author} {\bibfnamefont {J.}~\bibnamefont
  {{Roman-Duval}}}, \bibinfo {author} {\bibfnamefont {J.~M.}\ \bibnamefont
  {{Jackson}}}, \bibinfo {author} {\bibfnamefont {M.}~\bibnamefont {{Heyer}}},
  \bibinfo {author} {\bibfnamefont {A.}~\bibnamefont {{Johnson}}}, \bibinfo
  {author} {\bibfnamefont {J.}~\bibnamefont {{Rathborne}}}, \bibinfo {author}
  {\bibfnamefont {R.}~\bibnamefont {{Shah}}}, \ and\ \bibinfo {author}
  {\bibfnamefont {R.}~\bibnamefont {{Simon}}},\ }\href {\doibase
  10.1088/0004-637X/699/2/1153} {\bibfield  {journal} {\bibinfo  {journal}
  {\apj}\ }\textbf {\bibinfo {volume} {699}},\ \bibinfo {pages} {1153}
  (\bibinfo {year} {2009})},\ \Eprint {http://arxiv.org/abs/0905.0723}
  {arXiv:0905.0723 [astro-ph.GA]} \BibitemShut {NoStop}%
\bibitem [{\citenamefont {Kolpak}\ \emph {et~al.}(2003)\citenamefont {Kolpak},
  \citenamefont {Jackson}, \citenamefont {Bania}, \citenamefont {Clemens},\
  and\ \citenamefont {Dickey}}]{0004-637X-582-2-756}%
  \BibitemOpen
  \bibfield  {author} {\bibinfo {author} {\bibfnamefont {M.~A.}\ \bibnamefont
  {Kolpak}}, \bibinfo {author} {\bibfnamefont {J.~M.}\ \bibnamefont {Jackson}},
  \bibinfo {author} {\bibfnamefont {T.~M.}\ \bibnamefont {Bania}}, \bibinfo
  {author} {\bibfnamefont {D.~P.}\ \bibnamefont {Clemens}}, \ and\ \bibinfo
  {author} {\bibfnamefont {J.~M.}\ \bibnamefont {Dickey}},\ }\href
  {http://stacks.iop.org/0004-637X/582/i=2/a=756} {\bibfield  {journal}
  {\bibinfo  {journal} {The Astrophysical Journal}\ }\textbf {\bibinfo {volume}
  {582}},\ \bibinfo {pages} {756} (\bibinfo {year} {2003})}\BibitemShut
  {NoStop}%
\bibitem [{\citenamefont {{Nakanishi}}\ and\ \citenamefont
  {{Sofue}}(2003)}]{NS}%
  \BibitemOpen
  \bibfield  {author} {\bibinfo {author} {\bibfnamefont {H.}~\bibnamefont
  {{Nakanishi}}}\ and\ \bibinfo {author} {\bibfnamefont {Y.}~\bibnamefont
  {{Sofue}}},\ }\href {\doibase 10.1093/pasj/55.1.191} {\bibfield  {journal}
  {\bibinfo  {journal} {\pasj}\ }\textbf {\bibinfo {volume} {55}},\ \bibinfo
  {pages} {191} (\bibinfo {year} {2003})},\ \Eprint
  {http://arxiv.org/abs/astro-ph/0304338} {astro-ph/0304338} \BibitemShut
  {NoStop}%
\bibitem [{\citenamefont {{Tavakoli}}(2012)}]{2012arXiv1207.6150T}%
  \BibitemOpen
  \bibfield  {author} {\bibinfo {author} {\bibfnamefont {M.}~\bibnamefont
  {{Tavakoli}}},\ }\href@noop {} {\bibfield  {journal} {\bibinfo  {journal}
  {ArXiv e-prints}\ } (\bibinfo {year} {2012})},\ \Eprint
  {http://arxiv.org/abs/1207.6150} {arXiv:1207.6150 [astro-ph.GA]} \BibitemShut
  {NoStop}%
\end{thebibliography}%

\begin{appendix}

\section{Interstellar Gas Distributions}
\label{app:gas_extra}
The axisymmetric gas model implemented in {\tt Galprop} (for propagation) is inadequate for distributing sources in the inner Galaxy.  Here we briefly discuss some of the caveats which are improved by the use of the `PEB' model. These issues are also important when generating $\gamma$-ray skymaps, where the gas and CR density must be convolved along the line-of-sight and should be considered in future studies of gas-correlated $\gamma$-rays emission near the Galactic center.

\begin{enumerate}
\item {\em Vanishing kinematic resolution}: Line surveys sample the gas temperature as a function of latitude, longitude, and velocity relative to the solar system. In the direction of the Galactic center and anti-center, gas is moving tangentially to the line-of-sight, producing vanishing kinematic resolution.  This leads to a distance degeneracy along lines of sight near Galactic longitudes $l\approx0$ and $l\approx180$. In the {\tt Galprop} models, each annulus is linearly interpolated from the sides (averaged over $\Delta l=5^\circ$) between $|l|<10^\circ$ and $|180-l|^\circ$. The central annulus lies completely within the interpolated region, requiring a different procedure.  For HI, it is assumed that the central annulus has 60\% more gas than the neighboring ring while for CO all high velocity emission is assigned to the central ring as defined in Ref.~\cite{fermi_diffuse}. Finally, each pixel in interpolated region is renormalized to the total survey column density.  

\item {\em Kinematic distance ambiguity}:  Two distances correspond to the same radial velocity in the entire inner Galaxy.  A velocity deconvolution alone (without additional model inputs or absorption measures) cannot unambiguously assign gas to the near or far tangent point. In order to help alleviate the KDA generally -- i.e. not only for resolving distances to specific molecular clouds~\cite{2009ApJ...699.1153R} or HII regions~\cite{0004-637X-582-2-756} -- one can incorporate additional model assumptions into the deconvolution procedure, using for example gas-flow simulations~\cite{PEB}, or models of the molecular gas disk height~\cite{NS}.

\item {\em Peculiar Motion and Velocity Dispersion}: Many gas clouds and the central Galactic bar have significant non-circular motion.  If one can obtain independent distance estimates from e.g. absorption measures~\cite{2009ApJ...699.1153R,0004-637X-582-2-756}, individual clouds can be repositioned in the gas survey. This requires a significant effort considering the large number of independent structures.  Alternatively, one can model gas flow~\cite{PEB} in the inner Galaxy in order to kinematically resolve features with non-circular motion. For large gas clouds, internal velocity dispersions of the gas can be a significant fraction of the bulk velocity and one needs to make assumptions about the shape and size of the cloud in order to assign the correct location.
\end{enumerate}

\medskip

The starting survey data for the PEB model is based on the Dame 2001 survey~\cite{Dame:2001}, and provides much more accurate and 3-dimensional structure compared with {\tt Galprop's} (axisymmetric) analytic gas density model.  The survey is velocity cube is converted to a spatial data cube using a sophistocated deconvolution procedure which iteratively determining the best fitting line-of-sight density distribution based on the gas flow model. We refer the Reader to Ref.~\cite{PEB} for details.  The resulting gas datacube (in $(x,y,z)$) is output at 100 pc resolution giving more than a factor 10 resolution improvement compared with the {\tt Galprop} analytic gas models, and correctly representing major axisymmetric structures of the Galaxy.

In addition to the PEB models, three additional three-dimensional survey deconvolutions are available in the literature. For HI and $\rm H_2$, Nakanishi \& Sofue~\cite{NS} assume circular motion, and resolve the kinematic-distance-ambiguity in the inner Galaxy by assuming a model of hydrodynamical equilibrium in order to describe the gas disk heights as a function of Galactic radius.  This is essentially the same procedure used by Ref.~\cite{2012arXiv1207.6150T} for HI.  While this does represent an improvement on the {\tt Galprop} model, each of these models still assumes circular motion in the inner Galaxy and requires masking in a large region behind the Galactic center where kinematic resolution vanishes.  This makes these maps unsuitable for $\gamma$-ray studies toward the Galactic center.

\section{\boldmath $X_{\rm CO}$ in the Inner Galaxy}
\label{sec:X_CO}

Previous studies of the Galactic center excess have neglected variations on the $\rm H_2\to CO$ conversion factor $X_{\rm CO}$.  However, understanding the radial \xco provides an important indicator of the hadronic cosmic-ray density and calorimetry toward the Galactic center.   Here, we (i) briefly comment on the empirical and theoretical understanding of \xco toward the centers of star-forming Galaxies similar to the Milky Way, (ii) show that for realistic cosmic-ray injection rates from the CMZ, the gas-correlated $\gamma$-ray emission is over-saturated (assuming an SNR-like injection ratio $q_{p}/q_{e}\approx10$), and (iii) demonstrate that advective outflows at the Galactic center help to alleviate this problem.

\begin{figure*}[tb]
  \centering
	  \includegraphics[width=.45\textwidth]{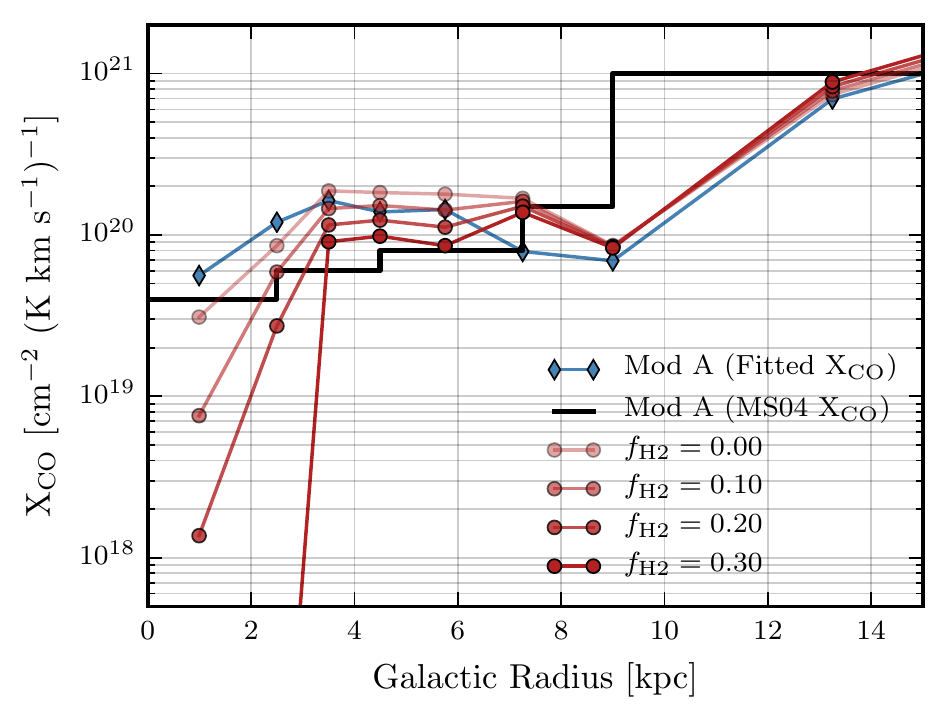}
    \includegraphics[width=.45\textwidth]{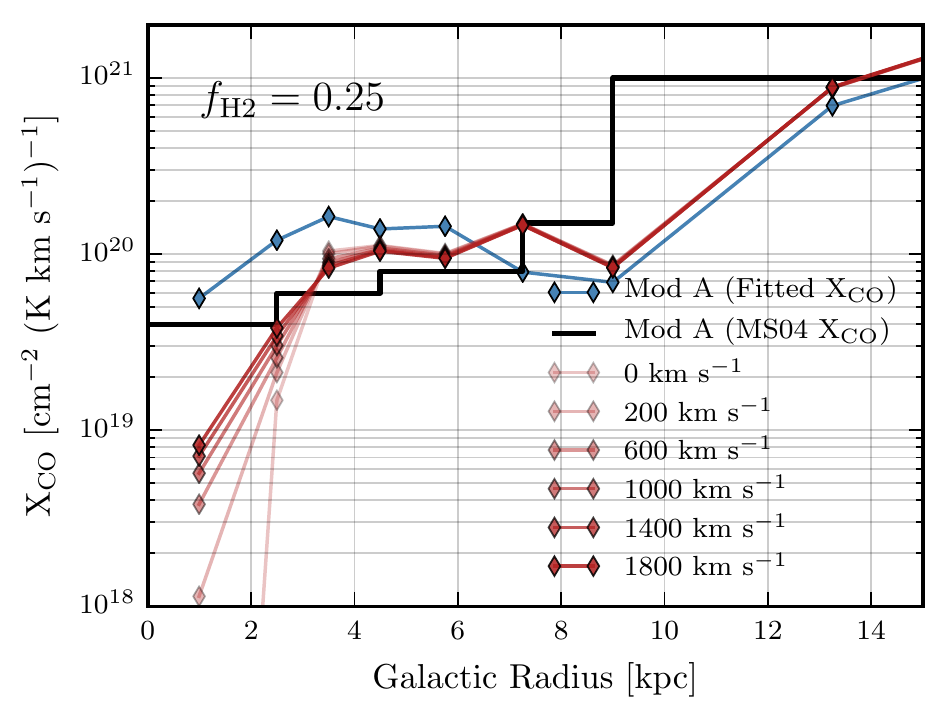}\\
      \includegraphics[width=.95\textwidth]{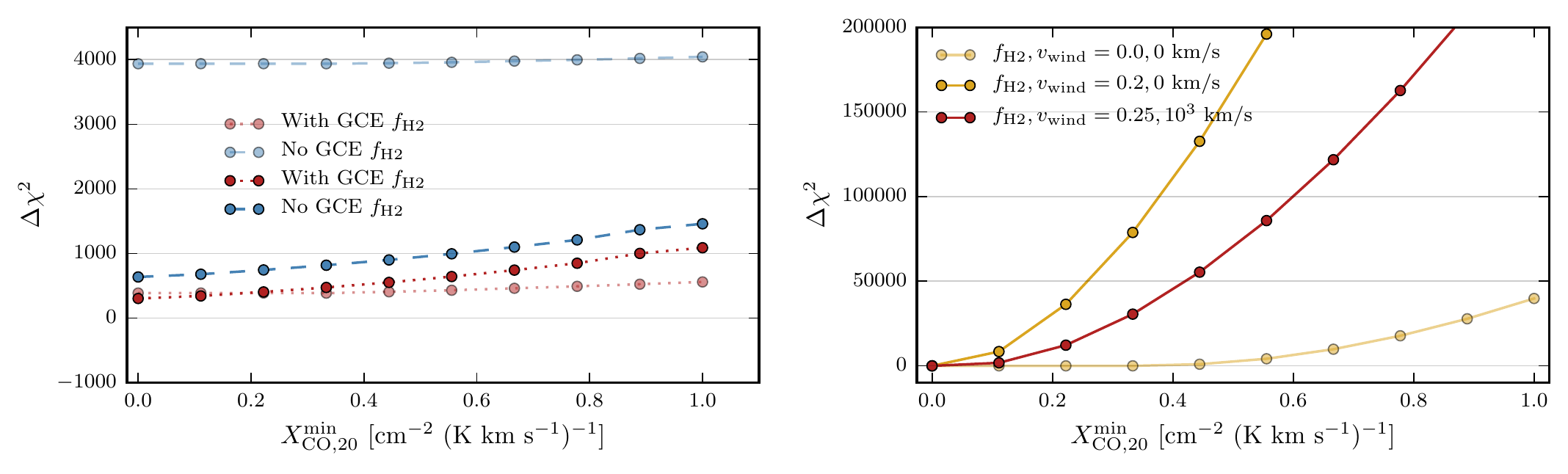}
\caption{{\bf Top-left:} Fitted values for \xco as \fh is increased.  {\bf Top-right:} \xco for $f_{\rm H2}=0.3$ as the Galactic center wind velocity $v_{\rm wind}$ is increased. Also shown our $X_{\rm CO}$-fitted version of Mod A, as well as the commonly used \xco profile from Ref.~\cite{MS:2004}. With the exception of the inner-most ring, all fitted values have statistical error bars $\lesssim 20\%$. {\bf Bottom-left:} Inner Galaxy $\Delta \chi^2$ for the traditional SNR and Canonical models when limiting minimal value of $X_{\rm CO}$ (in units of $10^{20} \rm\ cm^{-2} (K\ km\ s^{-1})^{-1}$). Traditional {\tt Galprop} models combined with $\gamma$-ray observations suggest a value around $X_{\rm CO}\approx 4\times 10^{19}\ \rm cm^{-2}\ (K\ km\ s^{-1})^{-1}$~\cite{galprop_x_co}.  The impact on the Galactic center excess spectrum and significance is negligible. {\bf Bottom-right:} Same, but for the global-inner analysis.  Here the disk is unmasked and the fit quickly degrades as the minimal \xco is increased.  Models which include winds help to alleviate this problem and allow for more realistic CMZ injection rates.}
 \label{fig:XCO}
\end{figure*}

Theoretical and observational results indicate that $X_{\rm CO}$ is subject to significant spatial and environmental variations, especially in the centers of local group star forming galaxies~\cite{Sandstrom:2012ni}. Comparisons of the total dust opacity to the CO-inferred gas density in local group spiral galaxies shows that $X_{\rm CO}$ is relatively flat throughout Galactic disks, but decreases on average by a factor more than 2 in the central regions.  In 3 of the 16 Galaxy cores surveyed in Ref.~\cite{Sandstrom:2012ni}, $X_{\rm CO}$ was at least a factor of 10 below the Milky Way average $X_{\rm CO}=2\times10^{20}~\rm cm^{-2}\ (K\ km\ s^{-1})^{-1}$.  $X_{\rm CO}$ was found to have a  strong inverse correlation with the interstellar radiation intensity and surface densities of star formation, stellar mass, and dust mass, all of which are much higher in the Milky Way center than the disk (see Sec.~\ref{sec:sources}).  $X_{\rm CO}$ is also believed to increase sharply with decreasing metallicity for $Z\lesssim Z_\odot/2$~\cite{Bolatto:2013}, which can severely alter the inferred gas density in chemically enriched regions of the Galaxy, such as the Galactic Center.  Finally, fitting EGRET~\cite{MS:2004} and Fermi~\cite{fermi_diffuse} data to models of the Milky-way's diffuse $\gamma$-ray emission requires $X_{\rm CO}$ in the inner 2 kpc to be a factor $\sim$4 lower than the rest of the Galaxy.  In addition to this, we have shown throughout this paper that {\em status quo} cosmic-ray models have dramatically underestimated the injection rate from the CMZ, implying that either \xco must be reduced even farther, or that transport in the inner Galaxy must be enhanced.

In the top-left panel of Figure~\ref{fig:XCO}, we show the fitted \xco profile as \fh is increased.  We also show the \xco profile from Moskalenko \& Strong (2004)~\cite{MS:2004} (MS04) which has previously been used for Ref.~\cite{Calore:2015}'s Mod A.  As cosmic-ray injection is enhanced toward the inner-Galaxy, \xco must simultaneously be reduced to maintain the same level of hadronic $\gamma$-ray emission.  Nowhere is this more important than the for the innermost Galactocentric ring ($r<1.8$ kpc) which, for $f_{\rm H2}=.2$, becomes suppressed by more two-orders of magnitude relative to the local value.  In the top-right panel, we add the advective wind discussed in Sec.~\ref{sec:winds} and show that while the outer Galaxy is relatively unaffected, as the wind velocity is increased the best-fitting \xco returns to potentially physical values.  All models remain biased to very high values in the outer Galaxy, consistent with previous studies~\cite{MS:2004,fermi_diffuse}.  We note that the MS04 profile was obtained from EGRET data, and does not employ point source masking. Interestingly, we find that our $X_{\rm CO}$ profiles inside the Solar circle are quite sensitive to the adopted point source masking, and that masking systematics should be understood for future studies of \xco utilizing $\gamma$-ray data.  

In all of the previous model fits, we have allowed the values of $X_{\rm CO}$ to float freely when performing global fits. Here we enforce a lower limit on $X_{\rm CO}$ in order to gauge the maximal allowed value.  In the bottom panels of Figure~\ref{fig:XCO} we show $\Delta\chi^2$ for the inner Galaxy and global-inner fits. The statistics of the inner Galaxy are hardly effected until \xco rises well above historical determinations~\cite{MS:2004,fermi_diffuse}.  This is to be expected in the inner Galaxy analysis where masking of the plane hides most of the $\rm H_2$ emission near the Galactic center.  For the global-inner fits, the plane is not masked and the fit rapidly worsens as the lower limit on $X_{\rm CO}$ is increased, which forces the inner H$_2$ rings to oversaturate.  For our Canonical model, physically likely values of \xco are very strongly disfavoured by the data, with statistical penalties outweighing even major changes to Galactic diffusion parameters.  We find that \xco {\em must} be substantially more than a factor 10 smaller than the Milky-way average in order to support the expected CMZ injection rates without Galactic center winds or much stronger vertical convection gradients. 

The gas-correlated $\gamma$-ray saturation toward the inner Galaxy is driven mostly by $\pi^0$ emission below a few GeV where the number of photons is largest.  Stronger winds blow low-energy protons and bremsstrahlung generating electrons away from the gas-rich CMZ.  This allows \xco to increase to more physical levels for wind velocities of at least several hundred km/s.  Alternative transport options are such as enhanced or anisotopric diffusion at the GC could also help alleviate the gas saturation.  However, none of these alternatives alone is likely to fully reconcile the expected CMZ injection rates with the observed level of $\pi^0$ emission.  We take this point as significant evidence in favor of strong GC winds.  Future modeling of gas in the CMZ should include combinations of improved molecular-line surveys tracing $\rm H_2$ -- such as the MOPRA CMZ survey~\cite{2012MNRAS.426.1972P} -- which are less prone to environmental variations than $\rm CO$.  Improved gas modelling is crucial in order to better constrain proton populations and winds at the GC, and can aid in discriminating between leptonic, hadronic, or mixed cosmic-ray injection models in the future.

\section{ROI Dependence of Fit Components}
\label{sec:scan_ROI}

\begin{figure*}[t!]
  \centering
  \includegraphics[width=.45\textwidth]{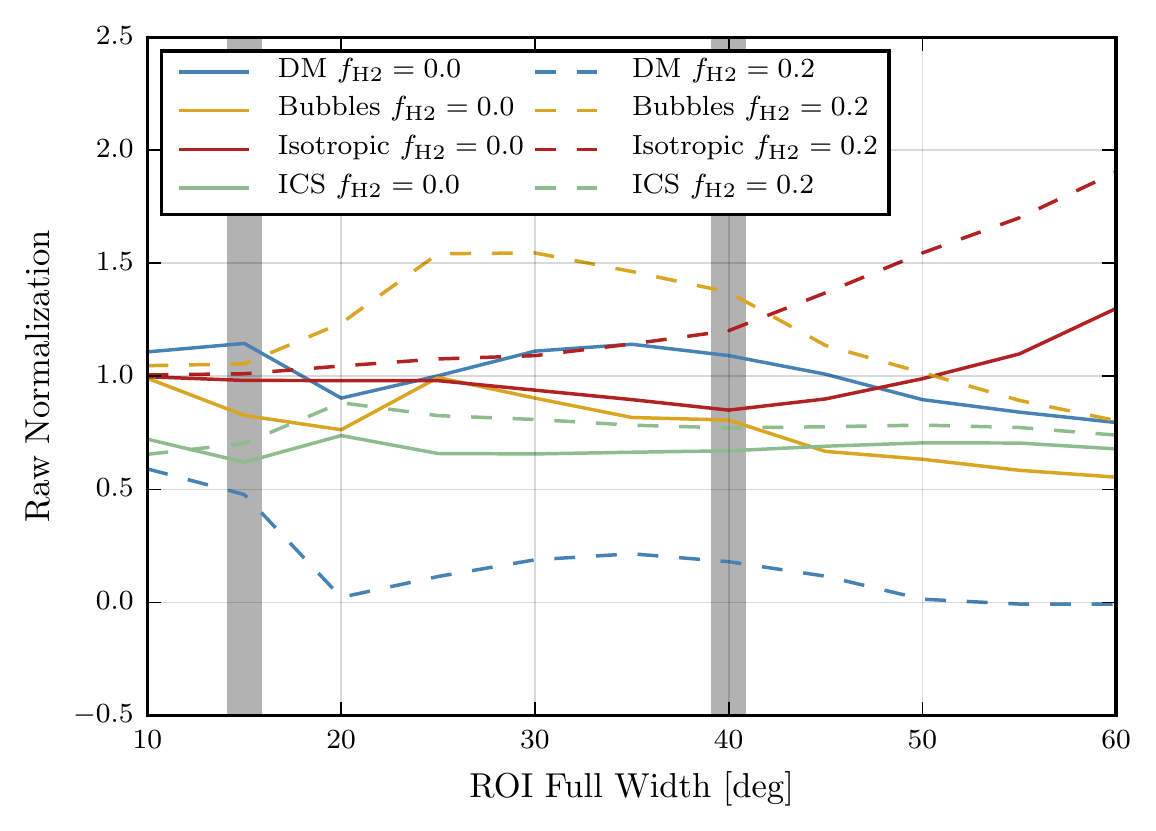}
  \includegraphics[width=.45\textwidth]{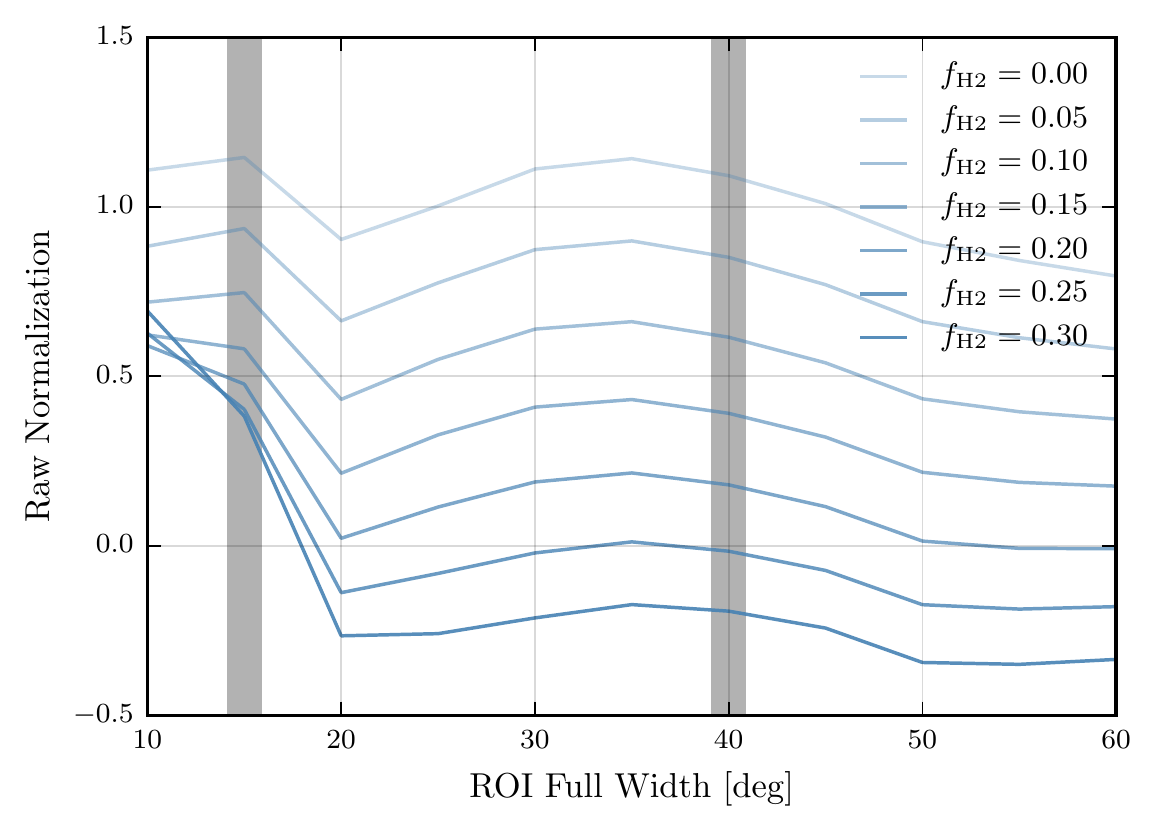}
  \caption{Dependence of fit components on choice of ROI for the inner Galaxy Analysis.  {\bf Left:} GDE and GCE component normalizations (averaged over 1-5 GeV) for $f_{\rm H2}=0$ and $f_{\rm H2}=0.2$ as the square ROI width is increased. All ROIs include a plane mask for $|b|<2^\circ$.  {\bf Right:} Same as left but for GCE template only.  Gray lines mark the Galactic center and inner Galaxy ROIs. }
 \label{fig:scan_ROI}
\end{figure*}

In Section~\ref{sec:gc} we showed that the GCE is not strongly reduced in the Galactic center analysis, compared with a strong reduction in the Inner Galaxy, and that the morphology of the GDE template in the IG and GC analyses differs.  The dependence of these results on the analysis windows implies that the GCE (or GDE templates) do not fully describe the residual data when cosmic-rays are added to the CMZ.  In Figure~\ref{fig:scan_ROI} we show the average raw normalizations, from 1-5 GeV, of each template in the inner Galaxy analysis (with a GCE template) as the fitting region is increased from $10^\circ\times10^\circ$ to $60^\circ\times60^\circ$.  We omit the $\pi^0$+bremsstrahlung template which remains flat to within $<10\%$ for all ROIs.  The Bubbles and isotropic spectrum normalizations are with respect to the original source spectra (see Sec.\ref{sec:gammarays}), while the DM spectrum is relative to Ref.~\cite{Calore:2015}.  The ICS normalization is relative to the {\tt Galprop} predictions of the respective models.  

As the ROI is expanded, the isotropic and ICS components remain flat until eventually the isotropic template increases by more than 50\% beyond $50^\circ$. The origin of this is not clear, but the smooth increase is a likely indication that the isotropic template prefers to be larger over all ROIs, but is constrained by the $\chi^2_{\rm ext}$, which has reduced weight as the ROI grows.  Because of the large latitudes involved, this is likely related to uncertainties in the foreground ICS rather than the GCE, which carries little weight over a 60$^\circ$ window.  The Fermi bubbles template used here is essentially isotropic over the central 5 degrees of the ROI before becoming bipolar at higher latitudes.  For $f_{\rm H2}=0.2$, windows between 20$^\circ$ and 50$^\circ$ result in a brightened bubbles template, but {\em only} in fits which include a GCE.  In fits that do not include the GC template, the bubbles template varies by less than $10$\% above 1 GeV, for all values of \fh (see Fig.~\ref{fig:spectra}).  In other words, the presence of the GCE template changes the inferred spectrum of the Fermi bubbles near the Galactic center (though the template morphology in this region is not well understood). 

The most dramatic change is the GCE template normalization (Fig.~\ref{fig:scan_ROI} right).  For $f_{\rm H2}=0$, the GCE spectrum is stable to within about 20\% over all ROIs. As we increase \fh2, overall normalization drops across all ROI sizes above $20^\circ$, eventually reducing near zero for $f_{\rm H2}=0.2-0.25$.  For ROIs under 20$^\circ$, however, the reduction saturates around $f_{\rm H2}=.1-.15$ at about half the original brightness.  The sharp decline of GCE template as the ROI increases indicates that the radial profile of the template is mismatched to the residual.  In particular, the ICS spike is too shallow to fully reproduce the inner few degrees. When the ROI is small, ICS is suppressed and the GCE template is bright.  As the ROI becomes larger, the ICS template brightens and the GCE becomes suppressed by the lack of excess emission far from the GC. This is in agreement with the findings of Sec.~\ref{sec:gc} and could relate to a number of factors including poor CMZ ISRF modeling, limited resolution of the simulations or injection gas maps, or be indicative of a very young injection of cosmic-rays as indicated by HESS observations of the Galactic center~\cite{2006Natur.439..695A}.

\section{Comparison with Gaussian CMZ Model}

Recently, Gaggero et al~\cite{Gaggero2015} noted that previous GDE models have neglected cosmic-ray sources at the Galactic center.  They proceeded to add a spherically symmetric Gaussian cosmic-ray source at the position of the Galactic center in order to account for the CMZ.  For cosmic-ray `spikes' of width 200-400 pc (constrained by the size of the CMZ), they found that such models could reproduce the bulk properties of the GCE, and strongly reduce the significance of the GCE. Our findings fully support this conclusion, but using a physical model and empirical inputs for the size and shape of the spike.  In addition, our star formation prescription correctly reproduces the normalization of the spike based on both the measured SNe rate of the CMZ {\em and} based on global fits to Fermi $\gamma$-ray data.  Their model is parametrized by a Gaussian width $\sigma$ and a normalization $\mathcal{N}$, which represents the fraction of total Galactic cosmic-ray injection contained in the CMZ ($r<300$ pc).  Here we reproduce this models for $\sigma=200$ pc and using Mod A (\xco fixed to MS04 profile) in order to make direct comparisons.

\begin{figure*}[thb]
  \centering
    \includegraphics[width=.45\textwidth]{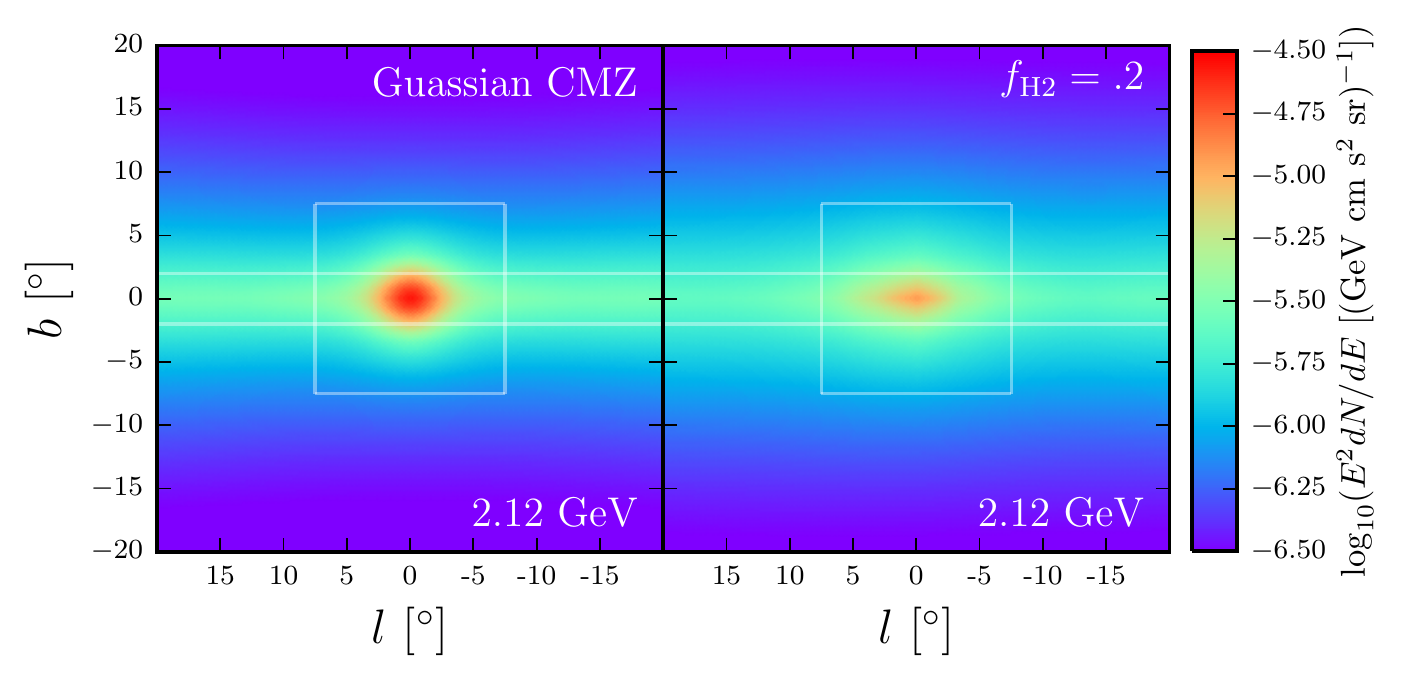}
  \includegraphics[width=.4\textwidth]{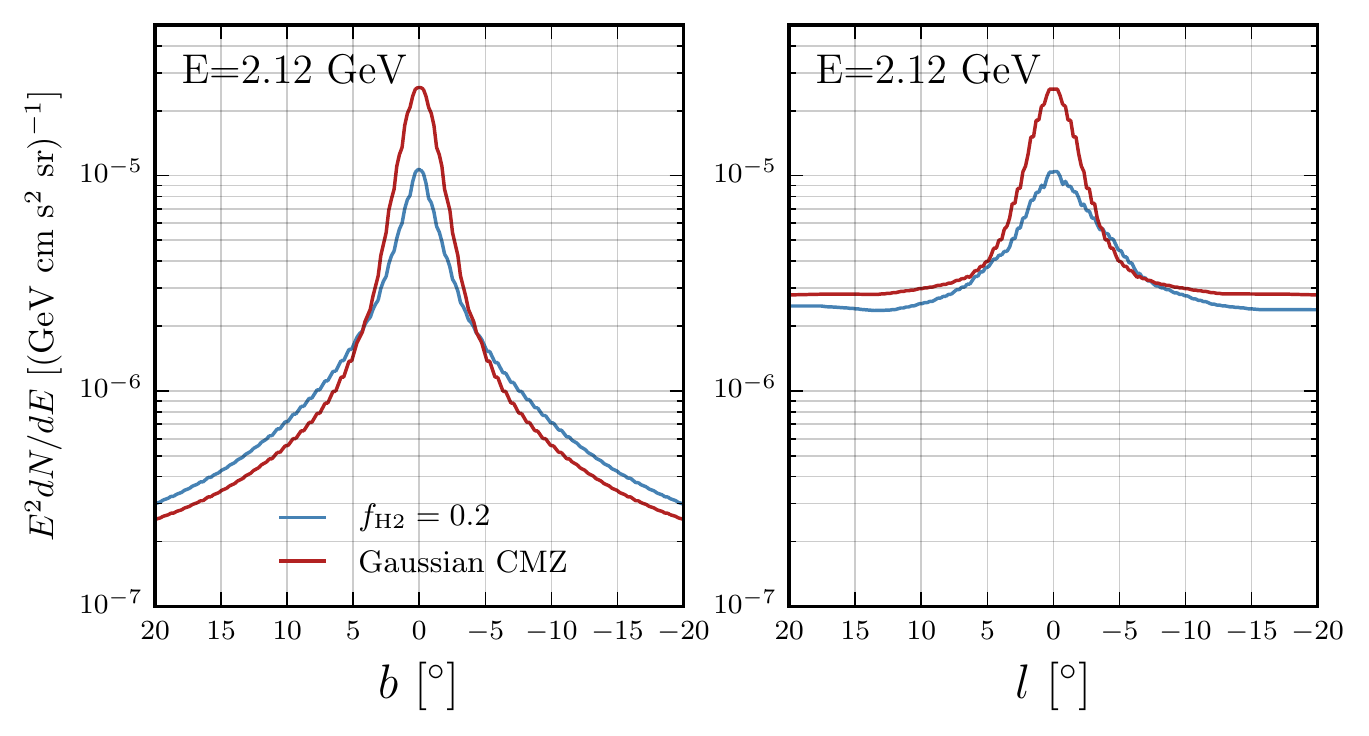} \\
    \includegraphics[width=.85\textwidth]{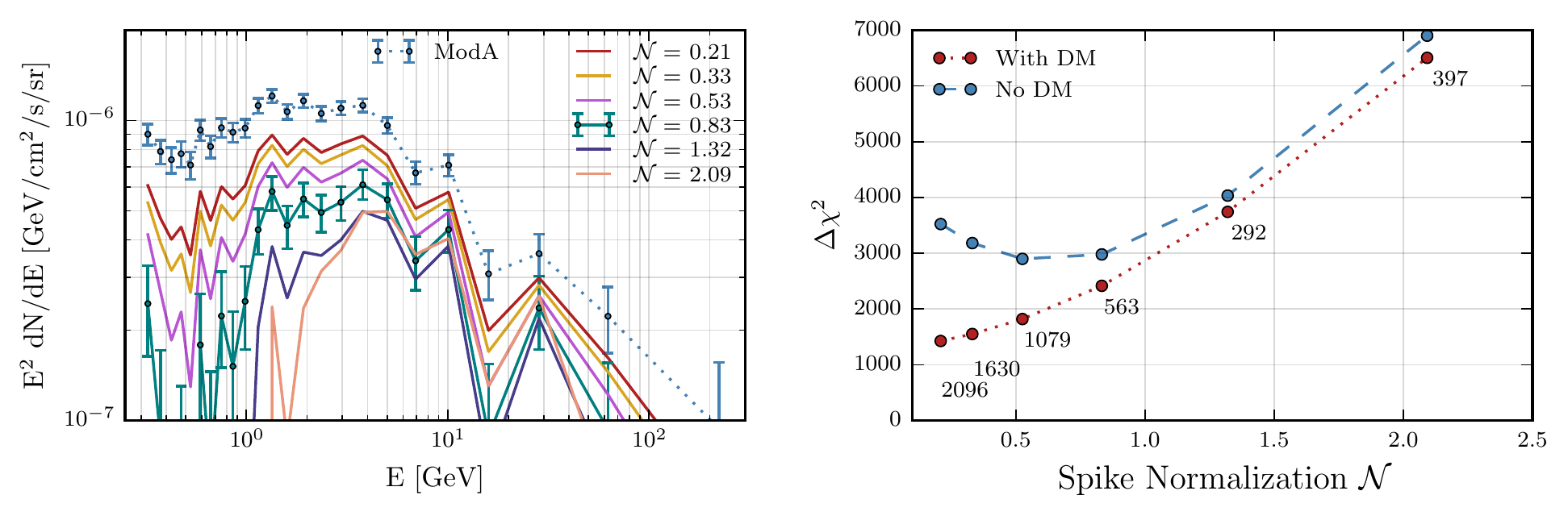}
  \caption{{\bf Top-left:} ICS emission over the Inner Galaxy ROI when adding a Gaussian cosmic-ray `spike' of width $\sigma=200$ and $\mathcal{N}=1.4\%$ versus our Canonical ($f_{\rm H2}=.2$) model.  {\bf Top-right:} Latitude and longitude profiles for ICS emission running through $l=0^\circ, b=0^\circ$ respectively. {\bf Bottom:} GCE Spectrum and fit statistics of the inner Galaxy ROI as the spike normalization $\mathcal{N}$ is varied in the Gaussian CMZ model.  To be compared with Figure~\ref{fig:vary_ns_fh2_Gal}.}
 \label{fig:spike_compare}
\end{figure*}

In the top-left panel of Figure~\ref{fig:spike_compare} we show the ICS emission for the $\mathcal{N}=1.4\%$ model and our Canonical star-formation based model over the inner Galaxy ROI at the peak GCE energy.  In the top-right panels, we show latitude and longitude profiles through $l=0$ and $b=0$, respectively.  Several features differentiate the two models.  First, the excess Gaussian CMZ spike is completely spherically symmetric (as is expected for an isotropic diffusion tensor), while the $f_{\rm H2}=0.2$ model is elongated along the plane (in reality, the height of the CMZ is very thin with FWHM$\approx$45 pc~\cite{Ferriere:2007}, and not 200 pc as taken in Ref~\cite{Gaggero2015}), and contains an axisymmetric structure from the bar which extends toward positive longitudes. Second, both the latitude and longitude profiles of the spike are substantially steeper than our Canonical model at 2 GeV, which may be driven by both the improved simulation resolution (thus injection resolution) afforded by the cylindrical simulations, or by the higher resolution of the gas model, which is an analytic model compared with our more limited $dx=dy=100$ pc empirical gas model.  We expect that our models produce a more realistic model of the CMZ shape and foreground emission as indicated by the much improved overall IG fit.

In the bottom panels of Figure~\ref{fig:spike_compare}, we show the GCE spectrum and statistics for the IG analysis.  Both quantities are remarkably similar to our findings in Section~\ref{sec:SFR_tuning}, and show that the GCE is highly degenerate with the new ICS profile.  The minimum TS for the added GCE template is comparable to our results.  A major difference is that the best-fitting model without a GCE template provides a much worse fit than our Canonical model.  This is not due to \xco fitting, or differences in diffusion parameters, but to axisymmetric and non-Gaussian features present in the injection morphology.  Our optimal no-DM fits not only provide a minimal GCE significance, but improve the IG fit by $\Delta \chi^2 \approx 4000$, compared to less than 600 here.  Finally, the fit improvement offered by the Gaussian CMZ model is limited to the inner 10-20 degrees, and does not improve the global $\gamma$-ray fits in a significant way, whereas our models provide an improved fit comparable to major changes to diffusion parameters in all regions except the outer Galaxy.

\section{The Morphology From Ten Sky Segments}
\label{sec:calore_regions}
\begin{figure*}[th!]
  \centering
  \includegraphics[width=1\textwidth]{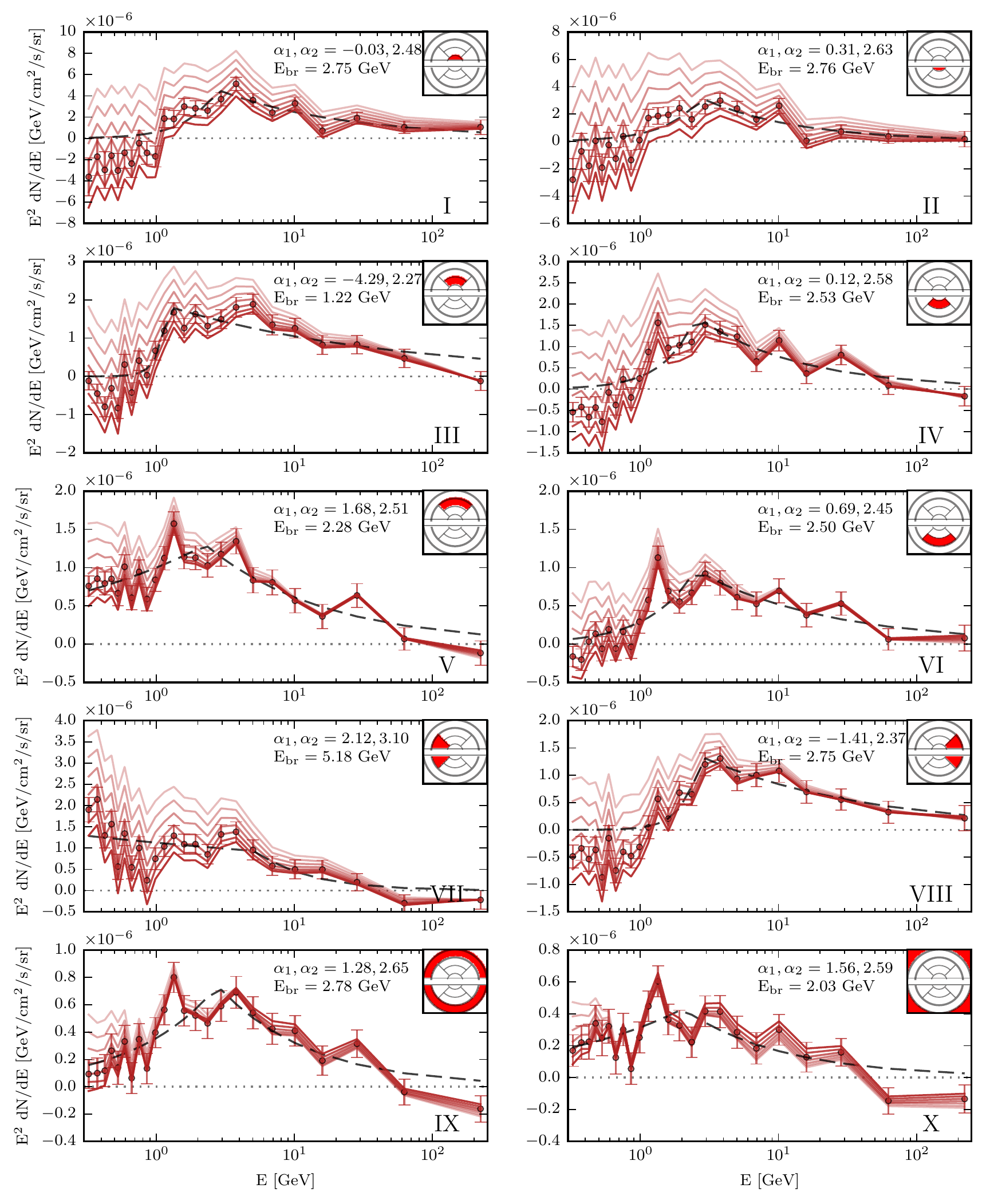}
  \caption{Spectrum of NFW$_{\gamma=1.05}$ template split into ten regions from Calore et al~\cite{Calore:2015}, with the active region indicated in red in the inset plots. Light to dark lines show the spectrum as \fh is increased from 0 in increments of 0.05, with markers highlighting the \fh=.2 case.  The black dashed lines indicate the best fit broken power-law spectrum with low (high) energy spectral index $alpha_1$ ($\alpha_2$), and a break energy $E_{\rm br}$.  Note the vertical scale in units of $10^{-6} \rm \ GeV\ cm^{-2}\ s^{-1}\ sr^{-1}$}
 \label{fig:calore_regions}
\end{figure*}

In Figure~\ref{fig:calore_regions} we divide the NFW$_{\alpha=1.05}$ GCE template into ten regions over the inner Galaxy ROI (defined in ref.~\cite{Calore:2015}) and show the spectrum as \fh is increased.  We also show the best fit broken power law spectrum (for $f_{\rm H2}=0.2$) with black-dashed lines, where the break energy and high/low energy spectral indices are allowed to vary.  The low energy spectrum is observed to very hard near the Galactic center ($\alpha_1 \lesssim 1$), with a gradual softening at larger radii. As we will see below, this low energy over-subtraction appears to be unphysical, and vanishes when the GCE template inner slope and radial profile are changed to their better fit values. A strong flattening and vertical elongation of the GCE profile is evident at all energies, with the sideband regions (regions VII and VIII) depleting much more significantly than the vertical regions (V \& VI).  The left sideband (fourth row left).  Note that the low energy spectrum of Region VII is heavily contaminated by the soft-spectrum residual from the Aquila Rift star-forming region.  In each region, we find that a broken power law provides an extremely good fit ($\chi^2/d.o.f.\approx 0.2$), even without including the larger systematic uncertainties.

\section{The Galactic Center Analysis with the 1FIG Catalog}
\label{appendix:1FIG}

\begin{figure*}[thb]
  \centering
  \includegraphics[width=1\textwidth]{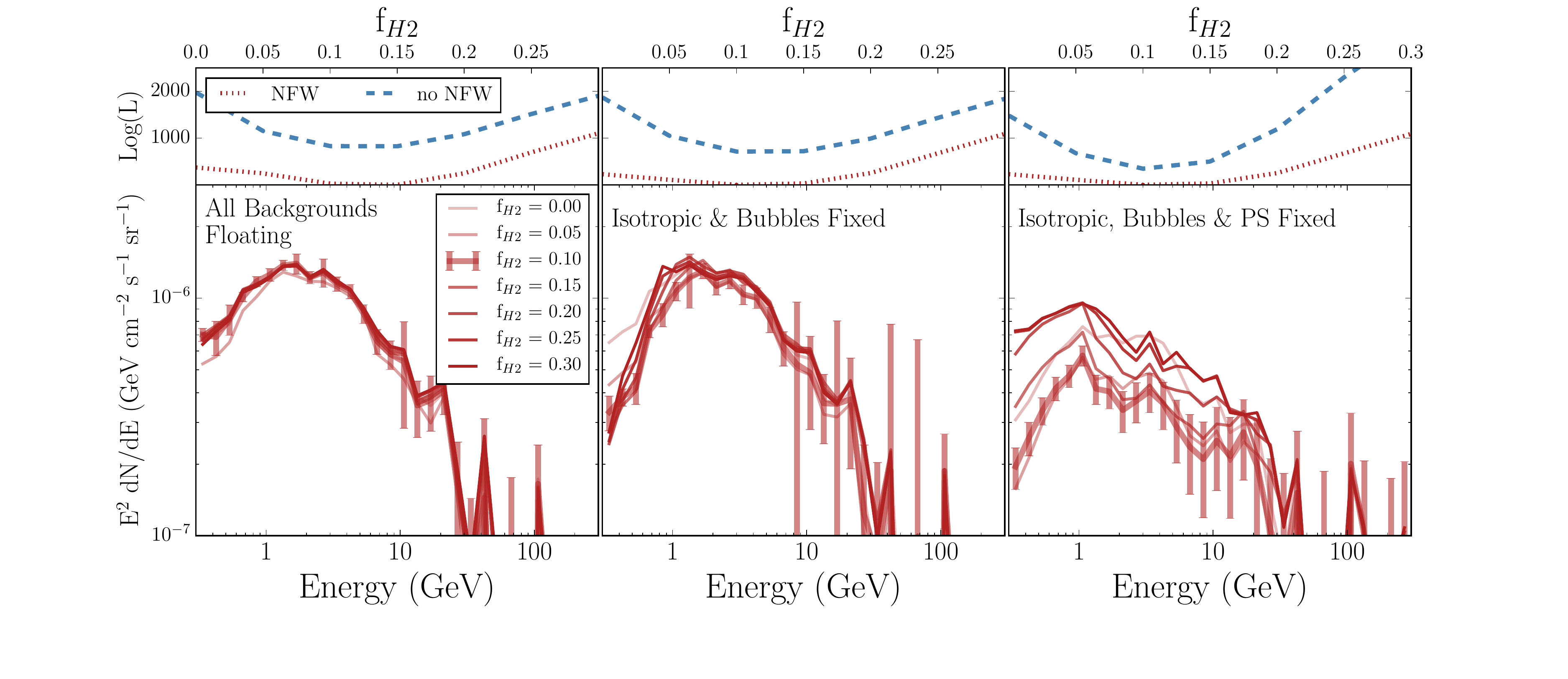}
\caption{Same as Figure~\ref{fig:gc_fits_no_mask} for an analysis using the 1FIG catalog produces by \citep{TheFermi-LAT:2015kwa}, rather than the standard 3FGL catalog.  The close comparison of these results with the results from the 3FGL point source population demonstrate the resilience of the GCE to changes in the $\gamma$-ray point source modeling.}
 \label{fig:gc_fits_no_mask_1FIG}
\end{figure*}

Recently, the Fermi-LAT collaboration produced an analysis of the 15$^\circ\times$15$^\circ$ region surrounding the Galactic center, utilizing new diffuse models and a new point source catalog to determine the properties of the Galactic Center excess~\citep{TheFermi-LAT:2015kwa}. While the diffuse emission models are not currently publicly available, the locations of 1FIG sources are. Using this source list, and again letting the normalization of all sources float independently in each energy bin (as in the case of 3FGL sources above). In Figure~\ref{fig:gc_fits_no_mask_1FIG} we show the resulting spectrum and intensity of the GCE component in models using the 1FIG catalog, finding that our results are in very close agreement with models utilizing the 3FGL catalog. Since there are substantial differences between the 1FIG and 3FGL catalogs, this result illustrates a point made previously by \cite{Daylan:2014rsa} among several other results --- that the observed $\gamma$-ray point source populations do not appear degenerate with the properties of the GCE component.

\end{appendix}

\end{document}